# CENTRAL FORCE OPTIMIZATION APPLIED TO THE PBM SUITE OF ANTENNA BENCHMARKS

**Richard A. Formato**

Registered Patent Attorney & Consulting Engineer
P.O. Box 1714, Harwich, MA 02645 USA
rf2@ieee.org

**Abstract** – Central Force Optimization (CFO) is a new nature-inspired deterministic multi-dimensional search and optimization metaheuristic based on the metaphor of gravitational kinematics. CFO is applied to the PBM antenna benchmark suite and the results compared to published performance data for other optimization algorithms. CFO acquits itself quite well. CFO's gradient-like nature is discussed, and it is speculated that a "generalized hyperspace derivative" might be defined for optimization problems as a new mathematical construct based on the Unit Step function. What appears to be a sufficient but not necessary condition for local trapping, oscillation in the probe average distance curve, is discussed in the context of the theory of gravitational "resonant returns" that gives rise to strikingly similar oscillatory curves. It is suggested that the theory may be applicable to CFO as an aid to understanding trapping and to developing effective mitigation techniques, possibly based on a concept of "energy" in CFO space. It also is suggested that CFO may be re-formulated as a "total energy" model by analogizing conservation of energy for orbiting masses in physical space.

## 1. INTRODUCTION

Central Force Optimization (CFO) is new nature-inspired, gradient-like metaheuristic for multi-dimensional search and optimization [1-4,21]. It comprises two simple *deterministic* equations based on the metaphor of gravitational kinematics. Because the law of gravity is deterministic, so too is CFO, unlike many nature-inspired algorithms that are fundamentally stochastic. The primary purpose of this paper is to report CFO's performance against a recognized suite of antenna benchmark problems. Its secondary purposes are (a) to describe the CFO metaheuristic and present its basic equations; (b) to suggest the possibility of a new mathematical construct, a "hyperspace directional derivative" based on the Unit step function; (c) to discuss the strikingly similar appearance of velocity curves for NEOs (Near Earth Objects) in gravitationally trapped close encounters with CFO's average probe distance curves; (d) to suggest the possibility of using probe total "energy" as a measure of local trapping; and (e) to suggest the possibility of re-formulating CFO using a conservation of energy model.

The original paper CFO [1] introduced CFO as a new metaheuristic and applied it to two problems in applied electromagnetics (EM): designing an equalizer for the canonical Fano load, and synthesizing a linear dipole array. CFO's results were compared to data from several other algorithms, and it also was tested against several analytic benchmark functions with known maxima. The results suggested that CFO merits further development as a promising, robust optimization methodology. CFO more recently has been used to synthesize linear and circular arrays [5], and it performed very well compared to the Quadrature Programming Method (QPM), to a Particle Swarm Optimization (PSO) algorithm, and to a Genetic Algorithm (GA). Even though CFO is a new metaheuristic that is not nearly as well developed as most of the widely used evolutionary algorithms (EAs), it nevertheless shows very robust performance at this early stage, and considerable promise for future development.

This paper provides further examples of CFO's effectiveness by applying it to the set of "real world" antenna benchmarks developed by Pantoja, Bretones, and Martin [6] ("PBM"). Their benchmark suite is designed to objectively evaluate the performance of EAs used to solve complex EM problems, representative examples being: PSO-based array synthesis [7]; swarm intelligence optimization of layered media [8]; GA antenna modeling [9]; array design using GAs, and memetic and tabu search algorithms [10]; crack detection using a finite-difference frequency domain/PSO methodology [11]; Vee-dipole optimization using a bacteria foraging algorithm [12]; and GA-optimized MRI coil design [13]. EAs solve these problems by "evolving" solutions that are unavailable analytically or numerically. The plethora of nature-inspired EAs makes comparison difficult, if not impossible, without a standardized set of benchmarks for testing. The PBM suite, which is based on the Numerical Electromagnetics Code (NEC), addresses this need for EAs that are applied to optimized antenna design. It therefore is a useful tool for evaluating CFO.

An EA's performance is measured by its *effectiveness* (locating global maxima accurately) and its *efficiency* (locating global maxima with minimum computational effort). An algorithm that fails to accurately locate global maxima fails to accomplish its intended purpose and consequently is ineffective. EAs that are, on a relative basis, computationally intensive are less desirable than ones requiring fewer calculations. This paper uses these performance measures to compare CFO to the four EAs discussed in [6]. The results are, for the most part, very encouraging. CFO performs very well compared to the PBM EAs, and in many cases much better. But CFO does suffer limitations, primarily a consequence of its being deterministic, which renders it prone to local trapping.

This paper is organized as follows: Section 2 provides an overview of the PBM benchmark suite and discusses NEC4 validation of the PBM data. Section 3 summarizes CFO's performance against the PBM benchmarks. Section 4 presents the CFO gravitational metaphor and its basic equations. It also discusses some interesting indirect evidence of the metaphor's validity drawn from the astronomy literature, the possibility of defining a CFO "energy" for trapping analysis or re-formulation of the algorithm, and it suggests the possibility of interpreting CFO as a "generalized gradient" methodology with the attendant possibility of defining a new Unit Step-based hyperspace derivative. Section 5 describes the PBM benchmark suite in detail, including antenna geometries and decision space landscapes. Section 6 discusses the CFO results in detail. Section 7 is the conclusion. Appendix A provides additional plots of the PBM benchmark topologies, including projections onto the principal planes, and sample NEC4 input/output files for each antenna. Appendix B contains pseudocode for the CFO implementation used here and Appendix C the Basic source code.

## 2. PBM BENCHMARKS – OVERVIEW AND NEC4 VALIDATION

### *2.1 The PBM Problems*

The PBM benchmark suite comprises five antenna problems designed to test an EA's effectiveness and efficiency. Table 1 lists their properties. In each case, the objective is to maximize the antenna's directivity ("fitness"). Four of the problems are two-dimensional (2D), while the fifth is $(N_{el}-1)$ D [ $N_{el}$ is the number of dipole elements in a collinear array]. The 2D problem topologies ("landscapes") are plotted in Section 5 to illustrate the nature and complexity of the decision space. Problems #1 and #4 are unimodal with a single global maximum. The first problem is "lumpy" with strong local maxima, whereas the fourth is "smooth." Problem #2 is "noisy" in a complex landscape with large amplitude nearby local maxima. The third problem's topology is extremely multimodal with four global maxima. Problem #5 is a unimodal high-dimensionality problem that optimizes the element spacing of an $N_{el}$-element collinear dipole array.

**Table 1. Properties of the PBM Benchmark Problems.**

| *PBM Benchmark #* | ***Problem Characteristics*** *(in each case objective is to maximize directivity)* |
|---|---|
| 1 | Variable length center-fed dipole. 2D, unimodal, single global maximum, strong local maxima. |
| 2 | Uniform 10-element array of center-fed $\frac{\lambda}{2}$-dipoles. 2D, added Gaussian noise, single global maximum, multiple strong local maxima. |
| 3 | 8-element circular array of center-fed $\frac{\lambda}{2}$-dipoles. 2D, highly multimodal, four global maxima. |
| 4 | Vee Dipole. 2D, unimodal, single global maximum, "smooth" landscape. |
| 5 | Collinear $N_{el}$-element array of center-fed $\frac{\lambda}{2}$-dipoles. $(N_{el}-1)$ D, unimodal, single global maximum. |

### *2.1 NEC4 Validation*

The PBM problems, which do not have analytic solutions, were solved numerically in [6] using NEC. The first step in evaluating CFO therefore is validation of the published PBM results. The five benchmark antennas were modeled using the Numerical Electromagnetics Code Version 4.1 Double Precision ("NEC4") [14], which may be a different version of NEC than the one used in [6]. The segmentation and wire radii used in [6] were replicated. NEC4 input and output files are included in Appendix A. Each PBM antenna was modeled using the coordinates for the maxima reported in [6]. In several cases they were estimated from graphical data and, as a consequence, necessarily are approximate. Tables 2(a) and (b) summarize the results. $x_1$ and $x_2$, respectively, refer to the abscissa and ordinate in the "Domain" column; λ is the wavelength; $N_d$ is the problem's dimensionality; and $D_{\max}^{PBM}$ and $D_{\max}^{NEC4}$ are the maximum directivities.

While the general conclusion is that NEC4 essentially recovers the PBM results, there are some noteworthy differences. For problems #1 and #2, NEC4's computed directivities are slightly less than PBM's. For problems #3 and #4, they are lower by a much wider margin. The best agreement is on problem #5 where the NEC4 and PBM data show very good agreement. What accounts for these discrepancies is not clear. There are several possible explanations, ranging from different versions of NEC to compiler differences in creating the executables to slight differences in the antenna models, for example, source modeling (NEC4 excitation was modeled following the guidelines for the "EX0" card [14, Part I, p. 48 *et seq.*]). These differences between the PBM results and NEC4 notwithstanding, the validation test shows that NEC4 can be used effectively to assess CFO's performance against the PBM benchmarks, and that CFO should recover maxima with similar amplitudes at about the same points in the decision space.

## 3. CFO PERFORMANCE SUMMARY

### 3.1 Effectiveness

This section summarizes CFO's performance (see §6 for detailed discussion of each run). CFO's *effectiveness* is measured by how accurately it locates the PBM maxima (coordinates and maxima values, that is, best fitnesses). Table 3 summarizes the coordinate results and Table 4 the fitnesses (Δ is the difference in values). Inspection of these tables shows that the agreement generally is quite good.

**Table 2(a). PBM and NEC4 Results for Benchmarks #1-4.**

| *Problem #* | *PBM* | | | | *NEC4* |
|---|---|---|---|---|---|
| | *Domain* | $x_1$ | $x_2$ | $D_{\max}^{PBM}$ *x* | $D_{\max}^{NEC4}$ |
| 1 | $0.5\lambda \le L \le 3\lambda$ <br> $0 \le \theta \le \frac{\pi}{2}$ | 2.58λ | 0.63 | 3.32 | 3.2 |
| 2 | $5\lambda \le d \le 15\lambda$ <br> $0 \le \theta \le \pi$ | ~5.85λ[1] | π/2 | ~18.3 | 18.11 |
| 3 | $0 \le \beta \le 4$ <br> $0 \le \theta \le \pi$ | 0.5, 1.5, 2.5, 3.5 | π/2 | ~7.05 | 6.15 |
| 4 | $0.5\lambda \le L_{arm} \le 1.5\lambda$ <br> $\frac{\pi}{18} \le \alpha \le \frac{\pi}{2}$ | 1.5λ | 0.834 | ~5.8 | 4.8 |

Note: [1] values marked with ~ are estimated from Figs. 6, 9 or 11 in [6].

**Table 2(b). PBM and NEC4 Results for Benchmark #5.**

| *Problem #* | *# Dipoles* | $N_d$ | *Domain* | $d_i$ | $D_{\max}^{PBM}$ | $D_{\max}^{NEC4}$ |
|---|---|---|---|---|---|---|
| 5 | 6 | 5 | $0.5\lambda \le d_i \le 1.5\lambda$ <br> $1 \le i \le 5$ | 0.99λ | ~11.25[1] | 11.22 |
| 5 | 10 | 9 | $0.5\lambda \le d_i \le 1.5\lambda$ <br> $1 \le i \le 9$ | 0.99λ | ~19 | 19.10 |
| 5 | 16 | 15 | $0.5\lambda \le d_i \le 1.5\lambda$ <br> $1 \le i \le 15$ | 0.99λ | ~31 | 30.97 |
| 5 | 24 | 23 | $0.5\lambda \le d_i \le 1.5\lambda$ <br> $1 \le i \le 23$ | 0.99λ | ~47 | 46.88 |

Note: [1] values marked with ~ are estimated from Fig. 13 in [6].

Turning to Table 3, for 2D problems #1 through #4, respectively, the coordinates agree to within (1.12%,1.9%), (1.26%,0.89%), (3.95%,0.16%), and (0.03%,14.75%) [relative to PBM's coordinates]. The only significant disagreements are in the abscissas for problem #3 and the ordinates for problem #4. In the case of problem #3, the location of the global maximum at ($\beta = 0.5, \theta = \frac{\pi}{2}$) is known analytically. No computed value was reported in [6], so that the true degree of agreement is not known. For problem #4, the discrepancy in the inner angle $\alpha$ may be a result of modeling differences, or possibly different versions of NEC. Of course, from a practical (read "engineering") point of view, agreement to with about 15% may be acceptable. For problem #5, the dipole separation for maximum directivity agreed to within 1% for all array sizes, which is quite good. If the 24-element array, for which $d_i = 1\lambda$, is excluded, then the agreement is even better, to within 0.7%.

**Table 3. Comparison of PBM and CFO maxima coordinates.**

| *PBM Problem #* | *PBM Coordinates* | | *CFO Coordinates* | | Δ | |
|---|---|---|---|---|---|---|
| | $x_1$ | $x_2$ | $x_1$ | $x_2$ | $x_1^{PBM} - x_1^{CFO}$ | $x_2^{PBM} - x_2^{CFO}$ |
| 1 | 2.58λ | 0.63 | 2.55088λ | 0.61805 | 0.02912λ | 0.01195 |
| 2a [no noise] | ~5.85λ | π/2 | 5.92359λ | 1.55685 | -0.07359λ | 0.01395 |
| 2b [noise] | nr(1) | nr | 6.93601λ | 1.54721 | - - - | - - - |
| 3 | 0.5 | π/2 | 0.48024 | 1.57327 | 0.01976 | -0.00247 |
| 4 | 1.5λ | 0.834 | 1.49520λ | 0.71098 | 0.00048λ | 0.12302 |
| - - - | $d_i, i = 1,...,N_{el} - 1$ | | $d_i, i = 1,...,N_{el} - 1$ | | $MAX(d_i^{PBM} - d_i^{CFO})$ | |
| 5 | 0.99λ | | 0.98310λ – 1λ | | 0.01λ | |

Notes: (1) not reported in [6]

Table 4 compares CFO's and PBM's best fitnesses as measured by the difference of directivities, $D_{\max}^{PBM} - D_{\max}^{CFO}$. The differences are 3.43%, 0.36%, 27.74%, and 1.47%, respectively, for problems #1 through #4 [relative to PBM's values]. The agreement is quite good, except for problem #3, which would be troublesome but for the data in Table 2(a). NEC4 returned a directivity 6.15 at the points $(\beta, \theta) = (i - 0.5, \frac{\pi}{2})$, $i = 1,...,4$, instead of 7.05 as reported in [6]. If, in fact, 6.15 is the correct value, then the difference decreases to a more modest 5.47%. At a minimum, there appears to be a question concerning the accuracy of the result reported in [6], and CFO's overall performance suggests that it accurately located the first global maximum for problem #3 after all. On problem #5 the directivity values are in excellent agreement across all array sizes. The differences range from only 0.08% ($N_{el} = 16$) to 0.52% ($N_{el} = 10$). In this case CFO recovered the PBM's suite best fitnesses with very high accuracy.

**Table 4. Comparison of PBM and CFO best fitnesses.**

| *PBM Problem #* | *CFO Results* | | | $D_{\max}^{PBM}$ | Δ $D_{\max}^{PBM} - D_{\max}^{CFO}$ |
|---|---|---|---|---|---|
| | $x_1$ | $x_2$ | $D_{\max}^{CFO}$ | | |
| 1 | 2.55088λ | 0.61805 | 3.20627 | 3.32 | 0.11373 |
| 2a (without noise)<br>2b (with noise) | 5.92359λ<br>6.93601λ | 1.55685<br>1.54721 | 18.3654<br>18.6880 | 18.3(1)<br>nr(2) | -0.0654<br>nr |
| 3 | 0.48024 | 1.57327 | 6.48634 | 7.05(1) | 0.56366 |
| 4 | 1.49520λ | 0.71098 | 5.71479 | 5.8(1) | 0.08521 |
| - - - | $d_i, i = 1,...,N_{el} - 1$ | | - - - | - - - | - - - |
| 5 (6 el) | 0.99105λ | | 11.2202 | ~11.25(3) | 0.0298 |
| 5 (7 el) | 0.98310λ | | 13.1826 | nr | - - - |
| 5 (10 el) | 0.99421λ | | 19.0985 | ~19(2) | -0.0985 |
| 5 (13 el) | 0.99629λ | | 25.0611 | nr | - - - |
| 5 (16 el) | 0.98958λ | | 30.9742 | ~31(2) | 0.0258 |
| 5 (24 el) | 1.00000λ | | 46.8813 | ~47(2) | 0.1187 |

Notes: (1) values marked with are estimated from the figures in [6].
(2) nr – not reported in [6].
(3) values marked with ~ are estimated from Fig. 13 in [6].

These data show that CFO essentially accurately recovered the global maximum for every one of the five PBM benchmark problems, with the caveat that it located only one of the four maxima for problem #3. This performance is better than any of the algorithms reported in [6]. None of the algorithms in that work achieved 100% effectiveness. CFO thus is very effective against the PBM antenna benchmarks.

### 3.2 Efficiency

An EA's efficiency is measured by how computationally intensive it is. As the detailed results in §6 show, CFO generally converges quickly, at least to the vicinity of a global maximum if trapping has not occurred. After its usually rapid rise, the best fitness often continues to increase, but much more slowly with only slight additional gains.

*Efficiency* is measured by the total number of calculations (objective function evaluations) required for the best fitness to saturate. The PBM paper examines four algorithms: ***GA-FPC***, ***μGA***, ***GA-RC***, and ***PSO*** (see [6] for details). The first three are GA variants, while the fourth is a particle swarm. Because these algorithms are stochastic, each one was run twenty times to develop statistics. The "mean hit time" in Table II in [6] is the average generation number at which the global maximum was located. Thus the *average* number of function evaluations is the product of mean hit time and the population size. Table 5 shows this measure for the PBM algorithms, along with the total number of objective function evaluations for saturation of CFO's best fitness [ $N_{eval} = (j^*+1)N_p$, where $j^*$ is the step at which the fitness saturated, and $N_p$ is the number of CFO "probes" (see §4.2)]. In Table 5 the best values are shown in bold **red** font, and the second best in bold **blue**.

On problem #1 CFO performed much better than all the other algorithms, requiring only 60 calculations to locate the global maximum compared to 1530 for the next best algorithm, ***PSO***. On problem #2, without and with noise, CFO did not perform as well as ***GA-FPC*** and ***μGA*** by a factor of about 2 to 3; it was on a par with ***GA-RC***; and it did substantially better than ***PSO***. For problem #3, CFO out-performed all the algorithms except ***PSO***, whose performance was about 17% better. Two different CFO initial probe distributions were used on problem #4, with the result that CFO's performance was comparable to ***GA-FPC***'s, and quite a bit better than ***μGA'***s and ***GA-RC'***s, by at least a factor of 2. Compared to ***PSO***, however, CFO did not do as well by a factor of about 3. Problems #5a and 5b in [6, Table II] are 7 and 13 element collinear arrays, respectively [6 @ p.1119]. For problem #5a CFO was nearly 15 times more efficient than the next most efficient algorithm, ***PSO***. On problem #5b, CFO was more than 12 times more efficient, again compared to ***PSO***.

Thus, CFO more than holds its own in terms of computational efficiency, especially for an algorithm that is not highly developed. Out of six problems, CFO performed the best on three, out-performing the other algorithms by a very wide margin. On problem #3, CFO turned in the second best result, but it was close to the best efficiency. On problem #4, CFO was the second best performer. And on problem #2 it was the third most efficient algorithm. Across the six problems, CFO was best on three and second best on two. ***PSO*** turned in the next best overall performance, placing first on two problems and second on three. On the one problem where CFO did not finish in the top two, it was the third best. Even in its infancy, CFO has proven to be very competitive in terms of efficiency.

**Table 5. CFO Efficiency (# function evaluations).**

| ***Problem #*** | ***CFO*** $N_{eval}$ | ***Results from PBM Paper [6], Table II*** (mean hit time x population) | | | |
|---|---|---|---|---|---|
| | | ***GA-FPC*** | ***μGA*** | ***GA-RC*** | ***PSO*** |
| 1 | **60** | 3140 | 5065 | 8920 | **1530** |
| 2a [no noise]<br>2b [with noise] | 1320<br>768 | **360** | **450** | 1400 | 2280 |
| 3 | **1050** | 1940 | 1685 | 5040 | **900** |
| 4 | 1488[(1)]<br>**1155** | 1300 | 3125 | 3800 | **330** |
| 5a | **72** | 1220 | 1700 | nr[(2)] | **1050** |
| 5b | **144** | 3480 | 5695 | nr | **1770** |

Notes: [(1)] different initial probe distributions.
[(2)] not reported in [6].

## 4. THE CFO METAPHOR

There are many nature-inspired search and optimization EAs, for example, metaheuristics based on animal and insect behaviour, human population behaviour, or physical processes such as water drop pooling. Most of these algorithms are fundamentally stochastic because the processes they analogize are inherently random. In fact, their ability to locate maxima derives from this inherent randomness. Without it, these EAs cannot function.

Central Force Optimization, by contrast, is completely deterministic because it is based on the metaphor of gravitational kinematics, the branch of physics that computes the motion of masses moving solely under the influence of gravity. Because gravity is deterministic, so too is CFO. This feature is a significant improvement over stochastic algorithms. Every CFO run with the same setup parameters returns precisely the same results. But, of course, as with all deterministic methodologies, the downside is a tendency for CFO to become trapped at local maxima. In fact, the two major issues in CFO's further development appear to be (1) finding effective ways of mitigating trapping and (2) establishing a precise methodology for specifying run parameters. This section describes CFO's basic equations and discusses these issues, in particular oscillation in $D_{avg}$ (defined below) as a harbinger of trapping and recent results in the astronomy literature that may be useful in understanding this phenomenon.

Even though CFO is fundamentally deterministic, the algorithm designer is free to inject some randomness if doing so improves performance. When local trapping occurs, stochastic perturbation of CFO's "probe" distribution may mitigate the trapping. The judicious injection of some measure of randomness has not been investigated here, but it has been used in other CFO implementations with very good results [5]. The author emphasizes CFO's deterministic nature because CFO is one of very few algorithms that provide true reproducibility and control over computed results. This characteristic may be especially useful in parameter-tuning implementations in which CFO's run parameters are adjusted using real-time feedback. Nevertheless, some CFO implementations may benefit from a stochastic component, and that possibility certainly merits consideration.

### 4.1 CFO, NEOs and Energy

Near Earth Objects (NEOs) approaching our planet can become gravitationally "trapped" for a while. As a general rule, in the absence of friction or some other energy dissipation mechanism, an object's trajectory will depend only upon its (constant) total energy. Recent work on gravitational "resonant returns" [15], however, shows that small objects closely approaching Earth can exchange energy with the planet without dissipation and, for some time at least, become gravitationally trapped in a tight orbit. A consequence of this type of encounter is the plot in Figure 2 of [16], which is reproduced below with permission. The curve plots the velocity change $\Delta v$ required to modify the trajectory of NEO asteroid Apophis to avoid a collision with Earth as a function of when the encounter occurs.

The Apophis $\Delta v$ plot is particularly interesting because of its unusual structure, the two plateau-like regions with superimposed oscillation separated by a rapid rise in $\Delta v$. The oscillatory behaviour is a direct result of actual gravitational trapping in the physical Universe. For comparison, Figure 32(b) from the original CFO paper [1] also is reproduced below. It plots the average distance, $D_{avg}$, between CFO's probe with the best fitness and all the other probes for the 2D Step Function.

Not only is the structural similarity between the Apophis $\Delta v$ plot and CFO's $D_{avg}$ self-evident, it is quite remarkable. It is difficult to imagine that the Apophis curve, which is based on actual NEO gravitational trapping, and the CFO $D_{avg}$ curve, which is based on the metaphor of gravitational kinematics, are not different manifestations of the same or, at a minimum, a very similar underlying phenomenon. Although admittedly speculative, the author believes that this similarity is compelling evidence of the validity of CFO's gravitational metaphor, and that it might point the way to techniques for mitigating trapping. In particular, the analysis of Valsecchi *et al.* [15] may be directly applicable to improving CFO so that it avoids local trapping.

Another possible application of conservative energy exchange between a small object and a larger gravitating mass in a trapping encounter is using the object's total energy, $E_{tot}$, as a measure of trapping. This suggestion, too, is speculative in nature, but it reflects what hopefully is strong correlation between real gravitational kinematics and CFO's metaphorical search methodology. If the effect of gravitational trapping is to modify $E_{tot}$, then with a suitable definition of energy in CFO space $E_{tot}$ may become a key factor in signalling trapping and possibly providing a mitigation mechanism simply by increasing its value. It may also be possible to re-formulate the basic CFO equations in terms of an energy model. Whether or not this approach has any merit is a question for researchers far more capable than the author, one that he hopes will be pursued.

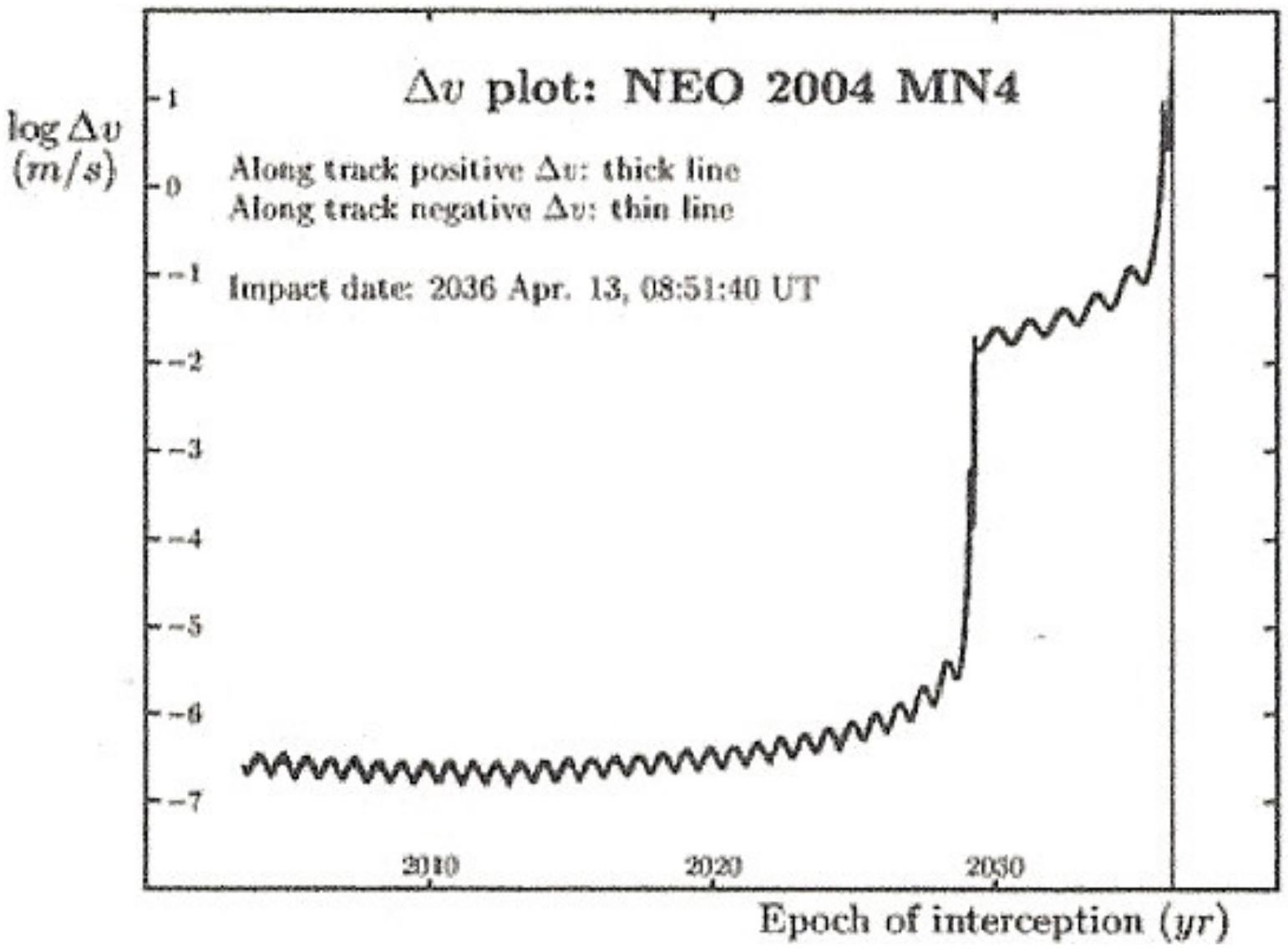

Reproduction of Figure 2 in Schweickart [16]

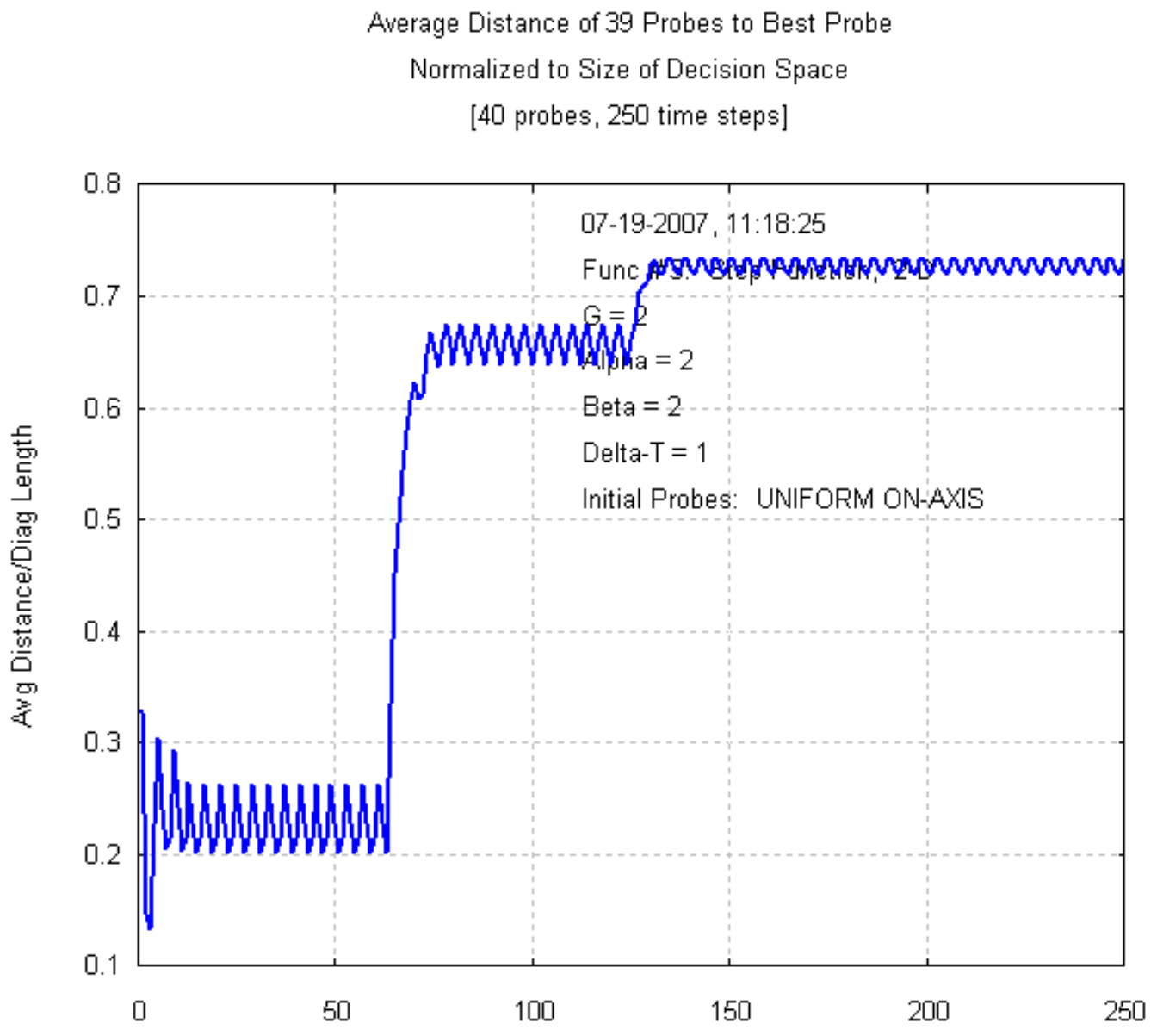

Reproduction of Figure 32(b) in the original CFO Paper [1].

### 4.2 The CFO Equations

CFO addresses the following problem: Locate the global maxima of the objective function $f(x_1, x_2, ..., x_{N_d})$ defined on the decision space $x_i^{\min} \le x_i \le x_i^{\max}$, $1 \le i \le N_d$, the $x_i$ being *decision variables*, where $i$ the coordinate number, and $N_d$ the space's dimensionality (number of coordinates). *Fitness* refers to the value of $f(\vec{x})$ at the point $\vec{x}$. The objective function's topology, which is unknown, may be continuous or discontinuous, highly multimodal or unimodal, and possibly subject to a set of constraints $\mathbf{\Omega}$ among the decision variables. While many optimization algorithms search for minima, CFO searches for maxima [which, of course, is the same as minimizing $-f(\vec{x})$].

CFO "flies" a group of "probes" through the decision space. Their trajectories are computed from the *equations of motion* for each probe's acceleration and position vector. These equations are derived by generalizing Newton's universal law of gravitation and law of motion from three-dimensional physical space to metaphorical $N_d$-dimensional "CFO space." The resulting equations are [1-4]:

$$\vec{a}_{j-1}^{\,p} = G \sum_{\substack{k=1 \\ k \ne p}}^{N_p} U\left(M_{j-1}^{k} - M_{j-1}^{p}\right) \cdot \left(M_{j-1}^{k} - M_{j-1}^{p}\right)^{\alpha} \times \frac{\left(\vec{R}_{j-1}^{k} - \vec{R}_{j-1}^{p}\right)}{\left\| \vec{R}_{j-1}^{k} - \vec{R}_{j-1}^{p} \right\|^{\beta}} \qquad (1)$$

$$\vec{R}_j^{p} = \vec{R}_{j-1}^{p} + \frac{1}{2} \vec{a}_{j-1}^{\,p} \Delta t^2, \quad j \ge 1 \qquad (2)$$

The probe number is $1 \le p \le N_p$, and the time step (iteration) number is $0 \le j \le N_t$, where $N_p$ and $N_t$ are the *total* number of probes and the *total* number of time steps, respectively. $\vec{a}_{j-1}^{\,p}$ is "probe" $p$'s acceleration at time step (iteration) $j-1$. $\vec{R}_j^{p} = \sum_{k=1}^{N_d} x_k^{p,j} \hat{e}_k$ is its position vector at step $j$, where the $x_k^{p,j}$ are probe $p$'s coordinates at step $j$, and $\hat{e}_k$ is the unit vector along the $x_k$ axis. Note that the version of eq. (2) in [1] contains a "velocity" term not included in the CFO formulation in [2]. The velocity term was dropped because including it actually impeded convergence (reasons unknown). In [1], however, this term already had been set to zero as a matter of convenience, so that from the outset it never was required, and consequently is no longer included.

CFO "flies" its $N_p$ probes through the decision space as a function of time (iteration number) along trajectories calculated using equations (1) and (2). A new probe distribution is computed at each step, and at the location of each probe the objective function's fitness is computed. For example, at time step $j-1$ at probe $p$'s location, the fitness is $M_{j-1}^{p} = f(x_1^{p,j-1}, x_2^{p,j-1}, ..., x_{N_d}^{p,j-1})$. Note that $M$ does *not* mean "mass" (see [1] for a discussion of the symbology). Each of the other probes has a fitness $M_{j-1}^{k}, k = 1,...,p-1, p+1,...,N_p$, associated with it at step $j-1$.

The function $U(\cdot)$ in eq. (1) is the Unit Step function, $U(z) = \begin{Bmatrix} 1, & z \ge 0 \\ 0, & otherwise \end{Bmatrix}$. Following standard notation, the magnitude of vector $\vec{B}$ is $\|\vec{B}\| = \left( \sum_{i=1}^{N_d} b_i^2 \right)^{\frac{1}{2}}$, where the $b_i$ are its scalar components. $G$ is CFO's "gravitational constant," and $\Delta t$ the "time" interval between steps during which the acceleration is constant. It is important to remember that this terminology is chosen solely to reflect CFO's gravitational metaphor, as is the factor $\frac{1}{2}$ in eq. (2), and beyond that there is no significance to the nomenclature. $\alpha$ and $\beta$ are the CFO exponents. While $G$ and $\Delta t$ have direct analogs in the equations of motion for gravitationally controlled real masses moving through physical space, there is no analog in Nature for $\alpha$ and $\beta$. They are included to provide added flexibility in how CFO is implemented by giving the algorithm designer the freedom to change how CFO's "gravity" varies with mass or distance, or both, in order to achieve a more effective algorithm. Of course, $\alpha$ and $\beta$ are perfectly acceptable parameters in metaphorical "CFO space," because there gravity can be whatever the designer wants it to be.

The Unit Step function in eq. (1) is extremely important because it creates positive-definite CFO mass. Under the gravitational analogy on which CFO is based objects in CFO space have "mass" just as they do in the physical Universe. But unlike real objects, mass in CFO space is a *user-defined* function of the objective function's fitness (not necessarily the fitness value itself). The problem is, depending on how the user defines that function, that the mass can be positive or negative. And negative mass leads to serious problems!

In this paper CFO "mass," by definition, is $U\left(M_{j-1}^{k} - M_{j-1}^{p}\right) \cdot \left(M_{j-1}^{k} - M_{j-1}^{p}\right)^{\alpha}$, that is, the difference in fitness values raised to the $\alpha$ power multiplied by the Unit Step. The *difference* of fitnesses intuitively seems to be a good measure of how much "gravitational" influence the value of the objective function at one point in the decision space should have on a probe at another point in the space, but other choices certainly are possible. The exponent $\alpha$ is optional, giving the algorithm designer added flexibility as discussed above. In marked contrast, with the difference of fitnesses definition, the Unit Step is an essential element, because without it the mass could be negative depending on which fitness is greater. In the real world, gravity always is attractive because mass is positive. Negative mass creates a repulsive gravitational force that flies probes away from maxima, instead of the positive gravitational force required to fly probes toward them. This is just the opposite of what CFO is intended to do. The Unit Step solves the negative mass problem by creating positive-definite CFO mass and, consequently, a positive gravitational force that always is attractive.

A frequently used measure of how well CFO's probes have clustered at a maximum is the average distance between the probe with the best fitness and all the other probes, normalized to the size of the decision space (length of the principal diagonal), that is,

$D_{avg} = \frac{1}{L \cdot (N_p - 1)} \sum_{p=1}^{N_p} \sqrt{\sum_{i=1}^{N_d} (x_i^{p,j} - x_i^{p^*,j})^2}$ where $p^*$ is the number of the probe with the best fitness, $L = \sqrt{\sum_{i=1}^{N_d} (x_i^{\max} - x_i^{\min})^2}$ is the diagonal length, and $\vec{R}_j^p = \sum_{k=1}^{N_d} x_k^{p,j} \hat{e}_k$ is probe $p$'s position vector at step $j$. Plotting $D_{avg}$ as a function of the time step is helpful in showing how the probe distribution evolves. As pointed out in §4.1, oscillation in $D_{avg}$ appears to be a sufficient but not necessary condition for local trapping of CFO's probe with the best fitness.

### 4.3 A New "Hyperspace Directional Derivative"?

The Unit Step introduced in the definition of mass in §4.2 may be important in another, purely mathematical way. It may be the basis for defining a new type of "hyperspace directional derivative." In Nature, gravity is a conservative vector force field that produces "action at a distance." While its fundamental equations are vector in nature, gravity also can be calculated as the gradient of a scalar potential function, just as the electric field is derived from an electric potential. The general notion of a "gradient" also applies in CFO space, but to some degree its utility depends on how mass is defined. For example, in this CFO implementation with $\alpha = \beta = 2$, the difference in fitness values divided by the distance between probes is essentially the square of the average slope, that is, the square of a derivative. When the Unit Step is included, this term, in a general sense, may be thought of as a sort of "generalized derivative." However, with entirely different CFO mass definitions, or the present definition with different exponent values, this generalization may not be as apparent or compelling; but in this case it seems both clear and appropriate. This observation suggests the possibility of defining a new mathematical construct, a "hyperspace directional derivative," that includes the Unit Step so that it always points towards an objective function's maxima. The author makes this suggestion entirely on a speculative basis. As with the "energy" notions discussed in §4.1, whether or not this suggestion makes any sense is a question left to the mathematicians who hopefully will pursue CFO's further development.

### 4.4 CFO Parameters

The CFO implementation described in this paper uses seven run parameters: $\alpha$, $\beta$, $G$, $\Delta t$, $N_t$, $N_p$, and $F_{rep}$. They are all defined above, except $F_{rep}$, the "repositioning factor." [Note that probe initial acceleration could be used as an eighth parameter, but it is zero for all runs reported here.] $F_{rep}$ is an algorithmic element and has nothing to do with fundamental CFO theory. For the specific CFO implementation used in this paper, $F_{rep}$ addresses the problem of errant probes. When equations (1) and (2) in §4.2 are used to compute a probe distribution, some "errant" probes may end up with coordinates outside the decision space domain. When that happens the question is, how to get them back? There are many possibilities. For example, the decision space boundary could be made "reflecting" (probes bouncing back into the decision space) or "absorbing" (probes extinguished and replaced with a new ones), and so on. A simple empirical scheme is used here.

On a coordinate-by-coordinate basis, probes flying out of the decision space are placed a fraction, $F_{rep}$, of the distance between the probe's starting coordinate and the corresponding boundary coordinate. But because there is no obvious way to assign a value to $0 \le F_{rep} \le 1$, the following simple, deterministic (somewhat clumsy, but workable) approach was taken. At each step, the current and previous 4 fitness values were stored in a 5-element array. $F_{rep}$ started at a value of 0.5 and was incremented by 0.005 whenever the absolute value of the difference between 5$^{th}$ array element and the average value of elements 3, 4, and 5 differed by less than 0.0005. If incrementing $F_{rep}$ in this manner resulted in $F_{rep} \ge 1$, then it was reset to the starting value, and this procedure was repeated with the then current probe distribution. If the data reported here are used for validation, it is important to note that the 5 saved fitness values are not necessarily sequential because of how the array index was computed (see pseudocode in Appendix B for details), which then affects when $F_{rep}$ is incremented. $F_{rep}$'s starting value and increment, and the fitness tolerance were determined empirically, that is, on a trial and error basis, as are all CFO parameters used in this and the other CFO papers.

Of all the CFO parameters, $F_{rep}$ and $N_p$ seem to have the greatest effect by far on the algorithm's performance. The sensitivity to $F_{rep}$ is what led to the empirical scheme described above. Changing $F_{rep}$ generally improves convergence, because doing so avoids or terminates local trapping (this effect discovered empirically). The number of probes, $N_p$, determines how well the decision space is sampled, that is, how much information CFO has about the objective function at the start of a run.

Numerical testing shows that CFO's convergence frequently is a sensitive function of how many initial probes are deployed. Another related non-numerical factor is where the initial probes are placed. As some of the cases in §6 show, changing the initial probe distribution may make the difference between capturing the global maximum or missing it.

The question of how best to assign values to CFO's run parameters is extremely important, yet it remains unanswered because CFO is in its infancy. At this point there is no methodology. Like most EAs, actually using the algorithm for a specific problem or class of problems, that is, setting up and making many runs, results in the user's learning how to make best use of the algorithm. Some parameter values work better than others, and the experience gained by making many runs is reflected in the user's choice of parameters in subsequent ones. Of course, the result is that the user *de facto* becomes a part of the algorithm because every run includes a degree of subjectivity. Needless to say, EAs should be implemented using an objective methodology to determine run parameters, but there presently is none for CFO, and, for that matter, for most EAs. One advantage CFO offers is its being deterministic. Results that are reproducible from one run to the next, coupled with CFO's frequently rapid convergence, even if to a local maximum, lend CFO to implementations that perform real-time parameter adjustments in response to a feedback mechanism. This sort of *Reactive Search* is proposed in [17], and it may be particularly appropriate for CFO because the algorithm is fast and deterministic.

## 5. PBM BENCHMARKS – ANTENNA GEOMETRIES & LANDSCAPE PLOTS

This section describes each PBM benchmark antenna and, for the 2D cases, its decision space using a perspective view of the topology. Additional perspective plots and projections onto the decision space's principal planes are included in Appendix A, as are samples of the NEC4 input/output files used to generate the plots and to validate NEC4.

### *5.1 Benchmark #1: Variable Length Center-Fed Dipole*

The antenna geometry for Problem #1 is shown in Fig. 1. The objective is to maximize a center-fed dipole's directivity, $D$, as a function of its total length, $L$, and the polar angle, $\theta$. A perspective view of the 2D landscape appears in Fig.2, with additional plots in Appendix A. The topology is smoothly varying with a single global maximum and two local maxima of similar amplitude.

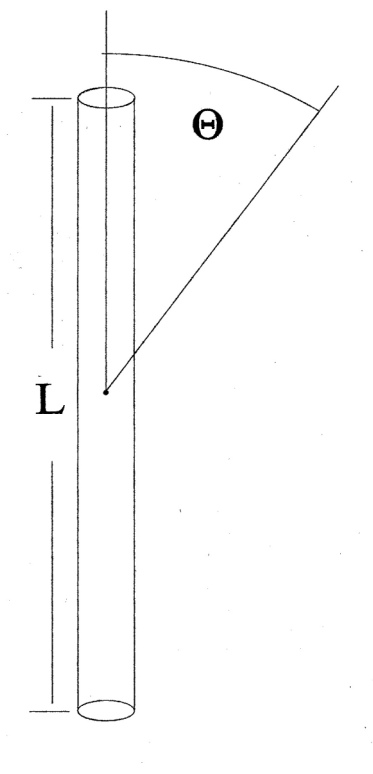

Fig. 1. Dipole.

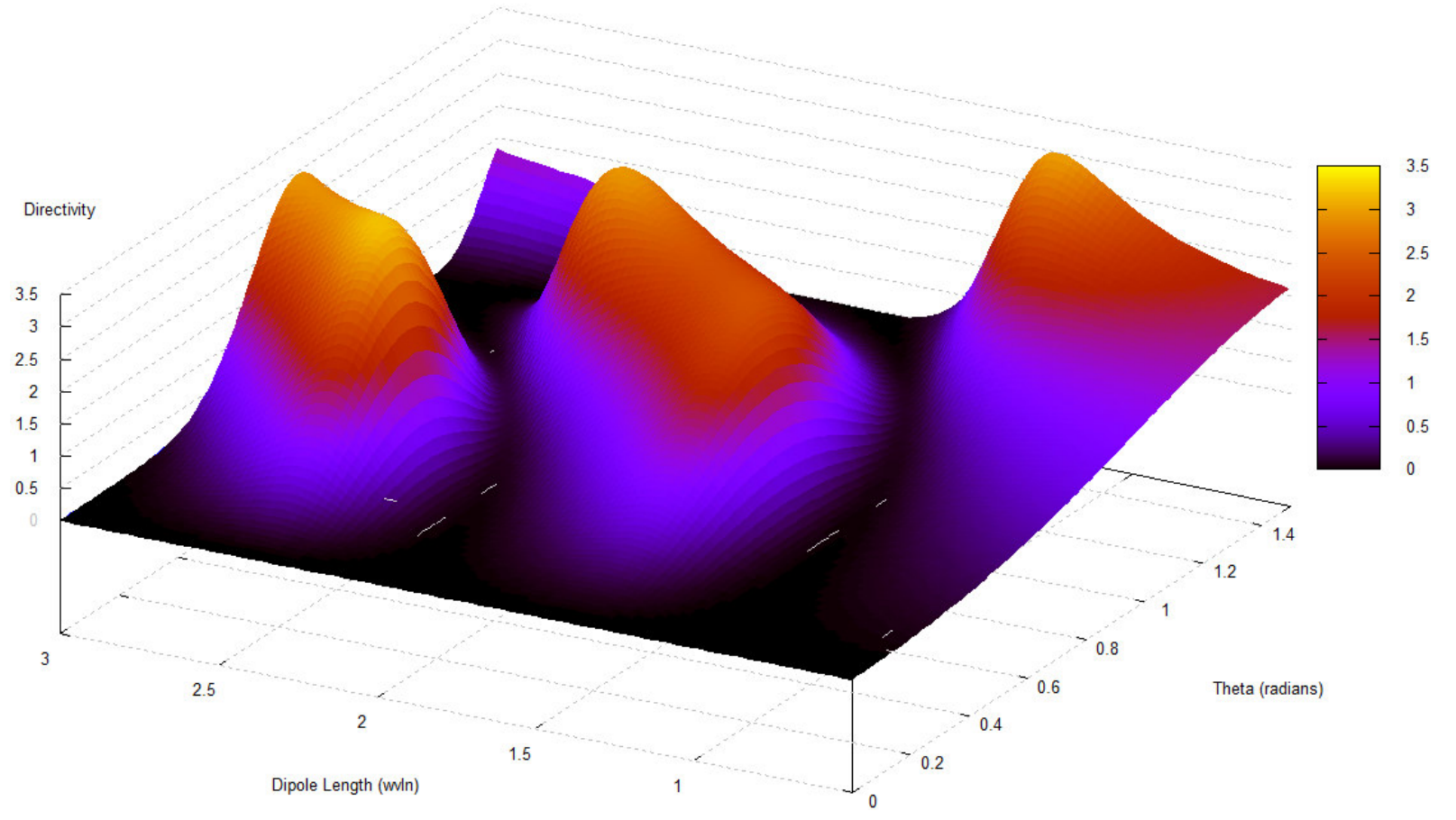

Fig. 2. Benchmark #1 topology, perspective view.

### *5.2. Benchmark #2: Uniform Dipole Array*

The problem #2 antenna is the uniform array of half-wave dipoles shown in Fig. 3. All elements are center-fed with in-phase equal amplitude sources. The standard right-handed Cartesian coordinate system used by NEC4 also is shown, as are the polar angle $\theta$ and azimuth angle $\phi$. The objective is to maximize directivity $D(d,\theta)$ in the plane $\phi = 90°$ as a function of element separation $d$ and polar angle $\theta$ in the presence of additive Gaussian noise. Fig. 4(a) shows the landscape without noise, and Fig. 4(b) with it. Following [6], noise is generated by adding to the NEC4-computed directivity a normally distributed zero-mean, 0.2-variance random variable $z$, here computed using the Box Muller method [18,19] as: $z = \mu + \sigma\sqrt{-2\ln(s)}\cos(2\pi t)$, where $\mu$ and $\sigma$, respectively, are the mean (zero) and standard deviation (0.4472), and $s$ and $t$ are random variables uniformly distributed on [0,1]. $s$ and $t$ are generated using the compiler's internal RND function [20] seeded with the CFO run's start time.

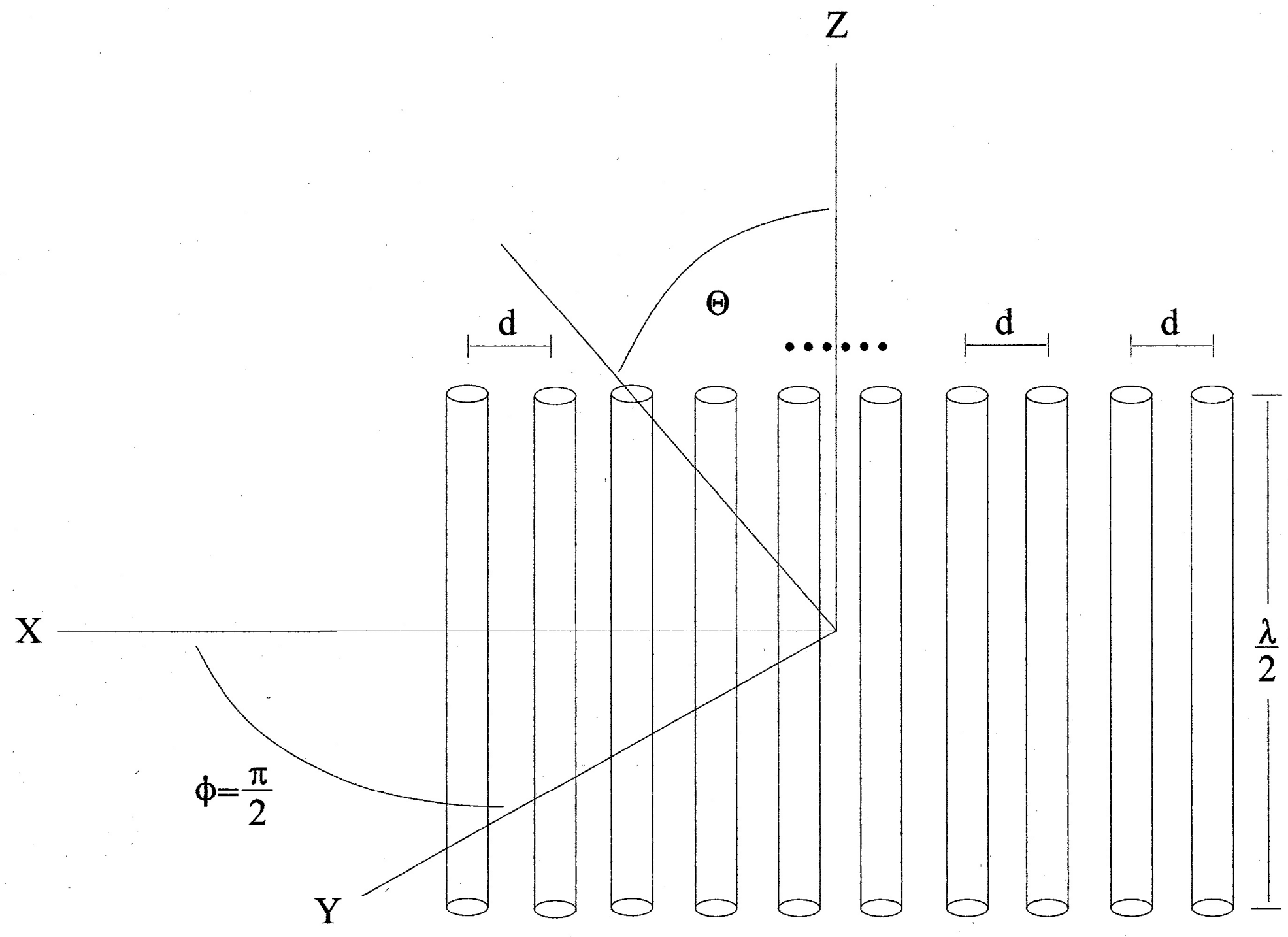

Fig. 3. Uniform Array of Half-Wave Center-Fed In-Phase Dipoles.

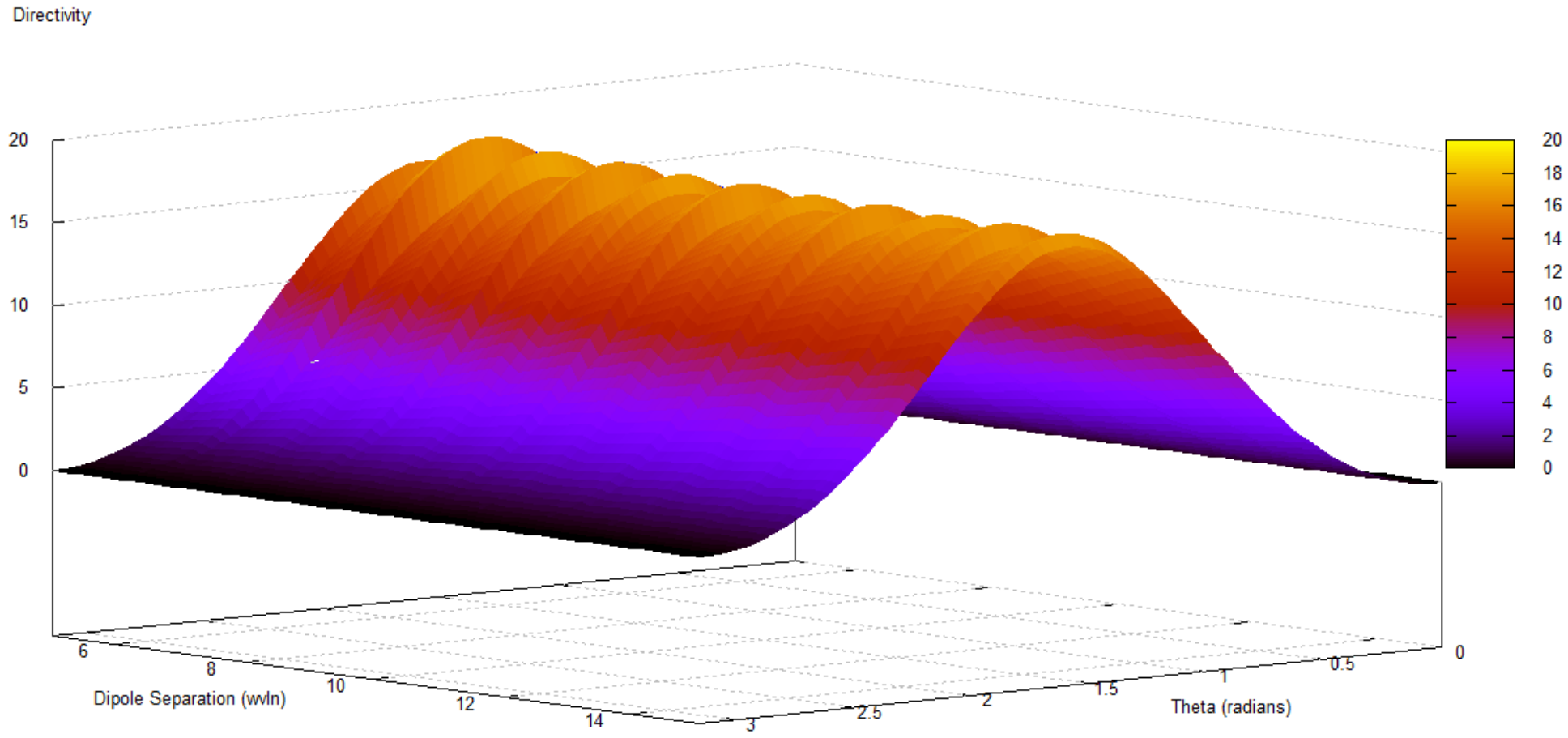

Fig. 4(a). Uniform half-wave dipole array without noise, perspective view.

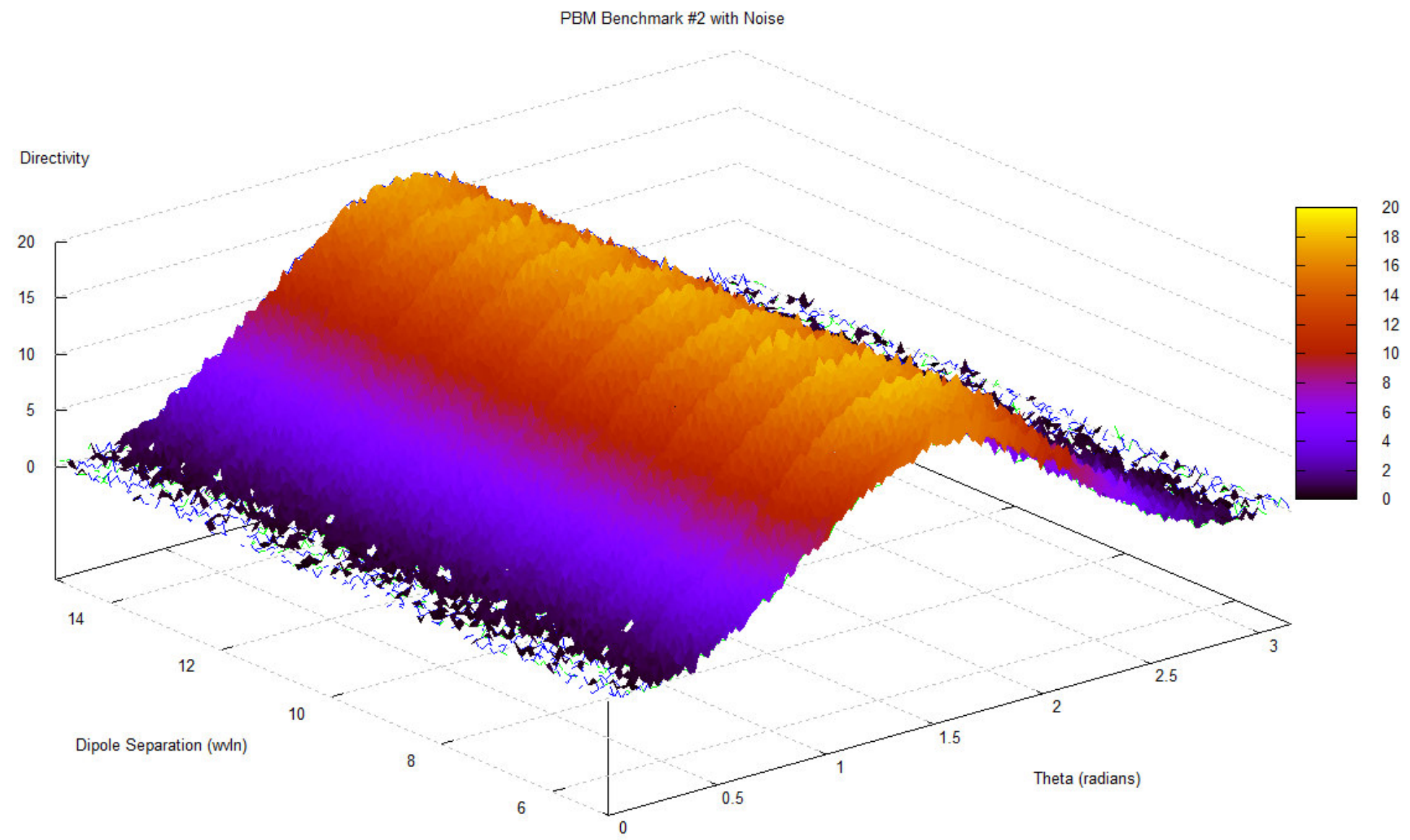

Fig. 4(a). Uniform dipole array with additive Gaussian noise, perspective view.

### *5.3 Benchmark #3: Circular Array of Half-Wave Dipoles*

The antenna for problem #3 is the circular array of half-wave dipoles shown in Fig. 5. The array comprises eight dipoles parallel to the z-axis uniformly deployed on a one-wavelength radius circle. All elements are center-fed by equal-amplitude sources; but, following [6], the phase varies as $\alpha_n = -\cos\left[2\pi\beta(n-1)\right]$, $n = 1,\ldots,8$. The unit-amplitude excitation therefore is $V_n = \cos\alpha_n + j\sin\alpha_n$. The objective is to maximize the directivity $D(\beta,\theta)$ in the plane $\phi = 0°$ as a function of the dimensionless phase parameter $0 \le \beta \le 4$ and the polar angle $\theta$. The range for $\beta$ produces the four global maxima at $(\beta_i = i - 0.5, i = 1,\ldots,4; \theta = \frac{\pi}{2})$ seen in the perspective topology plot in Fig. 6.

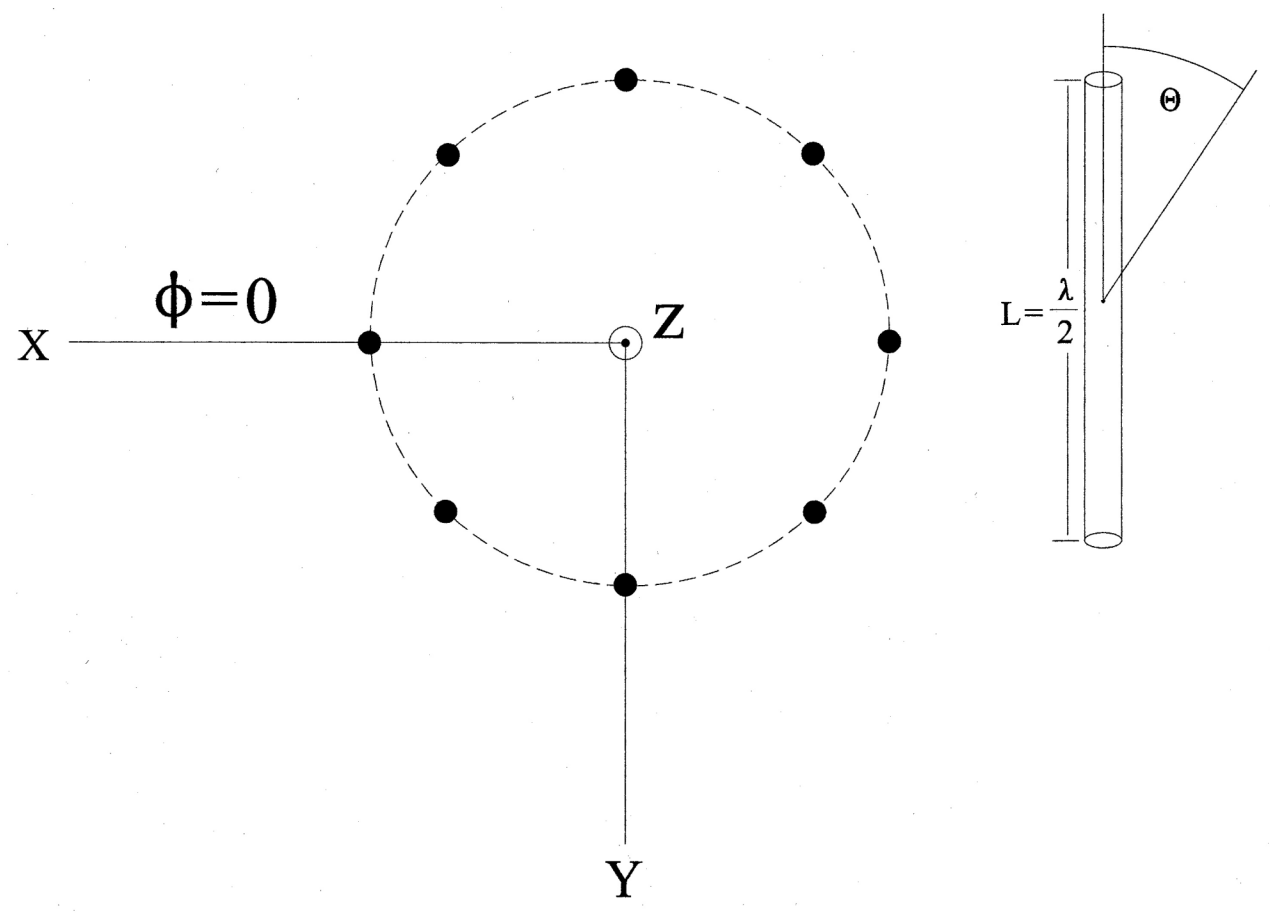

Fig. 5. Circular Array of Half-Wave Dipoles (1λ radius)

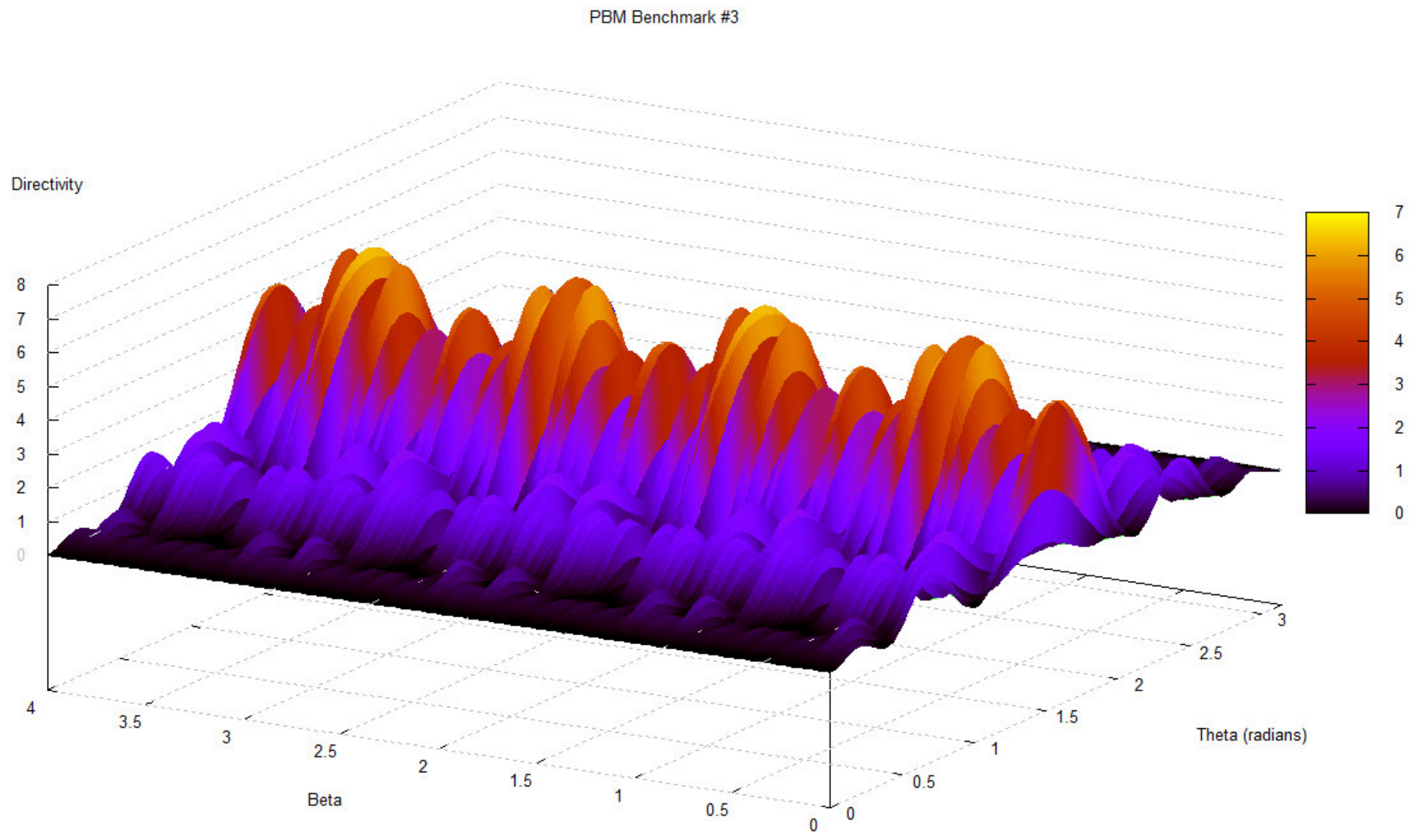

Fig. 6. Circular array landscape, perspective view.

### *5.4. Benchmark #4: Vee Dipole*

The Vee-dipole antenna for benchmark #4 is shown in Fig. 7. The antenna comprises two arms of length $L_{arm}$ with inner angle $2\alpha$ connected by a feed segment of length $2L_{feed}$ fed at its midpoint. The objective is to maximize the directivity $D(L_{total},\alpha)$ along the +X-axis as a function of the total dipole length $0.5\lambda \le L_{total} = 2L_{arm} + 2L_{feed} \le 1.5\lambda$ and the inner half-angle $\frac{\pi}{18} \le \alpha \le \frac{\pi}{2}$ with $L_{feed} = 0.01\lambda$. Topology of benchmark #4's decision space appears in Fig. 8. This objective function is unimodal with a single global maximum at $D(L_{total}, \alpha) = (1.5\lambda, 0.834)$. The surface is smoothly varying without pronounced local maxima.

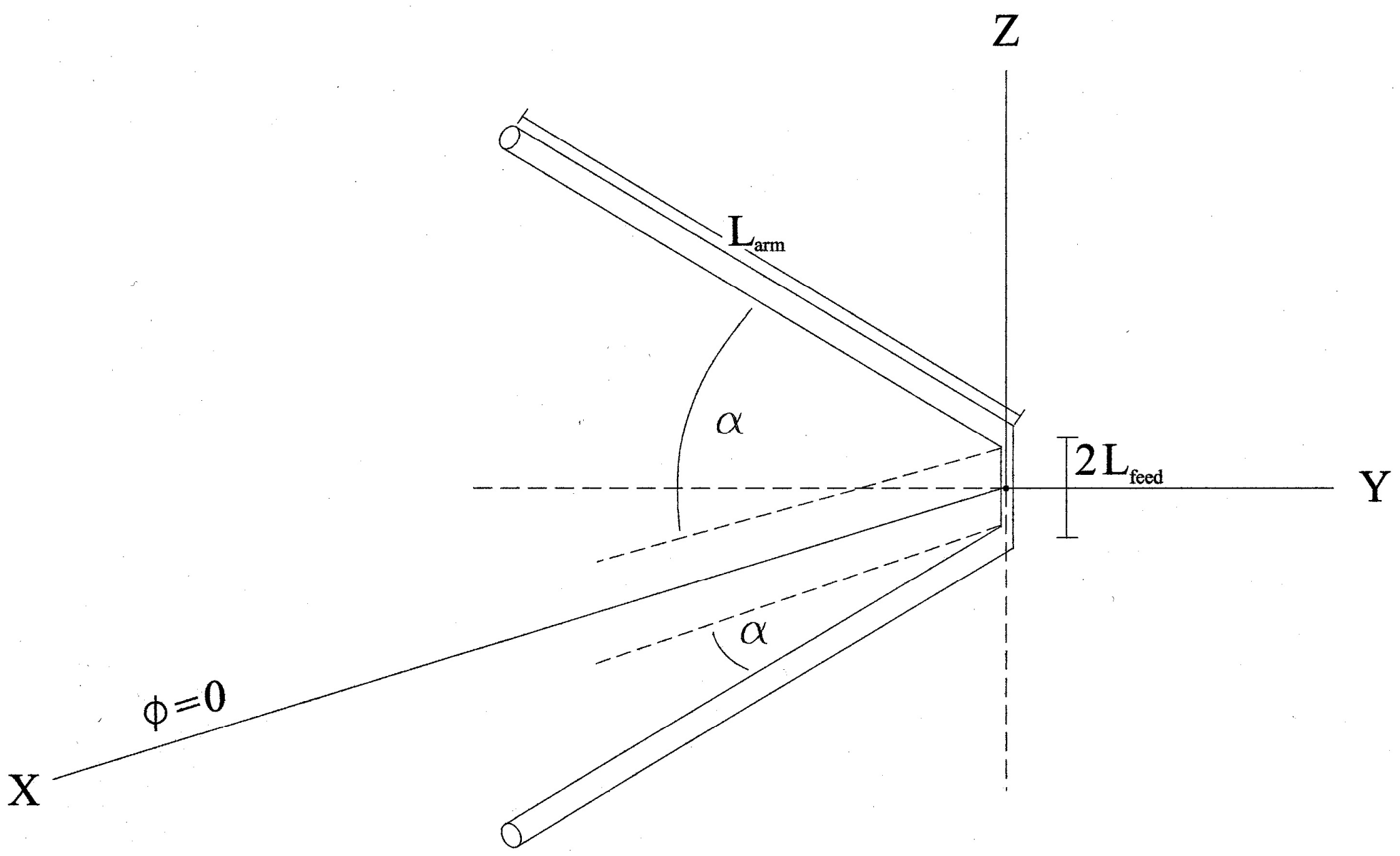

Fig. 7. Vee Dipole

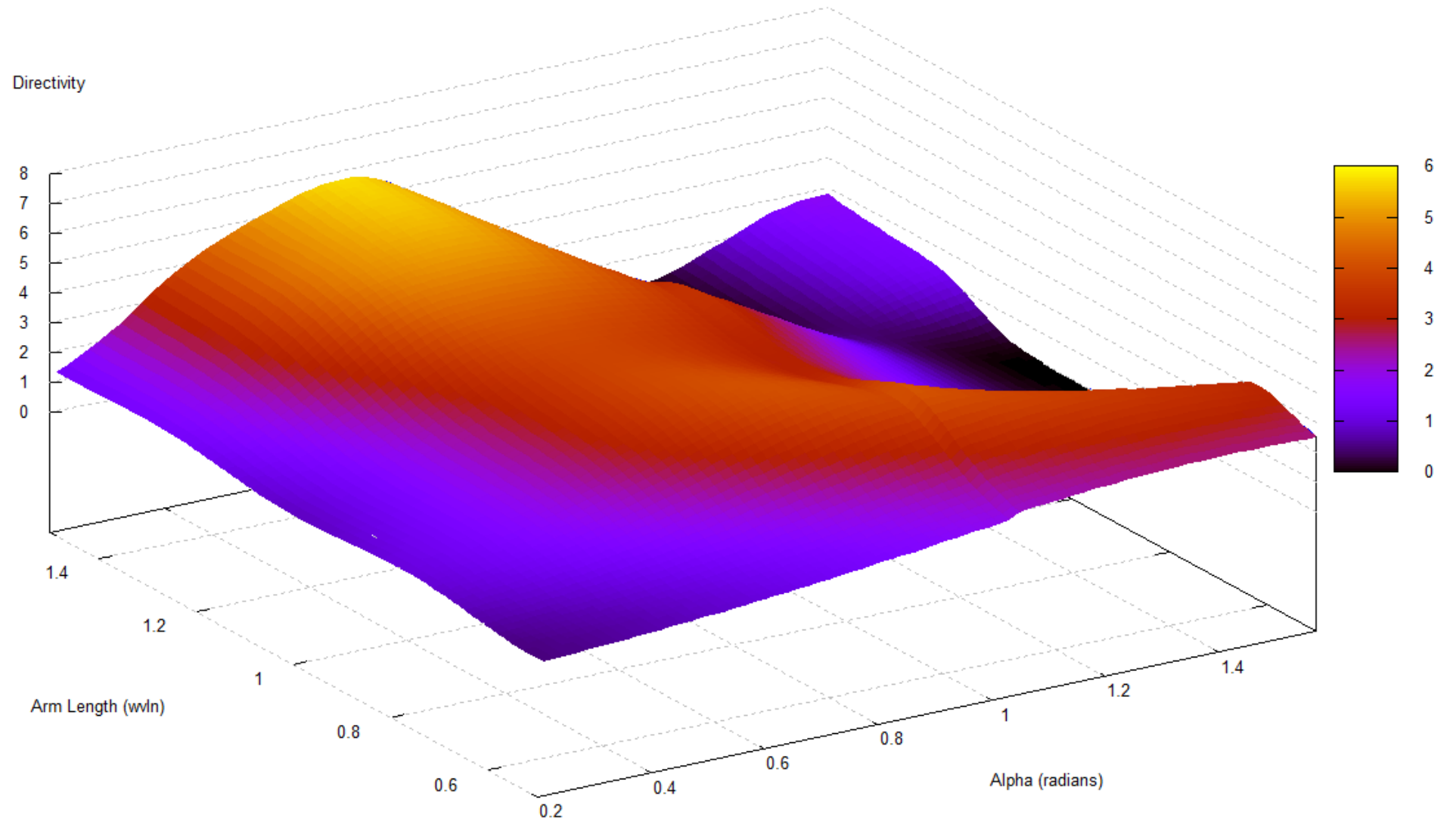

Fig. 8. Vee dipole decision space topology, perspective view.

### *5.5. Benchmark #5: N-element Collinear Dipole Array*

Benchmark #5 is a collinear array of $N_{el}$ half-wave dipoles as shown in Fig. 9. All elements are center-fed in-phase with equal amplitudes sources. The objective is to maximize directivity $D(d_i, i = 1,..., N_{el} - 1)$ in the plane $\phi = 0°$ as a function of the element center-to-center spacings $0.5\lambda \le d_i \le 1.5\lambda$. Because there are $N_{el} - 1$ spacings in an $N_{el}$ array, the dimensionality of this problem is $(N_{el} - 1)$ D, unlike the previous four benchmarks each of which is 2D. As discussed at length in [6], maximum directivity occurs at $d_i = 0.99\lambda, \forall i$ independent of the number of elements, that is, with all dipoles spaced $0.99\lambda$ regardless of the array size. Of course, the value of the directivity does depend on the array size, increasing approximately in proportion to the length.

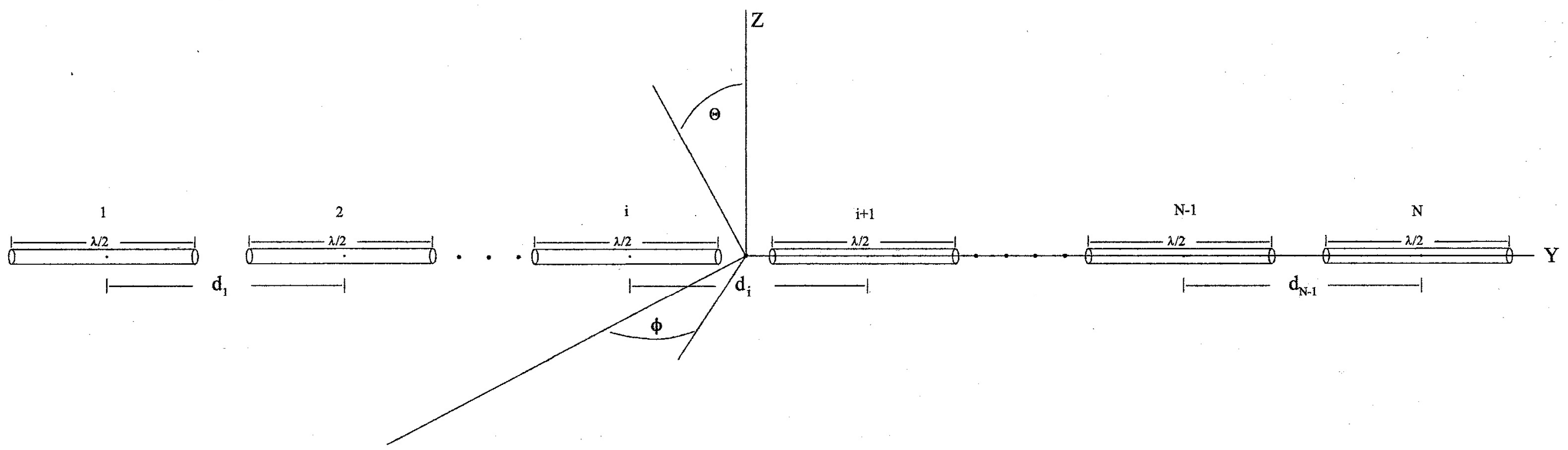

Fig. 9. ***N**<sub></sub>*$N_{el}$-element collinear dipole array.

## 6. CFO PERFORMANCE AGAINST PBM BENCHMARKS IN DETAIL

This section discusses in detail CFO's performance against the PBM benchmark suite. The following parameter values were used for all runs reported here: $\alpha = 2$, $\beta = 2$, $G = 2$, $\Delta t = 1$. The numbers of time steps and probes, $N_t$ and $N_p$, respectively, vary from run to run. And the important repositioning factor, $F_{rep}$, is variable as described in §4.4 and Appendix B. *All CFO run parameters are determined empirically because at this time there is no methodology for assigning parameter values.*

### *6.1 Benchmark #1: Variable Length, Center-Fed Dipole*

The CFO run was made with $N_t = 100$ and $N_p = 4$. The initial probes were located symmetrically in the $(L, \theta)$-plane at $(1.333\lambda, \frac{\pi}{4})$, $(2.167\lambda, \frac{\pi}{4})$, $(1.75\lambda, \frac{\pi}{6})$, $(1.75\lambda, \frac{\pi}{3})$ as shown in Fig. 10. The best fitness value of 3.2062693 was returned at step #14 at the point $(L, \theta) = (2.55088\lambda, 0.618046)$. While the PBM paper [6] reports a maximum directivity of 3.32 at $(L, \theta) = (2.58\lambda, 0.63)$, the directivity computed at that point by NEC4 is 3.2. CFO's returned value thus appears to be the actual maximum value that is computed by NEC4.

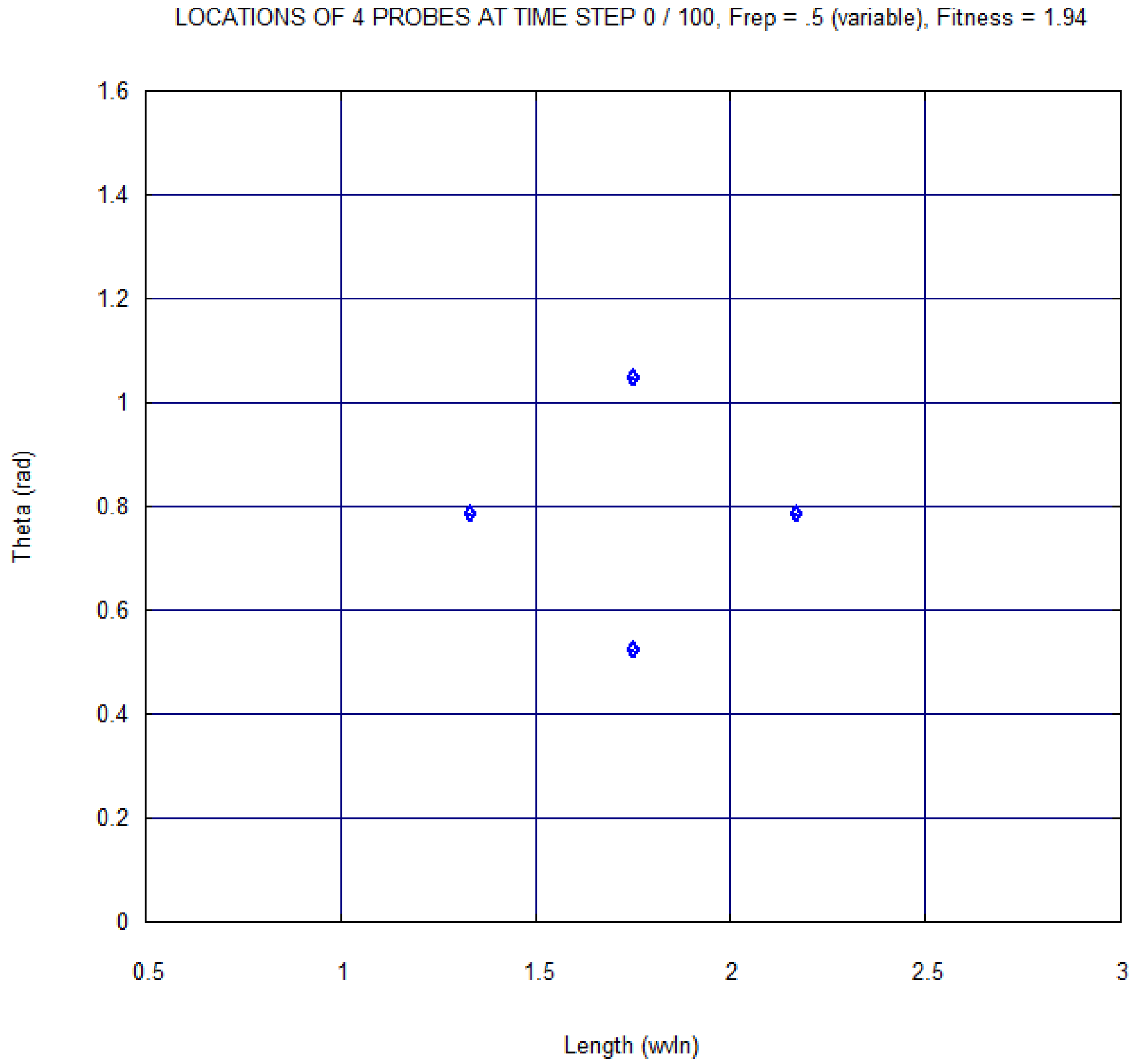

Fig. 10. Initial probe distribution for benchmark #1.

Table 6 shows the best fitness values at time steps when they changed. Saturation occurred at step #14, after only $N_{eval} = 60$ function evaluations [$N_{eval} = (j+1)N_p$ where $j$ is the time step number]. The run was terminated at step #100, and there was no change in fitness through that step. Figures 11(a) and (b) plot as a function of time step the best fitness and average distance between the probe with the best fitness and all other probes. In this case the best fitness increases in a step-wise fashion, whereas in many other CFO runs the increase is smoother. The plateau between steps #0 and 1 is a consequence of zero initial acceleration (it also will appear in all subsequent plots). Because the acceleration is zero, the probe distribution does not change. The $D_{avg}$ curve shows an erratic gradual decrease with a spike near step #70 followed by what visually appears to be complete coalescence of the probes thereafter. Inspection of the plot data, which are not tabulated here, shows that $D_{avg}$ is, in fact, constant at 0.001255 between steps #84 and 100, inclusive. Thus, all four probes have converged on the maximum, and it is reasonable to expect that the probe distribution at that point is static.

**Table 6. Best fitnesses for benchmark #1.**

| ***Step #*** | ***Fitness*** | $N_{eval}$ |
|---|---|---|
| 0 | 1.9364220 | 4 |
| 1 | 1.9364220 | 8 |
| 3 | 2.7352687 | 16 |
| 7 | 3.0338912 | 32 |
| 14 | 3.2062693 | 60 |
| 100 | 3.2062693 | 404 |

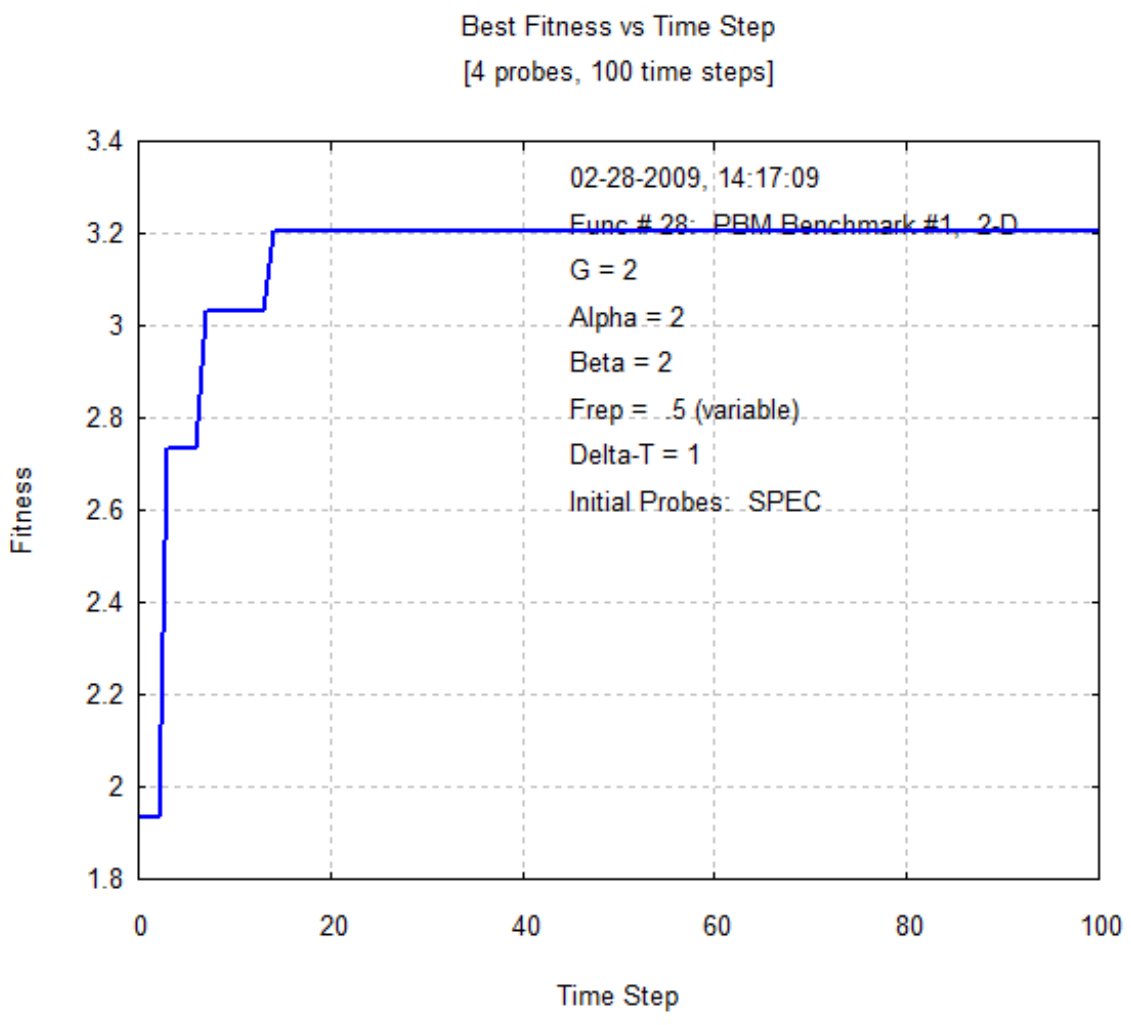

Fig. 11(a). Prob. #1 best fitness

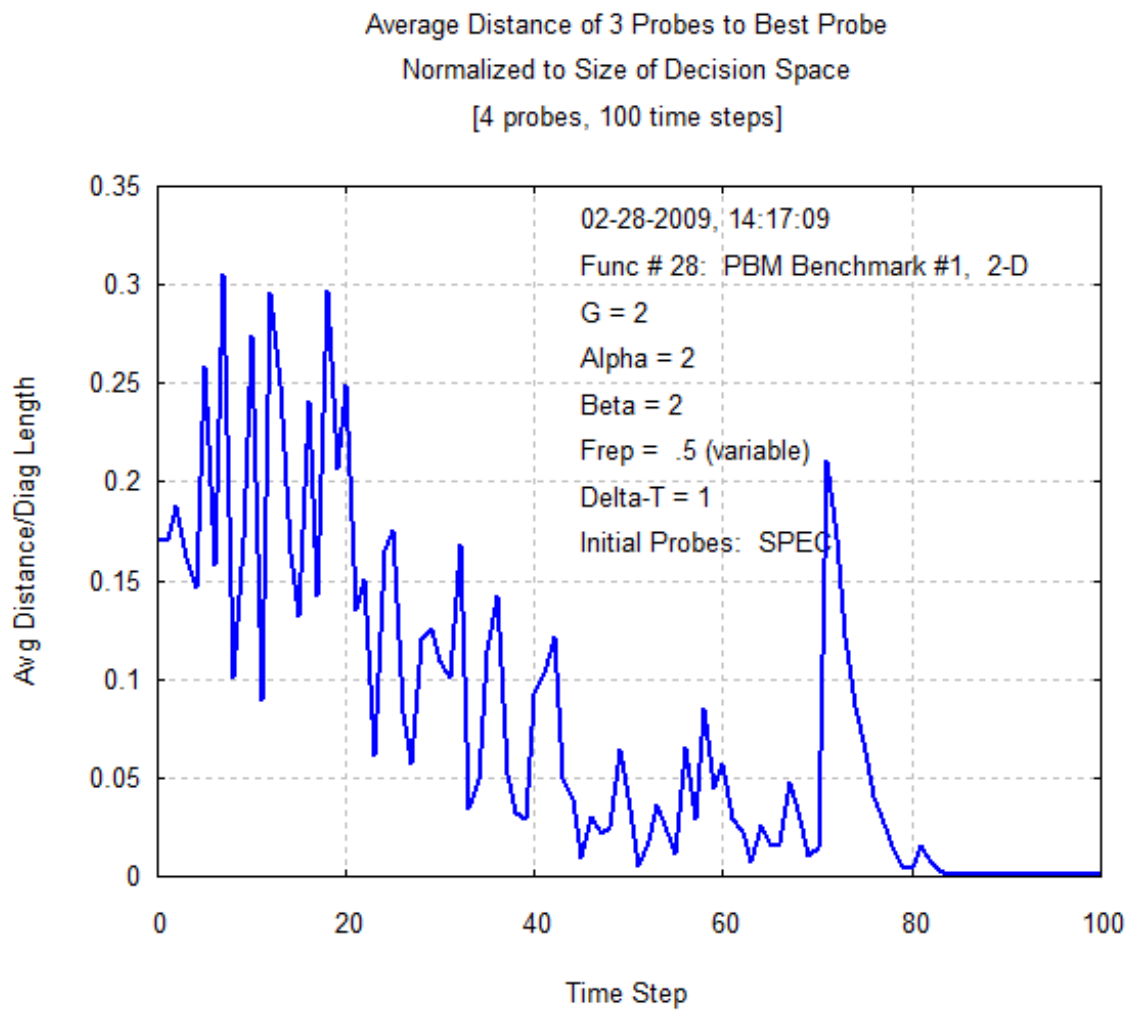

Fig. 11(b). Prob. #1 $D_{avg}$

### *6.2. Benchmark #2: Uniform Dipole Array, Added Noise*

Two CFO runs were made for problem #2, the first without noise, the second with additive Gaussian noise (see §5.2). $N_p = 24$ for both runs with initial probes deployed in the $(d,\theta)$-plane in a 6x4 grid as shown in Fig. 12. $N_t = 250$ for the noiseless run. Figs. 14(a) and (b) plot the evolution of the best fitness and $D_{avg}$, respectively. The fitness values at steps where they changed appear in Table 7. The best fitness was 18.3653834 at step #207 at $(d,\theta) = (5.92359\lambda, 1.55685)$. It increased very quickly through step 10 and slowly thereafter. The probe distribution at step #250 appears in Fig. 13, and it shows that by then almost all the probes have coalesced on the maximum. But at least one probe appears to be stuck at $(0.5\lambda, 0)$. The reason for this behavior (which has occurred in other CFO runs) is not understood. The $D_{avg}$ curve in Fig. 14(b ) clearly exhibits the oscillatory behavior that seems to be characteristic of local trapping, this time at the global maximum. Comparing these results to the PBM data show that CFO has successfully located the global maximum when the landscape is quiet.

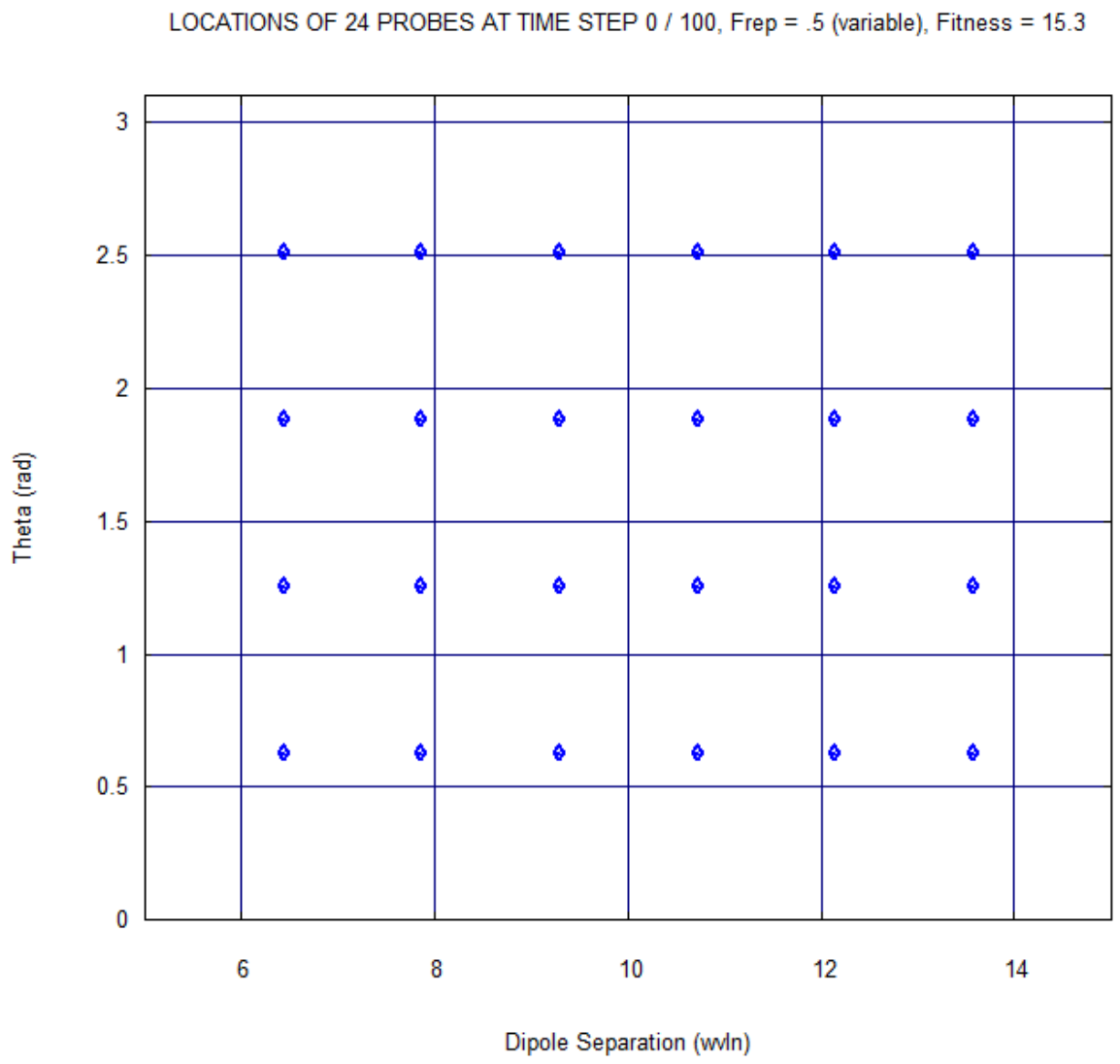

Fig. 12. Prob. #2 initial probes.

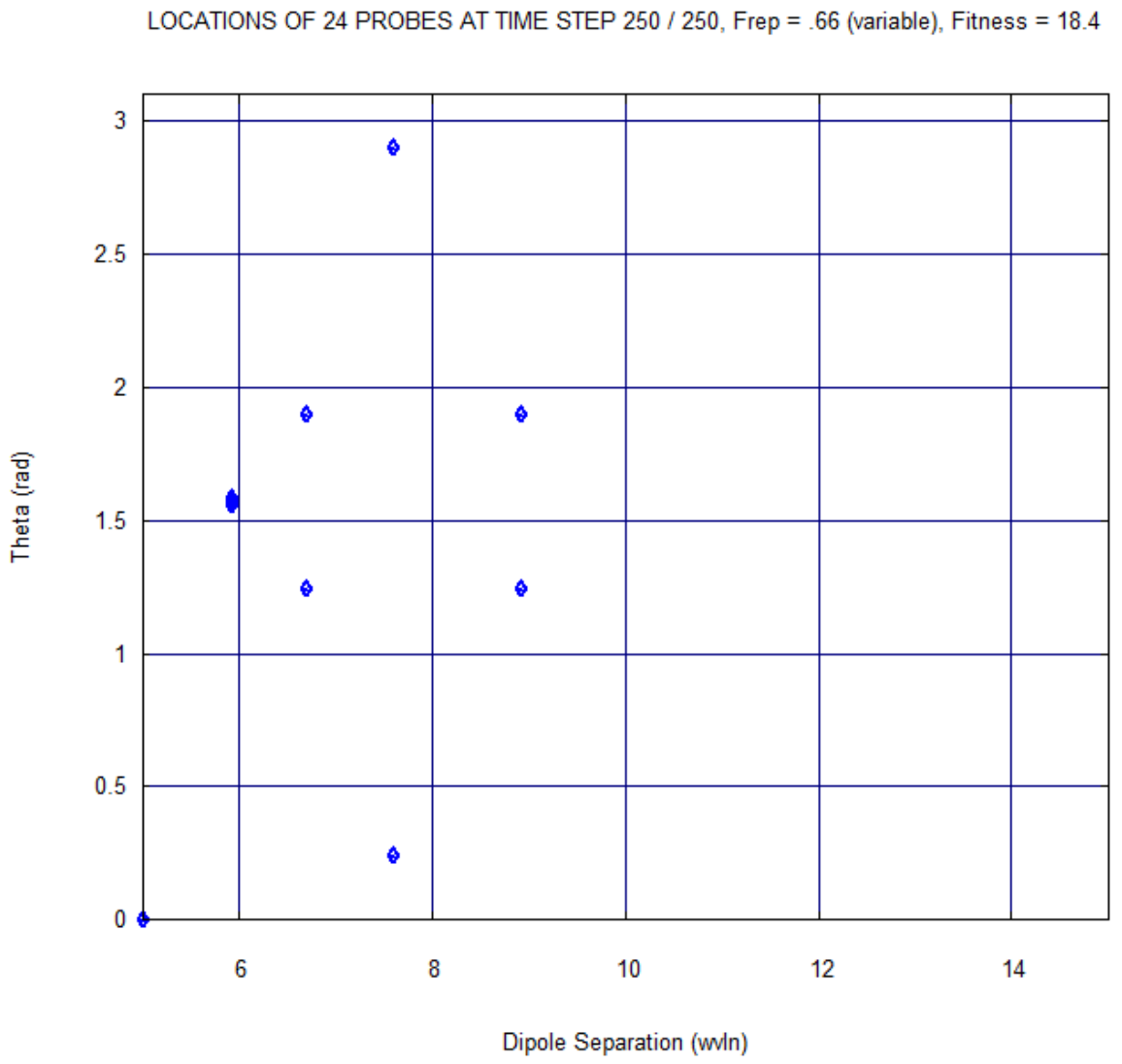

Fig.13. Probes at step #250, no noise.

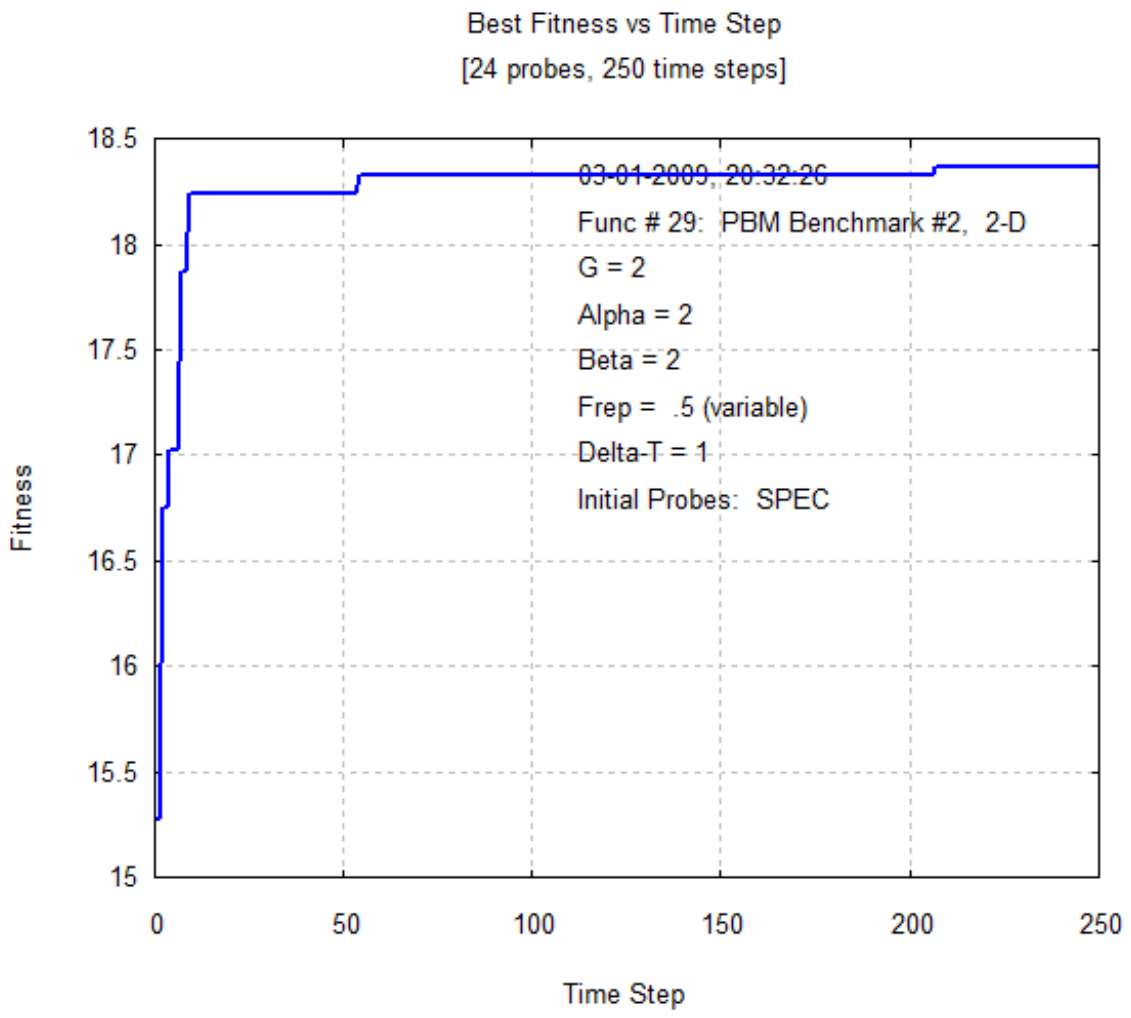

Fig. 14(a). Prob. #2 best fitness, no noise.

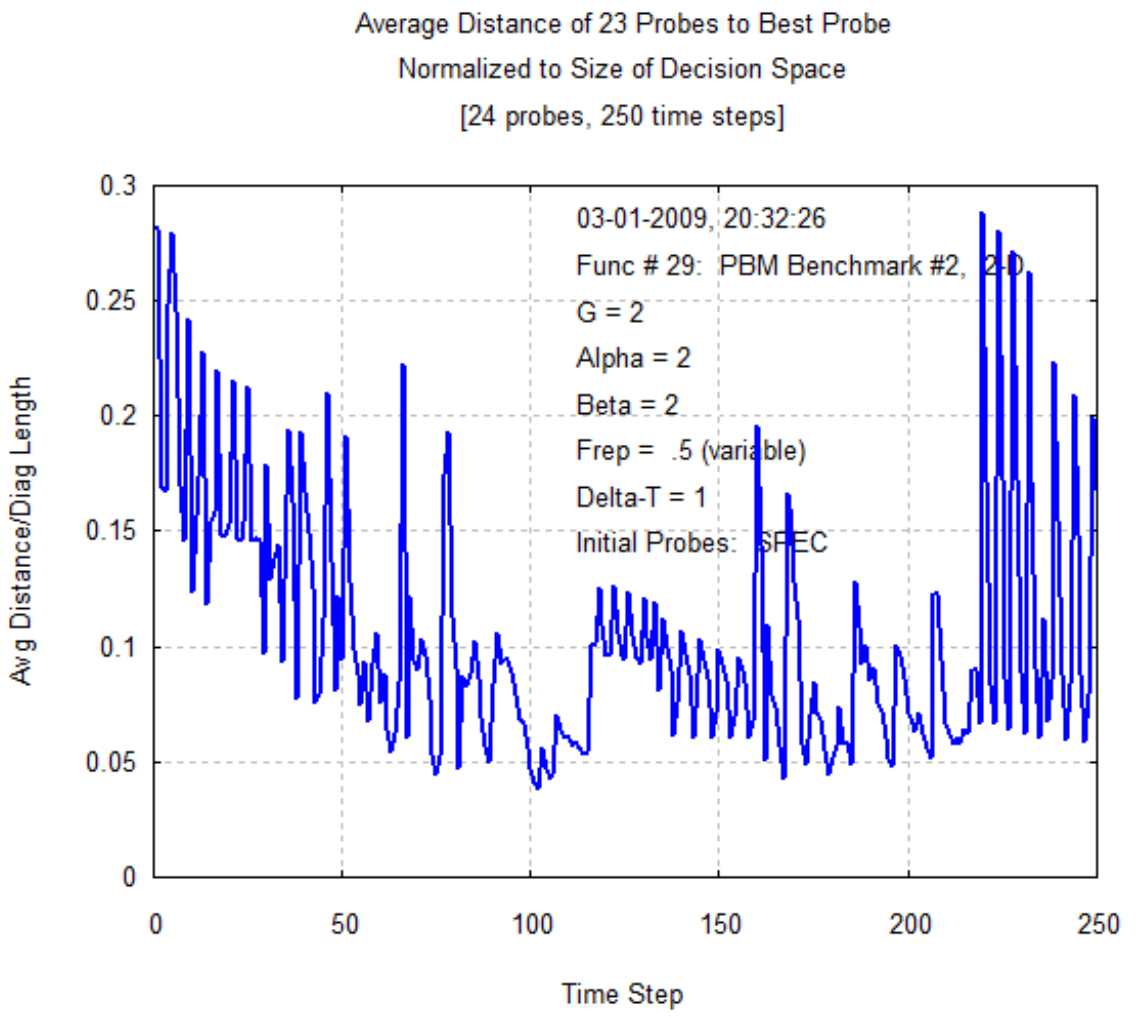

Fig. 14(b). Prob. #2 $D_{avg}$ , no noise.

**Table 7. Best fitnesses for benchmark #2 without noise.**

| ***Step #*** | ***Fitness*** | $N_{eval}$ $\ell$ |
|---|---|---|
| 0 | 15.2756606 | 24 |
| 1 | 15.2756606 | 48 |
| 2 | 16.7494288 | 72 |
| 4 | 17.0215851 | 120 |
| 7 | 17.8648757 | 192 |
| 9 | 18.2389570 | 240 |
| 54 | 18.3231442 | 1320 |
| 207 | 18.3653834 | 4992 |
| 250 | 18.3653834 | 6024 |

The second, noisy run was made with $N_t = 100$. Because of the additive noise, the best fitness changed at every time step. Inspection of the tabulated data (not shown) reveals that no two successive fitness values were the same. This result is expected because the normally distributed noise was added to the returned fitness at each calculation of the objective function's value. The best fitness occurred at step #31 with a value of 18.6880533 at $(d,\theta) = (6.93601\lambda, 1.54721)$. Consistent with the noisy landscape, the probe distributions fluctuate from step to step. Figs. 15(a) and (b) show the probe locations in the $(d,\theta)$-plane at steps 10 and 100, respectively, and neither plot has probes clustered near the maximum. Interestingly, the probe "stuck" at $(0.5\lambda, 0)$ is still evident. The evolution of the best fitness and $D_{avg}$ are plotted in Figs. 16(a) and (b), respectively.

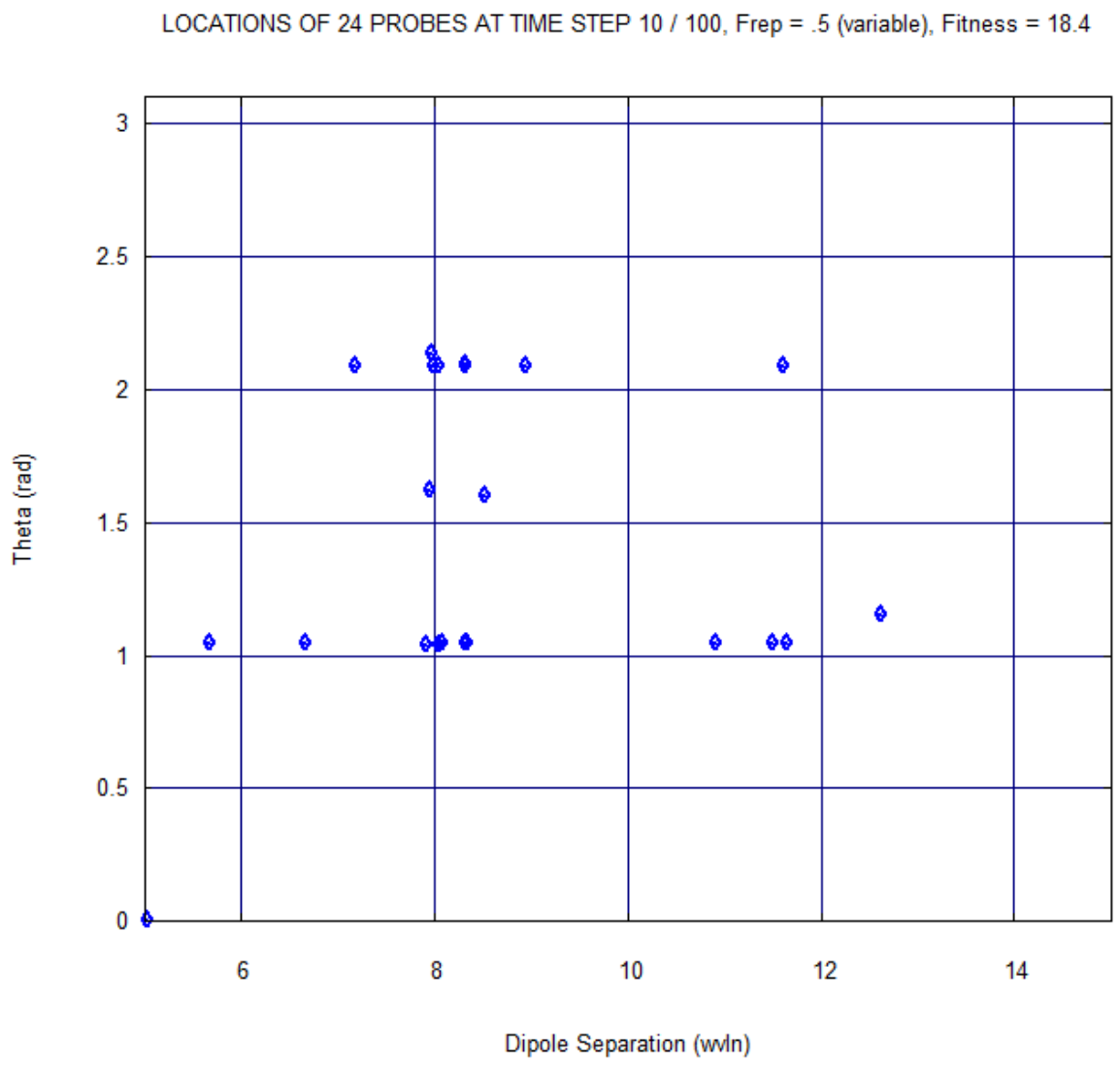

Fig. 15(a). Prob. #2 step 10 probes, noise.

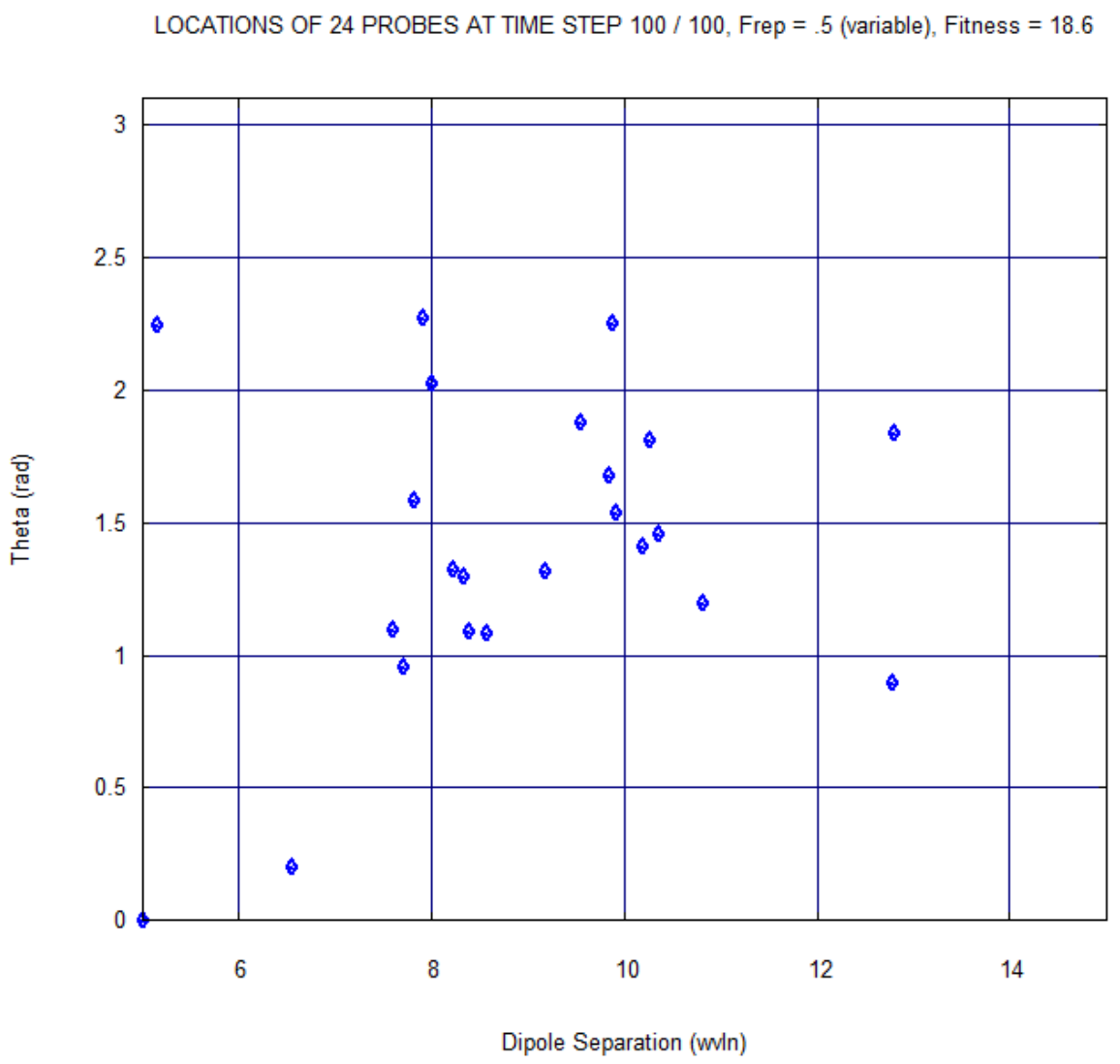

Fig. 15(b). Prob. #2 step 100 probes, noise.

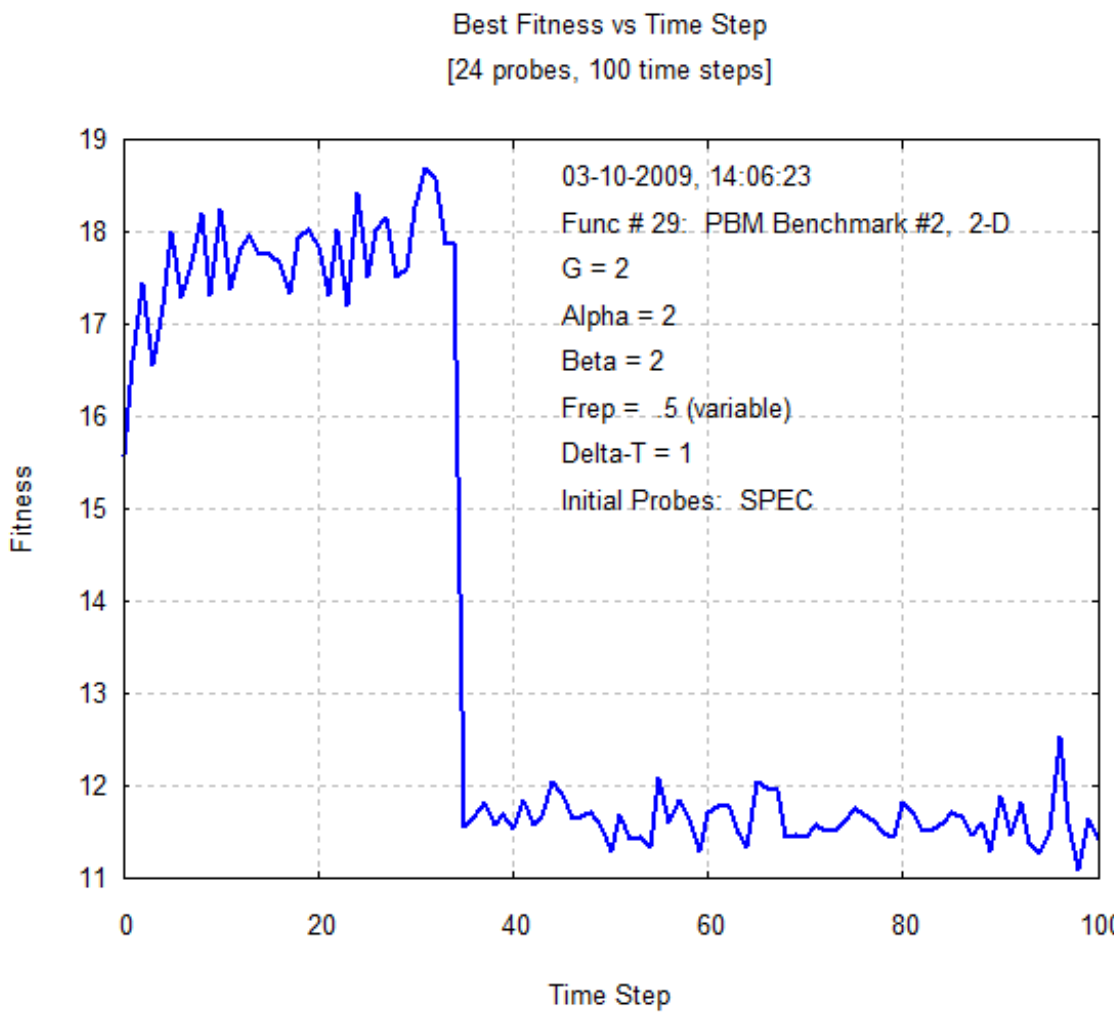

Fig. 16(a). Best fitness noisy prob. #2.

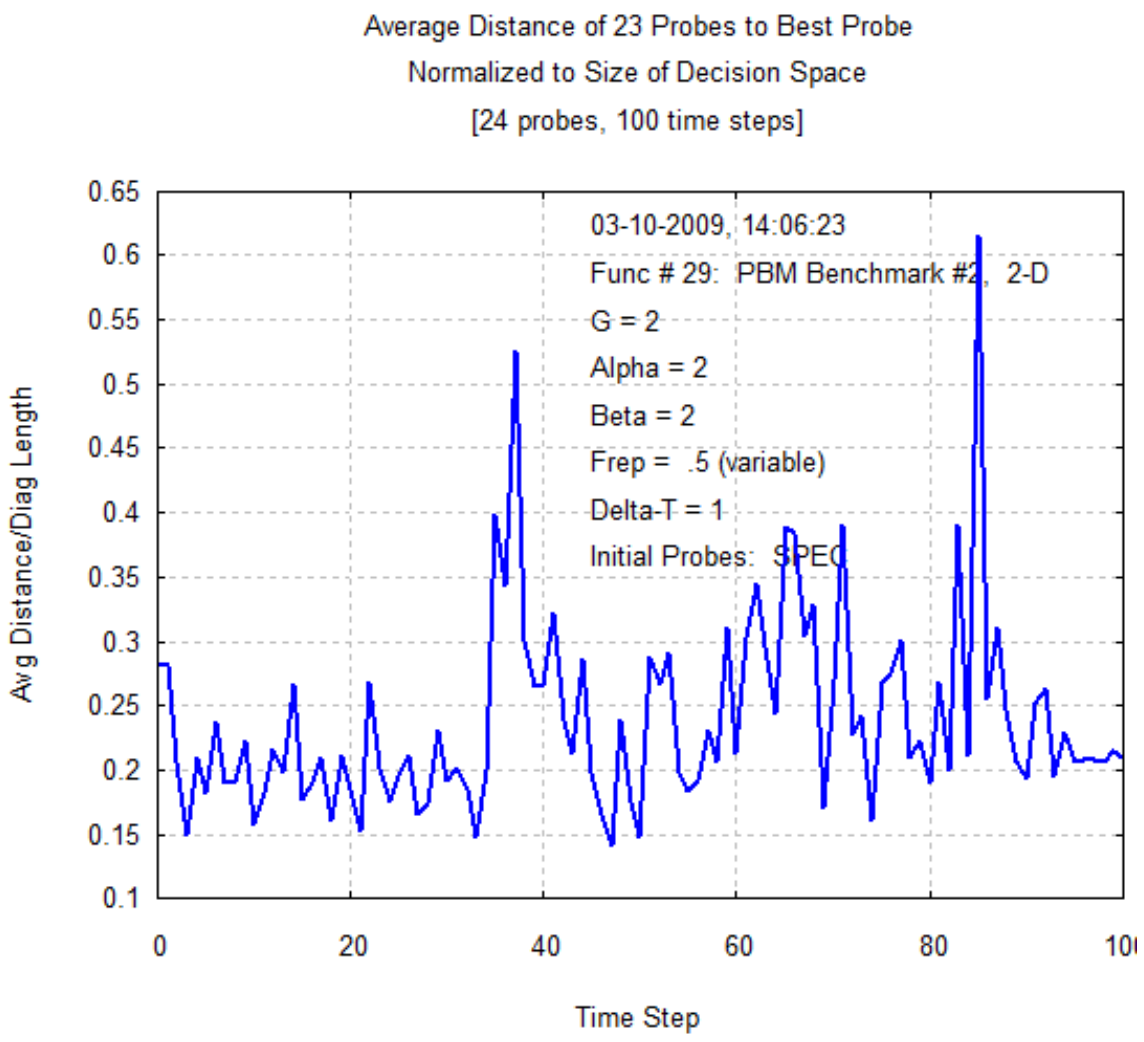

Fig. 16(b). $D_{avg}$ noisy prob. #2.

The fitness curve is quite interesting. There is an erratic increase to the global maximum at step #31, but there appears to be a clear upward trend in the data. Within a few steps of locating the maximum, the fitness drops precipitously, followed by an erratic variation around a value of about 11.5. The $D_{avg}$ curve also fluctuates erratically. Through step #31 the variation is around an "eyeball average" just below 0.2, and after that the variability increases substantially with no "average" value being evident. These behaviors in the best fitness and $D_{avg}$ do not have obvious explanations. Nevertheless, it is clear that CFO did successfully locate the global maximum even in the noisy decision space.

### *6.3 Benchmark #3: Circular Array of Half-Wave Dipoles*

The CFO run for problem #3 was made with $N_t = 300$ and $N_p = 10$ probes initially placed on the $\beta, \theta$-axes as shown in Fig. 17(a). Five probes were equally spaced on each of the decision space coordinate axes. Clustering of the probes at step #300 is shown in Fig. 17(b). One of the four global maxima was found at step #104 with a value of 6.4863443 at $(\beta, \theta) = (0.480235, 1.57327)$. The returned coordinates are close to the known first maximum's location of $(0.5, \frac{\pi}{2})$. But CFO did not locate the other three global maxima, a result that perhaps is not too surprising in view of the small number of probes and their initial deployment.

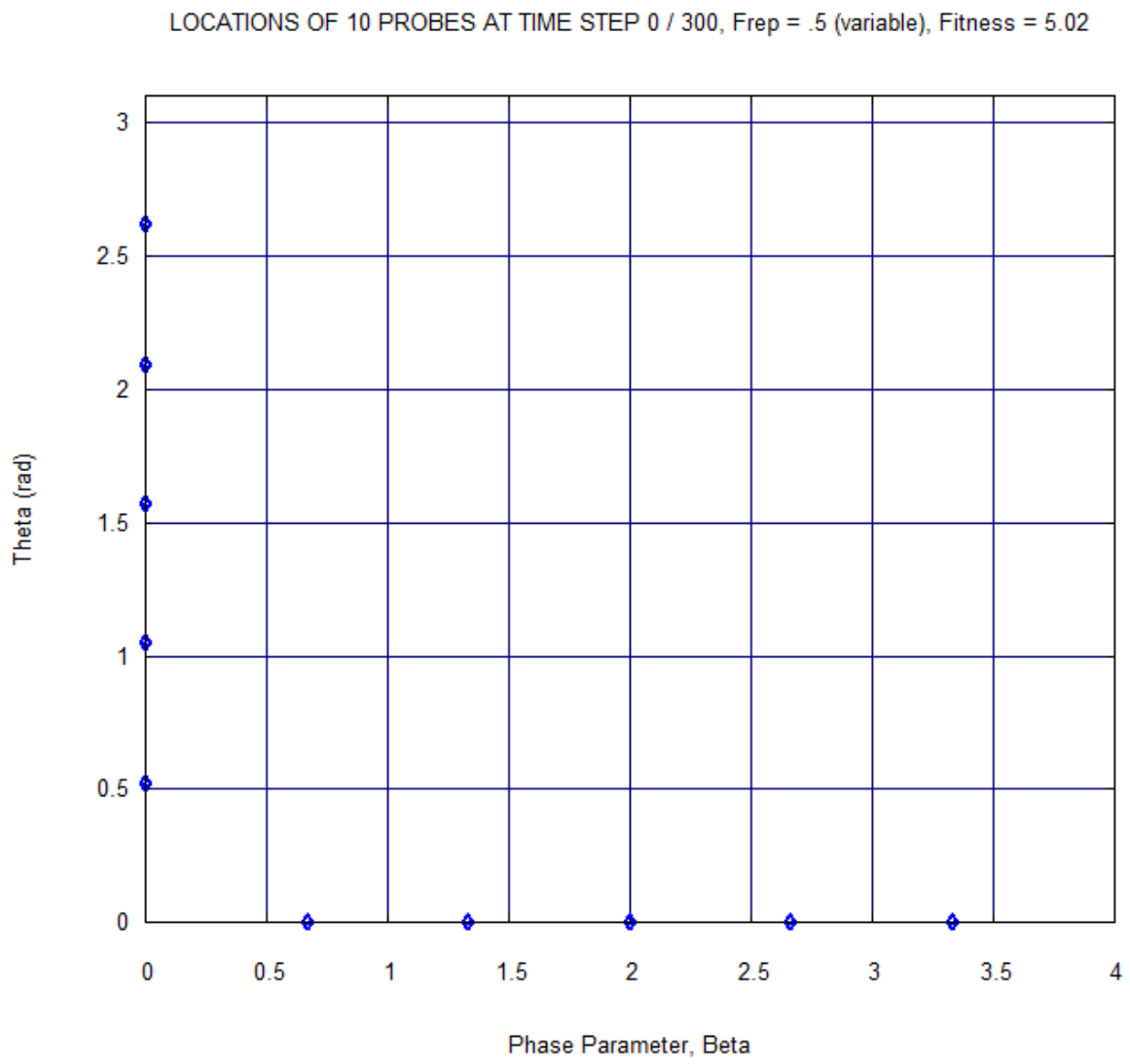

Fig. 17(a). Prob. #3 initial probes.

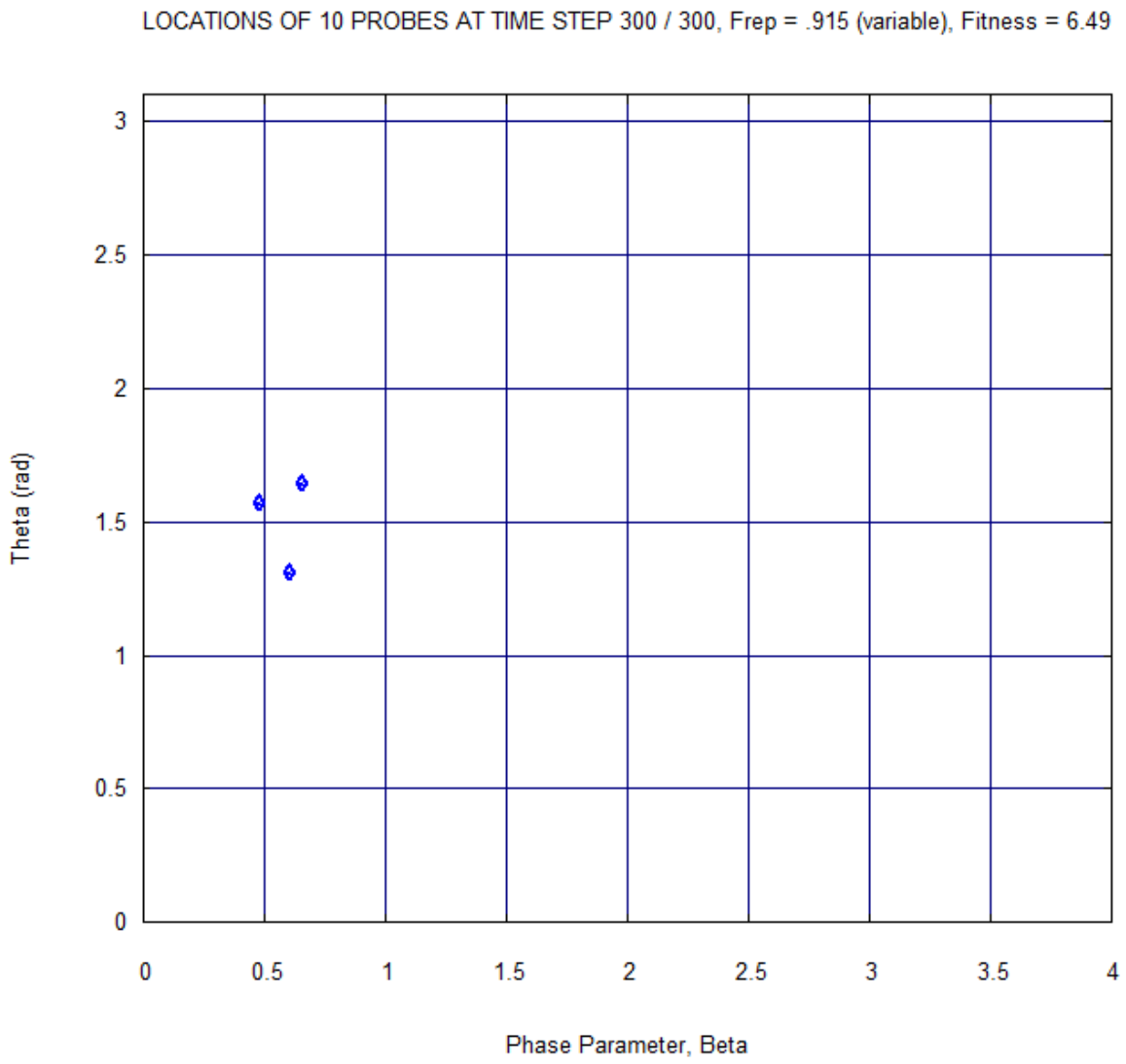

Fig. 17(b). Prob. #3 probes at step #300.

Figs. 18(a) and (b) plot the fitness and $D_{avg}$ evolution, respectively, and Table 8 lists the fitness values at time steps when they changed. As is often the case, CFO's best fitness increased very quickly through the first four steps, thereafter increasing much more slowly. The fitness saturated at step #104. The $D_{avg}$ curve is a good example of the oscillatory behavior that signals local trapping. Several distinct regions are evident in the plot, and each one presumably corresponds to a more or less stable oscillatory probe distribution. But the regions do not exhibit any obvious correlation to the plateaus in the fitness curve.

This problem provides a good example of how important the initial probe distribution can be in adequately sampling the objective function's topology. Because only one of the four global maxima was found, a second run was made with a much denser initial probe distribution, 100 probes instead of 10, deployed in a uniform 10x10 grid with $(\Delta\beta = 0.3636, \Delta\theta = \frac{\pi}{11})$, with the other run parameters unchanged. However, the run length was necessarily short, $N_t = 20$, because of excessive runtime. CFO failed to locate a global maximum. It returned a best fitness of 5.7543994 at the point $(\beta, \theta) = (1.63636, 1.68534)$ after only three steps. It appears that the initial probe distribution trapped CFO at a local maximum almost from the beginning of the run, and that only a very much longer run would determine whether or not a better fitness would evolve.

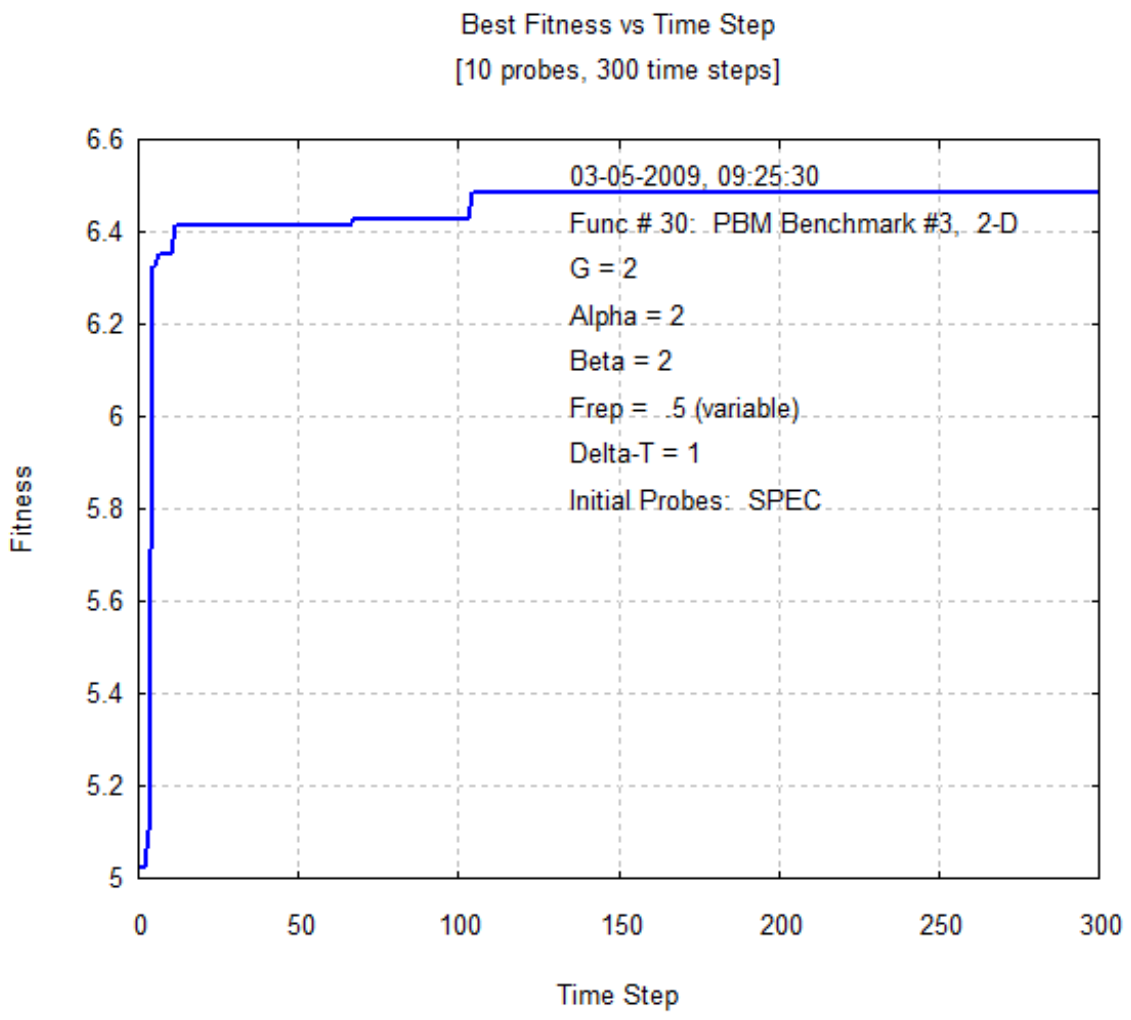

Fig. 18(a). Best fitness, prob. #3.

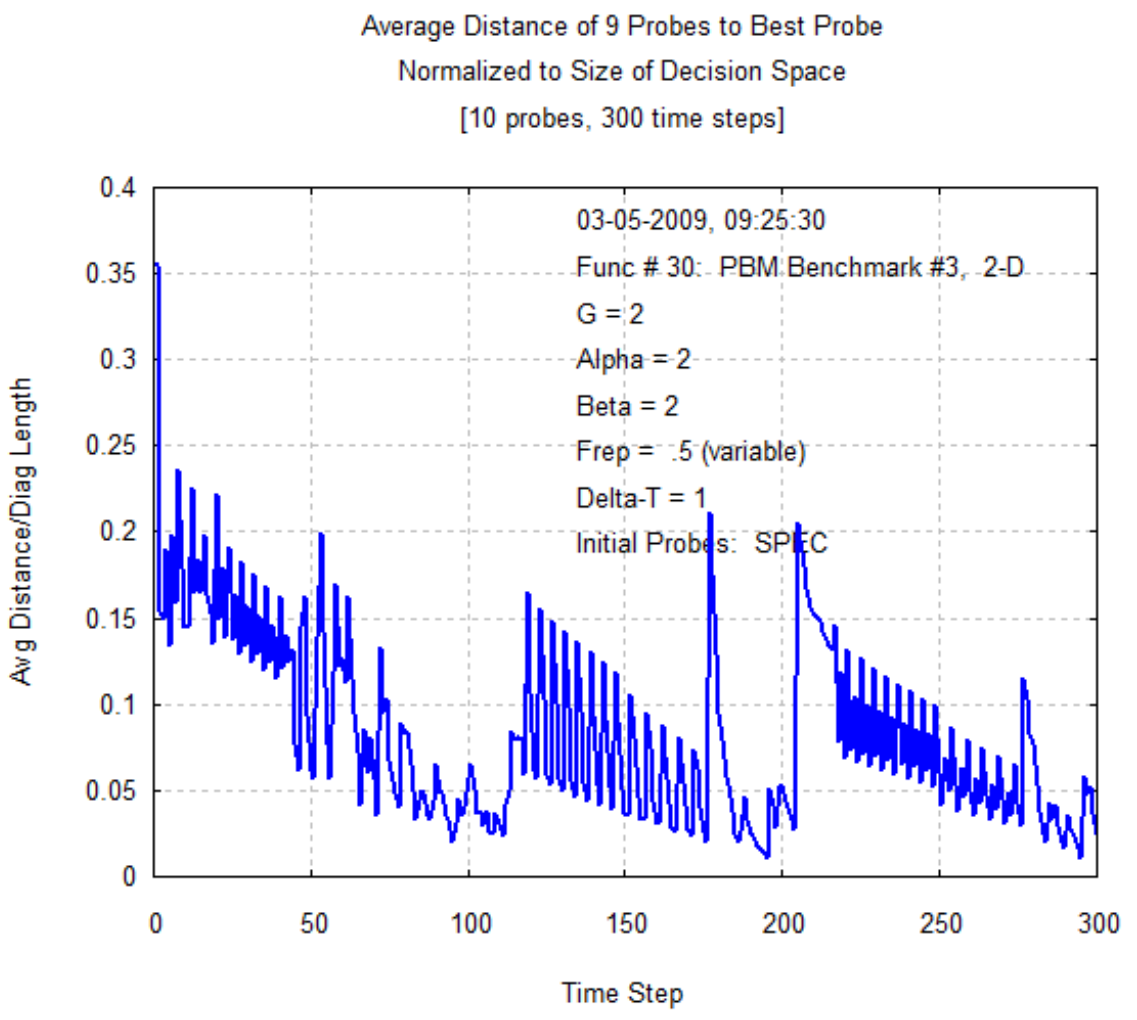

Fig. 18(b). $D_{avg}$, prob. #3.

**Table 8. Best fitnesses for benchmark #3.**

| ***Step #*** | ***Fitness*** | $N_{eval}$ |
|---|---|---|
| 0 | 5.0234259 | 10 |
| 1 | 5.0234259 | 20 |
| 3 | 5.1050500 | 40 |
| 4 | 6.3241185 | 50 |
| 6 | 6.3533093 | 70 |
| 11 | 6.4120958 | 120 |
| 67 | 6.4268772 | 680 |
| 104 | 6.4863443 | 1050 |
| 300 | 6.4863443 | 3000 |

### *6.4. Benchmark #4: Vee Dipole*

Problem #4 provides another interesting example of the importance of the initial probe distribution. Two runs were made with different distributions as shown below. Both converged on the single global maximum, but one was slightly more accurate. The first run was made with $N_t = 250$ and $N_p = 12$. The initial 3x4 probe grid and the probes at step #250 appear in Figs. 19(a) and (b), respectively. The global maximum was located at the point $(L_{arm}, \alpha) = (1.48104\lambda, 0.708983)$ with a value of 5.6885293 saturating at step #61 (see Table 9 for fitness values when they changed). The fitness evolution plot in Fig. 20(a) shows a very rapid initial increase through step #3 followed by step-wise increases through step #61. The $D_{avg}$ plot in Fig. 20(b) is quite erratic and shows not sign of the oscillation associated with trapping.

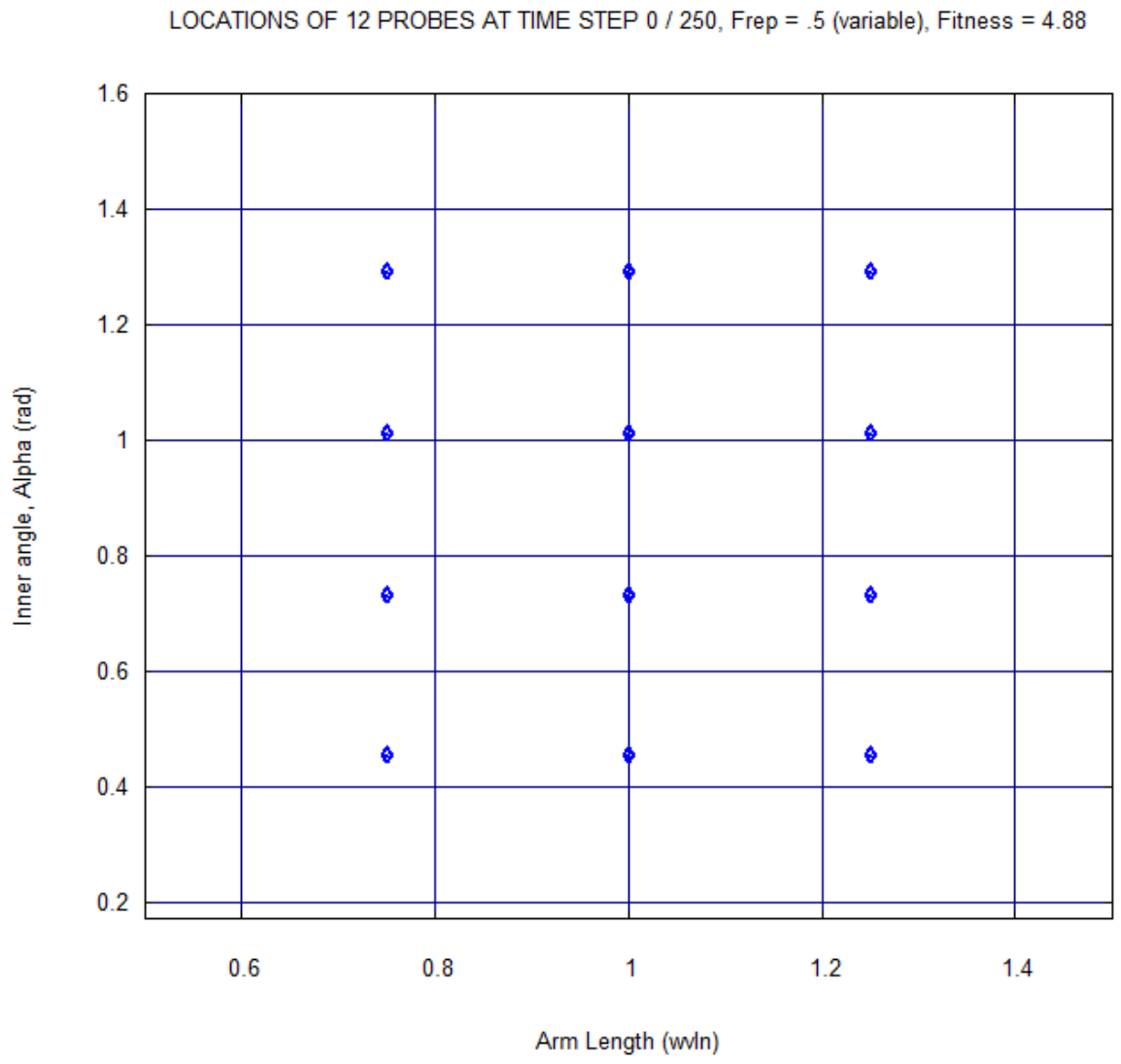

Fig. 19(a). Prob. #4 initial probes.

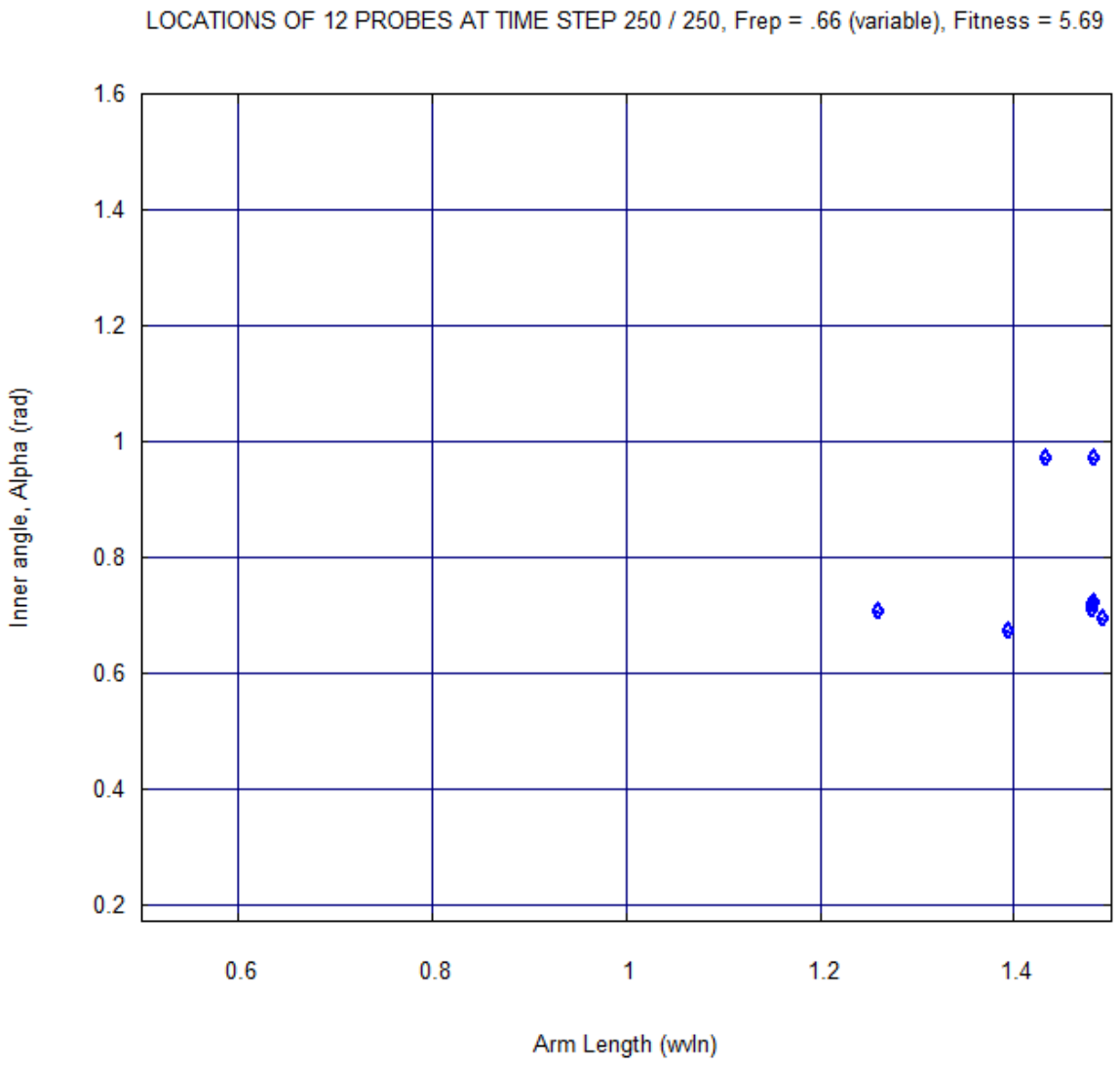

Fig. 19(b). Prob. #4 probes at step #250.

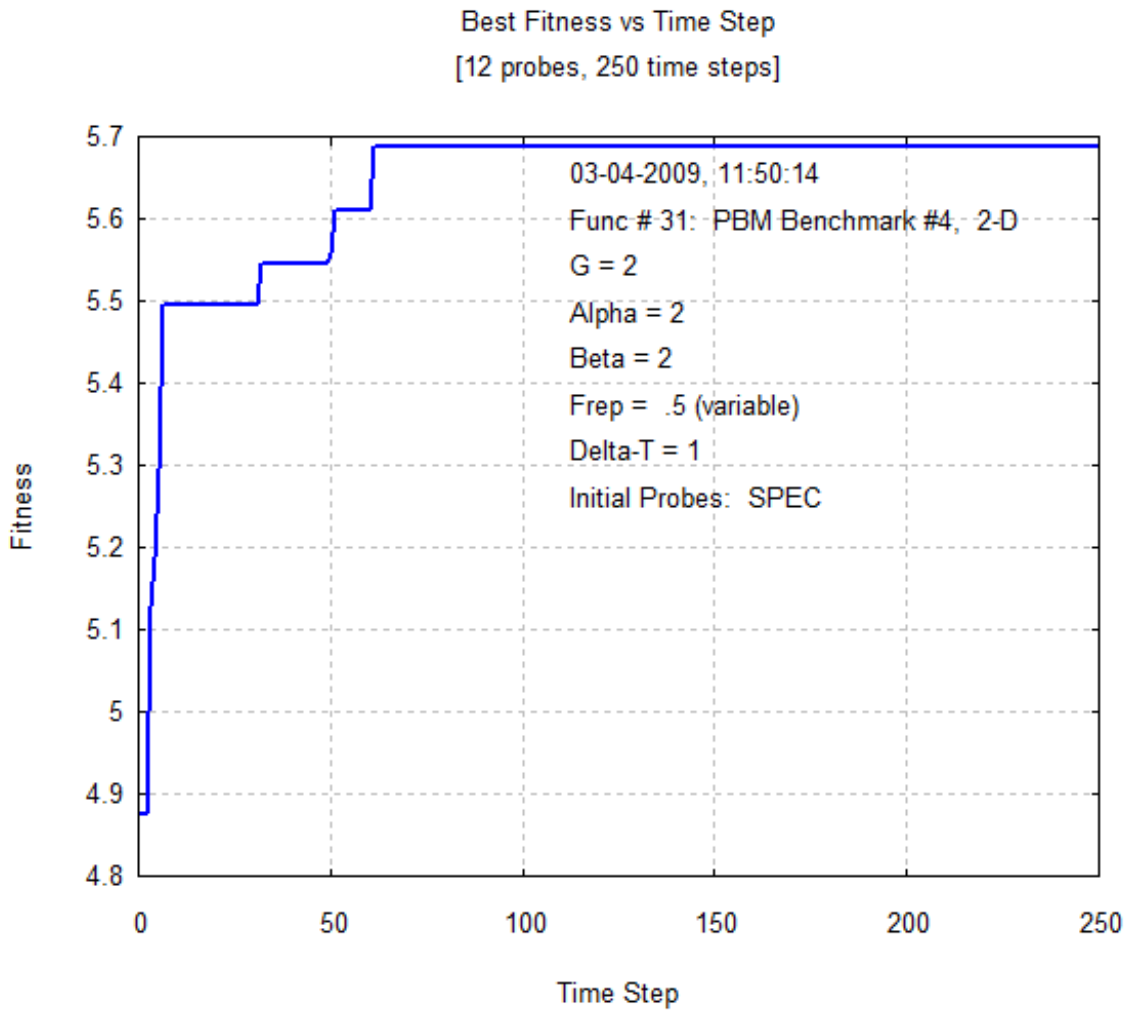

Fig. 20(a). Best fitness, prob. #4.

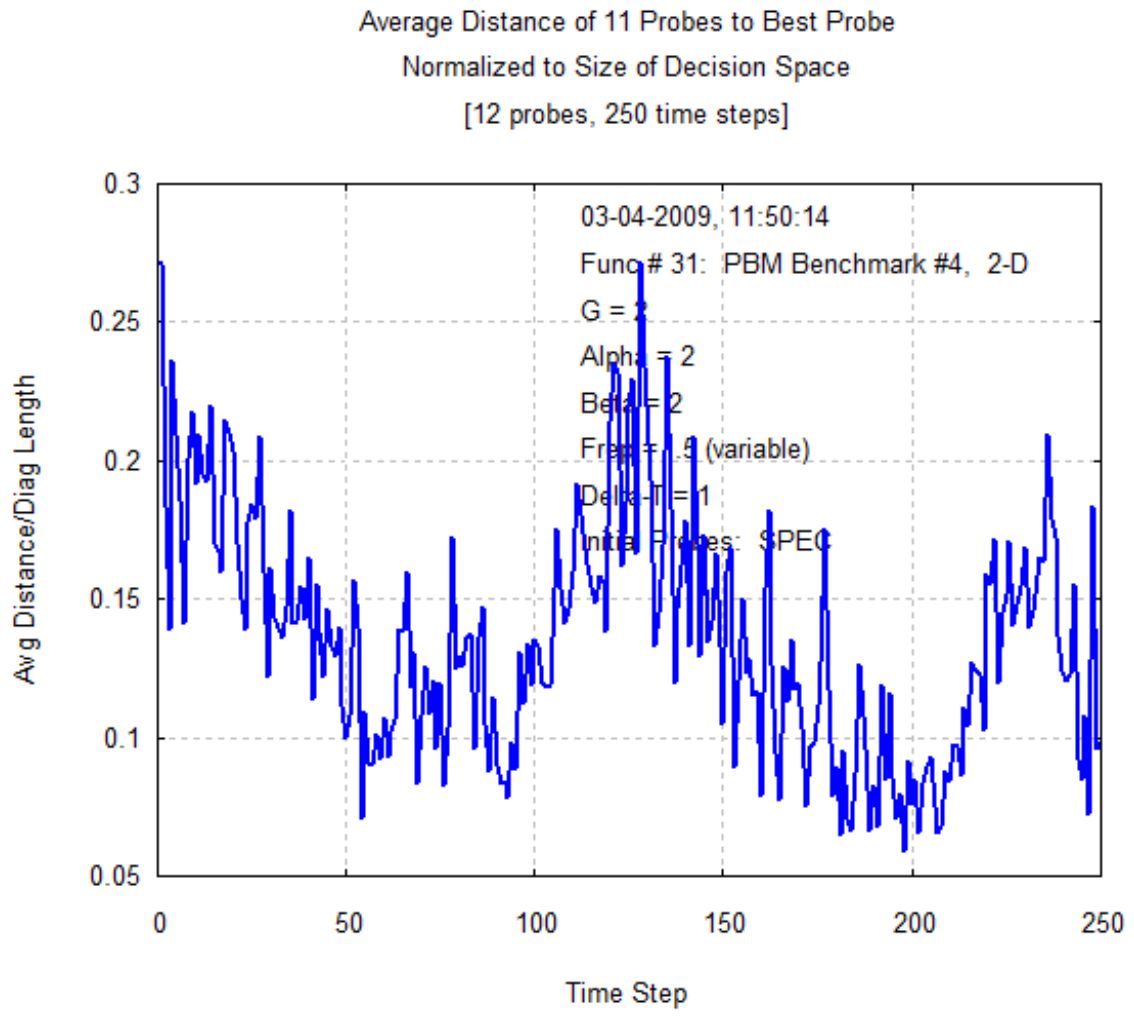

Fig. 20(b). $D_{avg}$ , prob. #4.

**Table 9 Best fitnesses for benchmark #4, run 1.**

| ***Step #*** | ***Fitness*** | $N_{eval}$ |
|---|---|---|
| 0 | 4.8752849 | 12 |
| 1 | 4.8752849 | 24 |
| 3 | 5.1286138 | 96 |
| 4 | 5.1880004 | 120 |
| 5 | 5.2966344 | 144 |
| 6 | 5.4954087 | 168 |
| 32 | 5.5462571 | 792 |
| 50 | 5.5590426 | 1224 |
| 51 | 5.6104798 | 1248 |
| 61 | 5.6885293 | 1488 |
| 250 | 5.6885293 | 6024 |

The second run was made for the same number of time steps ($N_t = 250$), but with fewer probes ($N_p = 7$). Three probes were placed on the $L_{arm}$-axis, and four on the $\alpha$-axis as shown in Fig. 21(a). The final probe distribution appears in Fig. 21(b), and, interestingly, it includes a probe "stuck" in the lower left corner at $(L_{arm}, \alpha) = (0.5\lambda, \frac{\pi}{18})$. The global maximum was somewhat greater than the previous run's at 5.7147684 located at the point $(L_{arm}, \alpha) = (1.4952\lambda, 0.710984)$. Convergence on the maximum was slower with saturation at step #164 (Table 10 lists fitness values when they changed), which seems reasonable in view of the smaller number of probes initially placed very far from the maximum. Figs. 22(a) and (b) plot the best fitness and $D_{avg}$ vs. time step. The fitness increase is very rapid through step #5, followed by a slower step-wise increase as seen in the first run. The $D_{avg}$ curve is less erratic than in the first run. Visually it seems to contain an oscillatory tendency, but the oscillation is not well-developed as in previous cases.

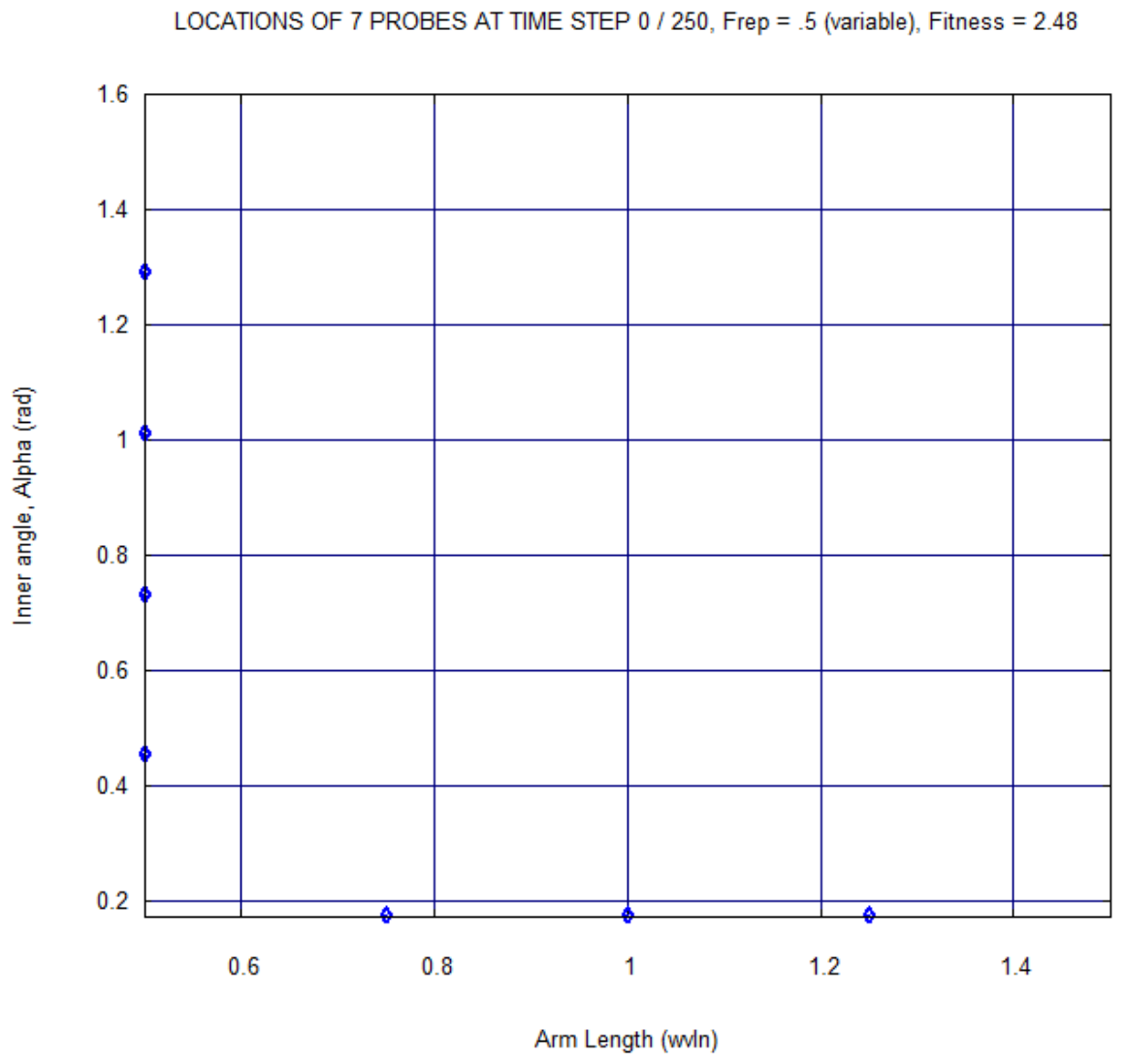

Fig. 21(a). Initial probes, run 2.

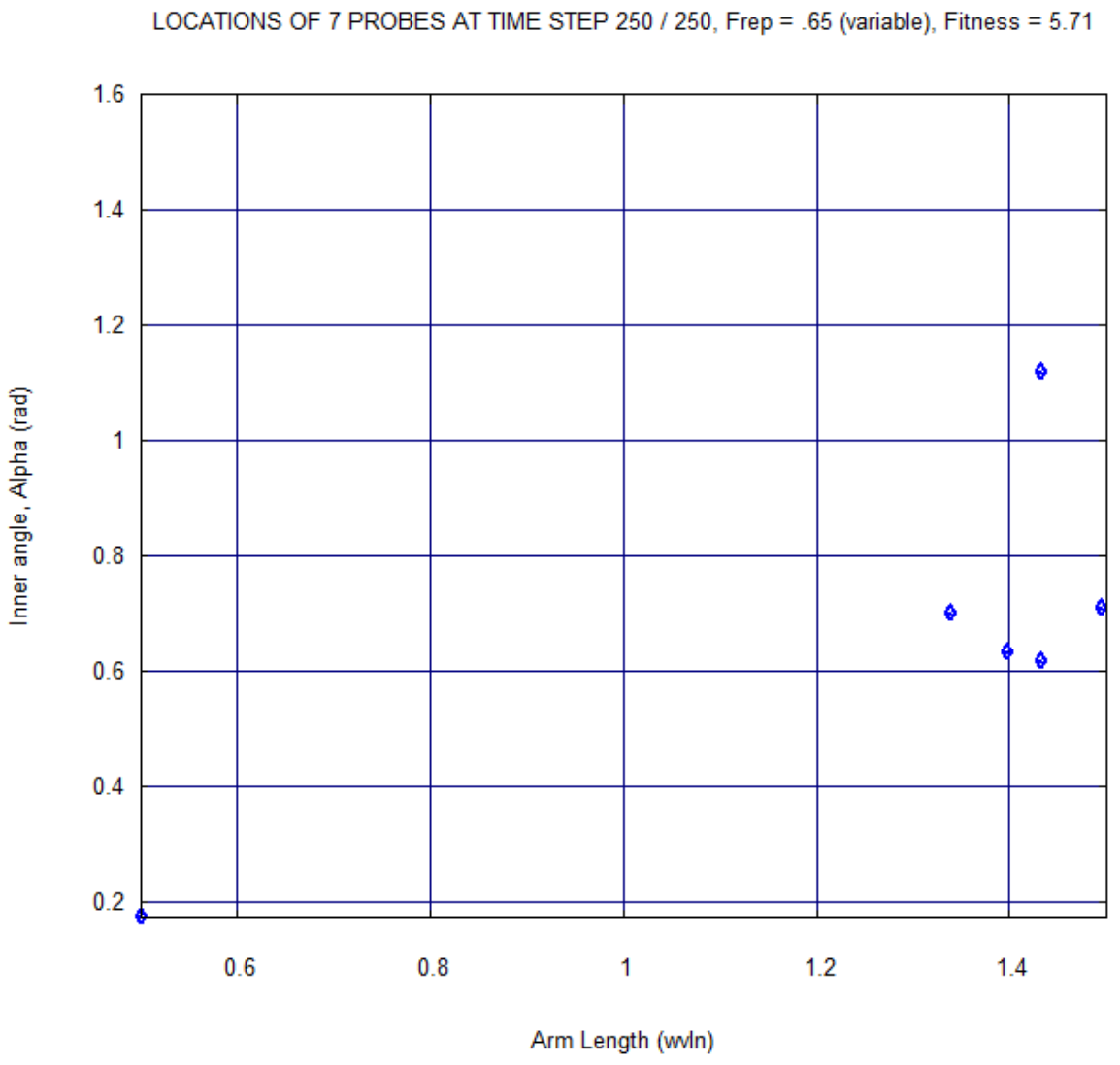

Fig. 21(b). Probes at step #250, run 2.

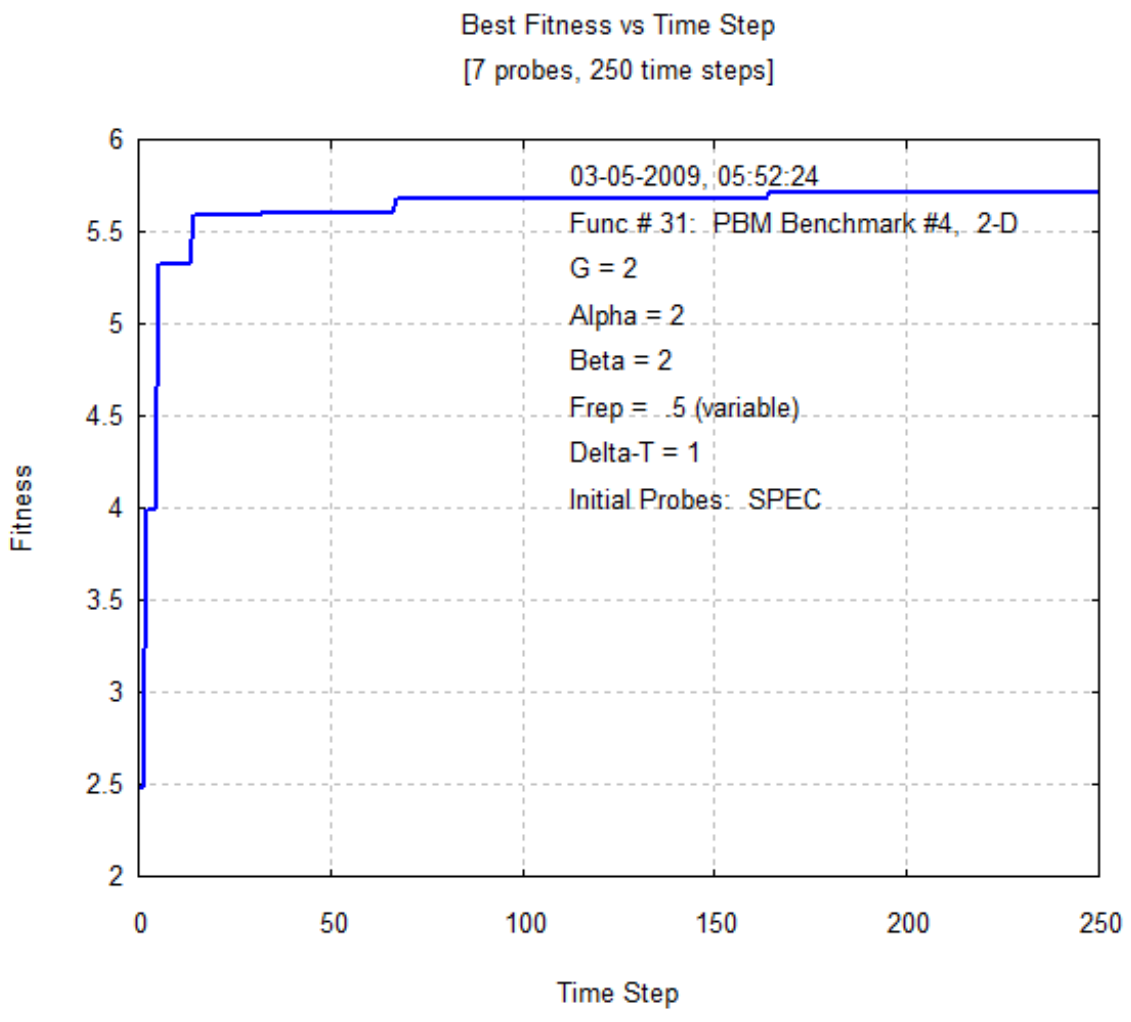

Fig. 22(a). Best fitness, run 2.

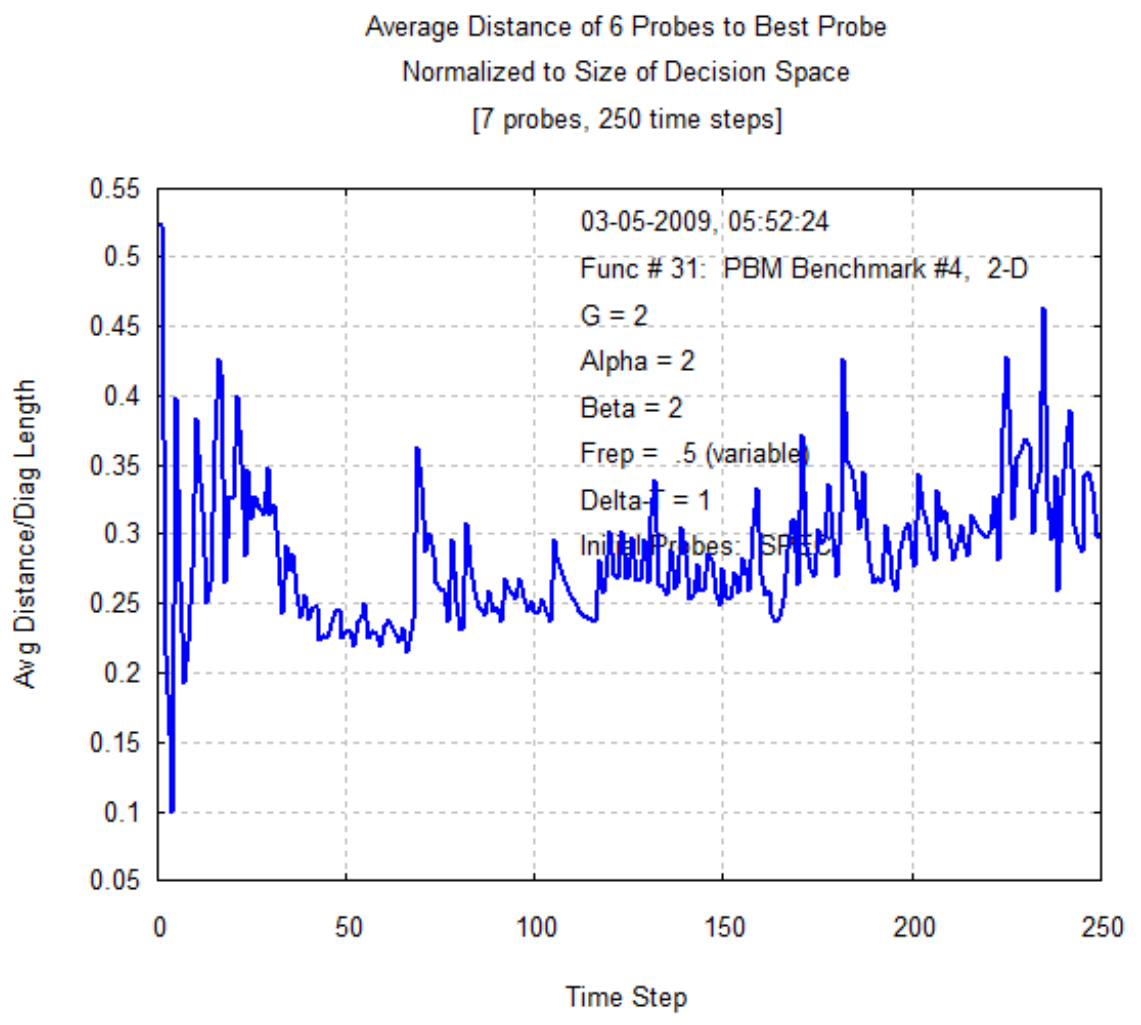

Fig. 22(b). $D_{avg}$, run 2.

**Table 10 Best fitnesses for benchmark #4, run 2.**

| *Step #* | *Fitness* | $N_{eval}$ |
|---|---|---|
| 0 | 2.4831331 | 7 |
| 1 | 2.4831331 | 14 |
| 2 | 3.9902490 | 21 |
| 5 | 5.3210826 | 42 |
| 14 | 5.5847019 | 105 |
| 32 | 5.5975760 | 231 |
| 67 | 5.6754461 | 476 |
| 164 | 5.7147864 | 1155 |
| 250 | 5.7147864 | 1757 |

### *6.5. Benchmark #5: N-element Collinear Dipole Array*

Benchmark #5 is an $N_{el}$-element array of collinear $\frac{\lambda}{2}$ dipoles. The problem is run with several array sizes, from 6 to 24 elements following [6], in order to create decision spaces of high dimensionality, $N_d = N_{el} - 1$. As discussed in [6], maximum directivity occurs when the dipole center-to-center spacing is $d_i = 0.99\lambda, i = 1,...,N_d$, $0.5\lambda \le d_i \le 1.5\lambda$, that is, all dipoles are uniformly separated by just less than one wavelength. The objective of this benchmark is to recover that spacing over a range of array sizes. CFO runs were made with $N_p = 2N_d$ initial probes on the principal diagonal at $\vec{R}_0^p = \sum_{i=1}^{N_d} (0.5 + \frac{p}{3}) \hat{e}_i$, $1 \le p \le N_p$ (note that "axis" in Figs. 23-29 refers to the diagonal, not coordinate axes).

CFO's performance is summarized in Table 11. In each case except $N_{el}$ = 24, 30, CFO recovers the PBM maximum as expected. The returned values of $d_i$ are all the same and range from a minimum of 0.983095 λ to a maximum of 0.996288λ. For $N_{el}$ = 24, however, CFO returns an optimum spacing of precisely $1\lambda$, and an explanation of this result is not apparent. The step at which fitness saturation takes place appears in column 5, and in every case it is extremely rapid, never more than 6 steps.

**Table 11. Best fitness coordinates and values for benchmark #5.**

| *# Dipoles* $N_{el}$ | $N_d$ | $N_p$ | $N_t$ | *Step # for Saturation* | $N_{eval}$ | $d_i$ $(i = 1,...,N_d)$ | $D_{\max}$ |
|---|---|---|---|---|---|---|---|
| 6 | 5 | 10 | 100 | 6 | 70 | 0.991050λ | 11.2202 |
| 7 | 6 | 12 | 10 | 5 | 72 | 0.983095λ | 13.1826 |
| 10 | 9 | 18 | 50 | 5 | 108 | 0.994210λ | 19.0985 |
| 13 | 12 | 24 | 16 | 5 | 144 | 0.996288λ | 25.0611 |
| 16 | 15 | 30 | 30 | 3 | 120 | 0.989583λ | 30.9742 |
| 24 | 23 | 46 | 10 | 3[(4)] | 184 | 1.000000λ | 46.8813 |
| 30 | 29 | 58 | 10 | 4[(4)] | 290 | 1.000000λ | 58.8844 |

Notes:
1. For all runs: probes on-diagonal, 2 probes per dimension.
2. Same value was returned for all coordinates of global maximum
3. 7 & 13 element arrays are cases 5a & 5b in [6, Table II].
4. Best fitness decreases at step #6

Figs. 23 through 29 plot the evolution of the best fitnesses [(a) plots] and $D_{avg}$ [(b) plots] for each of the arrays in Table 11. The results are consistent with other typical CFO results. In each case except $N_{el}=24$, the fitness increases very quickly followed by saturation and a plateau. The $D_{avg}$ curves exhibit correspondingly rapid decreases followed by essentially flat plateaus indicating a stable, tight probe distribution. There is no sign of the oscillation that indicates local trapping. The fitness curve for the $N_{el}=24$ array [Fig. 28(a)] is somewhat unusual because the best fitness plateaus at its maximum between step #3 and #5, decreases at step #6, and then begins to gradually increase. This behavior is not understood. While this behavior appears to be a rare occurrence, a similar precipitous drop in fitness was seen in the 2D Goldstein-Price function in [4]. But in that case it was accompanied by a corresponding increase in $D_{avg}$, while none is seen here. In order to rule of the possibility of an artifact in the 24-element case, another run was made with 30-elements (not one of the PBM problems) as shown in Table 11. The returned optimal dipole spacing was the same as for the 24-element array, *viz.*, $1\lambda$ (all 29 coordinates). The best fitness and $D_{avg}$ curves (Fig. 29) are similar. The fitness also decreases at step 6, but the drop is not as great. This case suggests that there is not artifact peculiar to the 24-element array, and that the results in that case are correct.

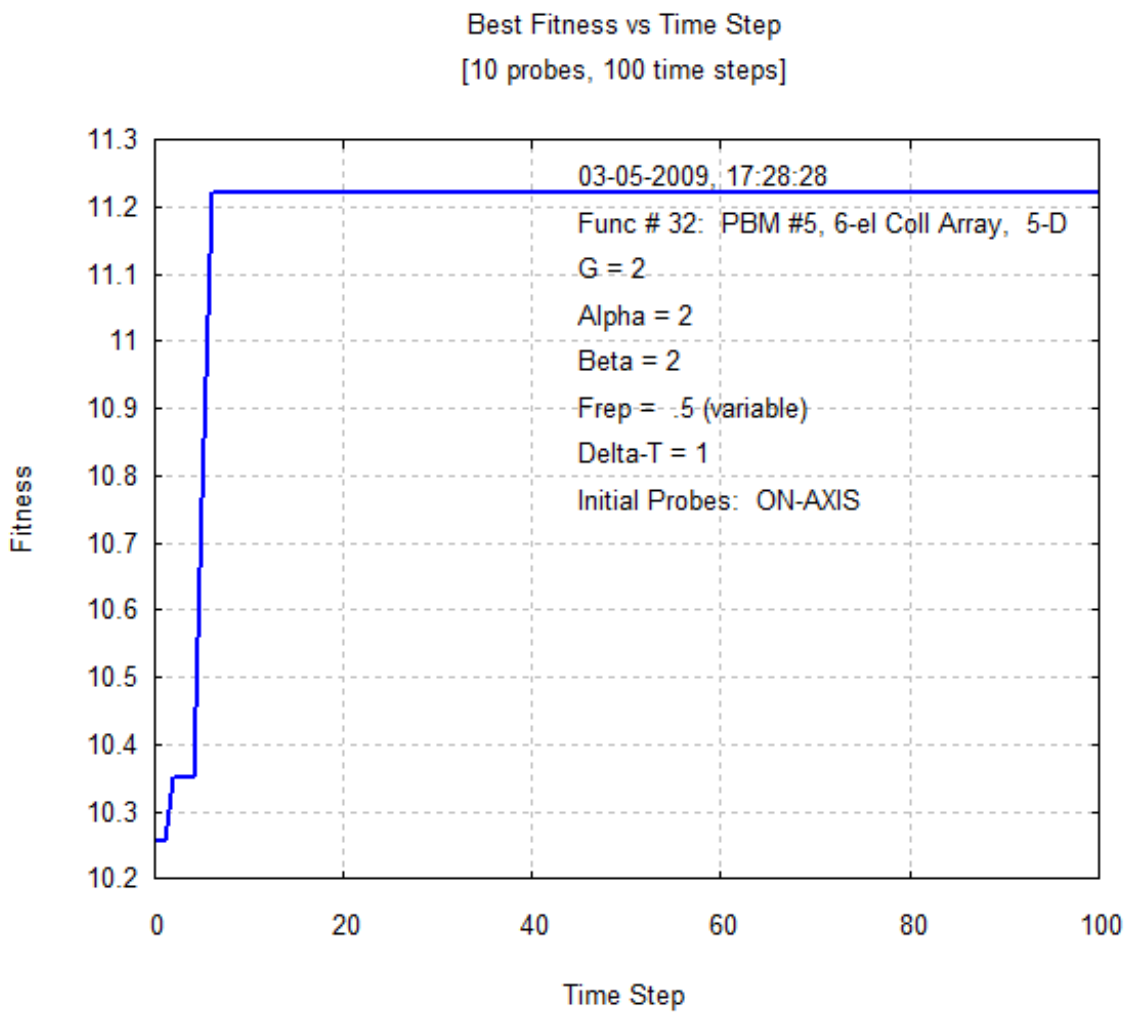

Fig. 23(a). Best fitness, 6-el array.

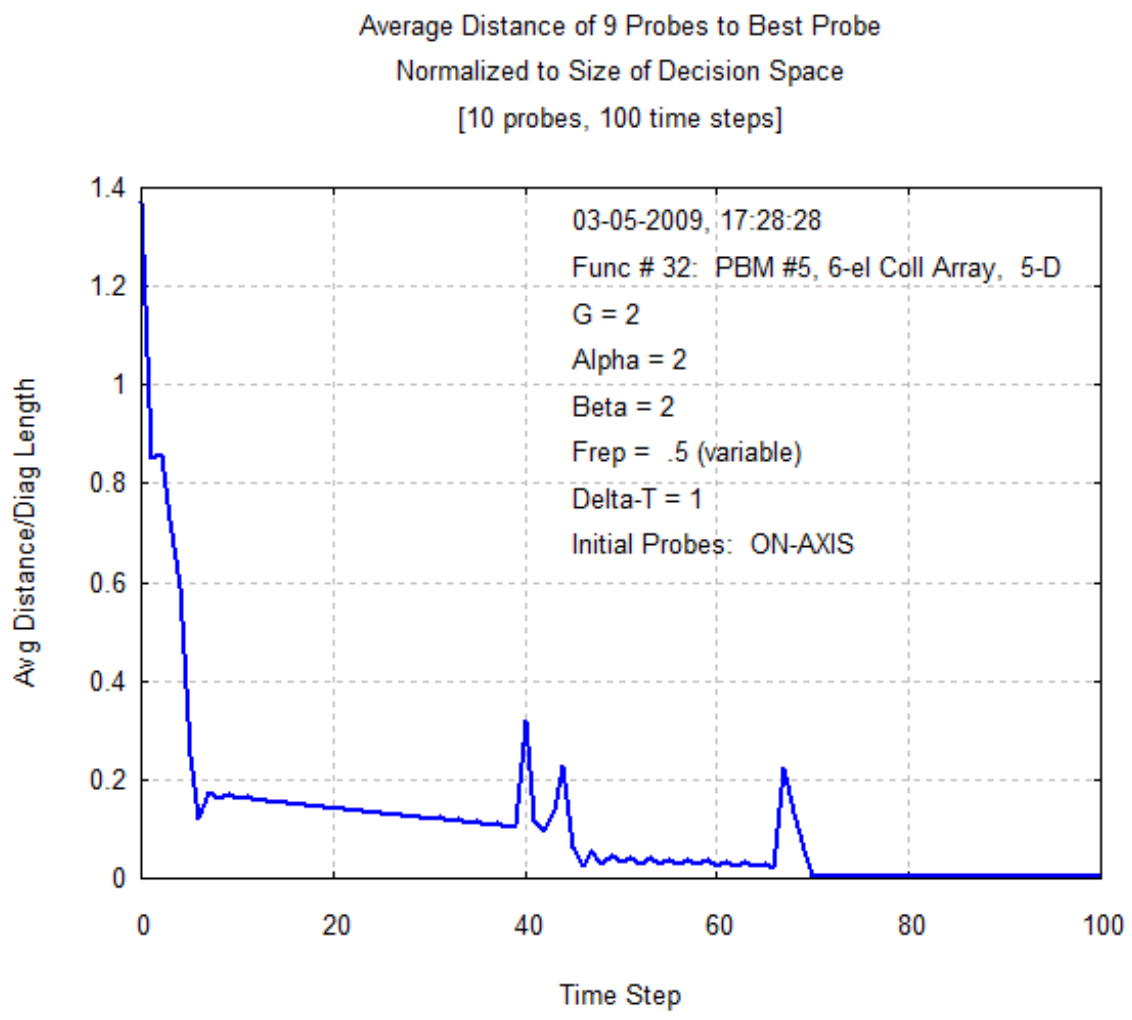

Fig. 23(b). $D_{avg}$, 6-el array.

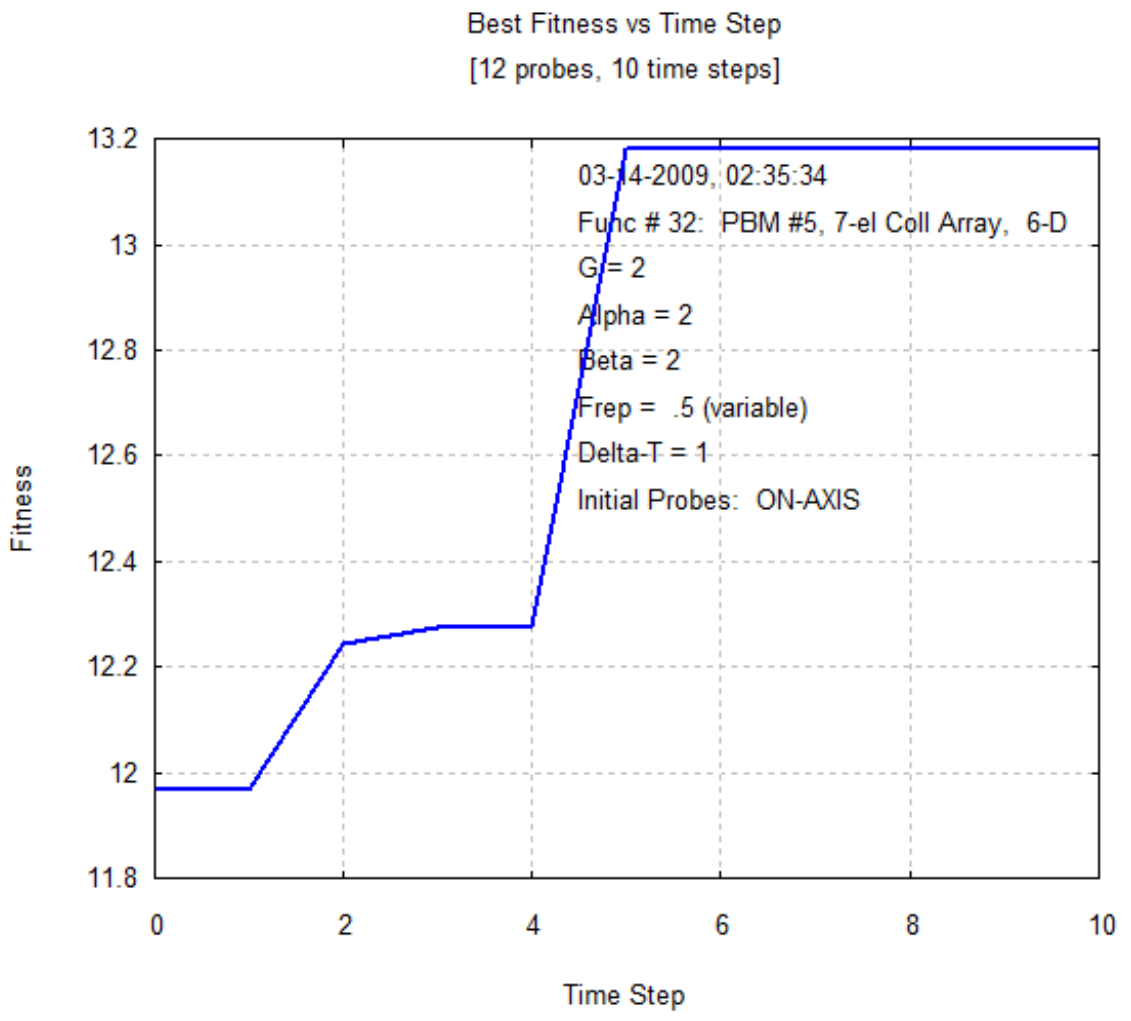

Fig. 24(a). Best fitness, 7-el array.

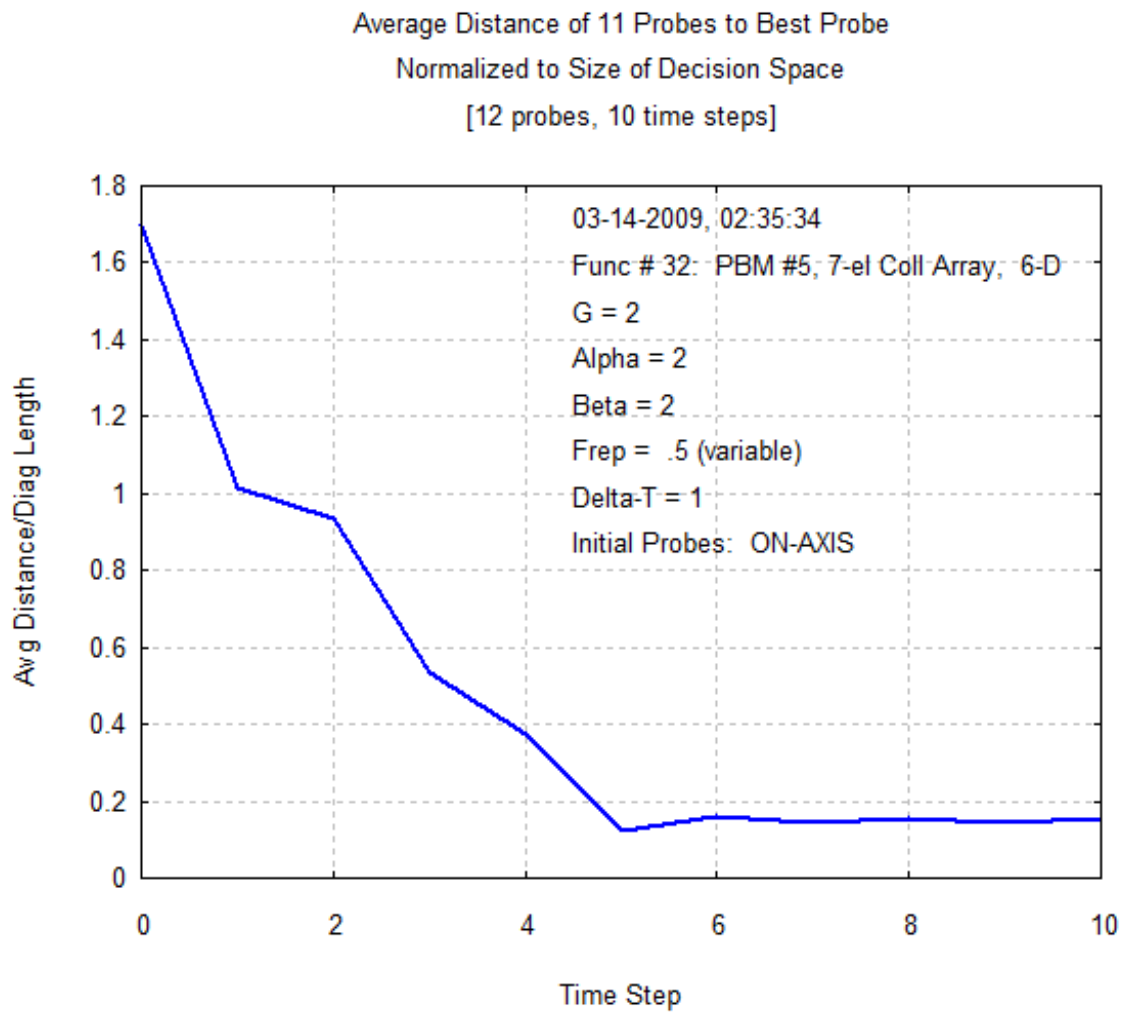

Fig. 24(b). $D_{avg}$, 7-el array.

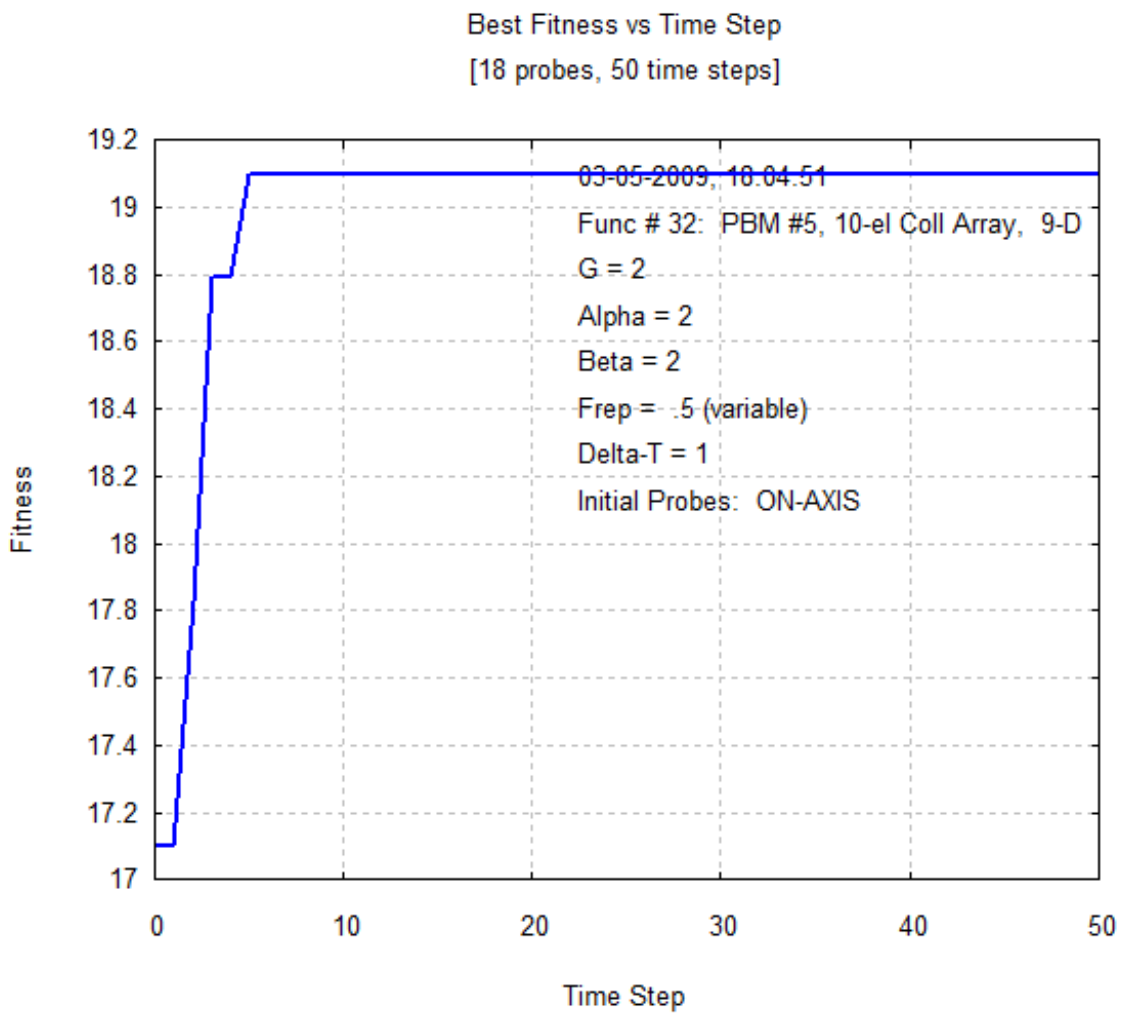

Fig. 25(a). Best fitness, 10-el array.

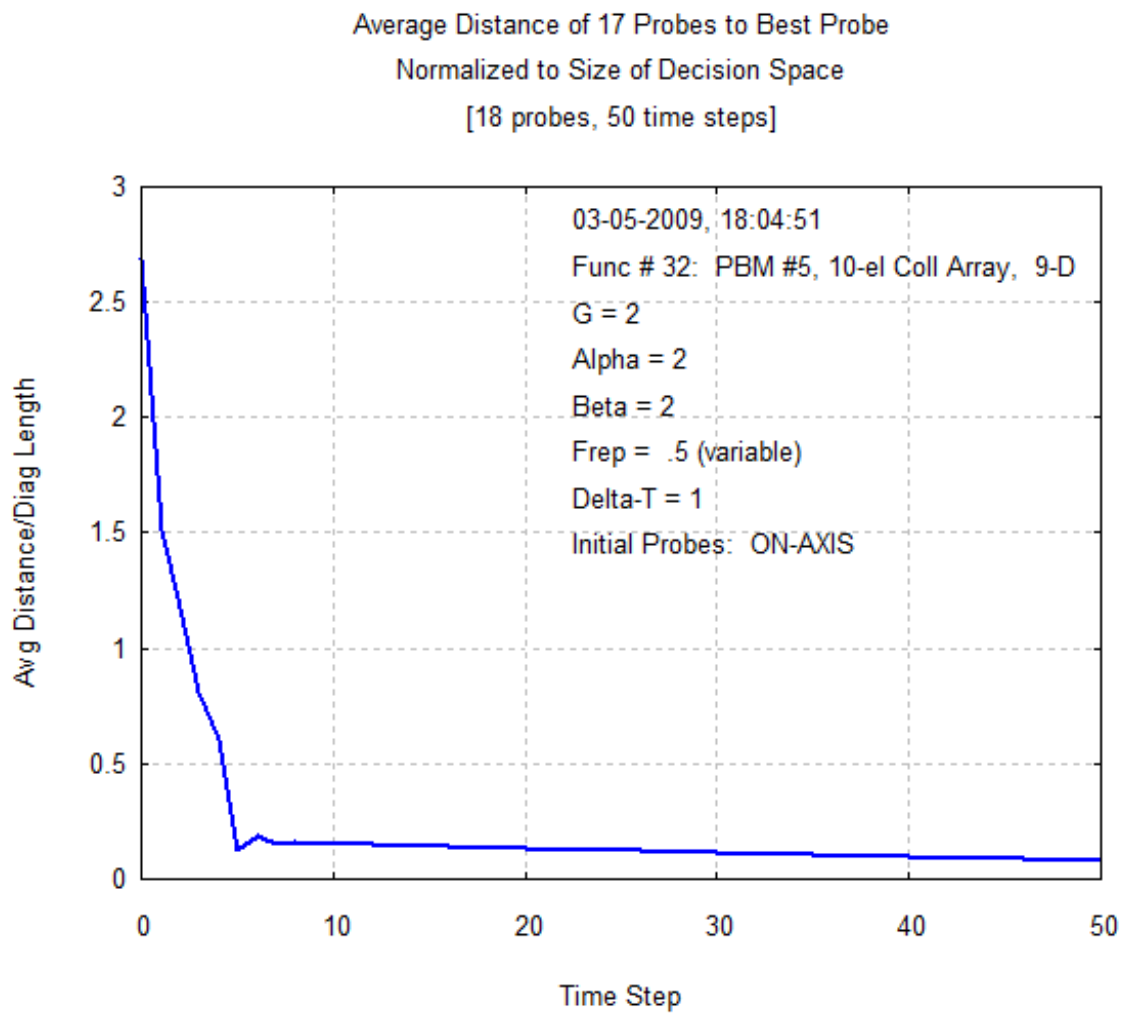

Fig. 25(b). $D_{avg}$ , 10-el array.

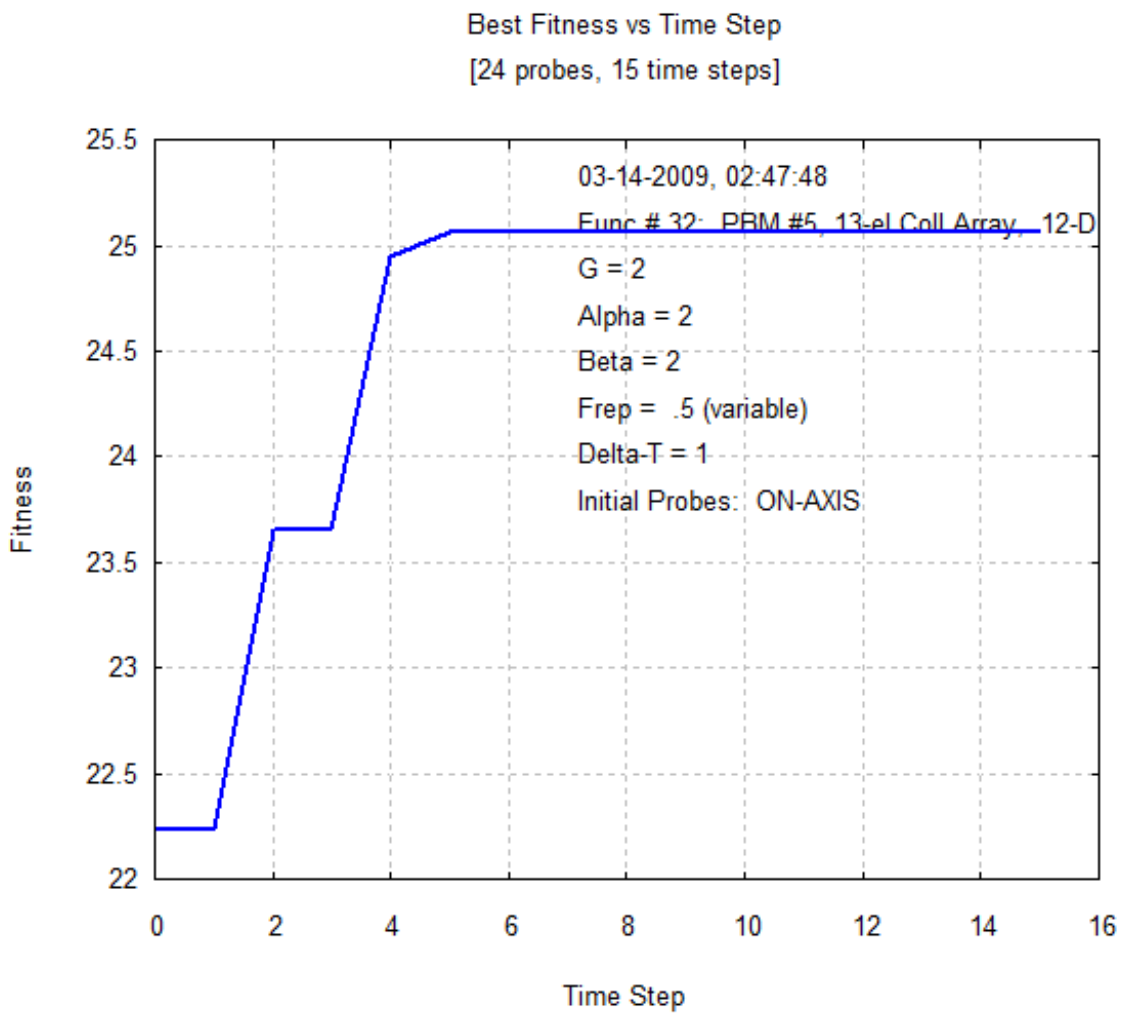

Fig. 26(a). Best fitness, 13-el array.

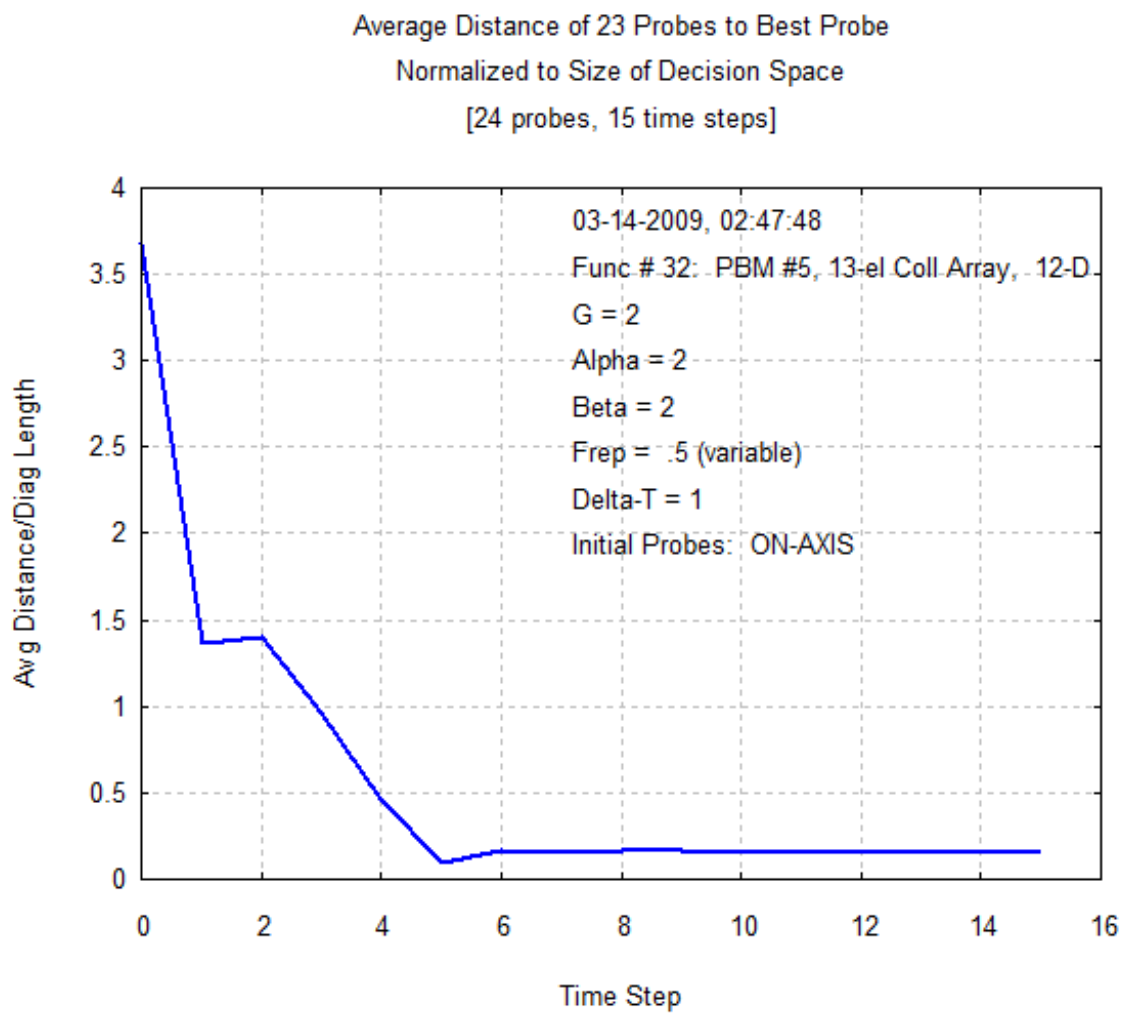

Fig. 26(b). $D_{avg}$ , 13-el array.

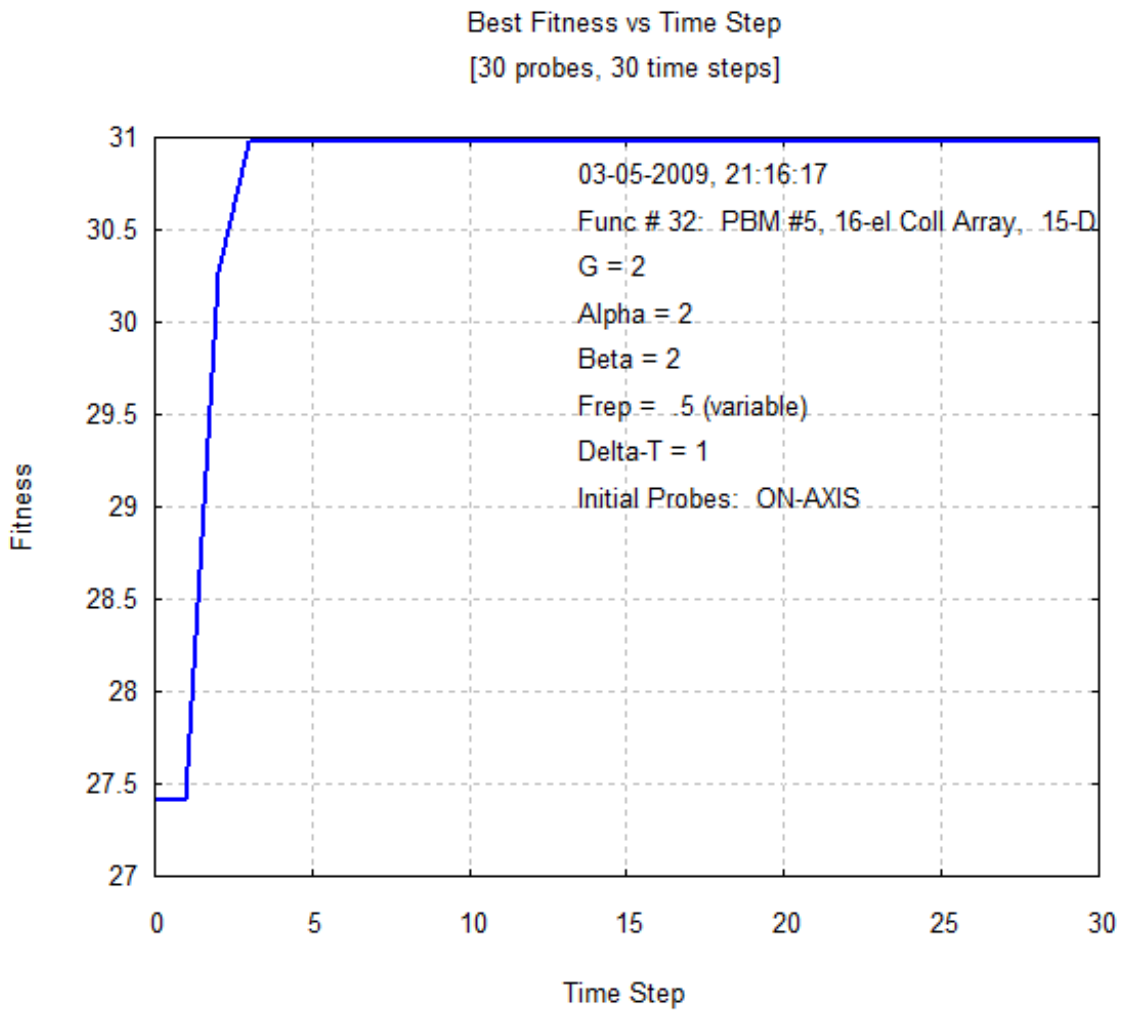

Fig. 27(a). Best fitness, 16-el array.

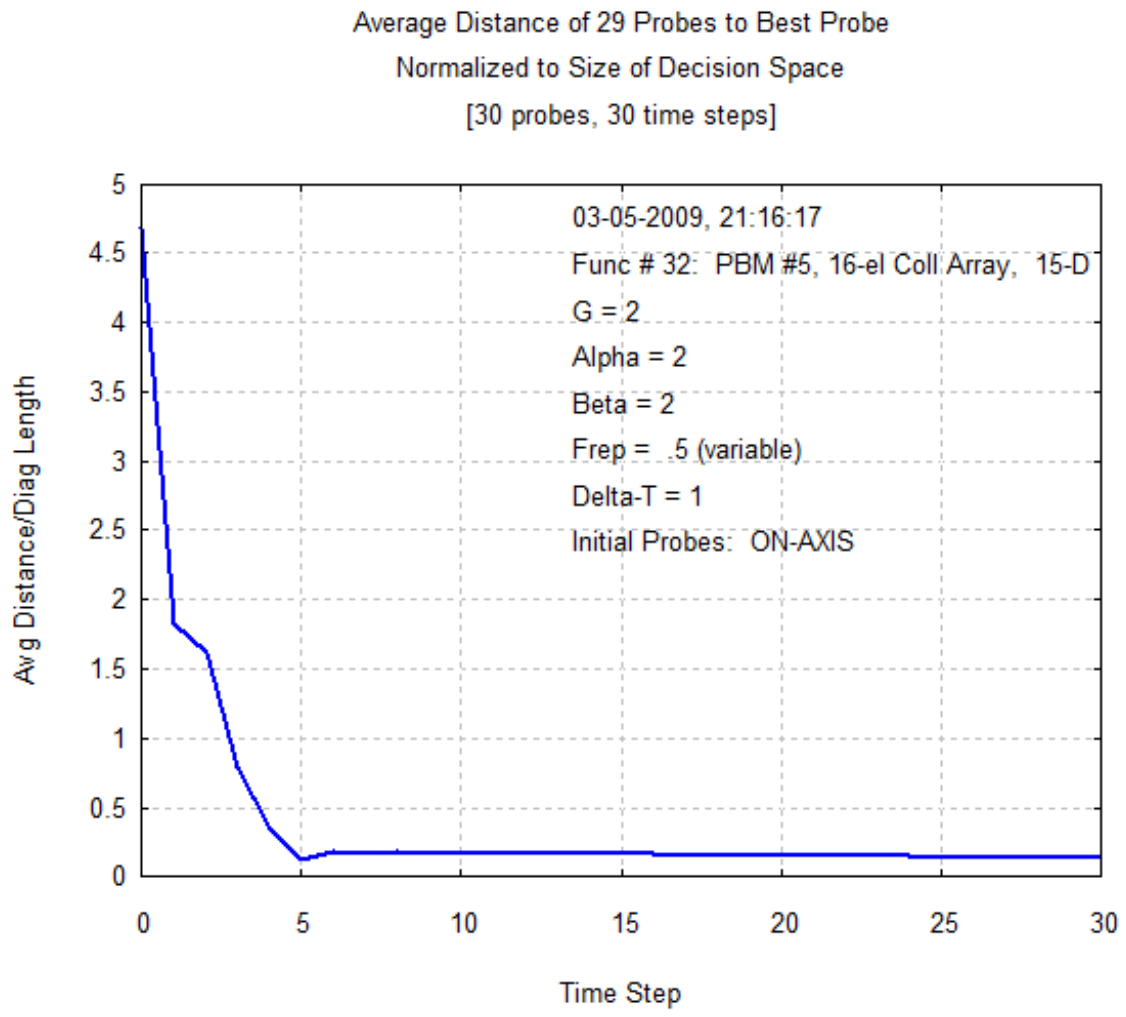

Fig. 27(b). $D_{avg}$ , 16-el array.

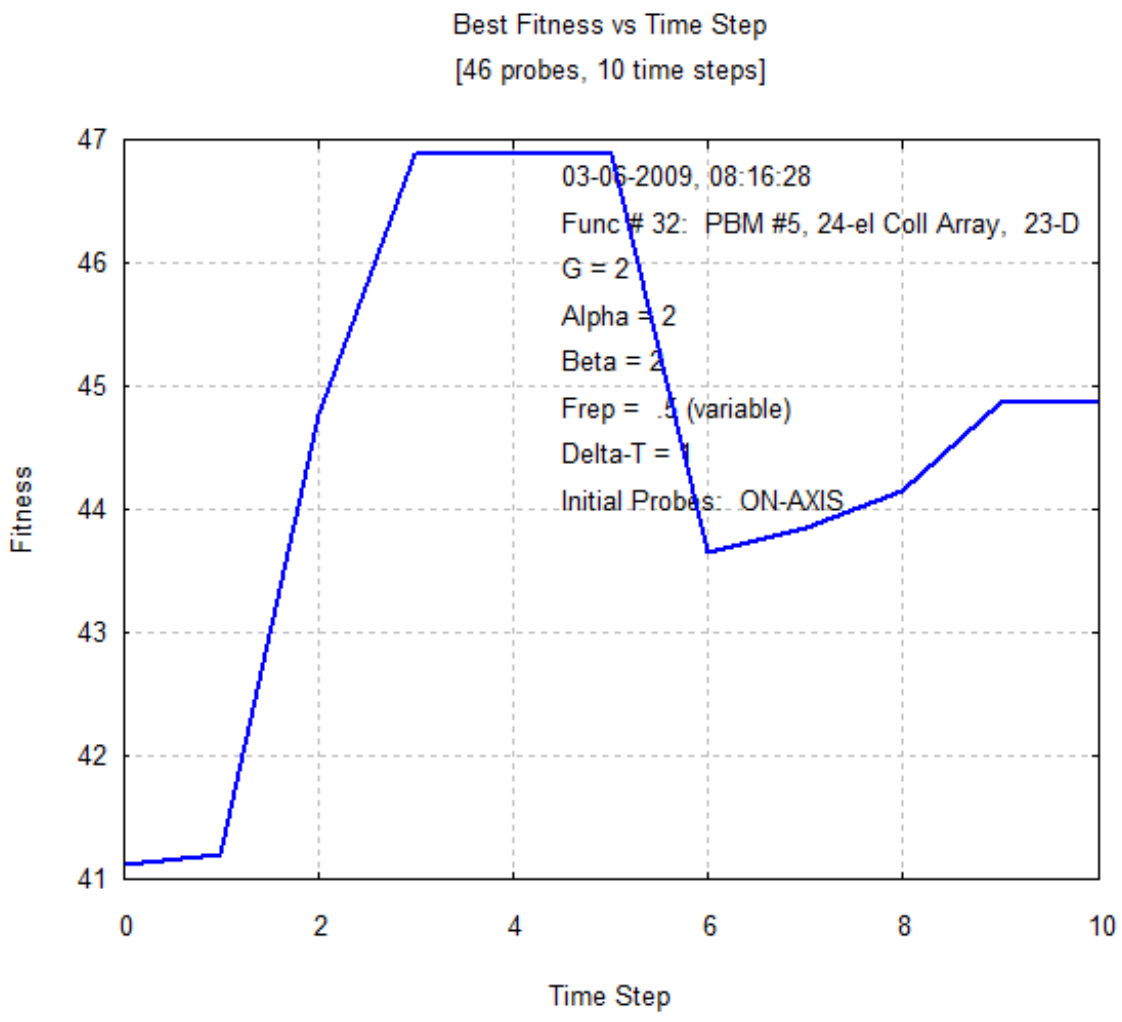

Fig. 28(a). Best fitness, 24-el array.

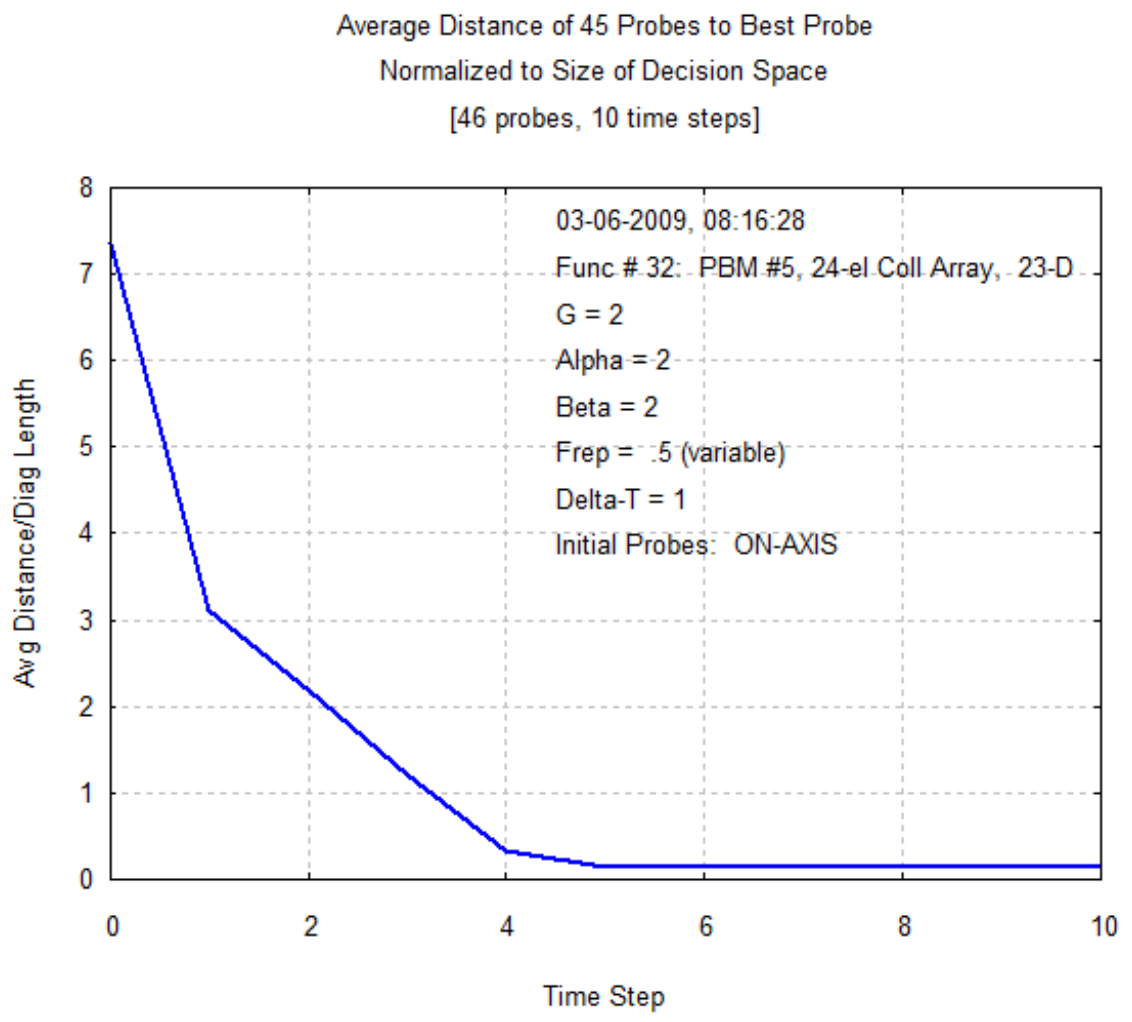

Fig. 28(b). $D_{avg}$ , 24-el array.

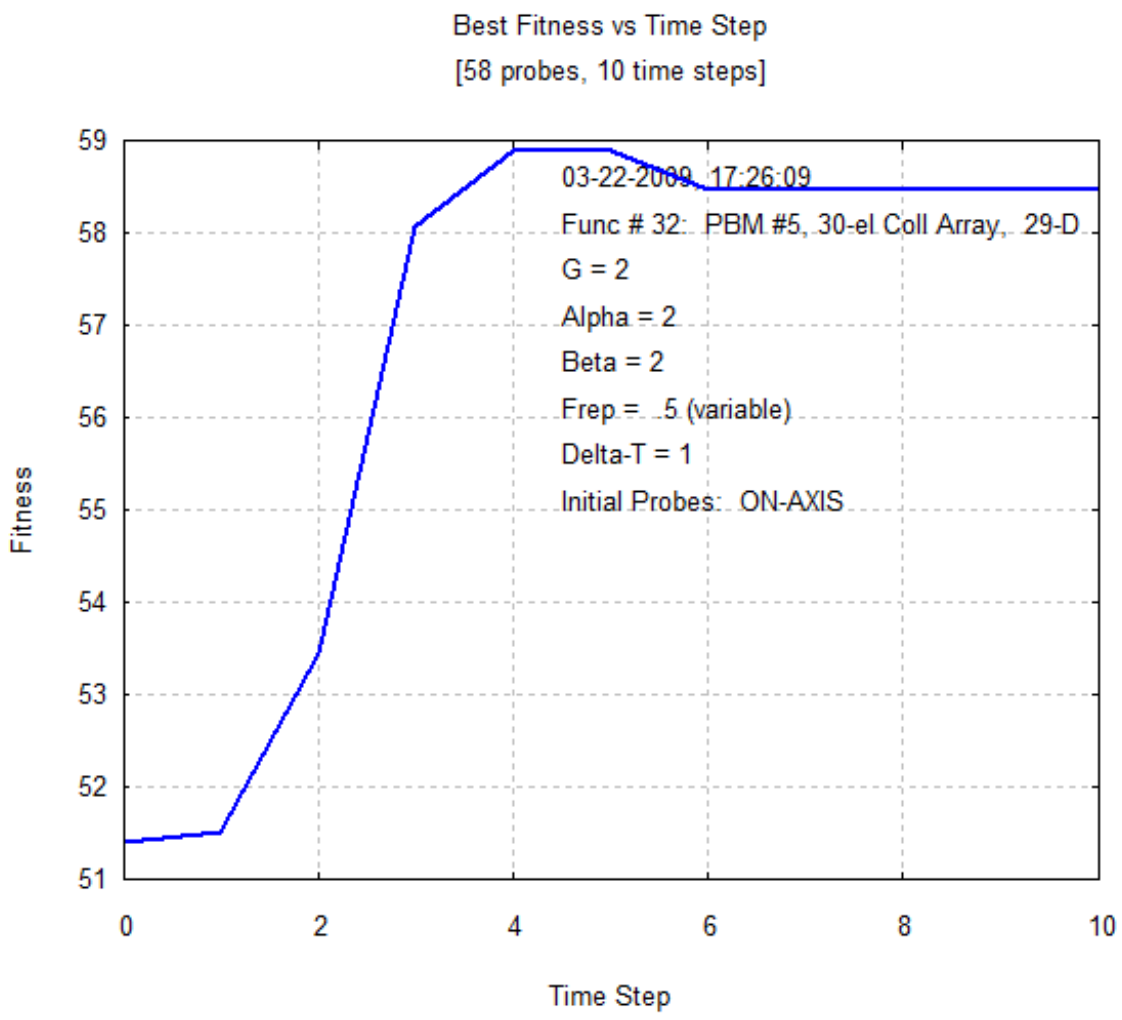

Fig. 29(a). Best fitness, 30-el array.

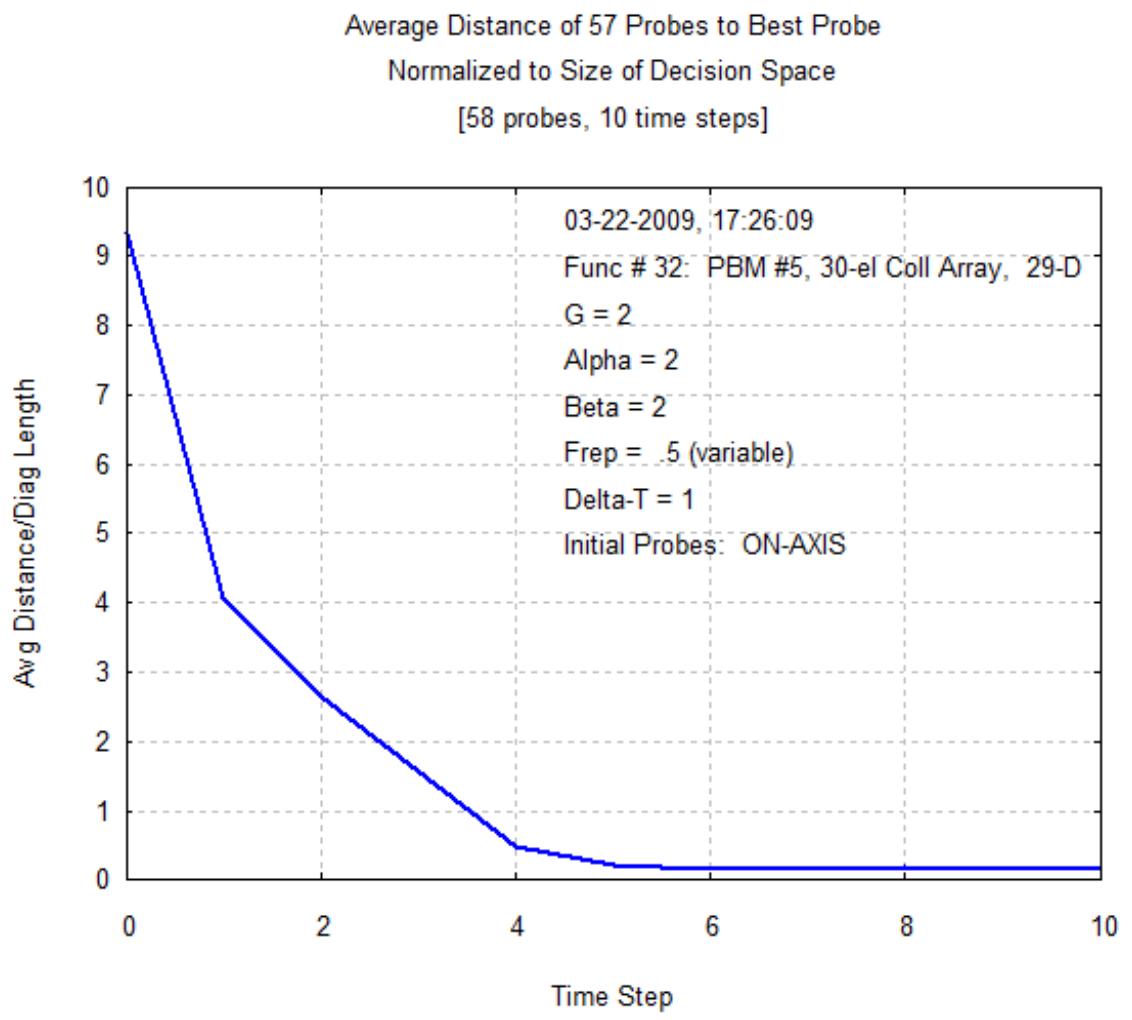

Fig. 29(b). $D_{avg}$ , 30-el array.

## 7. CONCLUSION

This paper describes Central Force Optimization as a new, deterministic, Nature-inspired search and optimization metaheuristic that embraces the metaphor of gravitational kinematics. It reports the results of testing CFO against the PBM suite of antenna benchmarks. CFO's performs quite well, in many respects better than the more established algorithms analyzed in PBM. CFO located all global maxima, even in a noisy decision landscape, and it did so with excellent efficiency. CFO generally performed better than the algorithms reported by PBM. This paper also highlights the striking similarity between velocity curves for Near Earth Objects in close encounters with planet Earth and CFO's average distance curves in the presence of local trapping. The nature of these plots suggests that oscillation is a sufficient but not necessary condition for local trapping, and that the recently developed theory of gravitational "resonant returns" might be of use in better understanding and perhaps mitigating trapping. The concept of "energy" in the gravitational metaphor was discussed, and suggestions made for its use as a test for local trapping and possibly a trapping mitigation mechanism. Also discussed was the possibility that CFO could be re-formulated on the basis of a "total energy" model. Finally, this paper suggests the possibility of a new mathematical construct, a hyperspace "directional derivative" based on the Unit Step function that always points towards a function's maxima.

## ACKNOWLEDGMENT

The author wishes to acknowledge and express his gratitude to the following individuals who in various ways have helped bring CFO to the attention of the optimization community. If CFO realizes what appears to be its considerable potential, in no small measure it will be due to their interest and encouragement: Professor Mario Pavone, Department of Mathematics, University of Catania, Sicily, Italy; Professor Zhihua Cui, Taiyuan University of Science and Technology, Division of System Simulation and Computer Application, Taiyuan, Shanxi, China; Professor Kusum Deep, Department of Mathematics, Indian Institute of Technology, Roorkee, India; Professor Nihad Dib, Department of Electrical Engineering, Jordan University of Science and Technology, Irbid, Jordan. If, on the other hand, CFO fails to meet this expectation, then the responsibility is entirely the author's.

***25 March 2009, Brewster, Massachusetts; Revised 16 December 2009.***

## APPENDIX A. ADDITIONAL BENCHMARK PLOTS & SAMPLE NEC4 FILES

This Appendix includes additional topology plots for each of the 2D benchmarks. In some cases additional perspective views are included, and in all cases projections of the objective function's surface onto the decision space principal planes are included. Sample NEC4 input and output files using the maxima coordinates reported in or estimated from [6] are included after the plots. To save space, the lengthy NEC4 output files are truncated to show only relevant data.

### A.1 <u>Benchmark #1</u>

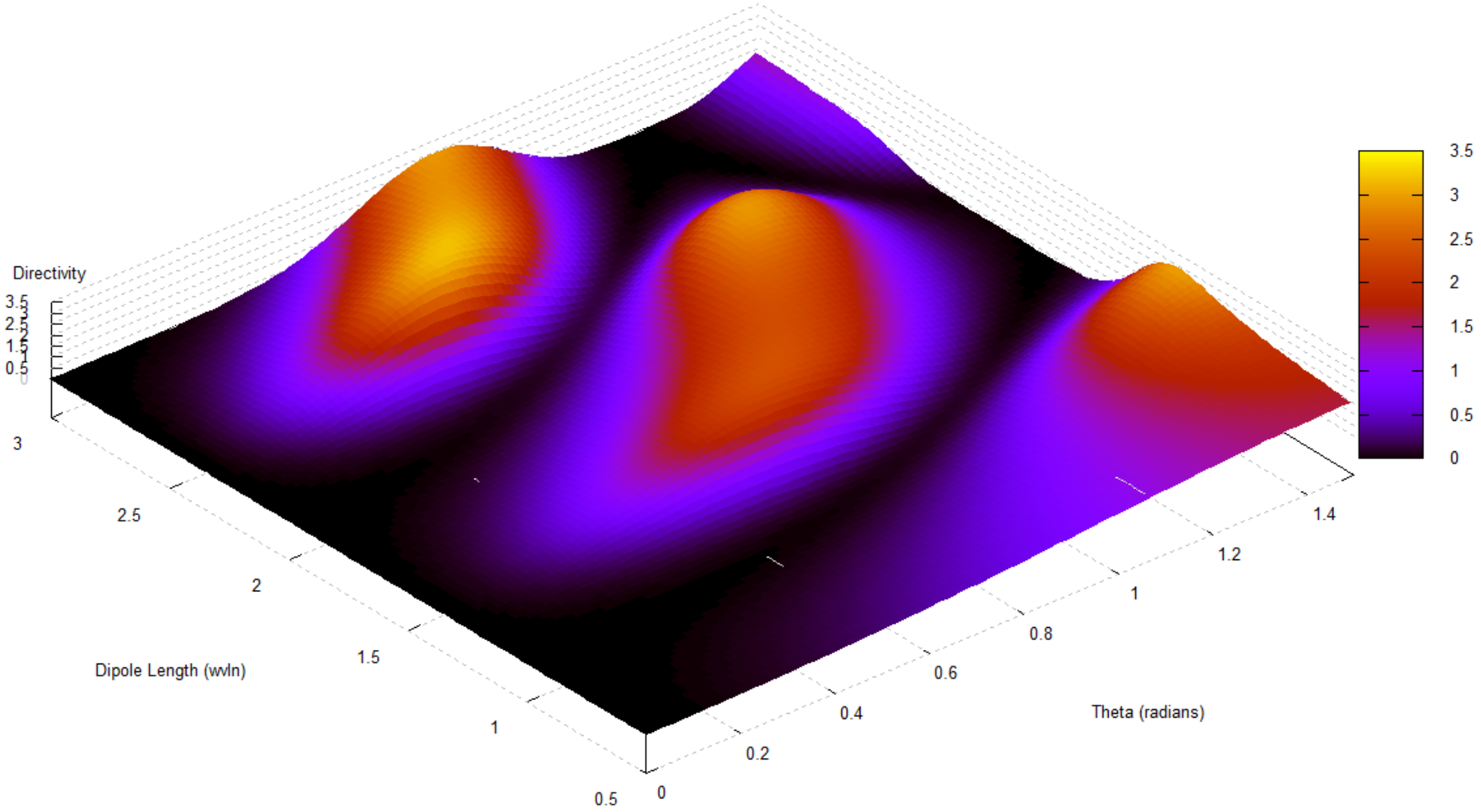

Fig. A1(a). Benchmark #1 perspective view.

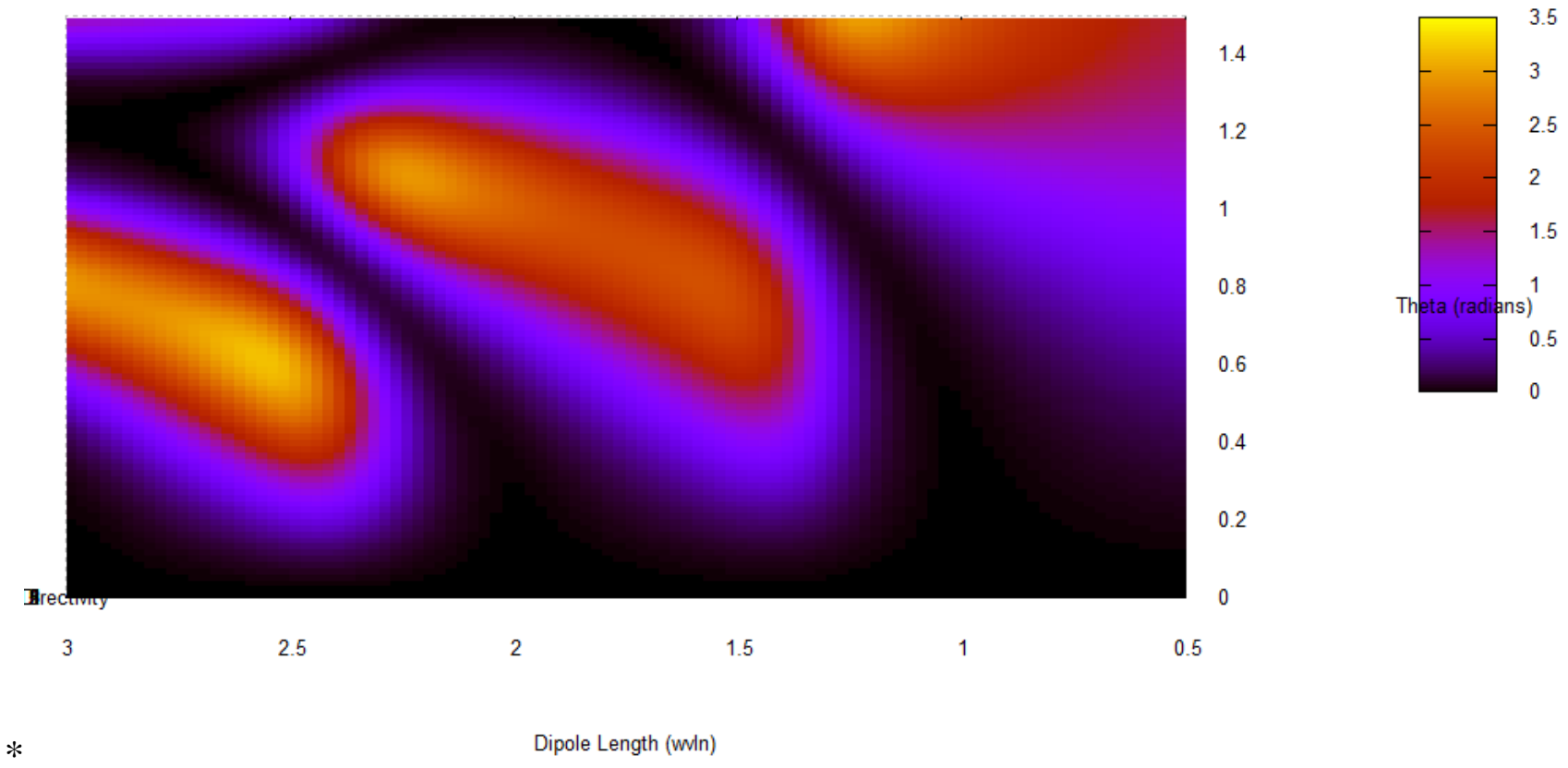

Fig. A1(b). Benchmark #1 projected onto ***L-θ*** plane.

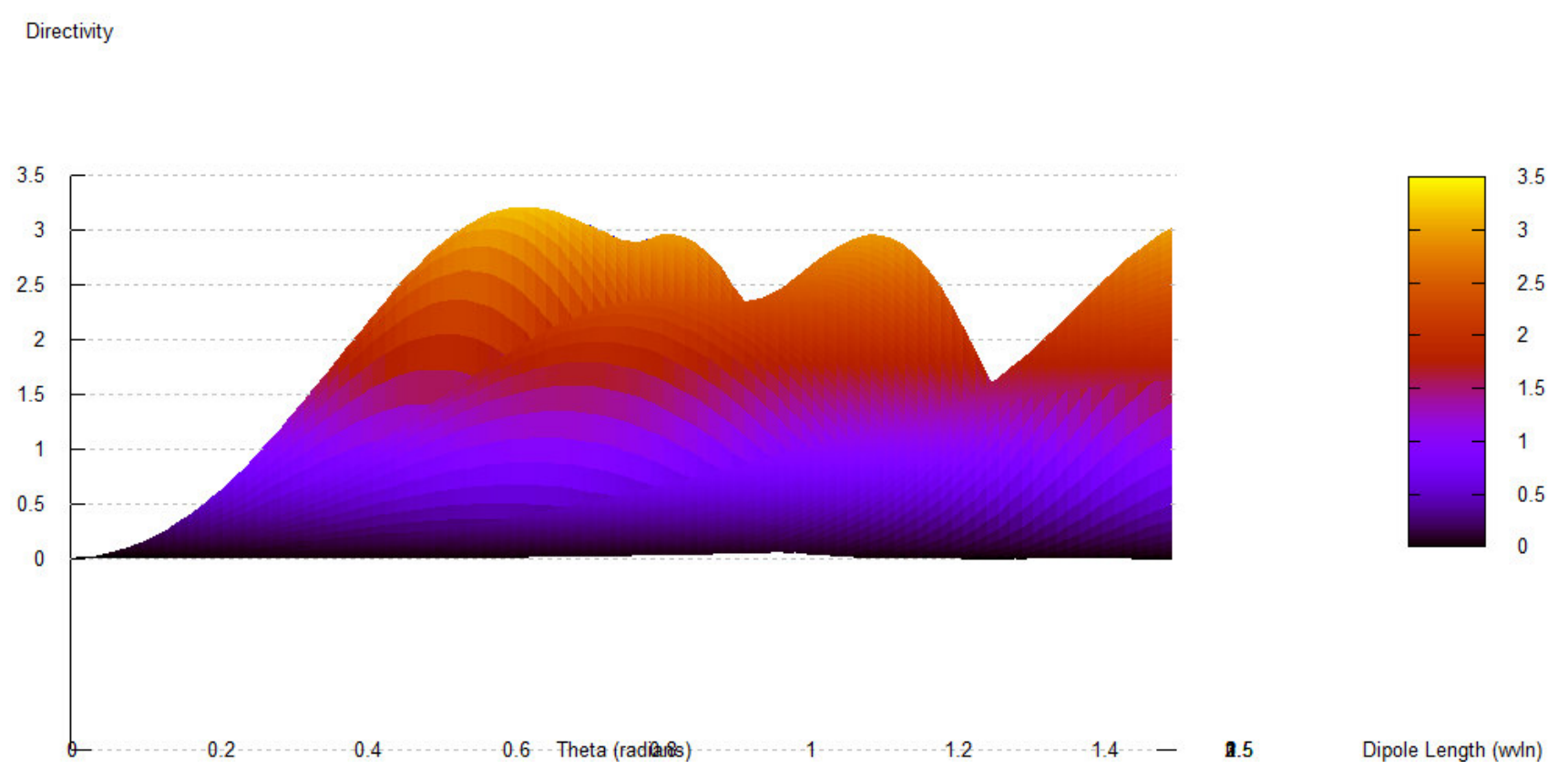

Fig. A1(c). Benchmark #1 projected onto ***θ-D*** plane.

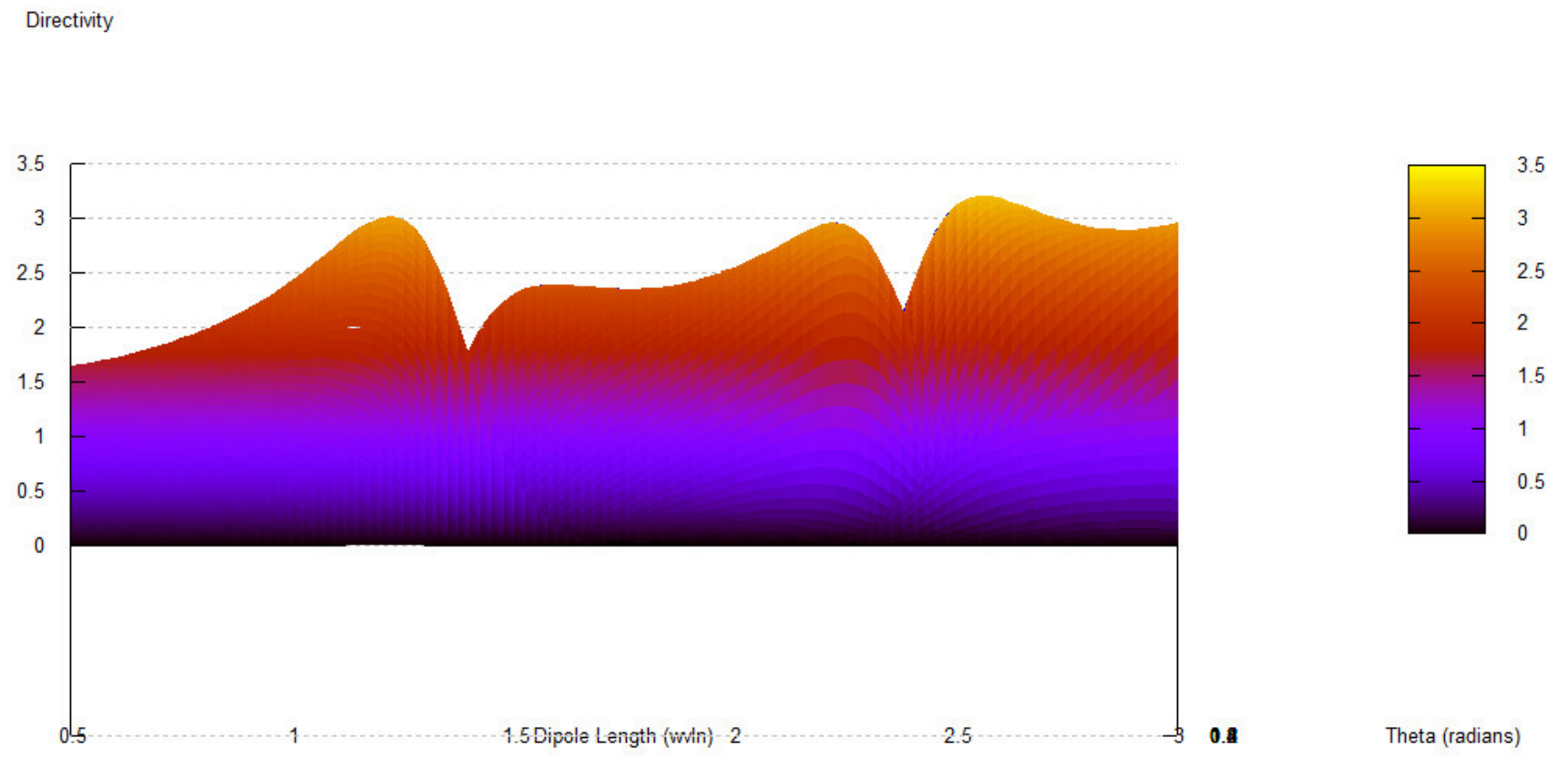

Fig. A1(d). Benchmark #1 projected onto ***L-D*** plane.

**PBM #1 NEC INPUT/OUTPUT FILES**

```
CM Input File: PBM1.NEC
CM Run ID 03-12-2009 18:42:53
CM R(p,1,j)= 2.58, R(p,2,j)= .63
CE
GW 1,259,0.,0.,-1.29,0.,0.,1.29,0.001
GE
EX 0,1,130,0,1.,0.
FR 0,1,0,0,299.79564,0.
RP 0,1,1,1001,36.1,0.,0.,0.,1000.
XQ
EN
```

```
                           ***********************************************************
                           *                                                         *
                           *  NUMERICAL ELECTROMAGNETICS CODE (NEC-4.1D, 23 MAR 05)  *
                           *                                                         *
                           *               RUN ID: 20090312 184253.969               *
                           *                                                         *
                           ***********************************************************

                      **********************************************************
                          Input File: PBM1.NEC
                          Run ID 03-12-2009 18:42:53
                          R(p,1,j)= 2.58, R(p,2,j)= .63

                      **********************************************************

                                 - - - STRUCTURE SPECIFICATION - - -
  WIRE                                                                                  NO. OF     FIRST  LAST      TAG
  NO.        X1         Y1         Z1          X2         Y2         Z2       RADIUS     SEG.      SEG.   SEG.      NO.
     1    0.00000    0.00000   -1.29000     0.00000    0.00000    1.29000    0.00100     259          1    259        1

   TOTAL SEGMENTS USED=   259      NO. SEG. IN A SYMMETRIC CELL=   259      SYMMETRY FLAG=  0


                                 - - - - - - FREQUENCY - - - - - -
                                    FREQUENCY= 2.9980E+02 MHZ
                                    WAVELENGTH= 1.0000E+00 METERS

                                  - - - ANTENNA ENVIRONMENT - - -
                                            FREE SPACE
```

```
                              - - - ANTENNA INPUT PARAMETERS - - -
  TAG   SEG.    VOLTAGE (VOLTS)         CURRENT (AMPS)         IMPEDANCE (OHMS)        ADMITTANCE (MHOS)      POWER
  NO.   NO.    REAL         IMAG.       REAL         IMAG.       REAL         IMAG.       REAL         IMAG.      (WATTS)
   1    130 1.00000E+00 0.00000E+00 2.29632E-03-2.25258E-03 2.21927E+02 2.17700E+02 2.29632E-03-2.25258E-03 1.14816E-03

                                - - - POWER BUDGET - - -
                                   INPUT POWER    = 1.1482E-03 WATTS
                                   RADIATED POWER= 1.1482E-03 WATTS
                                   WIRE LOSS      = 0.0000E+00 WATTS
                                   EFFICIENCY     = 100.00 PERCENT

                                        - - - RADIATION PATTERNS - - -
                                              RANGE= 1.000000E+03 METERS
                                              EXP(-JKR)/R= 1.00000E-03 AT PHASE   0.00 DEGREES

 - - ANGLES - -           - POWER GAINS -           - - - POLARIZATION - - -      - - - E(THETA) - - -       - - - E(PHI) - - -
  THETA      PHI         VERT.    HOR.    TOTAL       AXIAL      TILT   SENSE     MAGNITUDE     PHASE        MAGNITUDE     PHASE
 DEGREES   DEGREES        DB       DB       DB        RATIO      DEG.               VOLTS/M    DEGREES        VOLTS/M    DEGREES
   36.10      0.00        5.05 -999.99     5.05     0.00000      0.00  LINEAR    4.69508E-04    31.56      0.00000E+00      0.00
```

## A.2 Benchmark #2

***Without Noise***

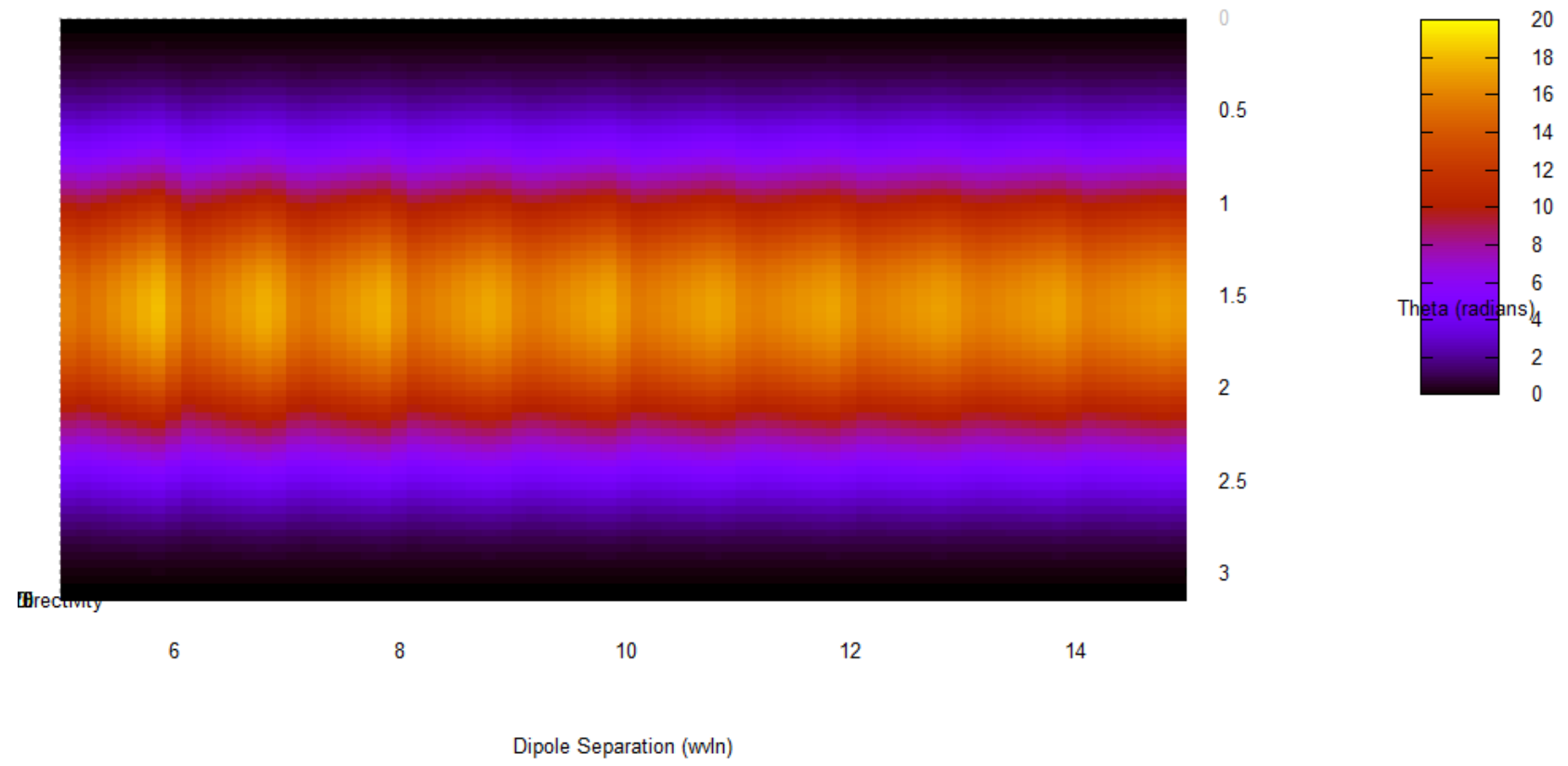

Fig. A2(a). Benchmark #2 without noise projected onto ***d-θ*** plane.

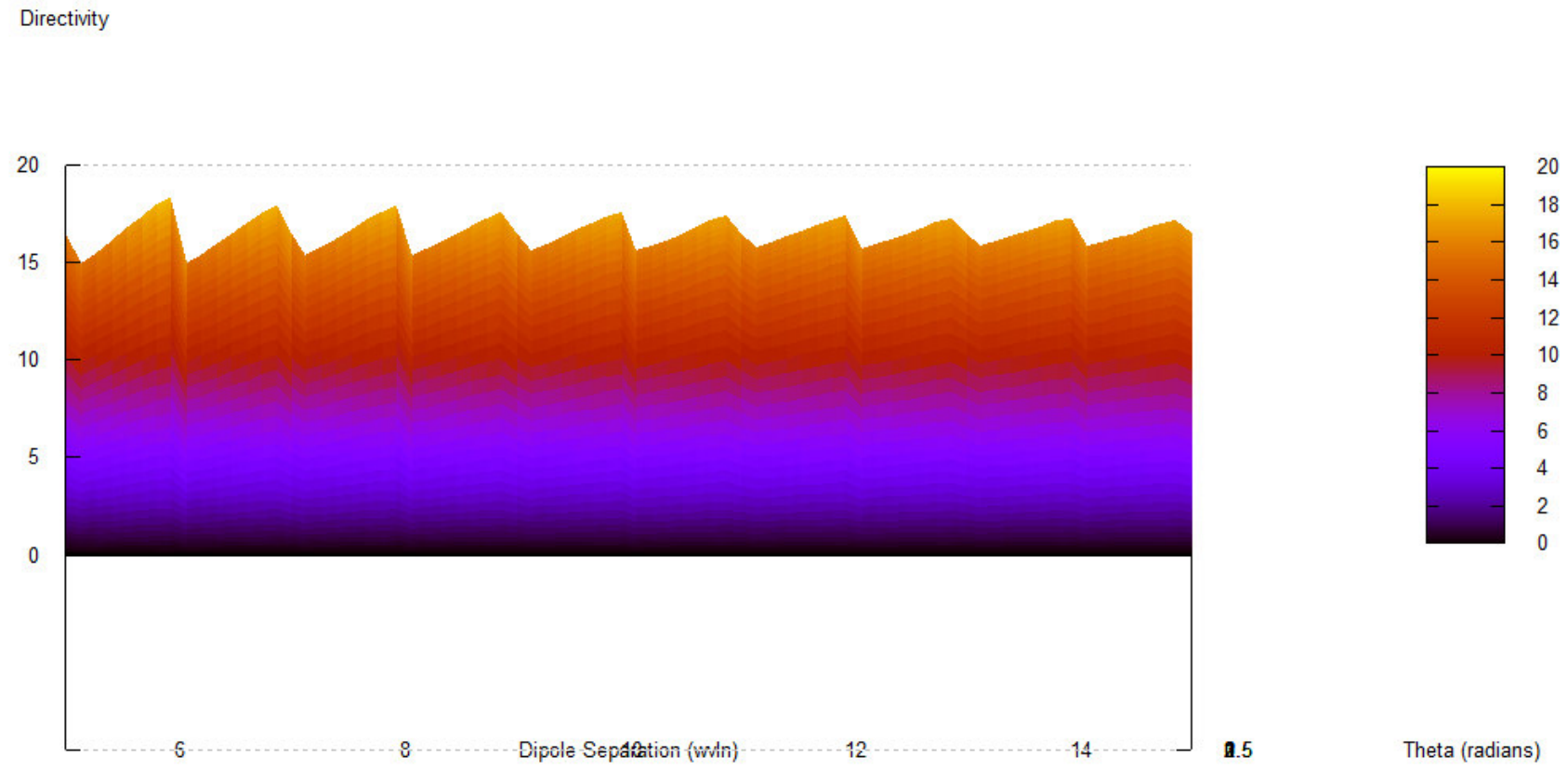

Fig. A2(b). Benchmark #2 without noise projected onto ***d-D*** plane.

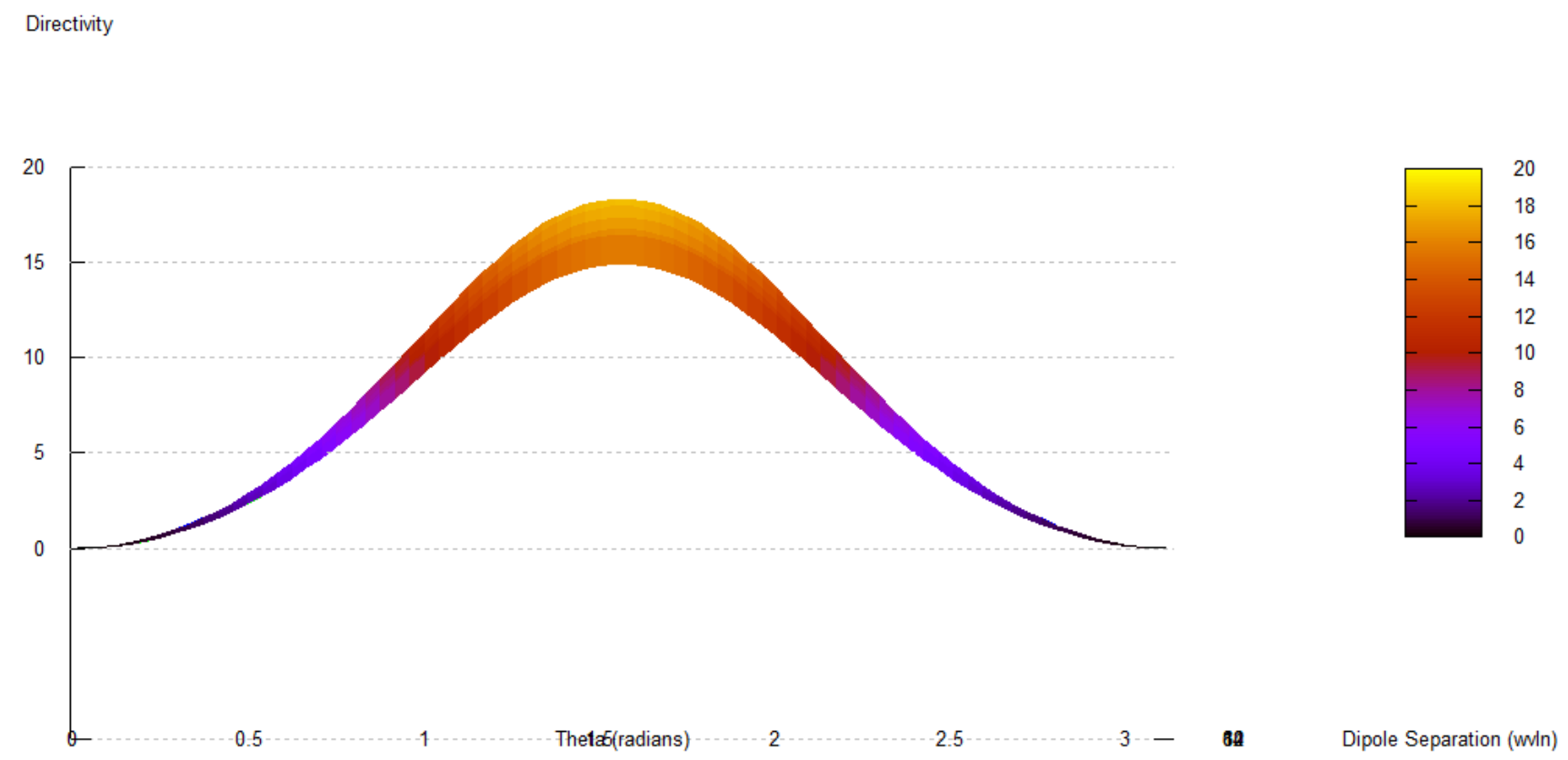

Fig. A2(c). Benchmark #2 without noise projected onto ***θ-D*** plane.

***With Noise***

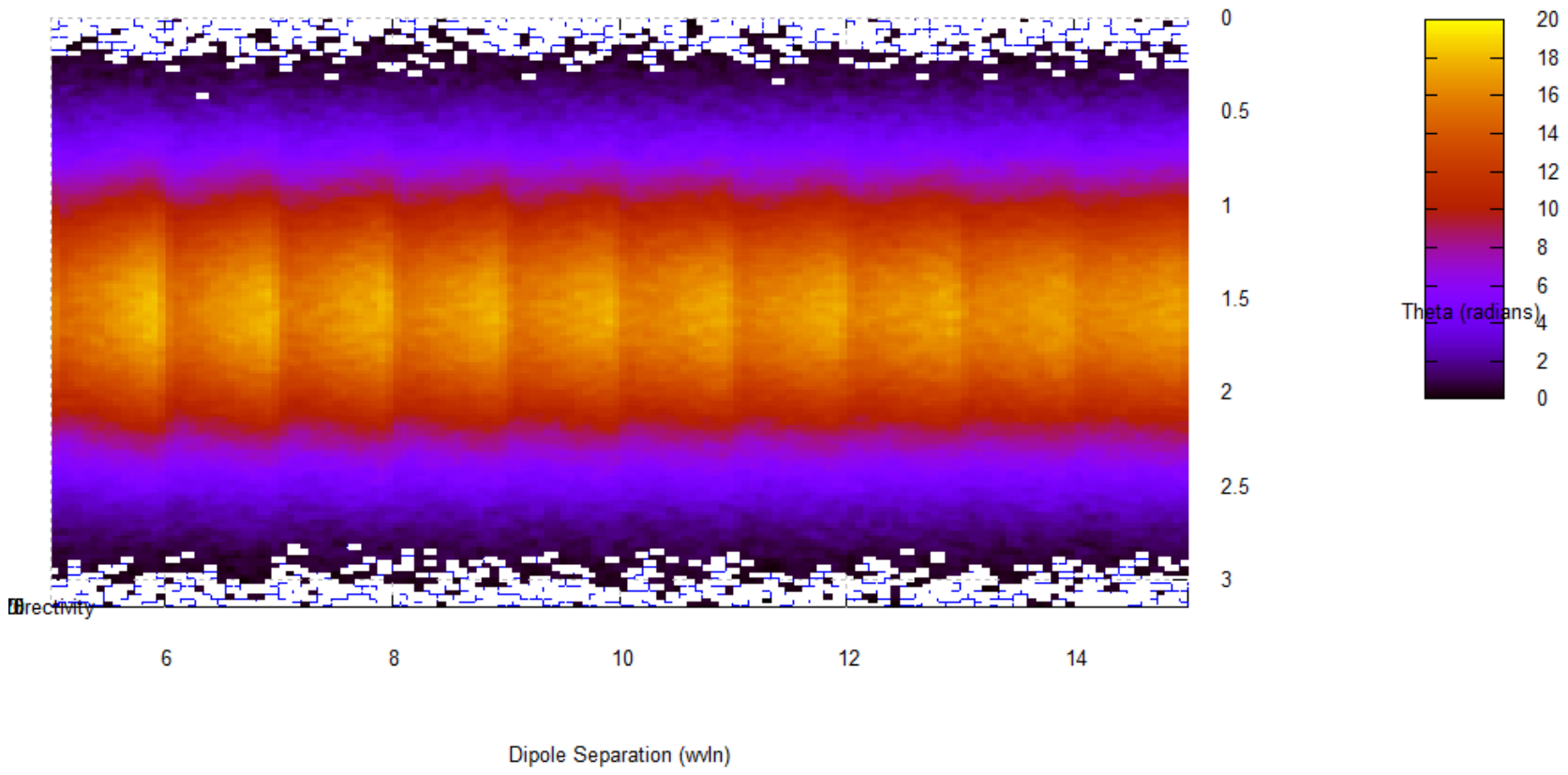

Fig. A2(d). Benchmark #2 with noise projected onto ***d-θ*** plane.

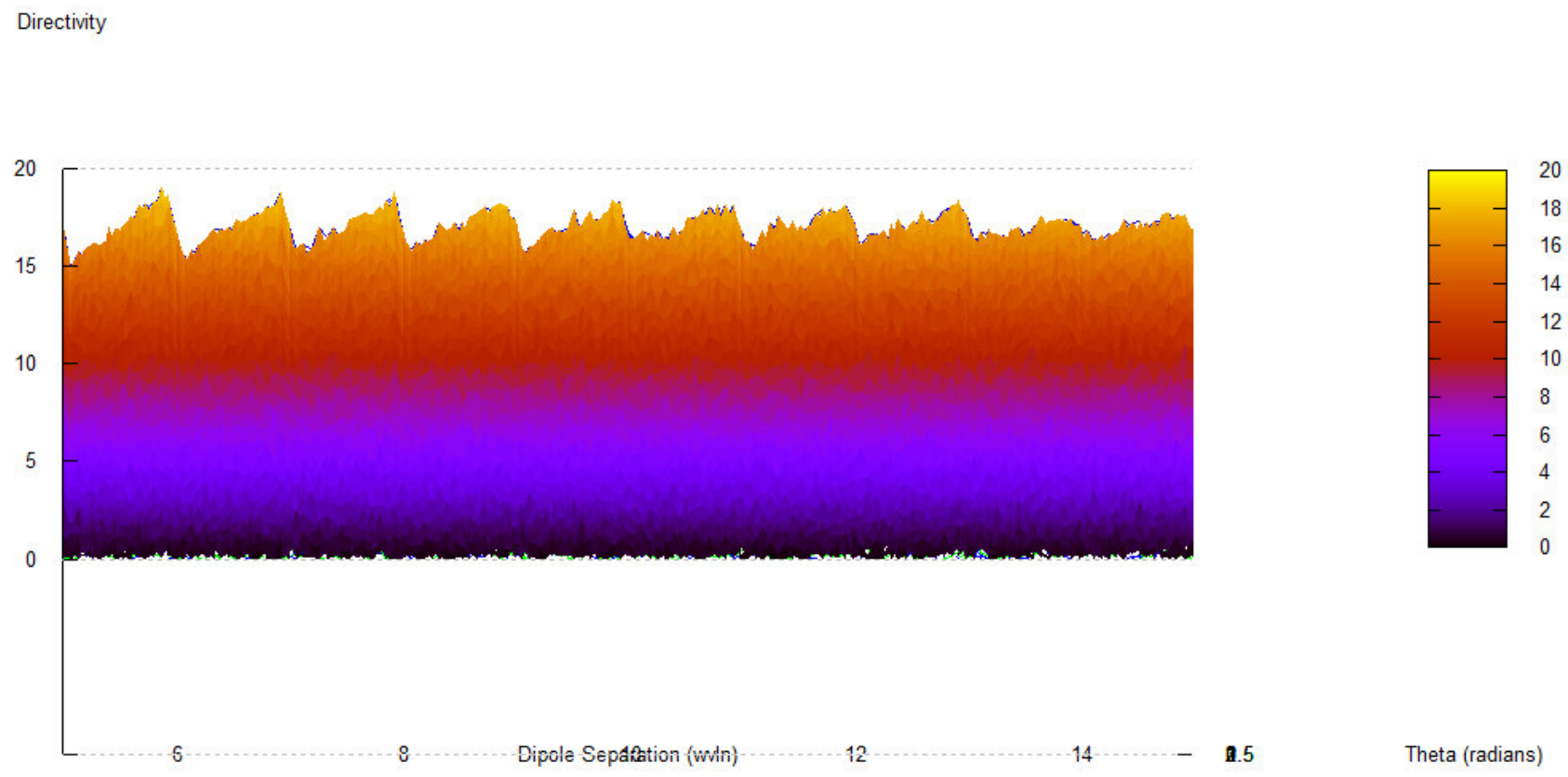

Fig. A2(e). Benchmark #2 with noise projected onto ***d-D*** plane.

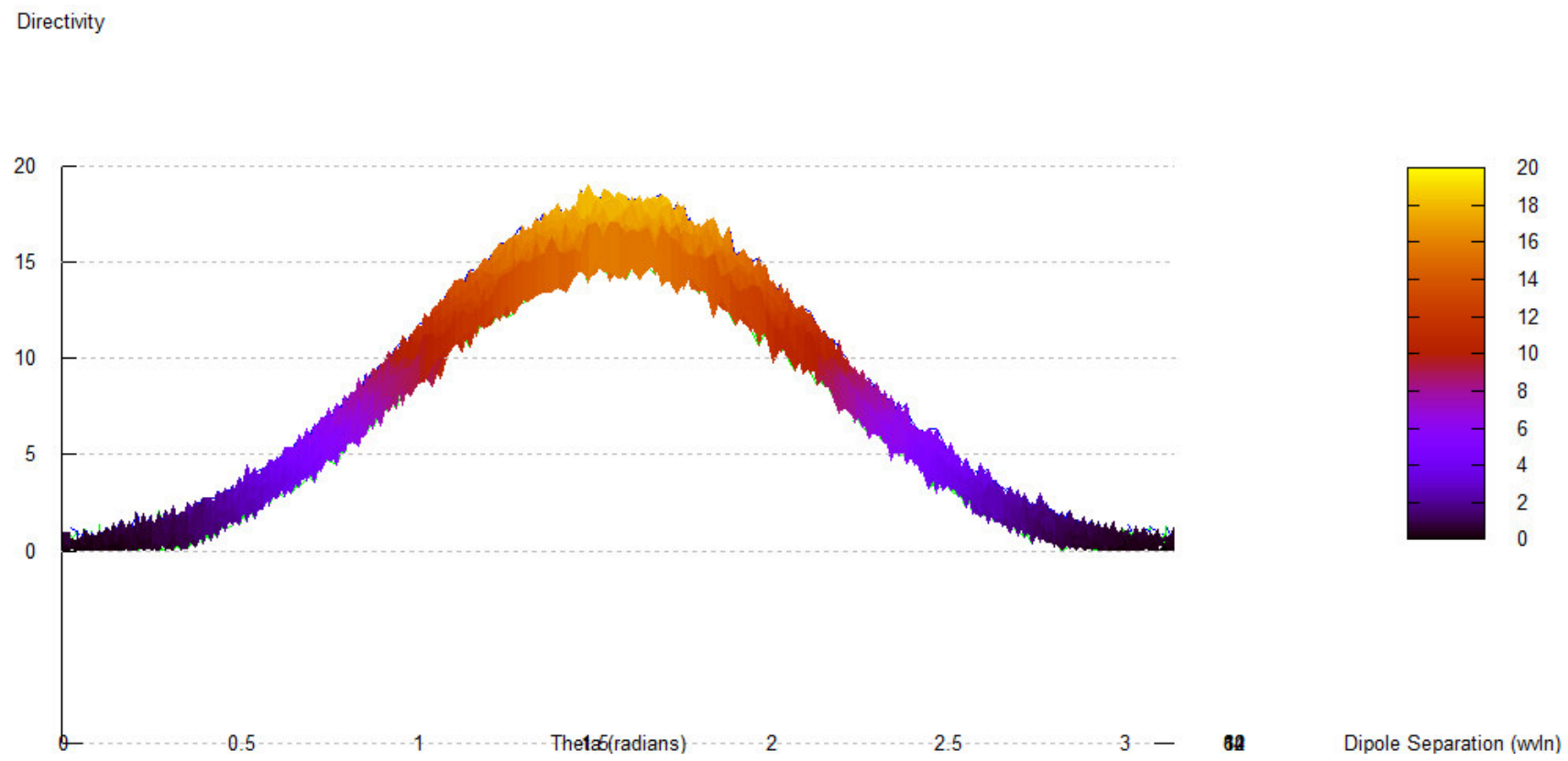

Fig. A2(f)). Benchmark #2 with noise projected onto $\theta$-$D$ plane.

## PBM #2 NEC INPUT/OUTPUT FILES

```
CM Input File: PBM2.NEC
CM Run ID 03-12-2009 18:42:57
CM R(p,1,j)= 5.85, R(p,2,j)= 1.5707963267949
CE
GW 1,49,-26.325,0.,-0.25,-26.325,0.,0.25,0.001
GW 2,49,-20.475,0.,-0.25,-20.475,0.,0.25,0.001
GW 3,49,-14.625,0.,-0.25,-14.625,0.,0.25,0.001
GW 4,49,-8.775,0.,-0.25,-8.775,0.,0.25,0.001
GW 5,49,-2.925,0.,-0.25,-2.925,0.,0.25,0.001
GW 6,49,2.925,0.,-0.25,2.925,0.,0.25,0.001
GW 7,49,8.775,0.,-0.25,8.775,0.,0.25,0.001
GW 8,49,14.625,0.,-0.25,14.625,0.,0.25,0.001
GW 9,49,20.475,0.,-0.25,20.475,0.,0.25,0.001
GW 10,49,26.325,0.,-0.25,26.325,0.,0.25,0.001
GE
EX 0,1,25,0,1.,0.
EX 0,2,25,0,1.,0.
EX 0,3,25,0,1.,0.
EX 0,4,25,0,1.,0.
EX 0,5,25,0,1.,0.
EX 0,6,25,0,1.,0.
EX 0,7,25,0,1.,0.
EX 0,8,25,0,1.,0.
EX 0,9,25,0,1.,0.
EX 0,10,25,0,1.,0.
FR 0,1,0,0,299.79564,0.
RP 0,1,1,1001,90.,90.,0.,0.,1000.
XQ
EN
```

```
                           ***********************************************************
                           *                                                         *
                           *  NUMERICAL ELECTROMAGNETICS CODE (NEC-4.1D, 23 MAR 05)  *
                           *                                                         *
                           *               RUN ID: 20090312 184257.992               *
                           *                                                         *
                           ***********************************************************


                      **********************************************************
                          Input File: PBM2.NEC
                          Run ID 03-12-2009 18:42:57
                          R(p,1,j)= 5.85, R(p,2,j)= 1.5707963267949

                      **********************************************************
```

- - - STRUCTURE SPECIFICATION - - -

| WIRE NO. | X1 | Y1 | Z1 | X2 | Y2 | Z2 | RADIUS | NO. OF SEG. | FIRST SEG. | LAST SEG. | TAG NO. |
|---|---|---|---|---|---|---|---|---|---|---|---|
| 1 | -26.32500 | 0.00000 | -0.25000 | -26.32500 | 0.00000 | 0.25000 | 0.00100 | 49 | 1 | 49 | 1 |
| 2 | -20.47500 | 0.00000 | -0.25000 | -20.47500 | 0.00000 | 0.25000 | 0.00100 | 49 | 50 | 98 | 2 |
| 3 | -14.62500 | 0.00000 | -0.25000 | -14.62500 | 0.00000 | 0.25000 | 0.00100 | 49 | 99 | 147 | 3 |
| 4 | -8.77500 | 0.00000 | -0.25000 | -8.77500 | 0.00000 | 0.25000 | 0.00100 | 49 | 148 | 196 | 4 |
| 5 | -2.92500 | 0.00000 | -0.25000 | -2.92500 | 0.00000 | 0.25000 | 0.00100 | 49 | 197 | 245 | 5 |
| 6 | 2.92500 | 0.00000 | -0.25000 | 2.92500 | 0.00000 | 0.25000 | 0.00100 | 49 | 246 | 294 | 6 |
| 7 | 8.77500 | 0.00000 | -0.25000 | 8.77500 | 0.00000 | 0.25000 | 0.00100 | 49 | 295 | 343 | 7 |
| 8 | 14.62500 | 0.00000 | -0.25000 | 14.62500 | 0.00000 | 0.25000 | 0.00100 | 49 | 344 | 392 | 8 |
| 9 | 20.47500 | 0.00000 | -0.25000 | 20.47500 | 0.00000 | 0.25000 | 0.00100 | 49 | 393 | 441 | 9 |
| 10 | 26.32500 | 0.00000 | -0.25000 | 26.32500 | 0.00000 | 0.25000 | 0.00100 | 49 | 442 | 490 | 10 |

```
 TOTAL SEGMENTS USED=   490      NO. SEG. IN A SYMMETRIC CELL=   490      SYMMETRY FLAG=  0

                                 - - - - - - FREQUENCY - - - - - -
                                    FREQUENCY= 2.9980E+02 MHZ
                                    WAVELENGTH= 1.0000E+00 METERS

                                  - - - ANTENNA ENVIRONMENT - - -
```

FREE SPACE

- - - ANTENNA INPUT PARAMETERS - - -

| TAG NO. | SEG. NO. | VOLTAGE (VOLTS) REAL | VOLTAGE (VOLTS) IMAG. | CURRENT (AMPS) REAL | CURRENT (AMPS) IMAG. | IMPEDANCE (OHMS) REAL | IMPEDANCE (OHMS) IMAG. | ADMITTANCE (MHOS) REAL | ADMITTANCE (MHOS) IMAG. | POWER (WATTS) |
|---|---|---|---|---|---|---|---|---|---|---|
| 1 | 25 | 1.00000E+00 | 0.00000E+00 | 8.86468E-03 | -5.47654E-03 | 8.16456E+01 | 5.04401E+01 | 8.86468E-03 | -5.47654E-03 | 4.43234E-03 |
| 2 | 74 | 1.00000E+00 | 0.00000E+00 | 8.70441E-03 | -5.84089E-03 | 7.92154E+01 | 5.31556E+01 | 8.70441E-03 | -5.84089E-03 | 4.35221E-03 |
| 3 | 123 | 1.00000E+00 | 0.00000E+00 | 8.81941E-03 | -5.96110E-03 | 7.78297E+01 | 5.26057E+01 | 8.81941E-03 | -5.96110E-03 | 4.40971E-03 |
| 4 | 172 | 1.00000E+00 | 0.00000E+00 | 8.99606E-03 | -5.94423E-03 | 7.73769E+01 | 5.11274E+01 | 8.99606E-03 | -5.94423E-03 | 4.49803E-03 |
| 5 | 221 | 1.00000E+00 | 0.00000E+00 | 9.11672E-03 | -5.89753E-03 | 7.73289E+01 | 5.00234E+01 | 9.11672E-03 | -5.89753E-03 | 4.55836E-03 |
| 6 | 270 | 1.00000E+00 | 0.00000E+00 | 9.11672E-03 | -5.89753E-03 | 7.73289E+01 | 5.00234E+01 | 9.11672E-03 | -5.89753E-03 | 4.55836E-03 |
| 7 | 319 | 1.00000E+00 | 0.00000E+00 | 8.99606E-03 | -5.94423E-03 | 7.73769E+01 | 5.11274E+01 | 8.99606E-03 | -5.94423E-03 | 4.49803E-03 |
| 8 | 368 | 1.00000E+00 | 0.00000E+00 | 8.81941E-03 | -5.96110E-03 | 7.78297E+01 | 5.26057E+01 | 8.81941E-03 | -5.96110E-03 | 4.40971E-03 |
| 9 | 417 | 1.00000E+00 | 0.00000E+00 | 8.70441E-03 | -5.84089E-03 | 7.92154E+01 | 5.31556E+01 | 8.70441E-03 | -5.84089E-03 | 4.35221E-03 |
| 10 | 466 | 1.00000E+00 | 0.00000E+00 | 8.86468E-03 | -5.47654E-03 | 8.16456E+01 | 5.04401E+01 | 8.86468E-03 | -5.47654E-03 | 4.43234E-03 |

- - - POWER BUDGET - - -

INPUT POWER = 4.4501E-02 WATTS
RADIATED POWER= 4.4501E-02 WATTS
WIRE LOSS = 0.0000E+00 WATTS
EFFICIENCY = 100.00 PERCENT

- - - RADIATION PATTERNS - - -

RANGE= 1.000000E+03 METERS
EXP(-JKR)/R= 1.00000E-03 AT PHASE 0.00 DEGREES

| - - ANGLES - - THETA DEGREES | PHI DEGREES | - POWER GAINS - VERT. DB | HOR. DB | TOTAL DB | - - - POLARIZATION - - - AXIAL RATIO | TILT DEG. | SENSE | - - - E(THETA) - - - MAGNITUDE VOLTS/M | PHASE DEGREES | - - - E(PHI) - - - MAGNITUDE VOLTS/M | PHASE DEGREES |
|---|---|---|---|---|---|---|---|---|---|---|---|
| 90.00 | 90.00 | 12.58 | -999.99 | 12.58 | 0.00000 | 0.00 | LINEAR | 6.95826E-03 | 52.71 | 0.00000E+00 | 0.00 |

## A.3 Benchmark #3

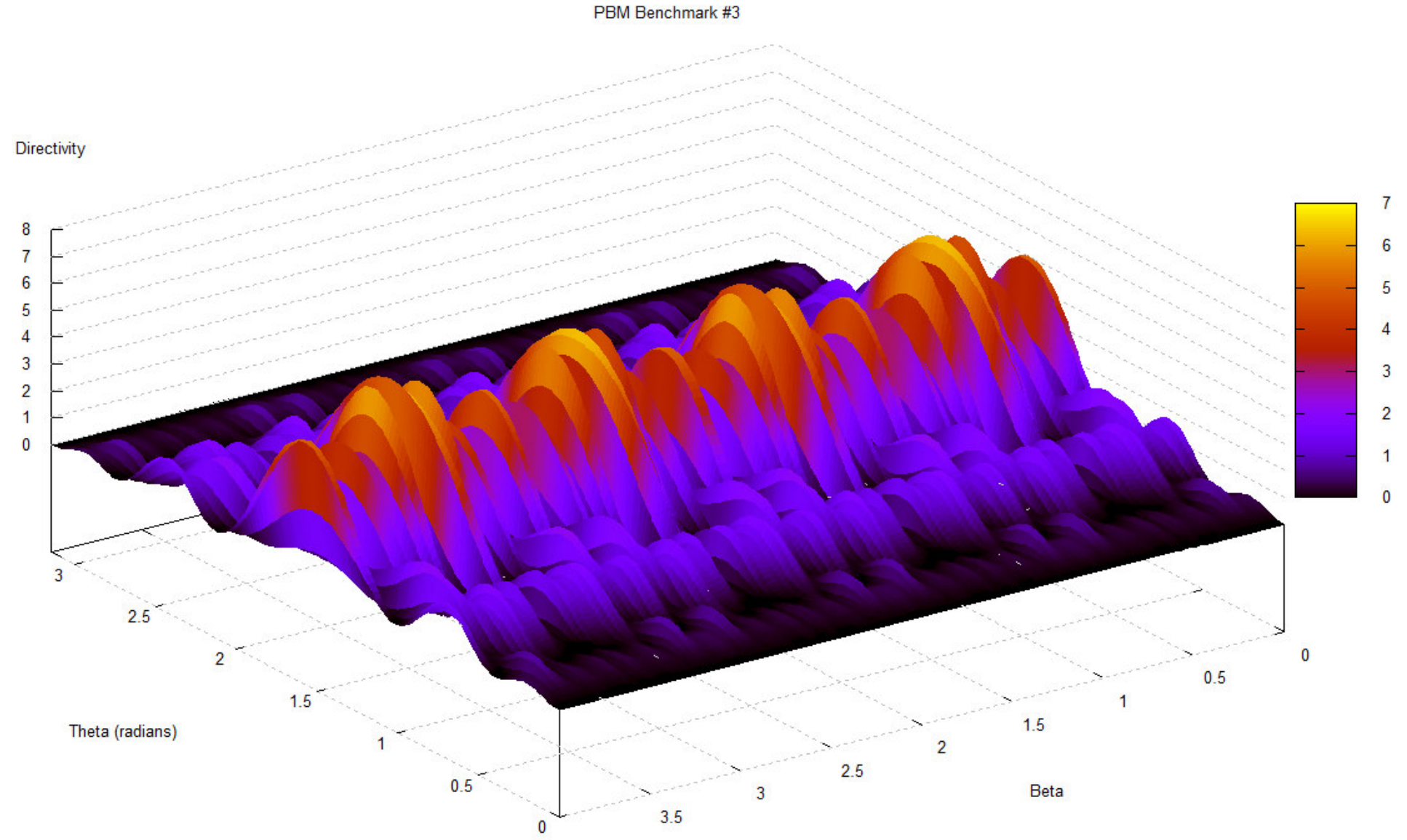

Fig. A3(a). Benchmark #3 perspective view.

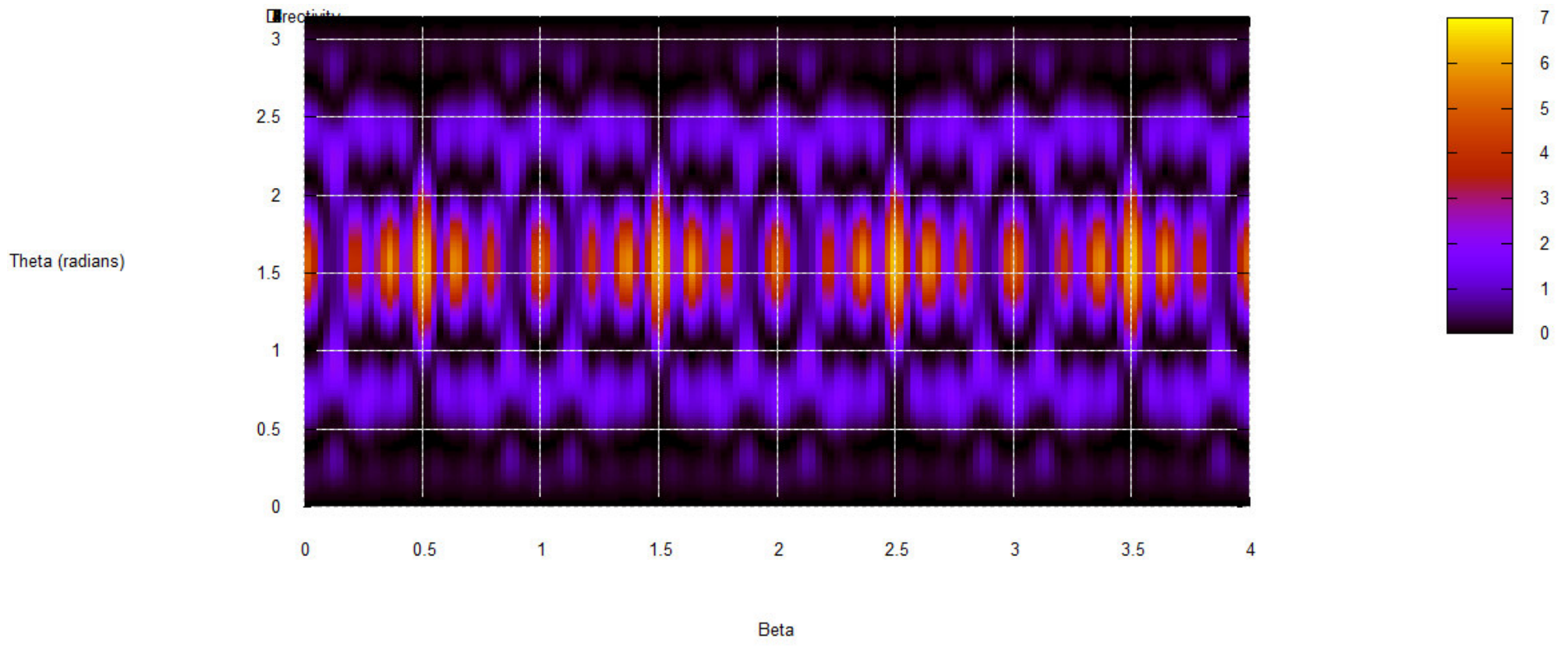

Fig. A3(b). Benchmark #3 projected onto $\beta$-$\theta$ plane.

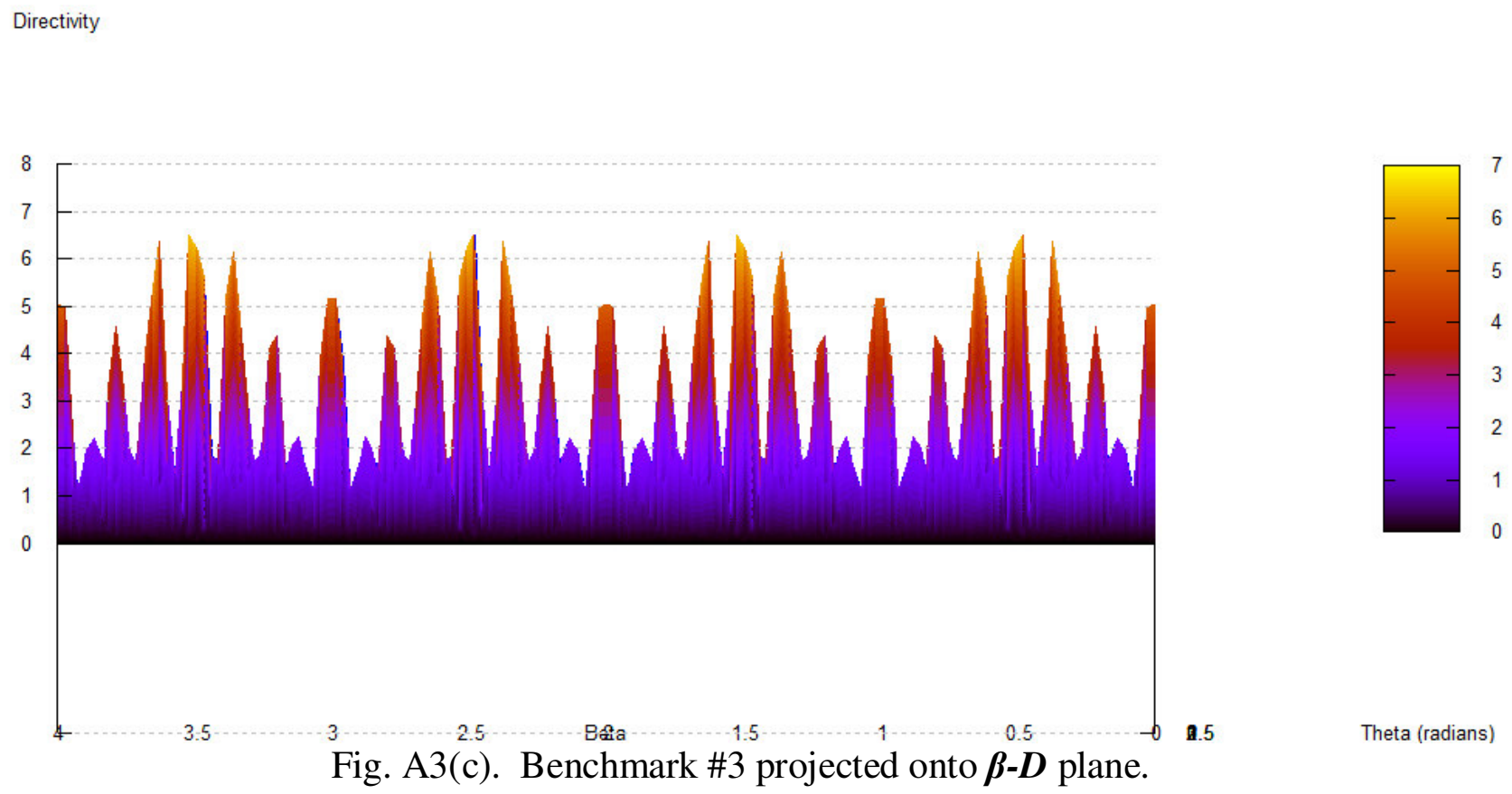

Fig. A3(c). Benchmark #3 projected onto ***β-D*** plane.

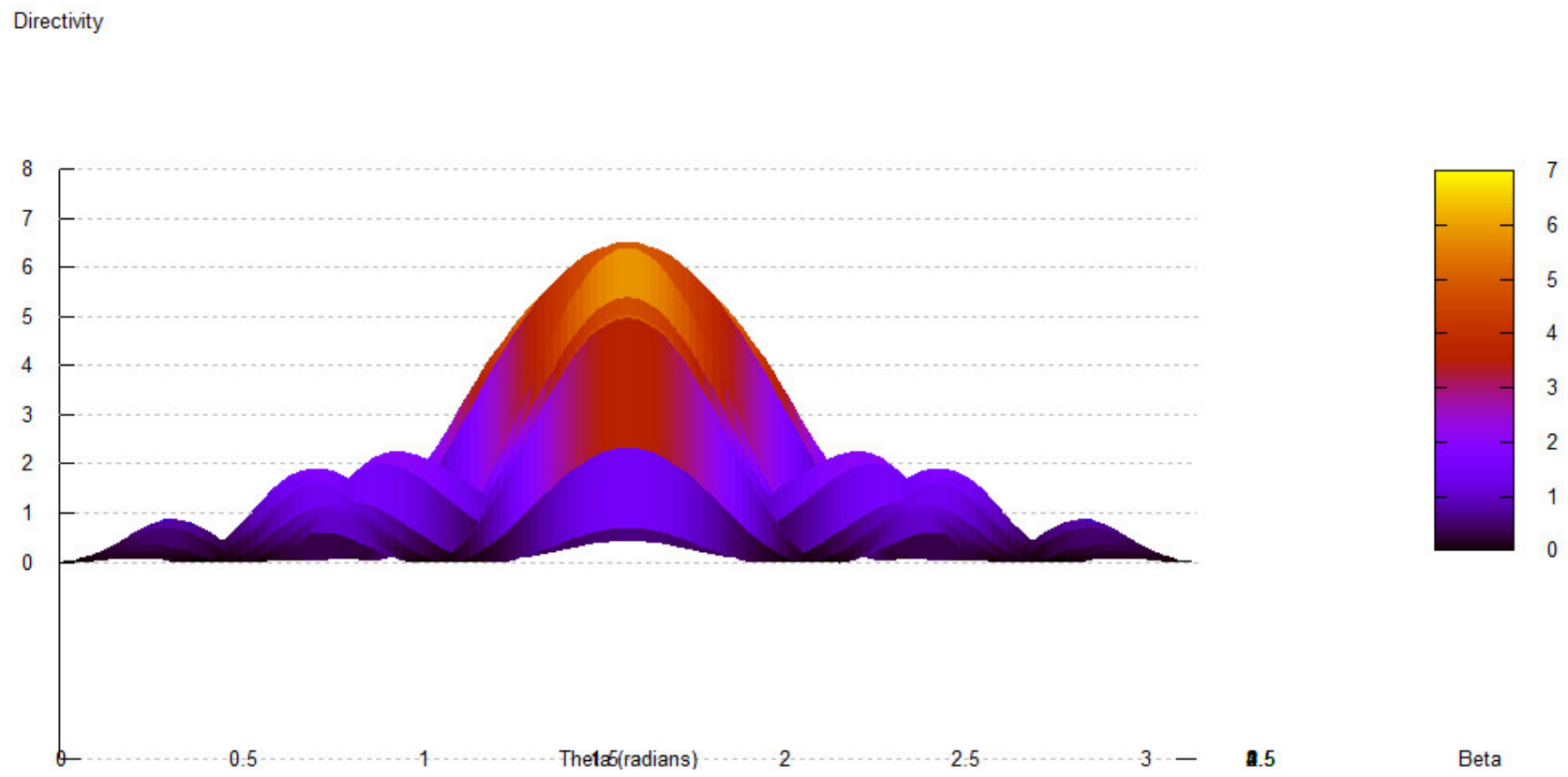

Fig. A3(d). Benchmark #3 projected onto ***θ-D*** plane.

**PBM #3 NEC INPUT/OUTPUT FILES**

```
CM Input File: PBM3.NEC
CM Run ID 03-12-2009 18:43:00
CM R(p,1,j)= 3.5, R(p,2,j)= 1.5707963267949
CE
GW 1,49,1,0,-0.25,1,0,0.25,0.001
GW 2,49,0.70711,0.70711,-0.25,0.70711,0.70711,0.25,0.001
GW 3,49,0,1,-0.25,0,1,0.25,0.001
GW 4,49,-0.70711,0.70711,-0.25,-0.70711,0.70711,0.25,0.001
GW 5,49,-1,0,-0.25,-1,0,0.25,0.001
GW 6,49,-0.70711,-0.70711,-0.25,-0.70711,-0.70711,0.25,0.001
GW 7,49,0,-1,-0.25,0,-1,0.25,0.001
GW 8,49,0.70711,-0.70711,-0.25,0.70711,-0.70711,0.25,0.001
GE
EX 0,1,25,0,.540302,-.841471
EX 0,2,25,0,.540302,.841471
EX 0,3,25,0,.540302,-.841471
EX 0,4,25,0,.540302,.841471
EX 0,5,25,0,.540302,-.841471
EX 0,6,25,0,.540302,.841471
EX 0,7,25,0,.540302,-.841471
EX 0,8,25,0,.540302,.841471
FR 0,1,0,0,299.79564,0.
RP 0,1,1,1001,90.,0.,0.,0.,1000.
XQ
EN
                           ***********************************************************
                           *                                                         *
                           *  NUMERICAL ELECTROMAGNETICS CODE (NEC-4.1D, 23 MAR 05)  *
                           *                                                         *
                           *             RUN ID: 20090312 184300.601                 *
                           *                                                         *
                           ***********************************************************

                      **********************************************************
                          Input File: PBM3.NEC
                          Run ID 03-12-2009 18:43:00
                          R(p,1,j)= 3.5, R(p,2,j)= 1.5707963267949

                      **********************************************************

                                 - - - STRUCTURE SPECIFICATION - - -
  WIRE                                                                              NO. OF     FIRST  LAST      TAG
```

| NO. | X1 | Y1 | Z1 | X2 | Y2 | Z2 | RADIUS | SEG. | SEG. | SEG. | NO. |
|---|---|---|---|---|---|---|---|---|---|---|---|
| 1 | 1.00000 | 0.00000 | -0.25000 | 1.00000 | 0.00000 | 0.25000 | 0.00100 | 49 | 1 | 49 | 1 |
| 2 | 0.70711 | 0.70711 | -0.25000 | 0.70711 | 0.70711 | 0.25000 | 0.00100 | 49 | 50 | 98 | 2 |
| 3 | 0.00000 | 1.00000 | -0.25000 | 0.00000 | 1.00000 | 0.25000 | 0.00100 | 49 | 99 | 147 | 3 |
| 4 | -0.70711 | 0.70711 | -0.25000 | -0.70711 | 0.70711 | 0.25000 | 0.00100 | 49 | 148 | 196 | 4 |
| 5 | -1.00000 | 0.00000 | -0.25000 | -1.00000 | 0.00000 | 0.25000 | 0.00100 | 49 | 197 | 245 | 5 |
| 6 | -0.70711 | -0.70711 | -0.25000 | -0.70711 | -0.70711 | 0.25000 | 0.00100 | 49 | 246 | 294 | 6 |
| 7 | 0.00000 | -1.00000 | -0.25000 | 0.00000 | -1.00000 | 0.25000 | 0.00100 | 49 | 295 | 343 | 7 |
| 8 | 0.70711 | -0.70711 | -0.25000 | 0.70711 | -0.70711 | 0.25000 | 0.00100 | 49 | 344 | 392 | 8 |

TOTAL SEGMENTS USED= 392 NO. SEG. IN A SYMMETRIC CELL= 392 SYMMETRY FLAG= 0

- - - - - - FREQUENCY - - - - - -
FREQUENCY= 2.9980E+02 MHZ
WAVELENGTH= 1.0000E+00 METERS

- - - ANTENNA ENVIRONMENT - - -
FREE SPACE

- - - ANTENNA INPUT PARAMETERS - - -

| TAG NO. | SEG. NO. | VOLTAGE (VOLTS) REAL | VOLTAGE (VOLTS) IMAG. | CURRENT (AMPS) REAL | CURRENT (AMPS) IMAG. | IMPEDANCE (OHMS) REAL | IMPEDANCE (OHMS) IMAG. | ADMITTANCE (MHOS) REAL | ADMITTANCE (MHOS) IMAG. | POWER (WATTS) |
|---|---|---|---|---|---|---|---|---|---|---|
| 1 | 25 | 5.40302E-01 | -8.41471E-01 | 3.32705E-03 | -1.17420E-02 | 7.84067E+01 | 2.37983E+01 | 1.16781E-02 | -3.54460E-03 | 5.83907E-03 |
| 2 | 74 | 5.40302E-01 | 8.41471E-01 | 1.99518E-03 | -6.13888E-04 | 1.28839E+02 | 4.61395E+02 | 5.61428E-04 | -2.01057E-03 | 2.80714E-04 |
| 3 | 123 | 5.40302E-01 | -8.41471E-01 | 3.32705E-03 | -1.17420E-02 | 7.84067E+01 | 2.37983E+01 | 1.16781E-02 | -3.54460E-03 | 5.83907E-03 |
| 4 | 172 | 5.40302E-01 | 8.41471E-01 | 1.99518E-03 | -6.13888E-04 | 1.28839E+02 | 4.61395E+02 | 5.61428E-04 | -2.01057E-03 | 2.80714E-04 |
| 5 | 221 | 5.40302E-01 | -8.41471E-01 | 3.32705E-03 | -1.17420E-02 | 7.84067E+01 | 2.37983E+01 | 1.16781E-02 | -3.54460E-03 | 5.83907E-03 |
| 6 | 270 | 5.40302E-01 | 8.41471E-01 | 1.99518E-03 | -6.13888E-04 | 1.28839E+02 | 4.61395E+02 | 5.61428E-04 | -2.01057E-03 | 2.80714E-04 |
| 7 | 319 | 5.40302E-01 | -8.41471E-01 | 3.32705E-03 | -1.17420E-02 | 7.84067E+01 | 2.37983E+01 | 1.16781E-02 | -3.54460E-03 | 5.83907E-03 |
| 8 | 368 | 5.40302E-01 | 8.41471E-01 | 1.99518E-03 | -6.13888E-04 | 1.28839E+02 | 4.61395E+02 | 5.61428E-04 | -2.01057E-03 | 2.80714E-04 |

- - - POWER BUDGET - - -
INPUT POWER = 2.4479E-02 WATTS
RADIATED POWER= 2.4479E-02 WATTS
WIRE LOSS = 0.0000E+00 WATTS
EFFICIENCY = 100.00 PERCENT

- - - RADIATION PATTERNS - - -
RANGE= 1.000000E+03 METERS
EXP(-JKR)/R= 1.00000E-03 AT PHASE 0.00 DEGREES

| - - ANGLES - - THETA DEGREES | PHI DEGREES | - POWER GAINS - VERT. DB | HOR. DB | TOTAL DB | - - - POLARIZATION - - - AXIAL RATIO | TILT DEG. | SENSE | - - - E(THETA) - - - MAGNITUDE VOLTS/M | PHASE DEGREES | - - - E(PHI) - - - MAGNITUDE VOLTS/M | PHASE DEGREES |
|---|---|---|---|---|---|---|---|---|---|---|---|
| 90.00 | 0.00 | 7.89 | -999.99 | 7.89 | 0.00000 | 0.00 | LINEAR | 3.00526E-03 | 8.47 | 0.00000E+00 | 0.00 |

## A.4 Benchmark #4

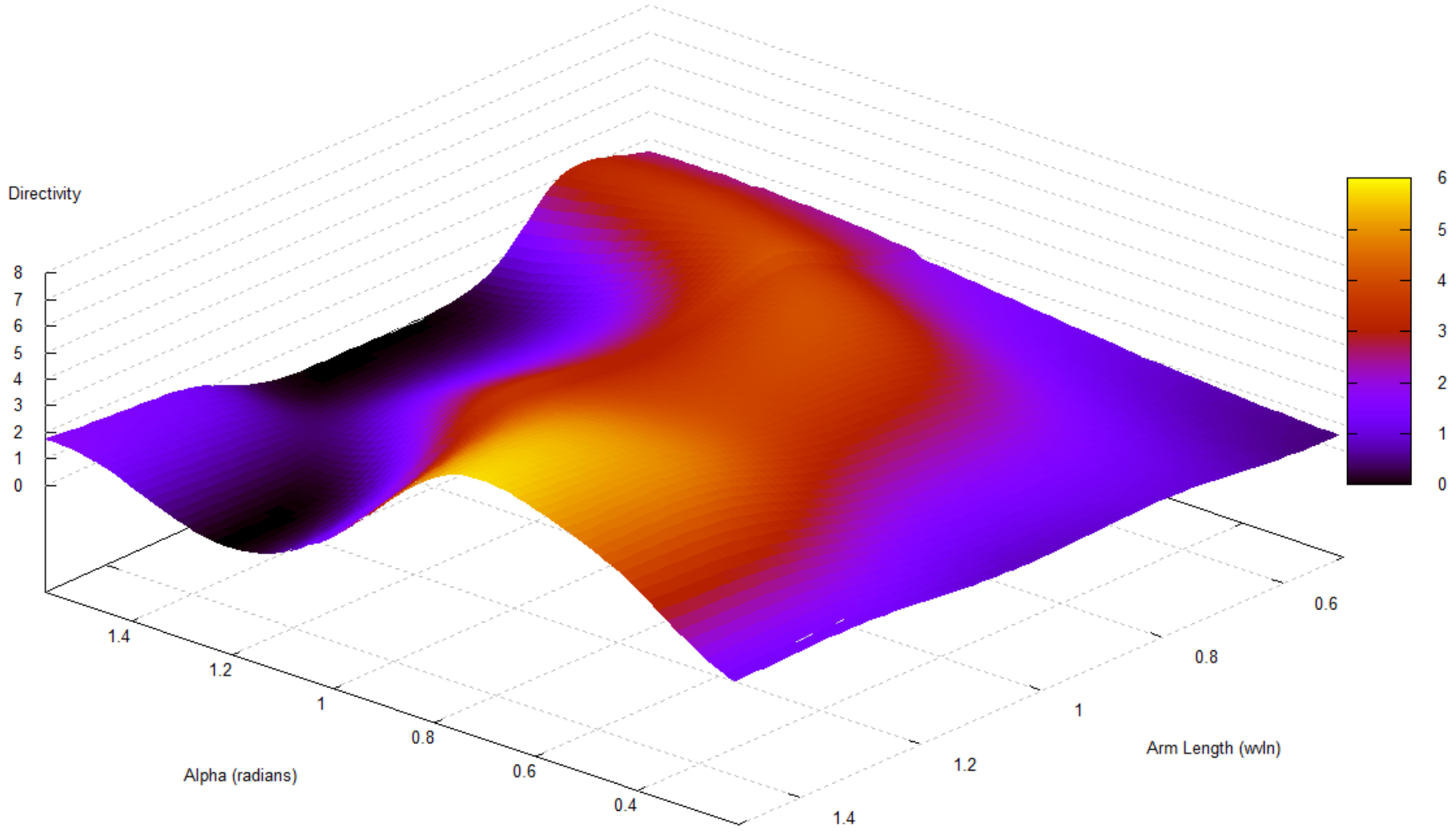

Fig. A4(a). Benchmark #4 perspective view.

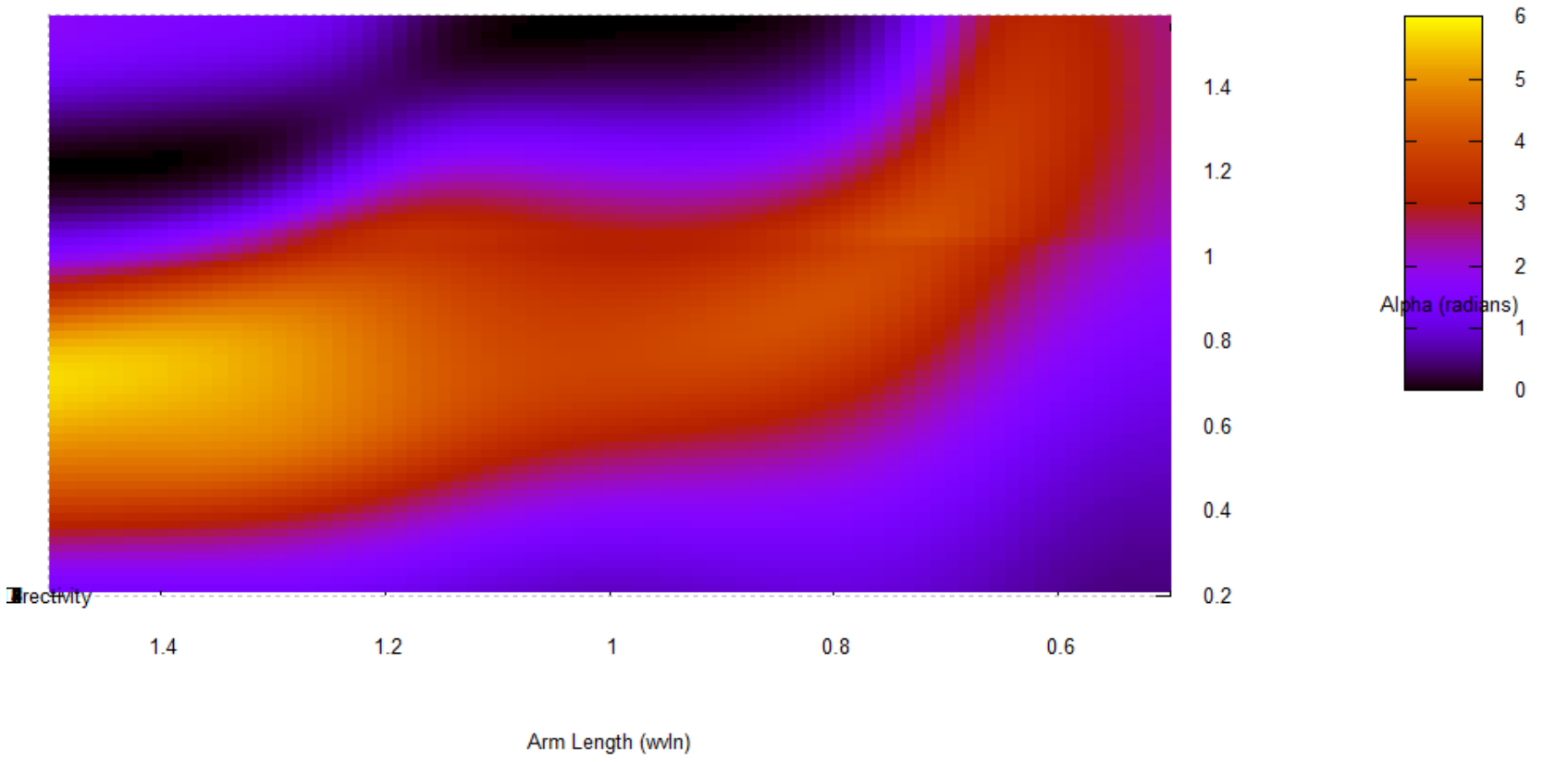

Fig. A4(b). Benchmark #4 projected onto ***L-α*** plane.

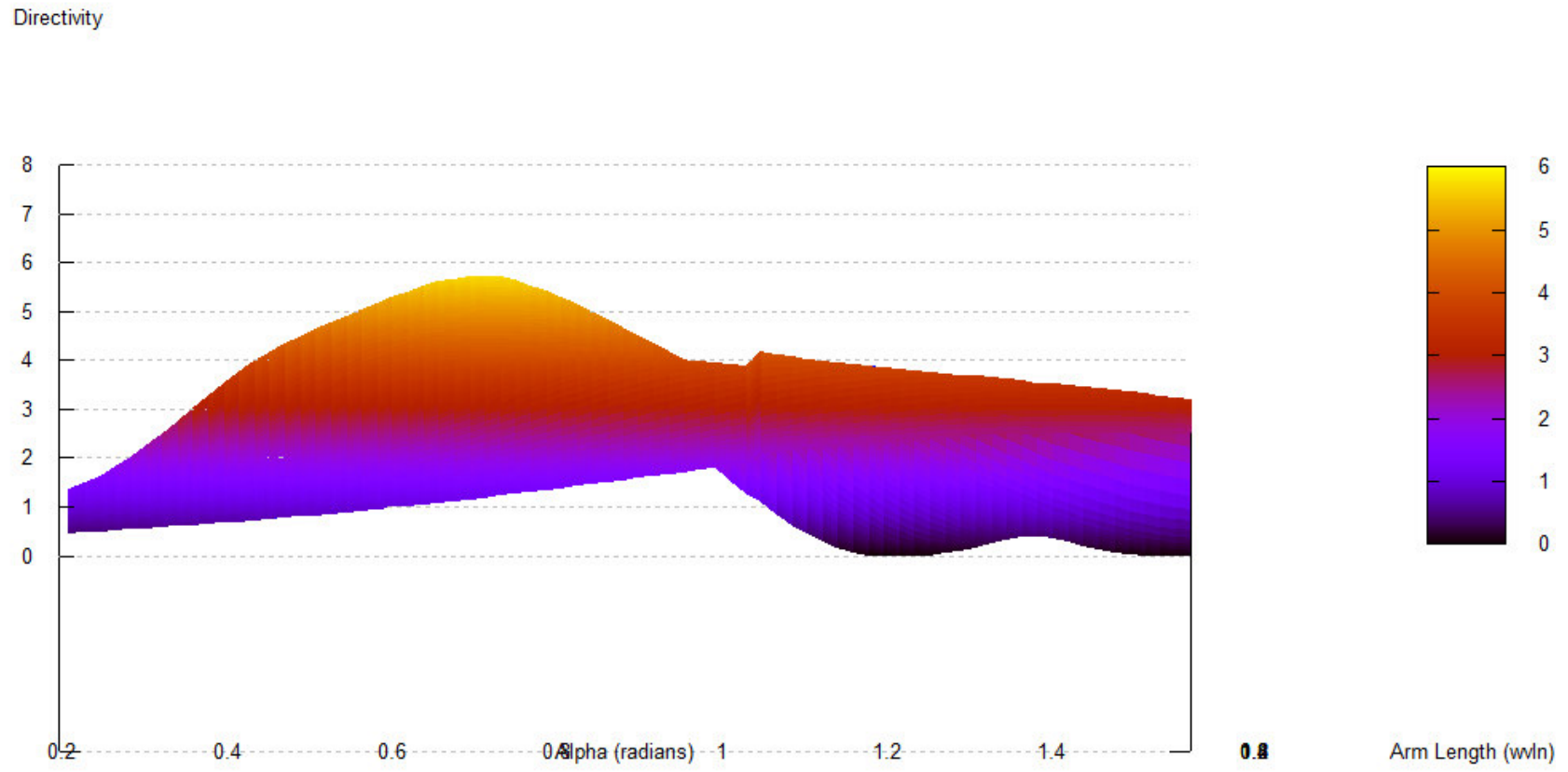

Fig. A4(c). Benchmark #4 projected onto ***α-D*** plane.

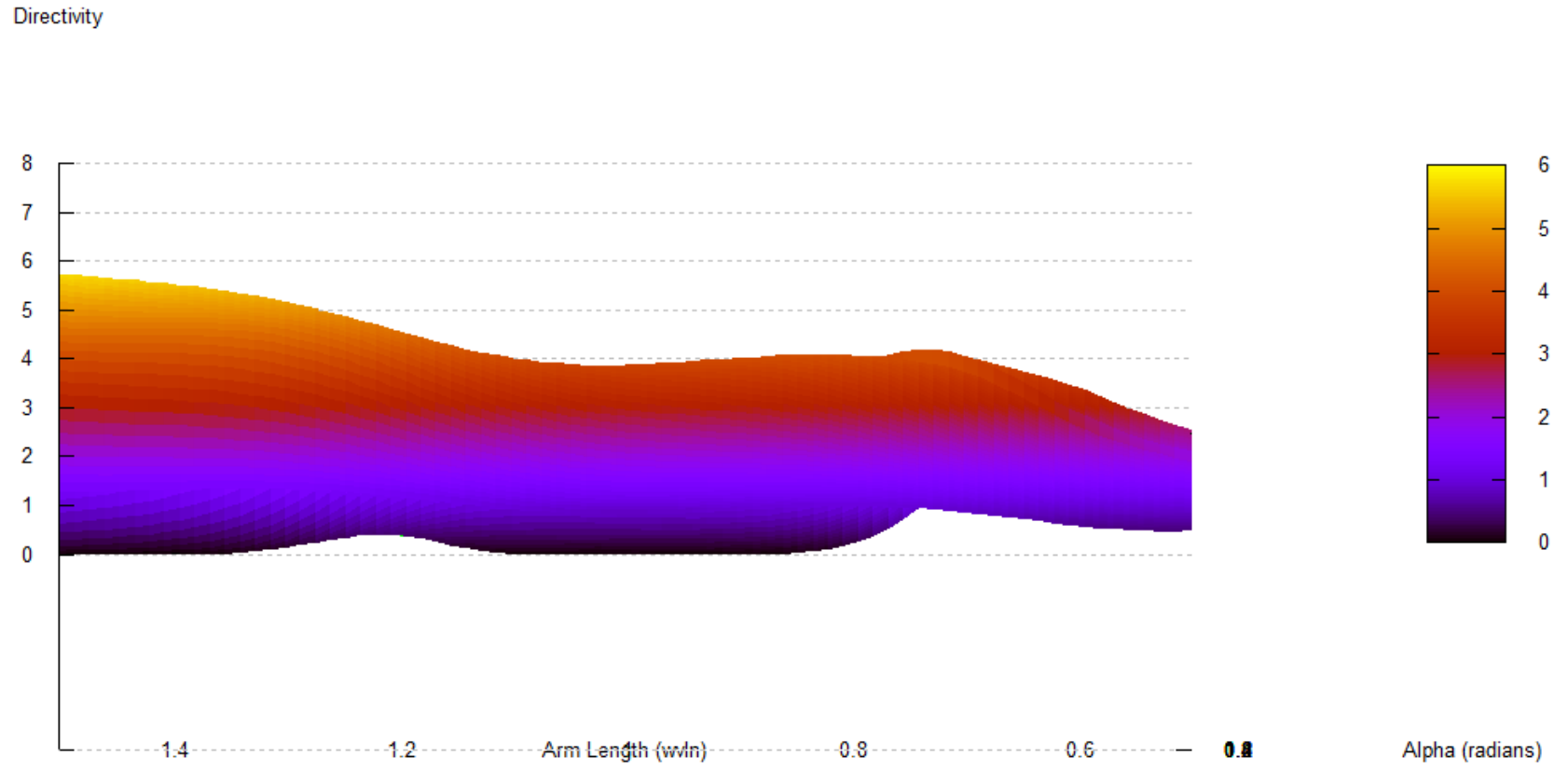

Fig. A4(d). Benchmark #4 projected onto ***L-D*** plane.

**PBM #4 NEC INPUT/OUTPUT FILES**

```
CM File: Input PBM4.NEC
CM Run ID 03-12-2009 18:43:01
CM R(p,1,j)= 1.5, R(p,2,j)= .834
CE
GW 1,5,0.,0.,-.01,0.,0.,.01,0.001
GW 2,150,0.,0.,.01,1.001159,0.,1.113531,0.001
GW 3,150,0.,0.,-.01,1.001159,0.,-1.113531,0.001
GE
EX 0,1,3,0,1.,0.
FR 0,1,0,0,299.79564,0.
RP 0,1,1,1001,90.,0.,0.,0.,1000.
XQ
EN
```

```
                           **********************************************************
                           *                                                        *
                           *  NUMERICAL ELECTROMAGNETICS CODE (NEC-4.1D, 23 MAR 05)  *
                           *                                                        *
                           *              RUN ID: 20090312 184301.906               *
                           *                                                        *
                           **********************************************************

                      **********************************************************
                          File: Input PBM4.NEC
                          Run ID 03-12-2009 18:43:01
                          R(p,1,j)= 1.5, R(p,2,j)= .834

                      **********************************************************

                                 - - - STRUCTURE SPECIFICATION - - -
 WIRE                                                                                 NO. OF    FIRST  LAST     TAG
 NO.        X1         Y1         Z1          X2         Y2         Z2      RADIUS   SEG.     SEG.   SEG.     NO.
    1    0.00000    0.00000   -0.01000     0.00000    0.00000    0.01000    0.00100      5        1     5       1
    2    0.00000    0.00000    0.01000     1.00116    0.00000    1.11353    0.00100    150        6   155       2
    3    0.00000    0.00000   -0.01000     1.00116    0.00000   -1.11353    0.00100    150      156   305       3

  TOTAL SEGMENTS USED=  305     NO. SEG. IN A SYMMETRIC CELL=  305     SYMMETRY FLAG=  0

                                 - - - - - - FREQUENCY - - - - - -
                                    FREQUENCY= 2.9980E+02 MHZ
                                    WAVELENGTH= 1.0000E+00 METERS

                                  - - - ANTENNA ENVIRONMENT - - -
                                          FREE SPACE

                                          - - - ANTENNA INPUT PARAMETERS - - -
  TAG   SEG.    VOLTAGE (VOLTS)         CURRENT (AMPS)         IMPEDANCE (OHMS)        ADMITTANCE (MHOS)      POWER
  NO.   NO.    REAL        IMAG.       REAL        IMAG.       REAL        IMAG.       REAL        IMAG.     (WATTS)
   1     3 1.00000E+00 0.00000E+00 4.80537E-04 7.09370E-04 6.54574E+02-9.66284E+02 4.80537E-04 7.09370E-04 2.40268E-04

                                       - - - POWER BUDGET - - -
                                          INPUT POWER   = 2.4027E-04 WATTS
                                          RADIATED POWER= 2.4027E-04 WATTS
                                          WIRE LOSS     = 0.0000E+00 WATTS
                                          EFFICIENCY    = 100.00 PERCENT

                                               - - - RADIATION PATTERNS - - -
                                                     RANGE= 1.000000E+03 METERS
                                                     EXP(-JKR)/R= 1.00000E-03 AT PHASE   0.00 DEGREES

 - - ANGLES - -          - POWER GAINS -        - - - POLARIZATION - - -    - - - E(THETA) - - -    - - - E(PHI) - - -
  THETA     PHI        VERT.   HOR.    TOTAL       AXIAL     TILT   SENSE     MAGNITUDE    PHASE      MAGNITUDE    PHASE
 DEGREES  DEGREES       DB      DB      DB         RATIO     DEG.               VOLTS/M    DEGREES      VOLTS/M    DEGREES
  90.00     0.00       6.81 -999.99    6.81    0.00000     0.00  LINEAR    2.62925E-04    10.31    0.00000E+00     0.00
```

**A.5 Benchmark #5**

**PBM #5 (10 elements) NEC INPUT/OUTPUT FILES**

```
CM File: Input PBM5_10.NEC
CM Run ID 03-12-2009 18:43:03
CM Nd= 9, p= 1, j= 0
CM R(p,i,j)= .99, i=1-9
CE
GW 1,49,0.,-4.705,0.,0.,-4.205,0.,0.001
GW 2,49,0.,-3.715,0.,0.,-3.215,0.,0.001
GW 3,49,0.,-2.725,0.,0.,-2.225,0.,0.001
GW 4,49,0.,-1.735,0.,0.,-1.235,0.,0.001
GW 5,49,0.,-.745,0.,0.,-.245,0.,0.001
GW 6,49,0.,.245,0.,0.,.745,0.,0.001
GW 7,49,0.,1.235,0.,0.,1.735,0.,0.001
GW 8,49,0.,2.225,0.,0.,2.725,0.,0.001
GW 9,49,0.,3.215,0.,0.,3.715,0.,0.001
GW 10,49,0.,4.205,0.,0.,4.705,0.,0.001
GE
EX 0,1,25,0,1.,0.
EX 0,2,25,0,1.,0.
EX 0,3,25,0,1.,0.
EX 0,4,25,0,1.,0.
EX 0,5,25,0,1.,0.
EX 0,6,25,0,1.,0.
EX 0,7,25,0,1.,0.
EX 0,8,25,0,1.,0.
EX 0,9,25,0,1.,0.
EX 0,10,25,0,1.,0.
FR 0,1,0,0,299.79564,0.
RP 0,1,1,1001,90.,0.,0.,0.,1000.
XQ
EN
                           **********************************************************
                           *                                                        *
                           *  NUMERICAL ELECTROMAGNETICS CODE (NEC-4.1D, 23 MAR 05)  *
                           *                                                        *
                           *              RUN ID: 20090312 184303.671               *
                           *                                                        *
                           **********************************************************
```

```
                    ********************************************************
                        File: Input PBM5_10.NEC
                        Run ID 03-12-2009 18:43:03
                        Nd= 9, p= 1, j= 0
                        R(p,i,j)= .99, i=1-9

                    ********************************************************

                               - - - STRUCTURE SPECIFICATION - - -
                                     WIRE
NO. OF    FIRST  LAST     TAG
  NO.       X1         Y1         Z1          X2         Y2          Z2      RADIUS   SEG.     SEG.   SEG.      NO.
    1    0.00000   -4.70500    0.00000    0.00000   -4.20500    0.00000    0.00100     49        1     49        1
    2    0.00000   -3.71500    0.00000    0.00000   -3.21500    0.00000    0.00100     49       50     98        2
    3    0.00000   -2.72500    0.00000    0.00000   -2.22500    0.00000    0.00100     49       99    147        3
    4    0.00000   -1.73500    0.00000    0.00000   -1.23500    0.00000    0.00100     49      148    196        4
    5    0.00000   -0.74500    0.00000    0.00000   -0.24500    0.00000    0.00100     49      197    245        5
    6    0.00000    0.24500    0.00000    0.00000    0.74500    0.00000    0.00100     49      246    294        6
    7    0.00000    1.23500    0.00000    0.00000    1.73500    0.00000    0.00100     49      295    343        7
    8    0.00000    2.22500    0.00000    0.00000    2.72500    0.00000    0.00100     49      344    392        8
    9    0.00000    3.21500    0.00000    0.00000    3.71500    0.00000    0.00100     49      393    441        9
   10    0.00000    4.20500    0.00000    0.00000    4.70500    0.00000    0.00100     49      442    490       10

   TOTAL SEGMENTS USED=  490      NO. SEG. IN A SYMMETRIC CELL=  490      SYMMETRY FLAG=  0

                               - - - - - - FREQUENCY - - - - - -
                                  FREQUENCY= 2.9980E+02 MHZ
                                  WAVELENGTH= 1.0000E+00 METERS

                                - - - ANTENNA ENVIRONMENT - - -
                                         FREE SPACE

                                        - - - ANTENNA INPUT PARAMETERS - - -
   TAG   SEG.    VOLTAGE (VOLTS)          CURRENT (AMPS)         IMPEDANCE (OHMS)        ADMITTANCE (MHOS)      POWER
   NO.   NO.     REAL         IMAG.       REAL         IMAG.       REAL         IMAG.       REAL         IMAG.     (WATTS)
    1     25 1.00000E+00 0.00000E+00 9.29297E-03-5.68215E-03 7.83251E+01 4.78916E+01 9.29297E-03-5.68215E-03 4.64649E-03
    2     74 1.00000E+00 0.00000E+00 9.61365E-03-6.12440E-03 7.39907E+01 4.71359E+01 9.61365E-03-6.12440E-03 4.80683E-03
    3    123 1.00000E+00 0.00000E+00 9.68606E-03-6.24915E-03 7.28978E+01 4.70315E+01 9.68606E-03-6.24915E-03 4.84303E-03
    4    172 1.00000E+00 0.00000E+00 9.71347E-03-6.29875E-03 7.24746E+01 4.69966E+01 9.71347E-03-6.29875E-03 4.85673E-03
    5    221 1.00000E+00 0.00000E+00 9.72396E-03-6.31782E-03 7.23131E+01 4.69830E+01 9.72396E-03-6.31782E-03 4.86198E-03
    6    270 1.00000E+00 0.00000E+00 9.72396E-03-6.31782E-03 7.23131E+01 4.69830E+01 9.72396E-03-6.31782E-03 4.86198E-03
    7    319 1.00000E+00 0.00000E+00 9.71347E-03-6.29875E-03 7.24746E+01 4.69966E+01 9.71347E-03-6.29875E-03 4.85673E-03
    8    368 1.00000E+00 0.00000E+00 9.68606E-03-6.24915E-03 7.28978E+01 4.70315E+01 9.68606E-03-6.24915E-03 4.84303E-03
    9    417 1.00000E+00 0.00000E+00 9.61365E-03-6.12440E-03 7.39907E+01 4.71359E+01 9.61365E-03-6.12440E-03 4.80683E-03
   10    466 1.00000E+00 0.00000E+00 9.29297E-03-5.68215E-03 7.83251E+01 4.78916E+01 9.29297E-03-5.68215E-03 4.64649E-03

                                     - - - POWER BUDGET - - -
                                        INPUT POWER   = 4.8030E-02 WATTS
                                        RADIATED POWER= 4.8030E-02 WATTS
                                        WIRE LOSS     = 0.0000E+00 WATTS
                                        EFFICIENCY    = 100.00 PERCENT

                                             - - - RADIATION PATTERNS - - -
                                                  RANGE= 1.000000E+03 METERS
                                                  EXP(-JKR)/R= 1.00000E-03 AT PHASE   0.00 DEGREES

 - - ANGLES - -          - POWER GAINS -         - - - POLARIZATION - - -     - - - E(THETA) - - -     - - - E(PHI) - - -
 THETA     PHI         VERT.   HOR.    TOTAL      AXIAL     TILT   SENSE     MAGNITUDE    PHASE       MAGNITUDE    PHASE
DEGREES  DEGREES        DB      DB      DB        RATIO     DEG.              VOLTS/M    DEGREES       VOLTS/M    DEGREES
  90.00     0.00     -999.99   12.81   12.81    0.00000   -90.00  LINEAR    0.00000E+00     0.00     7.41958E-03  -126.50
```

## APPENDIX B. CFO PSEUDOCODE

A CFO run requires seven user-specified parameters: $N_t$, $N_p$, $G$, $\alpha$, $\beta$, $\Delta t$, and $F_{rep}$, some of which are much more important than others. A possible eighth parameter is the initial probe acceleration, which is set to zero for all runs reported here and consequently not included in the parameter list. Constant values for $G$ and $\Delta t$ combine multiplicatively in a single coefficient, but there is the possibility that these parameters could be varied during a run. For the CFO runs reported here, $\alpha = \beta = 2$ because this value seems to provide good results for many test functions (it may, in fact, be a good default value). Note that in this paper there is no change in fitness value or average probe distance between steps 0 and 1 because the initial acceleration has been set to zero.

CFO is a simple algorithm that is easily implemented. The basic steps are: **(a)** compute initial probe positions and corresponding objective function fitnesses; assign initial accelerations; **(b)** successively compute each probe's new position using equation (2) §4.2 based on previously computed accelerations using equation (1) §4.2; **(c)** verify that each probe is located inside the decision space, making corrections as required; **(d)** update fitness at each new probe position; **(e)** compute accelerations for the next time step using new positions; and **(f)** loop over time steps or until a termination criterion is met.

The initial acceleration was set to zero for runs reported here; a non-zero value "flies" probes in the directions of the accelerations at step 0 which may or may not improve performance. Because CFO may "fly" probes outside the decision space domain into regions of unallowable solutions, any errant probes should be returned to the decision space. There are many possible retrieval schemes, and the one used here is simple and deterministic as described below and in §4.4. The repositioning factor, $F_{rep}$, which is used to return errant probes, in fact is one of CFO's most important parameters. When and how it should be changed are unanswered questions. The simple approach below seems to provide good results, and it was chosen solely for that reason.

Note that all CFO parameters presently are chosen on an empirical basis, that is, by "trial and error" because there is no methodology for specifying their values. This question remains one of the most pressing in CFO's further development. The parameter values that have been used provide good results for the purpose of demonstrating CFO's effectiveness and efficiency as an optimization metaheuristic, but these are not necessarily the best values.

CFO pseudocode is as follows (values for $N_{saved}$ , $TOL_{fit}$ , $F_{rep}^{init}$ and $\Delta F_{rep}$ are those used in this paper):

**(1)** ***<u>DATA STRUCTURES</u>***

(A) Create arrays $R(p,i,j)$ [position vectors], $A(p,i,j)$ [accelerations], $1 \le p \le N_p$ , $1 \le i \le N_d$ ,

$0 \le j \le N_t$ [probe #, coordinate #, time step #]

(B) Create fitness array $M(p,j) = f(R(p,i,j))$

(C) Create array of last $N_{saved}$ best fitnesses $M_{best}(q), 1 \le q \le N_{saved}$

**(2)** ***<u>INITIALIZATION</u>***

At time step $j = 0$:

(A)(i) <u>Uniform probes on each coordinate axis</u> (note: $\frac{N_p}{N_d}$, # probes/dimension, user-specified).

For $i = 1 \ to \ N_d$ , $n = 1 \ to \ \frac{N_p}{N_d}$:

$$p = n + \frac{(i-1)N_p}{N_d}, \ R(p,i,0) = x_i^{\min} + \frac{(n-1)\left(x_i^{\max} - x_i^{\min}\right)}{\frac{N_p}{N_d} - 1}.$$

(A)(ii) <u>Probes slightly off decision space diagonal</u>

For $p = 1 \ to \ N_p$, $i = 1 \ to \ N_d$ :

$$R(p,i,0) = x_i^{\min} + \frac{(n-1)\left(x_i^{\max} - x_i^{\min}\right)\left[N_d(p-1) + i - 1\right]}{N_p N_d - 1}.$$

(A)(iii) <u>Uniform 2D grid</u>

$$\Delta x_1 = \frac{x_1^{\max} - x_1^{\min}}{\frac{N_p}{N_d} - 1}, \ \Delta x_2 = \frac{x_2^{\max} - x_2^{\min}}{\frac{N_p}{N_d} - 1};$$ For $k = 1 \ to \ N_p / N_d$, $m = 1 \ to \ N_p / N_d$ :

$$p = \frac{N_p}{N_d}(k-1) + m\,,\;\; R(p,1,0) = x_1^{\min} + (k-1)\Delta x_1\,,\;\; R(p,2,0) = x_2^{\min} + (m-1)\Delta x_2$$

(A)(iv) Other user-specified probe distribution **[Note - This option used for this paper]**

(B) Initial acceleration

$$A(p,i,0) = 0, \quad 1 \le p \le N_p, \quad 1 \le i \le N_d$$

(C) Initial fitness

$$M(p,0) = f(R(p,i,0))\,,\; 1 \le p \le N_p,\; 1 \le i \le N_d\,,$$

$$Best\,Fitness = MAX\big(M(p,0), 1 \le p \le N_p\big)$$

(D) Saved Fitnesses & Fitness Tolerance

$$N_{saved} = 5\,,\; N_{sat} = 3\,,\; TOL_{fit} = 0.0005$$

(E) Initial Reposition Factor & Increment

$$F_{rep}^{init} = 0.5, \Delta F_{rep} = 0.005, F_{rep} = F_{rep}^{init}$$

**(3)** ***TIME LOOP*** $(0 \le j \le N_t)$

(A) *Saved Fitness Array Index:* $s = j\ MOD\ N_{saved}$; $If\ s = 0$ then $s = N_{saved}$

**IMPORTANT NOTE: Saved fitnesses are not necessarily sequential as a result of how *s* is calculated, which affects when $F_{rep}$ is incremented See discussion in §4.4.**

*New Probe Positions*

For $p = 1\ to\ N_p$, $i = 1\ to\ N_d$ :

(a) $$R(p,i,j) = R(p,i,j-1) + \frac{1}{2}A(p,i,j-1)\Delta t^2, \quad \Delta t^2 = 1$$

(b) *Retrieve errant probes as required* (Note: For Repositioning Factor, $0 \le F_{rep} \le 1$, see Formato [2-4])

$$If\ R(p,i,j) < x_i^{\min} \text{ then } R(p,i,j) = x_i^{\min} + F_{rep} \cdot (R(p,i,j-1) - x_i^{\min})$$

$$If\ R(p,i,j) > x_i^{\max} \text{ then } R(p,i,j) = x_i^{\max} - F_{rep} \cdot (x_i^{\max} - R(p,i,j-1))$$

(B) *Update Fitness Matrix*

For $p = 1\ to\ N_p$:

$$M(p,j) = f(R(p,i,j))$$

$$If\ M(p,j) \ge Best\,Fitness \text{ then } Best\,Fitness = M(p,j)\ \&\ M_{best}(s) = M(p,j)$$

(C) *Update Reposition Factor*

$$If\ \left| M_{best}(N_{saved}) - \frac{1}{N_{sat}} \sum_{q=s}^{N_{saved}} M_{best}(q) \right| \le TOL_{fit}\ , s = N_{saved} - N_{sat} + 1, \text{ then}$$

$$F_{rep} = F_{rep} + \Delta F_{rep}$$

$$If\ F_{rep} \ge 1 \text{ then } F_{rep} = F_{rep}^{init}$$

(D) <u>Update Accelerations</u>

For $p = 1\ to\ N_p, i = 1\ to\ N_d$ :

$$A(p,i,j) = G\sum_{\substack{k=1\\k\neq p}}^{N_p} U\big(M(k,j) - M(p,j)\big) \times$$

$$\big(M(k,j) - M(p,j)\big)^{\alpha} \times \frac{R(k,i,j) - R(p,i,j)}{\left|\vec{R}_j^k - \vec{R}_j^p\right|^{\beta}},$$

where $\left|\vec{R}_j^k - \vec{R}_j^p\right| = \sqrt{\sum_{m=1}^{N_d}\big(R(k,m,j) - R(p,m,j)\big)^2}$

(E) Increment $j \to j+1$, and repeat from (3)(A) until $j = N_t$ or other stopping criterion has been met.

## APPENDIX C. CFO SOURCE CODE (INCLUDING MODIFICATIONS THROUGH 3 DEC. 2009)

```
'NOTE: LINES BEGINNING WITH ' ARE COMMENT LINES.
'Program 'CFO_11-26-09.BAS' compiled with
'Power Basic/Windows Compiler 9.0
'LAST MOD 12-03-2009 ~0833 HRS EST
'NOTE: ALL PBM FUNCTIONS HAVE WIRE RADIUS SET TO 0.00001LAMBDA
'==============================================================
'THIS PROGRAM IMPLEMENTS A SIMPLE VERSION OF "CENTRAL
'FORCE OPTIMIZATION."  IT IS DISTRIBUTED FREE OF CHARGE
'TO INCREASE AWARENESS OF CFO AND TO ENCOURAGE EXPERI-
'MENTATION WITH THE ALGORITHM.
'CFO IS A MULTIDIMENSIONAL SEARCH AND OPTIMIZATION
'ALGORITHM THAT LOCATES THE GLOBAL MAXIMA OF A FUNCTION.
'UNLIKE MOST OTHER ALGORITHMS, CFO IS COMPLETELY DETERMIN-
'ISTIC, SO THAT EVERY RUN WITH THE SAME SETUP PRODUCES
'THE SAME RESULTS.
'Please email questions, comments, and significant
'results to: CFO_questions@yahoo.com.  Thanks!
'(c) 2006-2009 Richard A. Formato
'ALL RIGHTS RESERVED WORLDWIDE
'THIS PROGRAM IS FREEWARE.  IT MAY BE COPIED AND
'DISTRIBUTED WITHOUT LIMITATION AS LONG AS THIS
'COPYRIGHT NOTICE AND THE GNUPLOT AND REFERENCE
'INFORMATION BELOW ARE INCLUDED WITHOUT MODIFICATION,
'AND AS LONG AS NO FEE OR COMPENSATION IS CHARGED,
'INCLUDING "TIE-IN" OR "BUNDLING" FEES CHARGED FOR
'OTHER PRODUCTS.
'=======================================================
'THIS PROGRAM REQUIRES wgnuplot.exe TO DISPLAY PLOTS.
'Gnuplot is a copyrighted freeware plotting program
'available at http://www.gnuplot.info/index.html.
'=======================================================
'REFERENCES
'----------
'1. "Central Force Optimization: A New Metaheuristic with Applications in Applied Electromagnetics,"
'    Progress in Electromagnetics Research, PIER 77, 425-491, 2007 (online).
'Abstract - Central Force Optimization (CFO) is a new deterministic multi-dimensional search
'metaheuristic based on the metaphor of gravitational kinematics.  It models "probes" that "fly"
'through the decision space by analogy to masses moving under the influence of gravity.  Equations
'are developed for the probes' positions and accelerations using the analogy of particle motion
'in a gravitational field.  In the physical universe, objects traveling through three-dimensional
'space become trapped in close orbits around highly gravitating masses, which is analogous to
'locating the maximum value of an objective function.  In the CFO metaphor, "mass" is a user-defined
'function of the value of the objective function to be maximized.  CFO is readily implemented in a
'compact computer program, and sample pseudocode is presented.  As tests of CFO's effectiveness,
'an equalizer is designed for the well-known Fano load, and a 32-element linear array is synthesized.
'CFO results are compared to several other optimization methods.
'2. "Central Force Optimization: A New Nature Inspired Computational Framework for Multidimensional
'    Search and Optimization," Studies in Computational Intelligence (SCI), vol. 129, 221-238 (2008),
'    www.springerlink.com, Springer-Verlag Berlin Heidelberg 2008.
'Abstract - This paper presents Central Force Optimization, a novel, nature inspired, deterministic
'search metaheuristic for constrained multi-dimensional optimization.  CFO is based on the metaphor
'of gravitational kinematics.  Equations are presented for the positions and accelerations experienced
'by "probes" that "fly" through the decision space by analogy to masses moving under the influence of
'gravity.  In the physical universe, probe satellites become trapped in close orbits around highly
'gravitating masses.  In the CFO analogy, "mass" corresponds to a user-defined function of the value
'of an objective function to be maximized.  CFO is a simple algorithm that is easily implemented in
'a compact computer program.  A typical CFO implementation is applied to several test functions.
'CFO exhibits very good performance, suggesting that it merits further study.
'3. "Central Force Optimisation: A New Gradient-Like Metaheuristic for Multidimensional Search and
'    Optimisation," International Journal of Bio-Inspired Computation, vol. 1, no. 4, 217-238 (2009),
'Abstract: This paper introduces Central Force Optimization, a novel, Nature-inspired, deterministic
```

```
'search metaheuristic for constrained multidimensional optimization in highly multimodal, smooth, or
'discontinuous decision spaces.  CFO is based on the metaphor of gravitational kinematics.  The
'algorithm searches a decision space by “flying” its “probes” through the space by analogy to masses
'moving through physical space under the influence of gravity.  Equations are developed for the probes’
'positions and accelerations using the gravitational metaphor.  Small objects in our Universe can become
'trapped in close orbits around highly gravitating masses.  In “CFO space” probes are attracted to “masses”
'created by a user-defined function of the value of an objective function to be maximized.  CFO may be
'thought of in terms of a vector “force field” or, loosely, as a “generalized gradient” methodology
'because the force of gravity can be computed as the gradient of a scalar potential.  The CFO algorithm
'is simple and easily implemented in a compact computer program.  Its effectiveness is demonstrated by
'running CFO against several widely used benchmark functions.  The algorithm exhibits very good performance,
'suggesting that it merits further study.
'4. "Central Force Optimization: A New Gradient-Like Optimization Metaheuristic," OPSEARCH, Journal
'    of the Operational Research Society of India, vol. 46, no. 1, 25-51, 2009, Springer India.
'Abstract:   This paper introduces Central Force Optimization as a new, Nature-inspired metaheuristic
'for multidimensional search and optimization based on the metaphor of gravitational kinematics.  CFO is a
'“gradient-like” deterministic algorithm that explores a decision space by “flying” a group of “probes”
'whose trajectories are governed by equations analogous to the equations of gravitational motion in the
'physical Universe.  This paper suggests the possibility of creating a new “hyperspace directional derivative”
'using the Unit Step function t7o create positive-definite “masses” in “CFO space.”  A simple CFO implementation
'is tested against several recognized benchmark functions with excellent results, suggesting that CFO merits
'further investigation.
'5. "Synthesis of Antenna Arrays Using Central Force Optimization," Mosharaka International Conference
'    on Communications, Computers and Applications, MIC-CPE 2009.
'Abstract: Central force optimization (CFO) technique is a new deterministic multi-dimensional
'search metaheuristic based on an analogy to classical particle kinematics in a gravitational
'field. CFO is a simple technique and is still in its infancy. To enhance its global search
'ability while keeping its simplicity, a new selection part is introduced in this paper. CFO
'is briefly presented and applied in the design of linear antenna array and the modified CFO
'is applied to the design of circular array. The results are compared with those obtained using
'other evolutionary optimization techniques.
'6. "Central Force Optimization and NEOs - First Cousins?" Journal of Multiple Valued Logic and Soft
'    Computing (to be published)
'Abstract: Central Force Optimization is a new deterministic multidimensional search and optimization algorithm
'based on the metaphor of gravitational kinematics.  This paper describes CFO and suggests some possible
'directions for its future development.  Because CFO is deterministic, it is more computationally efficient
'than stochastic algorithms and may lend itself well to “parameter tuning” implementations.  But, like
'all deterministic algorithms, CFO is prone to local trapping.  Oscillation in CFO’s Davg curve appears to
'be a reliable harbinger of trapping.  And there seems to be a reasonable basis for believing that
'trapping can be handled deterministically using the theory of gravitationally trapped Near Earth Objects.
'Deterministic mitigation of local trapping would be a major step forward in optimization theory.  Finally,
'CFO may be thought of as a “gradient-like” algorithm utilizing the Unit Step function as a critical element,
'and it is suggested that a useful, new derivative-like mathematical construct might be defined based on the Unit
'Step.
'7. "Array Synthesis and Antenna Benchmark Performance Using Central Force Optimization," IET (UK)
'    (in review)
'Abstract: Central force optimization (CFO) is a new deterministic multi-dimensional search metaheuristic
'based on an analogy to classical particle kinematics in a gravitational field. CFO is a simple technique
'that is still in its infancy.  This paper evaluates CFO’s performance and provides further examples of its
'effectiveness by applying it to a set of “real world” antenna benchmarks and to pattern synthesis for linear
'and circular array antennas.  A new selection scheme is introduced that enhances CFO’s global search ability
'while maintaining its simplicity.  The improved CFO algorithm is applied to the design of a circular array
'with very good results.  CFO’s performance on the antenna benchmarks and the synthesis problems is compared
'to that of other evolutionary optimization techniques.
'==================================================================================================
#COMPILE EXE
#DIM ALL
%USEMACROS = 1
#INCLUDE "Win32API.inc"
DEFEXT A-Z
'------ EQUATES -----
%IDC_FRAME1     = 101
%IDC_FRAME2     = 102
%IDC_Function_Number1  = 121
%IDC_Function_Number2  = 122
%IDC_Function_Number3  = 123
%IDC_Function_Number4  = 124
%IDC_Function_Number5  = 125
%IDC_Function_Number6  = 126
%IDC_Function_Number7  = 127
%IDC_Function_Number8  = 128
%IDC_Function_Number9  = 129
%IDC_Function_Number10 = 130
%IDC_Function_Number11 = 131
%IDC_Function_Number12 = 132
%IDC_Function_Number13 = 133
%IDC_Function_Number14 = 134
%IDC_Function_Number15 = 135
%IDC_Function_Number16 = 136
%IDC_Function_Number17 = 137
%IDC_Function_Number18 = 138
%IDC_Function_Number19 = 139
%IDC_Function_Number20 = 140
%IDC_Function_Number21 = 141
%IDC_Function_Number22 = 142
%IDC_Function_Number23 = 143
%IDC_Function_Number24 = 144
%IDC_Function_Number25 = 145
%IDC_Function_Number26 = 146
%IDC_Function_Number27 = 147
%IDC_Function_Number28 = 148
%IDC_Function_Number29 = 149
%IDC_Function_Number30 = 150
%IDC_Function_Number31 = 151
%IDC_Function_Number32 = 152
```

```
%IDC_Function_Number33 = 153
%IDC_Function_Number34 = 154
%IDC_Function_Number35 = 155
%IDC_Function_Number36 = 156
%IDC_Function_Number37 = 157
%IDC_Function_Number38 = 158
%IDC_Function_Number39 = 159
%IDC_Function_Number40 = 160
%IDC_Function_Number41 = 161
%IDC_Function_Number42 = 162
%IDC_Function_Number43 = 163
%IDC_Function_Number44 = 164
%IDC_Function_Number45 = 165
%IDC_Function_Number46 = 166
%IDC_Function_Number47 = 167
%IDC_Function_Number48 = 168
%IDC_Function_Number49 = 169
%IDC_Function_Number50 = 170
'------------------------------ GLOBAL CONSTANTS & SYMBOLS ---------------------------
GLOBAL Aij() AS EXT 'array for Shekel's Foxholes function
GLOBAL EulerConst, Pi, Pi2, Pi4, TwoPi, FourPi, e, Root2 AS EXT 'mathematical constants
GLOBAL Alphabet$, Digits$, RunID$  'upper/lower case alphabet, digits 0-9 & Run ID
GLOBAL Quote$, SpecialCharacters$  'quotation mark & special symbols
GLOBAL Mu0, Eps0, c, eta0 AS EXT   'E&M constants
GLOBAL Rad2Deg, Deg2Rad, Feet2Meters, Meters2Feet, Inches2Meters, Meters2Inches AS EXT 'conversion factors
GLOBAL Miles2Meters, Meters2Miles, NautMi2Meters, Meters2NautMi AS EXT                 'conversion factors
GLOBAL ScreenWidth&, ScreenHeight& 'screen width & height
GLOBAL xOffset&, yOffset&          'offsets for probe plot windows
GLOBAL FunctionNumber%
GLOBAL AddNoiseToPBM2$
'----------------------------- TEST FUNCTION DECLARATIONS --------------------------------
DECLARE FUNCTION F1(R(),Nd%,p%,j&)              'F1 (n-D)
DECLARE FUNCTION F2(R(),Nd%,p%,j&)              'F2(n-D)
DECLARE FUNCTION F3(R(),Nd%,p%,j&)              'F3 (n-D)
DECLARE FUNCTION F4(R(),Nd%,p%,j&)              'F4 (n-D)
DECLARE FUNCTION F5(R(),Nd%,p%,j&)              'F5 (n-D)
DECLARE FUNCTION F6(R(),Nd%,p%,j&)              'F6 (n-D)
DECLARE FUNCTION F7(R(),Nd%,p%,j&)              'F7 (n-D)
DECLARE FUNCTION F8(R(),Nd%,p%,j&)              'F8 (n-D)
DECLARE FUNCTION F9(R(),Nd%,p%,j&)              'F9 (n-D)
DECLARE FUNCTION F10(R(),Nd%,p%,j&)             'F10 (n-D)
DECLARE FUNCTION F11(R(),Nd%,p%,j&)             'F11 (n-D)
DECLARE FUNCTION F12(R(),Nd%,p%,j&)             'F12 (n-D)
DECLARE FUNCTION F13(R(),Nd%,p%,j&)             'F13 (n-D)
DECLARE FUNCTION F14(R(),Nd%,p%,j&)             'F14 (n-D)
DECLARE FUNCTION F15(R(),Nd%,p%,j&)             'F15 (n-D)
DECLARE FUNCTION F16(R(),Nd%,p%,j&)             'F16 (n-D)
DECLARE FUNCTION F17(R(),Nd%,p%,j&)             'F17 (n-D)
DECLARE FUNCTION F18(R(),Nd%,p%,j&)             'F18 (n-D)
DECLARE FUNCTION F19(R(),Nd%,p%,j&)             'F19 (n-D)
DECLARE FUNCTION F20(R(),Nd%,p%,j&)             'F20 (n-D)
DECLARE FUNCTION F21(R(),Nd%,p%,j&)             'F21 (n-D)
DECLARE FUNCTION F22(R(),Nd%,p%,j&)             'F22 (n-D)
DECLARE FUNCTION F23(R(),Nd%,p%,j&)             'F23 (n-D)
DECLARE FUNCTION F24(R(),Nd%,p%,j&)             'F24 (n-D)
DECLARE FUNCTION F25(R(),Nd%,p%,j&)             'F25 (n-D)
DECLARE FUNCTION F26(R(),Nd%,p%,j&)             'F26 (n-D)
DECLARE FUNCTION F27(R(),Nd%,p%,j&)             'F27 (n-D)
DECLARE FUNCTION ParrottF4(R(),Nd%,p%,j&)       'Parrott F4 (1-D)
DECLARE FUNCTION SGO(R(),Nd%,p%,j&)             'SGO Function (2-D)
DECLARE FUNCTION GoldsteinPrice(R(),Nd%,p%,j&)  'Goldstein-Price Function (2-D)
DECLARE FUNCTION StepFunction(R(),Nd%,p%,j&)    'Step Function (n-D)
DECLARE FUNCTION Schwefel226(R(),Nd%,p%,j&)     'Schwefel Prob. 2.26 (n-D)
DECLARE FUNCTION Colville(R(),Nd%,p%,j&)        'Colville Function (4-D)
DECLARE FUNCTION Griewank(R(),Nd%,p%,j&)        'Griewank (n-D)
DECLARE FUNCTION Himmelblau(R(),Nd%,p%,j&)      'Himmelblau (2-D)
DECLARE FUNCTION PBM_1(R(),Nd%,p%,j&)           'PBM Benchmark #1
DECLARE FUNCTION PBM_2(R(),Nd%,p%,j&)           'PBM Benchmark #2
DECLARE FUNCTION PBM_3(R(),Nd%,p%,j&)           'PBM Benchmark #3
DECLARE FUNCTION PBM_4(R(),Nd%,p%,j&)           'PBM Benchmark #4
DECLARE FUNCTION PBM_5(R(),Nd%,p%,j&)           'PBM Benchmark #5
'----------------------------------- SUB DECLARATIONS ------------------------------------
DECLARE SUB GetTestFunctionNumber(FunctionName$)
DECLARE SUB FillArrayAij
DECLARE SUB ResetDecisionSpaceBoundaries(Nd%,XiMin(),XiMax(),StartingXiMin(),StartingXiMax())
DECLARE SUB Plot3DbestProbeTrajectories(NumTrajectories%,M(),R(),XiMin(),XiMax(),Np%,Nd%,LastStep&,FunctionName$)
DECLARE SUB Plot2DbestProbeTrajectories(NumTrajectories%,M(),R(),XiMin(),XiMax(),Np%,Nd%,LastStep&,FunctionName$)
DECLARE SUB Plot2DindividualProbeTrajectories(NumTrajectories%,M(),R(),XiMin(),XiMax(),Np%,Nd%,LastStep&,FunctionName$)
DECLARE SUB SelectTestFunction(FunctionName$)
DECLARE SUB
Show2Dprobes(R(),Np%,Nt&,j&,XiMin(),XiMax(),Frep,BestFitness,BestProbeNumber%,BestTimeStep&,FunctionName$,RepositionFac
tor$,Gamma)
DECLARE SUB StatusWindow(FunctionName$,StatusWindowHandle???)
DECLARE SUB
DisplayRunParameters(FunctionName$,Nd%,Np%,Nt&,G,DeltaT,Alpha,Beta,Frep,R(),A(),M(),PlaceInitialProbes$,InitialAccelera
tion$,RepositionFactor$,RunCFO$,ShrinkDS$,CheckForEarlyTermination$)
DECLARE SUB GetBestFitness(M(),Np%,StepNumber&,BestFitness,BestProbeNumber%,BestTimeStep&)
DECLARE SUB
Tabulate1DprobeCoordinates(Max1DprobesPlotted%,Nd%,Np%,LastStep&,G,DeltaT,Alpha,Beta,Frep,R(),M(),PlaceInitialProbes$,I
nitialAcceleration$,RepositionFactor$,FunctionName$,Gamma)
DECLARE SUB
GetPlotAnnotation(PlotAnnotation$,Nd%,Np%,Nt&,G,DeltaT,Alpha,Beta,Frep,M(),PlaceInitialProbes$,InitialAcceleration$,Rep
ositionFactor$,FunctionName$,Gamma)
```

```
DECLARE SUB
ChangeRunParameters(NumProbesPerDimension%,Np%,Nd%,Nt&,G,Alpha,Beta,DeltaT,Frep,PlaceInitialProbes$,InitialAcceleration
$,RepositionFactor$,FunctionName$)
DECLARE SUB CLEANUP
DECLARE SUB DisplayBestFitness(Np%,Nd%,LastStep&,M(),R(),BestFitnessProbeNumber%,BestFitnessTimeStep&,FunctionName$)
DECLARE SUB
Plot1DprobePositions(Max1DprobesPlotted%,Nd%,Np%,LastStep&,G,DeltaT,Alpha,Beta,Frep,R(),M(),PlaceInitialProbes$,Initial
Acceleration$,RepositionFactor$,FunctionName$,Gamma)
DECLARE SUB DisplayMmatrix(Np%,Nt&,M())
DECLARE SUB DisplayMmatrixThisTimeStep(Np%,j&,M())
DECLARE SUB DisplayAmatrix(Np%,Nd%,Nt&,A())
DECLARE SUB DisplayAmatrixThisTimeStep(Np%,Nd%,j&,A())
DECLARE SUB DisplayRmatrix(Np%,Nd%,Nt&,R())
DECLARE SUB DisplayRmatrixThisTimeStep(Np%,Nd%,j&,R())
DECLARE SUB DisplayXiMinMax(Nd%,XiMin(),XiMax())
DECLARE SUB
DisplayRunParameters2(FunctionName$,Nd%,Np%,Nt&,G,DeltaT,Alpha,Beta,Frep,PlaceInitialProbes$,InitialAcceleration$,Repos
itionFactor$)
DECLARE SUB
PlotBestProbeVsTimeStep(Nd%,Np%,LastStep&,G,DeltaT,Alpha,Beta,Frep,M(),PlaceInitialProbes$,InitialAcceleration$,Reposit
ionFactor$,FunctionName$,Gamma)
DECLARE SUB
PlotBestFitnessEvolution(Nd%,Np%,LastStep&,G,DeltaT,Alpha,Beta,Frep,M(),PlaceInitialProbes$,InitialAcceleration$,Reposi
tionFactor$,FunctionName$,Gamma)
DECLARE SUB
PlotAverageDistance(Nd%,Np%,LastStep&,G,DeltaT,Alpha,Beta,Frep,M(),PlaceInitialProbes$,InitialAcceleration$,RepositionF
actor$,FunctionName$,R(),DiagLength,Gamma)
DECLARE SUB Plot2Dfunction(FunctionName$,XiMin(),XiMax(),R())
DECLARE SUB Plot1Dfunction(FunctionName$,XiMin(),XiMax(),R())
DECLARE SUB
GetFunctionRunParameters(FunctionName$,Nd%,Np%,Nt&,G,DeltaT,Alpha,Beta,Frep,R(),A(),M(),XiMin(),XiMax(),StartingXiMin()
,StartingXiMax(),_
                                     DiagLength,PlaceInitialProbes$,InitialAcceleration$,RepositionFactor$)
DECLARE SUB InitialProbeDistribution(Np%,Nd%,Nt&,XiMin(),XiMax(),R(),PlaceInitialProbes$,Gamma)
DECLARE SUB
InitialProbeAccelerations(Np%,Nd%,A(),InitialAcceleration$,MaxInitialRandomAcceleration,MaxInitialFixedAcceleration)
DECLARE SUB RetrieveErrantProbes(Np%,Nd%,j&,XiMin(),XiMax(),R(),M(),RepositionFactor$,Frep)
DECLARE SUB
CFO(Nd%,Np%,Nt&,G,DeltaT,Alpha,Beta,Frep,R(),A(),M(),XiMin(),XiMax(),DiagLength,PlaceInitialProbes$,InitialAcceleration
$,_

RepositionFactor$,FunctionName$,LastStep&,CheckForEarlyTermination$,BestFitnessThisRun,RunNumber%,NumRuns%,Gamma,BestOv
erallFitness,ShrinkDS$)
DECLARE SUB ThreeDplot(PlotFileName$,PlotTitle$,Annotation$,xCoord$,yCoord$,zCoord$, _
                       XaxisLabel$,YaxisLabel$,ZaxisLabel$,zMin$,zMax$,GnuPlotEXE$,A$)
DECLARE SUB ThreeDplot2(PlotFileName$,PlotTitle$,Annotation$,xCoord$,yCoord$,zCoord$,XaxisLabel$,_
                        YaxisLabel$,ZaxisLabel$,zMin$,zMax$,GnuPlotEXE$,A$,xStart$,xStop$,yStart$,yStop$)
DECLARE SUB TwoDplot(PlotFileName$,PlotTitle$,xCoord$,yCoord$,XaxisLabel$,YaxisLabel$, _
                     LogXaxis$,LogYaxis$,xMin$,xMax$,yMin$,yMax$,xTics$,yTics$,GnuPlotEXE$,LineType$,Annotation$)
DECLARE SUB
TwoDplot2Curves(PlotFileName1$,PlotFileName2$,PlotTitle$,Annotation$,xCoord$,yCoord$,XaxisLabel$,YaxisLabel$, _
                            LogXaxis$,LogYaxis$,xMin$,xMax$,yMin$,yMax$,xTics$,yTics$,GnuPlotEXE$,LineSize)
DECLARE SUB
TwoDplot3curves(NumCurves%,PlotFileName1$,PlotFileName2$,PlotFileName3$,PlotTitle$,Annotation$,xCoord$,yCoord$,XaxisLab
el$,YaxisLabel$, _
                            LogXaxis$,LogYaxis$,xMin$,xMax$,yMin$,yMax$,xTics$,yTics$,GnuPlotEXE$)
DECLARE SUB CreateGNUplotINIfile(PlotWindowULC_X%,PlotWindowULC_Y%,PlotWindowWidth%,PlotWindowHeight%)
DECLARE SUB Delay(NumSecs)
DECLARE SUB MathematicalConstants
DECLARE SUB AlphabetAndDigits
DECLARE SUB SpecialSymbols
DECLARE SUB EMconstants
DECLARE SUB ConversionFactors
DECLARE SUB ShowConstants
'------ FUNCTION DECLARATIONS -------
DECLARE CALLBACK FUNCTION DlgProc
DECLARE FUNCTION HasFITNESSsaturated$(Nsteps&,j&,Np%,Nd%,M(),R(),DiagLength)
DECLARE FUNCTION HasDAVGsaturated$(Nsteps&,j&,Np%,Nd%,M(),R(),DiagLength)
DECLARE FUNCTION OscillationInDavg$(j&,Np%,Nd%,M(),R(),DiagLength)
DECLARE FUNCTION DavgThisStep(j&,Np%,Nd%,M(),R(),DiagLength)
DECLARE FUNCTION NoSpaces$(X,NumDigits%)
DECLARE FUNCTION FormatFP$(X,Ndigits%)
DECLARE FUNCTION FormatInteger$(M%)
DECLARE FUNCTION TerminateNowForSaturation$(j&,Nd%,Np%,Nt&,G,DeltaT,Alpha,Beta,R(),A(),M())
DECLARE FUNCTION MagVector(V(),N%)
DECLARE FUNCTION UniformDeviate(u&&)
DECLARE FUNCTION RandomNum(a,b)
DECLARE FUNCTION GaussianDeviate(Mu,Sigma)
DECLARE FUNCTION UnitStep(X)
DECLARE FUNCTION Fibonacci&&(N%)
DECLARE FUNCTION ObjectiveFunction(R(),Nd%,p%,j&,FunctionName$)
DECLARE FUNCTION UnitStep(X)
'====================================================================================================
'----- MAIN PROGRAM ------
FUNCTION PBMAIN () AS LONG
'   ------ CFO Parameters -----
    LOCAL Nd%, Np%, Nt&
    LOCAL G, DeltaT, Alpha, Beta, Frep AS EXT
    LOCAL PlaceInitialProbes$, InitialAcceleration$, RepositionFactor$
    LOCAL R(), A(), M() AS EXT    'position, acceleration & fitness matrices
    LOCAL XiMin(), XiMax(), StartingXiMin(), StartingXiMax() AS EXT 'decision space boundaries
    LOCAL FunctionName$           'name of objective function
    LOCAL DiagLength AS EXT
'   ------------ Miscellaneuous Setup Parameters --------------
    LOCAL StartingG, StartingDeltaT, StartingAlpha, StartingBeta, StartingFrep AS EXT
```

```
    LOCAL N%, i%, YN&
    LOCAL A$
    LOCAL NumRuns%, RunNumber%, BestRunNumber%, TotalFunctionEvaluations&&
    LOCAL Gamma, StartingGamma, StoppingGamma, BestFitnessThisRun, BestOverallFitness AS EXT
    LOCAL NumTrajectories%
    LOCAL Max1DprobesPlotted%
    LOCAL BestFitnessProbeNumber%, BestFitnessTimeStep&
    LOCAL RunCFO$
    LOCAL StatusWindowHandle???
    LOCAL LastStep&
    LOCAL CheckForEarlyTermination$ 'early termination checking? (YES/NO)
    LOCAL ShrinkDS$ 'adaptively shrink DS? (YES/NO)
'   -------------------- Global Constants ---------------------
    REDIM Aij(1 TO 2, 1 TO 25) '(GLOBAL array for Shekel's Foxholes function)
    CALL FillArrayAij
    CALL MathematicalConstants 'NOTE: Calling order is important!!
    CALL AlphabetAndDigits
    CALL SpecialSymbols
    CALL EMconstants
    CALL ConversionFactors        ': CALL ShowConstants 'to verify constants have been set
    xOffset& = 20 : yOffset& = 20 'offsets for successive probe position plots
'   -------------------------- General Setup ---------------------------
    RANDOMIZE TIMER  'seed random number generator with program start time
    DESKTOP GET SIZE TO ScreenWidth&, ScreenHeight&  'get screen size (global variables)
    IF DIR$("wgnuplot.exe") = "" THEN
        MSGBOX("WARNING!  'wgnuplot.exe' not found.  Run terminated.") : EXIT FUNCTION
    END IF
'   ------------------- NEC Files Required for PBM Antenna Benchmarks -------------------
    IF DIR$("n41_2k1.exe") = "" THEN
            MSGBOX("WARNING!  'n41_2k1.exe' not found.  Run terminated.") : EXIT FUNCTION
    END IF
    N% = FREEFILE : OPEN "FILE_MSG.DAT" FOR OUTPUT AS #N% : PRINT #N%, "YES"     : CLOSE #N%
    N% = FREEFILE : OPEN "FILE_MSG.DAT" FOR OUTPUT AS #N% : PRINT #N%, "NO"      : CLOSE #N%
    N% = FREEFILE : OPEN "NHSCALE.DAT"  FOR OUTPUT AS #N% : PRINT #N%, "0.00001" : CLOSE #N%
    IF DIR$("ProbeCoordinates.DAT") <> "" THEN KILL "ProbeCoordinates.DAT" 'get rid of this file to avoid confusion
    IF DIR$("Frep.DAT") <> "" THEN KILL "Frep.DAT" 'ditto
'   ---------------------------------------------------------------------------- CFO RUN PARAMETERS --------------------
--------------------------------------------------------
    Max1DprobesPlotted% = 15 : Max1DprobesPlotted% = MIN(15,Max1DprobesPlotted%) '15 MAX!
   ' CALL SelectTestFunction(FunctionName$)
    CALL GetTestFunctionNumber(FunctionName$)' : exit function 'DEBUG
    'MSGBOX("Test Function is #"+STR$(FunctionNumber%)+", "+FunctionName$)
    CALL
GetFunctionRunParameters(FunctionName$,Nd%,Np%,Nt&,G,DeltaT,Alpha,Beta,Frep,R(),A(),M(),XiMin(),XiMax(),StartingXiMin()
,StartingXiMax(),_
                              DiagLength,PlaceInitialProbes$,InitialAcceleration$,RepositionFactor$)
    StartingG = G : StartingDeltaT = DeltaT : StartingAlpha = Alpha : StartingBeta = Beta : StartingFrep = Frep
'   IMPORTANT NOTE: Arrays XiMin() & XiMax() are dimensioned (1 TO Nd%) in SUB GetFunctionRunParameters
    REDIM R(1 TO Np%, 1 TO Nd%, 0 TO Nt&), A(1 TO Np%, 1 TO Nd%, 0 TO Nt&), M(1 TO Np%, 0 TO Nt&) 'position,
acceleration & fitness matrices
'   ------------------------ PLOT 1D and 2D FUNCTIONS ON-SCREEN FOR VISUALIZATION ------------------------
    IF Nd% = 2 AND INSTR(FunctionName$,"PBM_") > 0 THEN MSGBOX("Begin computing plot of function "+FunctionName$+"?
May take a while - be patient...")
    SELECT CASE Nd%
        CASE 1 : CALL Plot1Dfunction(FunctionName$,XiMin(),XiMax(),R()) : REDIM R(1 TO Np%, 1 TO Nd%, 0 TO Nt&) 'erases
coordinate data in R()used to plot function
        CASE 2 : CALL Plot2Dfunction(FunctionName$,XiMin(),XiMax(),R()) : REDIM R(1 TO Np%, 1 TO Nd%, 0 TO Nt&) 'ditto
    END SELECT
'   ------------------------------------------------------------------- DISPLAY PARAMETERS & RUN CFO --------------------
---------------------------------------------------
    CALL
DisplayRunParameters(FunctionName$,Nd%,Np%,Nt&,G,DeltaT,Alpha,Beta,Frep,R(),A(),M(),PlaceInitialProbes$,InitialAccelera
tion$,RepositionFactor$,RunCFO$,ShrinkDS$,CheckForEarlyTermination$)
    IF RunCFO$ = "YES" THEN 'run CFO
        CALL StatusWindow(FunctionName$,StatusWindowHandle???)
        StartingGamma = 0## : StoppingGamma = 1##
'       StartingGamma = 0.5## : StoppingGamma = 0.5##  'USE THIS VALUE FOR SGO !!
        NumRuns% = 11 '2 '20 '25
'       ---------------- Output Data File & Header ----------------
        N% = FREEFILE : OPEN FunctionName$+".DAT" FOR OUTPUT AS #N%
        PRINT #N%, "Run ID: "+RunID$+CHR$(13)+CHR$(13)+"FUNCTION: "+UCASE$(FunctionName$)+CHR$(13)+CHR$(13)
        PRINT #N%, _
        "Run #     Gamma     Nt     Nd     Np      G     DelT   Alpha    Beta    #Steps    Neval       Frep
Fitness       Initial Probes" + CHR$(13) +_
        "-----    -------   ----   ----   ----   -----   ----   -----   ------   ------   -------   --------    -------
--------   ---------------"
        A$ = _
        "####     ###.###   ####   ####   ####   ###.#   ##.#   ##.##    ##.##   ######   #######   #.#####\\
########.########   \             \"
'
UNIFORM ON-AXIS
        PRINT #N%,
USING$(A$,0,StartingGamma,Nt&,Nd%,Np%,StartingG,StartingDeltaT,StartingAlpha,StartingBeta,0,0,StartingFrep,LEFT$(Reposi
tionFactor$,1),-9999,PlaceInitialProbes$) 'header
        PRINT #N%,_
        "------------------------------------------------------------------------------------------------------------
--------------------------"
'       ------- Loop on Gamma --------
        BestOverallFitness = -1E4200 'very large negative number...
        TotalFunctionEvaluations&& = 0
        FOR RunNumber% = 1 TO NumRuns%
            Gamma = StartingGamma + (RunNumber%-1)*(StoppingGamma-StartingGamma)/(NumRuns%-1)
            G = StartingG : DeltaT = StartingDeltaT : Alpha = StartingAlpha : Beta = StartingBeta : Frep = StartingFrep
            CALL ResetDecisionSpaceBoundaries(Nd%,XiMin(),XiMax(),StartingXiMin(),StartingXiMax())
```

```
            CALL
CFO(Nd%,Np%,Nt&,G,DeltaT,Alpha,Beta,Frep,R(),A(),M(),XiMin(),XiMax(),DiagLength,PlaceInitialProbes$,InitialAcceleration
$,_

RepositionFactor$,FunctionName$,LastStep&,CheckForEarlyTermination$,BestFitnessThisRun,RunNumber%,NumRuns%,Gamma,BestOv
erallFitness,ShrinkDS$)
            IF BestFitnessThisRun >= BestOverallFitness THEN
                BestOverallFitness = BestFitnessThisRun : BestRunNumber% = RunNumber%
            END IF
            TotalFunctionEvaluations&& = TotalFunctionEvaluations&& + (LastStep&+1)*Np%
            PRINT #N%, USING$(A$,RunNumber%,Gamma,Nt&,Nd%,Np%,G,DeltaT,Alpha,Beta,LastStep&,(Laststep&+1)*Np%,Frep,_
                              LEFT$(RepositionFactor$,1),BestFitnessThisRun,PlaceInitialProbes$)
        NEXT RunNumber%
        PRINT #N%, CHR$(13)+USING$("                                                       Total Function Evaluations:
###########",TotalFunctionEvaluations&&)+CHR$(13)
'       -------------------------------------------- Re-Run Best Run for Final Results ---------------------------------
----------------
        Gamma = StartingGamma + (BestRunNumber%-1)*(StoppingGamma-StartingGamma)/(NumRuns%-1)
        G = StartingG : DeltaT = StartingDeltaT : Alpha = StartingAlpha : Beta = StartingBeta : Frep = StartingFrep
        CALL ResetDecisionSpaceBoundaries(Nd%,XiMin(),XiMax(),StartingXiMin(),StartingXiMax())
        CALL
CFO(Nd%,Np%,Nt&,G,DeltaT,Alpha,Beta,Frep,R(),A(),M(),XiMin(),XiMax(),DiagLength,PlaceInitialProbes$,InitialAcceleration
$,_

RepositionFactor$,FunctionName$,LastStep&,CheckForEarlyTermination$,BestFitnessThisRun,BestRunNumber%,NumRuns%,Gamma,Be
stOverallFitness,ShrinkDS$)
        CALL ResetDecisionSpaceBoundaries(Nd%,XiMin(),XiMax(),StartingXiMin(),StartingXiMax())
        PRINT #N%,"--------------------------------------------------------------------------------------------------
------------------------------------"
        PRINT #N%, USING$(A$,BestRunNumber%,Gamma,Nt&,Nd%,Np%,G,DeltaT,Alpha,Beta,LastStep&,(Laststep&+1)*Np%,Frep,_
                          LEFT$(RepositionFactor$,1),BestFitnessThisRun,PlaceInitialProbes$)
        CLOSE #N%
    ELSE
        GOTO ExitPBMAIN
    END IF
'   -------------------------------- Display Best Fitness -----------------------------------------
    CALL DisplayBestFitness(Np%,Nd%,LastStep&,M(),R(),BestFitnessProbeNumber%,BestFitnessTimeStep&,FunctionName$)
'   ------------------------------------------------ PLOT EVOLUTION OF BEST FITNESS, AVG DISTANCE & BEST PROBE # -------
------------------------------------------------
    CALL
PlotBestFitnessEvolution(Nd%,Np%,LastStep&,G,DeltaT,Alpha,Beta,Frep,M(),PlaceInitialProbes$,InitialAcceleration$,Reposi
tionFactor$,FunctionName$,Gamma)
    CALL
PlotAverageDistance(Nd%,Np%,LastStep&,G,DeltaT,Alpha,Beta,Frep,M(),PlaceInitialProbes$,InitialAcceleration$,RepositionF
actor$,FunctionName$,R(),DiagLength,Gamma)
    CALL
PlotBestProbeVsTimeStep(Nd%,Np%,LastStep&,G,DeltaT,Alpha,Beta,Frep,M(),PlaceInitialProbes$,InitialAcceleration$,Reposit
ionFactor$,FunctionName$,Gamma)
'   ---------------------------------------------- PLOT TRAJECTORIES OF BEST PROBES FOR 2/3-D FUNCTIONS ---------------
------------------------------------
    IF Nd% = 2 THEN
        NumTrajectories% = 10 : CALL
Plot2DbestProbeTrajectories(NumTrajectories%,M(),R(),XiMin(),XiMax(),Np%,Nd%,LastStep&,FunctionName$)
        NumTrajectories% = 16 : CALL
Plot2DindividualProbeTrajectories(NumTrajectories%,M(),R(),XiMin(),XiMax(),Np%,Nd%,LastStep&,FunctionName$)
    END IF
    IF Nd% = 3 THEN
        NumTrajectories% = 4 : CALL
Plot3DbestProbeTrajectories(NumTrajectories%,M(),R(),XiMin(),XiMax(),Np%,Nd%,LastStep&,FunctionName$)
    END IF
'   ----------- For 1-D Objective Functions, Tabulate Probe Coordinates & if Np% =< Max1DprobesPlotted% Plot Evolution
of Probe Positions ------------
    IF Nd% = 1 THEN
        CALL
Tabulate1DprobeCoordinates(Max1DprobesPlotted%,Nd%,Np%,LastStep&,G,DeltaT,Alpha,Beta,Frep,R(),M(),PlaceInitialProbes$,I
nitialAcceleration$,RepositionFactor$,FunctionName$,Gamma)
        IF Np% =< Max1DprobesPlotted% THEN CALL
Plot1DprobePositions(Max1DprobesPlotted%,Nd%,Np%,LastStep&,G,DeltaT,Alpha,Beta,Frep,R(),M(),PlaceInitialProbes$,Initial
Acceleration$,RepositionFactor$,FunctionName$,Gamma)
        CALL CLEANUP 'delete probe coordinate plot files, if any
    END IF
ExitPBMAIN:
END FUNCTION 'PBMAIN()
'=================================================================================  CFO SUBROUTINE
=====================================================================================
SUB
CFO(Nd%,Np%,Nt&,G,DeltaT,Alpha,Beta,Frep,R(),A(),M(),XiMin(),XiMax(),DiagLength,PlaceInitialProbes$,InitialAcceleration
$,_

RepositionFactor$,FunctionName$,LastStep&,CheckForEarlyTermination$,BestFitnessThisRun,RunNumber%,NumRuns%,Gamma,BestOv
erallFitness,ShrinkDS$)
LOCAL p%, i%, j& 'Standard Indices: Probe #, Coordinate #, Time Step #
LOCAL k%, L%     'Dummy summation indices
LOCAL SumSQ, Denom, Numerator, MaxInitialRandomAcceleration, MaxInitialFixedAcceleration AS EXT
LOCAL StepNumber&, BestProbeNumber%, BestTimeStep&
LOCAL BestFitness AS EXT
LOCAL DavgOscillation$, DavgSaturation$, FitnessSaturation$
LOCAL Nsteps&
'STEP (A1) ------------- Compute Initial Probe Distribution (Step 0)----------------
    CALL InitialProbeDistribution(Np%,Nd%,Nt&,XiMin(),XiMax(),R(),PlaceInitialProbes$,Gamma)
    IF Nd% = 2 THEN 'plot 2-D initial probe distribution at Step 0
        CALL CreateGNUplotINIfile(0.1##*ScreenWidth&,0.25##*ScreenHeight&,0.6##*ScreenHeight&,0.6##*ScreenHeight&)
        CALL Show2Dprobes(R(),Np%,Nt&,0,XiMin(),XiMax(),Frep,0##,1,0,FunctionName$,RepositionFactor$,Gamma) 'show
initial probes for 2-D functions
'       msgbox("Press ENTER to continue.")
```

```
    END IF
'STEP (A2) ------------- Compute Initial Fitness Matrix (Step 0) -------------
    FOR p% = 1 TO Np% : M(p%,0) = ObjectiveFunction(R(),Nd%,p%,0,FunctionName$) : NEXT p%
'STEP (A3) ------------- Compute Initial Probe Accelerations (Step 0)---------------
    MaxInitialRandomAcceleration =  2## 'maximum value for random initial acceleration: Random[0-MAX]
    MaxInitialFixedAcceleration  = 10## 'maximum value for fixed initial acceleration: [0.001-MAX]
    CALL
InitialProbeAccelerations(Np%,Nd%,A(),InitialAcceleration$,MaxInitialRandomAcceleration,MaxInitialFixedAcceleration)
'   ======================================= LOOP ON TIME STEPS STARTING AT STEP #1
===========================================
    LastStep& = Nt& 'unless run is terminated earlier
    BestFitnessThisRun = M(1,0)
    FOR j& = 1 TO Nt&
'STEP (B) ----------- Compute Probe Position Vectors for this Time Step --------
        FOR p% = 1 TO Np% : FOR i% = 1 TO Nd% : R(p%,i%,j&) = R(p%,i%,j&-1) + 0.5##*A(p%,i%,j&-1)*DeltaT^2 : NEXT i% :
NEXT p%
'STEP (C) ----------- Retrieve Errant Probes ---------------
        CALL RetrieveErrantProbes(Np%,Nd%,j&,XiMin(),XiMax(),R(),M(),RepositionFactor$,Frep)
'STEP (D) ----------- Compute Fitness Matrix for Current Probe Distribution ---------
        FOR p% = 1 TO Np% : M(p%,j&) = ObjectiveFunction(R(),Nd%,p%,j&,FunctionName$) : NEXT p%
'STEP (E) ----------- Compute Accelerations Based on Current Probe Distribution & Fitnesses ----------------
        FOR p% = 1 TO Np%
            FOR i% = 1 TO Nd%
                A(p%,i%,j&) = 0
                FOR k% = 1 TO Np%
                    IF k% <> p% THEN
                        SumSQ = 0##
                        FOR L% = 1 TO Nd%  : SumSQ = SumSQ + (R(k%,L%,j&)-R(p%,L%,j&))^2 : NEXT L% 'dummy index
                        Denom = SQR(SumSQ) : Numerator = UnitStep((M(k%,j&)-M(p%,j&)))*(M(k%,j&)-M(p%,j&))
                        A(p%,i%,j&) = A(p%,i%,j&) + G*(R(k%,i%,j&)-R(p%,i%,j&))*Numerator^Alpha/Denom^Beta
                    END IF
                NEXT k% 'dummy index
            NEXT i% 'coord (dimension) #
        NEXT p% 'probe #
'   --------- Get Best Fitness & Corresponding Probe # and Time Step ---------
    CALL GetBestFitness(M(),Np%,j&,BestFitness,BestProbeNumber%,BestTimeStep&)
    IF BestFitness >= BestFitnessThisRun THEN BestFitnessThisRun = BestFitness
'   ----------------------------------------- Display Best Fitness/Probe#/Time Step and Plot Probe Positions for 2-D
Functions -------------------------------------------
'    IF Nd% = 2 AND ((j& MOD MAX(Nt&\16,4) = 0 OR (j& =< 16 AND j& MOD 2 = 0) OR j& = Nt&)) THEN 'plot 2D probes at
selected time steps
'        xOffset& = xOffset& - 20 : yOffset& = yOffset& + 10
'        CALL
CreateGNUplotINIfile(0.55##*ScreenWidth&+xOffset&,0.01##*ScreenHeight&+yOffset&,0.6##*ScreenHeight&,0.6##*ScreenHeight&
)
'        CALL
Show2Dprobes(R(),Np%,Nt&,j&,XiMin(),XiMax(),Frep,BestFitness,BestProbeNumber%,BestTimeStep&,FunctionName$,RepositionFac
tor$,Gamma) 'plot positions of probes in 2-D
'    END IF
'   ---------------------- Check for Davg OSC and Fitness/Davg SAT &  Display Run Status --------------------------
    DavgOscillation$   = OscillationInDavg$(j&,Np%,Nd%,M(),R(),DiagLength)           'check for oscillation in Davg
    Nsteps& = 50 '25 'for 100D tests 'used 50 for new paper '15 '# steps for averaging Davg & Best Fitness to test for
saturation
    DavgSaturation$    = HasDAVGsaturated$(Nsteps&,j&,Np%,Nd%,M(),R(),DiagLength)    'check for saturation of Davg
    FitnessSaturation$ = HasFITNESSsaturated$(Nsteps&,j&,Np%,Nd%,M(),R(),DiagLength) 'check for saturation of best
fitness
    GRAPHIC SET PIXEL (35,15) : GRAPHIC PRINT "Run #"    + REMOVE$(STR$(RunNumber%),ANY" ")     + "/"
+REMOVE$(STR$(NumRuns%),ANY" ") +_
                                              ", Step #" + REMOVE$(STR$(j&),ANY" ")             + "/" +
REMOVE$(STR$(Nt&),ANY" ")     +_
                                              ", Gamma=" + REMOVE$(STR$(ROUND(Gamma,4)),ANY" ") +_
                                              ", Frep="  + REMOVE$(STR$(Frep,4),ANY" ")         + STRING$(10," ")
    GRAPHIC SET PIXEL (35,35) : GRAPHIC PRINT "Best Fitness This Run = " + REMOVE$(STR$(ROUND(BestFitness,8)),ANY" ")
+_
                                              " @ Probe #" + REMOVE$(STR$(BestProbeNumber%),ANY" ") + ", Step #" +
REMOVE$(STR$(BestTimeStep&),ANY" ") + STRING$(15," ")
    IF RunNumber% > 1 THEN
        GRAPHIC SET PIXEL (35,55) : GRAPHIC PRINT "Best Fitness Overall = " +
REMOVE$(STR$(ROUND(BestOverallFitness,8)),ANY" ") + STRING$(15," ")
    ELSE
        GRAPHIC SET PIXEL (35,55) : GRAPHIC PRINT "Best Fitness Overall = N/A @ Run #1" + STRING$(15," ")
    END IF
    GRAPHIC SET PIXEL (35,75) : GRAPHIC PRINT "Osc Davg? " +  DavgOscillation$ + ",  Sat Davg? " +  DavgSaturation$ +",
Sat Fitness? " + FitnessSaturation$ + STRING$(10," "): GRAPHIC REDRAW
'   ------------------------------------- If Variable Frep, Adjust Value-------------------------------------------
    IF RepositionFactor$ = "VARIABLE" THEN
        Frep = Frep + 0.05##
        IF Frep > 1## THEN Frep = 0.05## 'keep Frep in range [0.05,1]
    END IF
    IF RepositionFactor$ = "RANDOM" THEN Frep = RandomNum(0##,1##) 'set new Frep value randomly
'   ------------------------------ If DavgOSC => Reposition Probes ------------------------------
'   ---------------------------- If DavgSauration => Reposition Probes ------------------------------
'   IF DavgSaturation$ = "YES" and j& < Nt& THEN
'   IF DavgOscillation$ = "YES" THEN
'        FOR p% = 1 TO Np%
'            If p% = BestProbeNumber% then
'                FOR i% = 1 TO Nd%
'                    R(p%,i%,j&) = 0.5##*R(p%,i%,j&) 'adjust each coordinate by 10%
'                    R(p%,i%,j&) = R(p%,i%,j&)*(1##+GaussianDeviate(0##,0.1##)) 'adjust each coordinate randomly
(mu,sigma)
'                   IF R(p%,i%,j&) < XiMin(i%) THEN R(p%,i%,j&) = XiMin(i%)
'                   IF R(p%,i%,j&) > XiMax(i%) THEN R(p%,i%,j&) = XiMax(i%)
'                NEXT i%
'            end if
'        NEXT p%
```

```
'        CALL RetrieveErrantProbes(Np%,Nd%,j&,XiMin(),XiMax(),R(),M(),RepositionFactor$,Frep)
'    END IF
'   ------------------------------------ Every 20 Steps Shrink Decision Space Around Best Probe -----------------------
-------------
    IF j& MOD 20 = 0 AND ShrinkDS$ = "YES" THEN
'    IF j& MOD 30 = 0 THEN
        FOR i% = 1 TO Nd%
            XiMin(i%) = XiMin(i%)+(R(BestProbeNumber%,i%,BestTimeStep&)-XiMin(i%))/2## : XiMax(i%) = XiMax(i%)-
(XiMax(i%)-R(BestProbeNumber%,i%,BestTimeStep&))/2##
        NEXT i%
'       LastStep& = j& : EXIT FOR  'time step loop
    END IF
'   ------------------------------------ If DavgOSC => Increase All Probes' Acceleration to Break Away from Trapping(?)
--------------------------------
'    IF DavgOscillation$ = "YES" THEN
 '       For p% = 1 to Np%
  '          For i% = 1 to Nd%
   '             A(p%,i%,j&) = 1##*A(p%,i%,j&) 'factor of XX increase
    '        next i%
     '   next p%
    'END IF
'STEP (F) -------------------- Check for Early Run Termination ---------------------
    IF CheckForEarlyTermination$ = "YES" THEN 'insert termination test
        IF FitnessSaturation = "YES" THEN 'terminate run
            LastStep& = j& : EXIT FOR  'time step loop
        END IF
    END IF
    NEXT j& 'END OF TIME STEP LOOP
END SUB 'CFO()
'=======================================================================================================================
===========================
SUB GetBestFitness(M(),Np%,StepNumber&,BestFitness,BestProbeNumber%,BestTimeStep&)
LOCAL p%, i&, A$
    BestFitness = M(1,0)
    FOR i& = 0 TO StepNumber&
        FOR p% = 1 TO Np%
            IF M(p%,i&) >= BestFitness THEN
                BestFitness = M(p%,i&) : BestProbeNumber% = p% : BestTimeStep& = i&
            END IF
        NEXT p%
    NEXT i&
END SUB
'============================================================= FUNCTION DEFINITIONS
=============================================================
SUB SelectTestFunction(FunctionName$)
LOCAL A$
    A$ = INPUTBOX$("Which Function?"+CHR$(13)+"1 - Parrott F4 (1-D)"+CHR$(13)+"2 - SGO (2-D)"+CHR$(13)+"3 - GP (2-
D)"+CHR$(13)+"4 - Step (n-D)"+CHR$(13)+ _
                   "5 - Schwefel 2.26 (n-D)"+CHR$(13)+"6 - Colville (4-D)"+CHR$(13)+"7 - RESERVED"+CHR$(13)+"8 -
more","SELECT OBJECTIVE FUNCTION","1")
    IF VAL(A$) < 1 OR VAL(A$) > 8 THEN A$ = "1"
    SELECT CASE VAL(A$)
        CASE 1 : FunctionName$ = "ParrottF4"
        CASE 2 : FunctionName$ = "SGO"
        CASE 3 : FunctionName$ = "GP"
        CASE 4 : FunctionName$ = "STEP"
        CASE 5 : FunctionName$ = "SCHWEFEL_226"
        CASE 6 : FunctionName$ = "COLVILLE"
        CASE 7 : FunctionName$ = "GRIEWANK"
        CASE 8 : FunctionName$ = "more"
    END SELECT
    IF FunctionName$ = "more" THEN
        A$ = INPUTBOX$("Which Function?"+CHR$(13)+"1 - F1"+CHR$(13)+"2 - F2"+CHR$(13)+"3 - F3"+CHR$(13)+"4 -
F4"+CHR$(13)+ _
                   "5 - F5"+CHR$(13)+"6 - F6"+CHR$(13)+"7 - F7"+CHR$(13)+"8 - more","SELECT OBJECTIVE FUNCTION","1")
        IF VAL(A$) < 1 OR VAL(A$) > 8 THEN A$ = "1"
        SELECT CASE VAL(A$)
            CASE 1 : FunctionName$ = "F1"
            CASE 2 : FunctionName$ = "F2"
            CASE 3 : FunctionName$ = "F3"
            CASE 4 : FunctionName$ = "F4"
            CASE 5 : FunctionName$ = "F5"
            CASE 6 : FunctionName$ = "F6"
            CASE 7 : FunctionName$ = "F7"
            CASE 8 : FunctionName$ = "more"
        END SELECT
    END IF
        IF FunctionName$ = "more" THEN
        A$ = INPUTBOX$("Which Function?"+CHR$(13)+"1 - F8"+CHR$(13)+"2 - F9"+CHR$(13)+"3 - F10"+CHR$(13)+"4 -
F11"+CHR$(13)+ _
                   "5 - F12"+CHR$(13)+"6 - F13"+CHR$(13)+"7 - F14"+CHR$(13)+"8 - more","SELECT OBJECTIVE FUNCTION","1")
        IF VAL(A$) < 1 OR VAL(A$) > 8 THEN A$ = "1"
        SELECT CASE VAL(A$)
            CASE 1 : FunctionName$ = "F8"
            CASE 2 : FunctionName$ = "F9"
            CASE 3 : FunctionName$ = "F10"
            CASE 4 : FunctionName$ = "F11"
            CASE 5 : FunctionName$ = "F12"
            CASE 6 : FunctionName$ = "F13"
            CASE 7 : FunctionName$ = "F14"
            CASE 8 : FunctionName$ = "more"
        END SELECT
    END IF
    IF FunctionName$ = "more" THEN
```

```
        A$ = INPUTBOX$("Which Function?"+CHR$(13)+"1 - F15"+CHR$(13)+"2 - F16"+CHR$(13)+"3 - F17"+CHR$(13)+"4 -
F18"+CHR$(13)+ _
                   "5 - F19"+CHR$(13)+"6 - F20"+CHR$(13)+"7 - F21"+CHR$(13)+"8 - more","SELECT OBJECTIVE FUNCTION","1")
        IF VAL(A$) < 1 OR VAL(A$) > 8 THEN A$ = "1"
        SELECT CASE VAL(A$)
            CASE 1 : FunctionName$ = "F15"
            CASE 2 : FunctionName$ = "F16"
            CASE 3 : FunctionName$ = "F17"
            CASE 4 : FunctionName$ = "F18"
            CASE 5 : FunctionName$ = "F19"
            CASE 6 : FunctionName$ = "F20"
            CASE 7 : FunctionName$ = "F21"
            CASE 8 : FunctionName$ = "more"
        END SELECT
    END IF
    IF FunctionName$ = "more" THEN
        A$ = INPUTBOX$("Which Function?"+CHR$(13)+"1 - F22"+CHR$(13)+"2 - F23"+CHR$(13)+"3 - F24"+CHR$(13)+"4 -
F25"+CHR$(13)+ _
                   "5 - F26"+CHR$(13)+"6 - F27"+CHR$(13)+"7 - F28"+CHR$(13)+"8 - more","SELECT OBJECTIVE FUNCTION","1")
        IF VAL(A$) < 1 OR VAL(A$) > 2 THEN A$ = "1"
        SELECT CASE VAL(A$)
            CASE 1 : FunctionName$ = "F22"
            CASE 2 : FunctionName$ = "F23"
            CASE 3 : FunctionName$ = "F24"
            CASE 4 : FunctionName$ = "F25"
            CASE 5 : FunctionName$ = "F26"
            CASE 6 : FunctionName$ = "F27"
            CASE 7 : FunctionName$ = "F28"
            CASE 8 : FunctionName$ = "more"
        END SELECT
    END IF
END SUB 'SelectTestFunction()
'------
FUNCTION ObjectiveFunction(R(),Nd%,p%,j&,FunctionName$) 'Objective function to be MAXIMIZED is defined here
    SELECT CASE FunctionName$
        CASE "ParrottF4"     : ObjectiveFunction = ParrottF4(R(),Nd%,p%,j&)      'Parrott F4 (1-D)
        CASE "SGO"           : ObjectiveFunction = SGO(R(),Nd%,p%,j&)            'SGO Function (2-D)
        CASE "GP"            : ObjectiveFunction = GoldsteinPrice(R(),Nd%,p%,j&) 'Goldstein-Price Function (2-D)
        CASE "STEP"          : ObjectiveFunction = StepFunction(R(),Nd%,p%,j&)   'Step Function (n-D)
        CASE "SCHWEFEL_226"  : ObjectiveFunction = Schwefel226(R(),Nd%,p%,j&)    'Schwefel Prob. 2.26 (n-D)
        CASE "COLVILLE"      : ObjectiveFunction = Colville(R(),Nd%,p%,j&)       'Colville Function (4-D)
        CASE "GRIEWANK"      : ObjectiveFunction = Griewank(R(),Nd%,p%,j&)       'Griewank Function (n-D)
        CASE "HIMMELBLAU"    : ObjectiveFunction = Himmelblau(R(),Nd%,p%,j&)     'Himmelblau Function (2-D)
'       -------------------------- GSO Paper Benchmark Functions ----------------------------
        CASE  "F1"           : ObjectiveFunction = F1(R(),Nd%,p%,j&)             'F1  (n-D)
        CASE  "F2"           : ObjectiveFunction = F2(R(),Nd%,p%,j&)             'F2  (n-D)
        CASE  "F3"           : ObjectiveFunction = F3(R(),Nd%,p%,j&)             'F3  (n-D)
        CASE  "F4"           : ObjectiveFunction = F4(R(),Nd%,p%,j&)             'F4  (n-D)
        CASE  "F5"           : ObjectiveFunction = F5(R(),Nd%,p%,j&)             'F5  (n-D)
        CASE  "F6"           : ObjectiveFunction = F6(R(),Nd%,p%,j&)             'F6  (n-D)
        CASE  "F7"           : ObjectiveFunction = F7(R(),Nd%,p%,j&)             'F7  (n-D)
        CASE  "F8"           : ObjectiveFunction = F8(R(),Nd%,p%,j&)             'F8  (n-D)
        CASE  "F9"           : ObjectiveFunction = F9(R(),Nd%,p%,j&)             'F9  (n-D)
        CASE "F10"           : ObjectiveFunction = F10(R(),Nd%,p%,j&)            'F10 (n-D)
        CASE "F11"           : ObjectiveFunction = F11(R(),Nd%,p%,j&)            'F11 (n-D)
        CASE "F12"           : ObjectiveFunction = F12(R(),Nd%,p%,j&)            'F12 (n-D)
        CASE "F13"           : ObjectiveFunction = F13(R(),Nd%,p%,j&)            'F13 (n-D)
        CASE "F14"           : ObjectiveFunction = F14(R(),Nd%,p%,j&)            'F14 (2-D)
        CASE "F15"           : ObjectiveFunction = F15(R(),Nd%,p%,j&)            'F15 (4-D)
        CASE "F16"           : ObjectiveFunction = F16(R(),Nd%,p%,j&)            'F16 (2-D)
        CASE "F17"           : ObjectiveFunction = F17(R(),Nd%,p%,j&)            'F17 (2-D)
        CASE "F18"           : ObjectiveFunction = F18(R(),Nd%,p%,j&)            'F18 (2-D)
        CASE "F19"           : ObjectiveFunction = F19(R(),Nd%,p%,j&)            'F19 (3-D)
        CASE "F20"           : ObjectiveFunction = F20(R(),Nd%,p%,j&)            'F20 (6-D)
        CASE "F21"           : ObjectiveFunction = F21(R(),Nd%,p%,j&)            'F21 (4-D)
        CASE "F22"           : ObjectiveFunction = F22(R(),Nd%,p%,j&)            'F22 (4-D)
        CASE "F23"           : ObjectiveFunction = F23(R(),Nd%,p%,j&)            'F23 (4-D)
'       ------------------------------ PBM Antenna Benchmarks ------------------------------
        CASE "PBM_1"         : ObjectiveFunction = PBM_1(R(),Nd%,p%,j&)          'PBM_1 (2-D)
        CASE "PBM_2"         : ObjectiveFunction = PBM_2(R(),Nd%,p%,j&)          'PBM_2 (2-D)
        CASE "PBM_3"         : ObjectiveFunction = PBM_3(R(),Nd%,p%,j&)          'PBM_3 (2-D)
        CASE "PBM_4"         : ObjectiveFunction = PBM_4(R(),Nd%,p%,j&)          'PBM_4 (2-D)
        CASE "PBM_5"         : ObjectiveFunction = PBM_5(R(),Nd%,p%,j&)          'PBM_5 (2-D)
    END SELECT
END FUNCTION 'ObjectiveFunction()
'-----------
SUB ResetDecisionSpaceBoundaries(Nd%,XiMin(),XiMax(),StartingXiMin(),StartingXiMax())
LOCAL i%
    FOR i% = 1 TO Nd% : XiMin(i%) = StartingXiMin(i%) : XiMax(i%) = StartingXiMax(i%) : NEXT i%
END SUB
'------
SUB
GetFunctionRunParameters(FunctionName$,Nd%,Np%,Nt&,G,DeltaT,Alpha,Beta,Frep,R(),A(),M(),XiMin(),XiMax(),StartingXiMin()
,StartingXiMax(),_
                             DiagLength,PlaceInitialProbes$,InitialAcceleration$,RepositionFactor$)
LOCAL i%, NumProbesPerDimension%, NN%, NumCollinearElements%
    SELECT CASE FunctionName$
        CASE "ParrottF4"
            Nd%                    = 1
            NumProbesPerDimension% = 3
            Np%                    = NumProbesPerDimension%*Nd%
            Nt&      = 500
            G        = 2##
            Alpha    = 2##
            Beta     = 2##
```

```
            DeltaT   = 1##
            Frep     = 0.9##
            PlaceInitialProbes$  = "UNIFORM ON-AXIS"
            InitialAcceleration$ = "ZERO"
            RepositionFactor$    = "FIXED"
            CALL
ChangeRunParameters(NumProbesPerDimension%,Np%,Nd%,Nt&,G,Alpha,Beta,DeltaT,Frep,PlaceInitialProbes$,InitialAcceleration
$,RepositionFactor$,FunctionName$)
            Nd% = 1 'cannot change dimensionality of Parrott F4 function!
            NumProbesPerDimension% = MAX(NumProbesPerDimension%,3) 'at least three for 1-D functions
            Np% = NumProbesPerDimension%*Nd%
            REDIM XiMin(1 TO Nd%), XiMax(1 TO Nd%) : XiMin(1) = 0## : XiMax(1) = 1##
            REDIM StartingXiMin(1 TO Nd%), StartingXiMax(1 TO Nd%) : FOR i% = 1 TO Nd% : StartingXiMin(i%) = XiMin(i%)
: StartingXiMax(i%) = XiMax(i%) : NEXT i%
        CASE "SGO"
            Nd%                    = 2
            NumProbesPerDimension% = 4 '10 '4
            Np%                    = NumProbesPerDimension%*Nd%
            Nt&      = 500
            G        = 2##
            Alpha    = 2##
            Beta     = 2##
            DeltaT   = 1##
            Frep     = 0.5##
            PlaceInitialProbes$  = "UNIFORM ON-AXIS" '"2D GRID"
            InitialAcceleration$ = "ZERO"
            RepositionFactor$    = "VARIABLE"
            CALL
ChangeRunParameters(NumProbesPerDimension%,Np%,Nd%,Nt&,G,Alpha,Beta,DeltaT,Frep,PlaceInitialProbes$,InitialAcceleration
$,RepositionFactor$,FunctionName$)
            Nd% = 2 'cannot change dimensionality of SGO function!
            Np% = NumProbesPerDimension%*Nd%
            REDIM XiMin(1 TO Nd%), XiMax(1 TO Nd%) : FOR i% = 1 TO Nd% : XiMin(i%) = -50## : XiMax(i%) = 50## : NEXT i%
            REDIM StartingXiMin(1 TO Nd%), StartingXiMax(1 TO Nd%) : FOR i% = 1 TO Nd% : StartingXiMin(i%) = XiMin(i%)
: StartingXiMax(i%) = XiMax(i%) : NEXT i%
            IF PlaceInitialProbes$ = "2D GRID" THEN
                Np% = NumProbesPerDimension%^2 : REDIM R(1 TO Np%, 1 TO Nd%, 0 TO Nt&) 'to create (Np/Nd) x (Np/Nd)
grid
            END IF
        CASE "GP"
            Nd%                    = 2
            NumProbesPerDimension% = 4 '10
            Np%                    = NumProbesPerDimension%*Nd%
            Nt&      = 500
            G        = 2##
            Alpha    = 2## '0.2##
            Beta     = 2##
            DeltaT   = 1##
            Frep     = 0.5## '0.8## '0.9##
            PlaceInitialProbes$  = "UNIFORM ON-AXIS" '"2D GRID"
            InitialAcceleration$ = "ZERO"
            RepositionFactor$    = "VARIABLE"
            CALL
ChangeRunParameters(NumProbesPerDimension%,Np%,Nd%,Nt&,G,Alpha,Beta,DeltaT,Frep,PlaceInitialProbes$,InitialAcceleration
$,RepositionFactor$,FunctionName$)
            Nd% = 2 'cannot change dimensionality of GP function!
            Np% = NumProbesPerDimension%*Nd%
            REDIM XiMin(1 TO Nd%), XiMax(1 TO Nd%) : FOR i% = 1 TO Nd% : XiMin(i%) = -100## : XiMax(i%) = 100## : NEXT
i%
            REDIM StartingXiMin(1 TO Nd%), StartingXiMax(1 TO Nd%) : FOR i% = 1 TO Nd% : StartingXiMin(i%) = XiMin(i%)
: StartingXiMax(i%) = XiMax(i%) : NEXT i%
            IF PlaceInitialProbes$ = "2D GRID" THEN
                Np% = NumProbesPerDimension%^2 : REDIM R(1 TO Np%, 1 TO Nd%, 0 TO Nt&) 'to create (Np/Nd) x (Np/Nd)
grid
            END IF
        CASE "STEP"
            Nd%                    = 2
            NumProbesPerDimension% = 4 '20
            Np%                    = NumProbesPerDimension%*Nd%
            Nt&      = 500
            G        = 2##
            Alpha    = 2##
            Beta     = 2##
            DeltaT   = 1##
            Frep     = 0.5##
            PlaceInitialProbes$  = "UNIFORM ON-AXIS"
            InitialAcceleration$ = "ZERO"
            RepositionFactor$    = "VARIABLE" '"FIXED"
            CALL
ChangeRunParameters(NumProbesPerDimension%,Np%,Nd%,Nt&,G,Alpha,Beta,DeltaT,Frep,PlaceInitialProbes$,InitialAcceleration
$,RepositionFactor$,FunctionName$)
            Np% = NumProbesPerDimension%*Nd%
            REDIM XiMin(1 TO Nd%), XiMax(1 TO Nd%) : FOR i% = 1 TO Nd% : XiMin(i%) = -100## : XiMax(i%) = 100## : NEXT
i%
'           REDIM XiMin(1 TO Nd%), XiMax(1 TO Nd%) : XiMin(1) = 72## : XiMax(1) = 78## : XiMin(2) = 27## : XiMax(2) =
33## 'use this to plot STEP detail
            REDIM StartingXiMin(1 TO Nd%), StartingXiMax(1 TO Nd%) : FOR i% = 1 TO Nd% : StartingXiMin(i%) = XiMin(i%)
: StartingXiMax(i%) = XiMax(i%) : NEXT i%
            IF PlaceInitialProbes$ = "2D GRID" THEN
                Np% = NumProbesPerDimension%^2 : REDIM R(1 TO Np%, 1 TO Nd%, 0 TO Nt&) 'to create (Np/Nd) x (Np/Nd)
grid
            END IF 'STEP
        CASE "SCHWEFEL_226"
            Nd%                    = 30
            NumProbesPerDimension% = 4
```

```
            Np%                     = NumProbesPerDimension%*Nd%
            Nt&      = 500
            G        = 2##
            Alpha    = 2##
            Beta     = 2##
            DeltaT   = 1##
            Frep     = 0.5##
            PlaceInitialProbes$  = "UNIFORM ON-AXIS"
            InitialAcceleration$ = "ZERO"
            RepositionFactor$    = "VARIABLE"
            CALL
ChangeRunParameters(NumProbesPerDimension%,Np%,Nd%,Nt&,G,Alpha,Beta,DeltaT,Frep,PlaceInitialProbes$,InitialAcceleration
$,RepositionFactor$,FunctionName$)
            Np% = NumProbesPerDimension%*Nd%
            REDIM XiMin(1 TO Nd%), XiMax(1 TO Nd%) : FOR i% = 1 TO Nd% : XiMin(i%) = -500## : XiMax(i%) = 500## : NEXT
i%
            REDIM StartingXiMin(1 TO Nd%), StartingXiMax(1 TO Nd%) : FOR i% = 1 TO Nd% : StartingXiMin(i%) = XiMin(i%)
: StartingXiMax(i%) = XiMax(i%) : NEXT i%
            IF PlaceInitialProbes$ = "2D GRID" THEN
                Np% = NumProbesPerDimension%^2 : REDIM R(1 TO Np%, 1 TO Nd%, 0 TO Nt&) 'to create (Np/Nd) x (Np/Nd)
grid
            END IF
        CASE "COLVILLE"
            Nd%                     = 4
            NumProbesPerDimension% = 4 '14
            Np%                     = NumProbesPerDimension%*Nd%
            Nt&      = 500
            G        = 2##
            Alpha    = 2##
            Beta     = 2##
            DeltaT   = 1##
            Frep     = 0.5##
            PlaceInitialProbes$  = "UNIFORM ON-AXIS"
            InitialAcceleration$ = "ZERO"
            RepositionFactor$    = "VARIABLE"
            CALL
ChangeRunParameters(NumProbesPerDimension%,Np%,Nd%,Nt&,G,Alpha,Beta,DeltaT,Frep,PlaceInitialProbes$,InitialAcceleration
$,RepositionFactor$,FunctionName$)
            Nd% = 4 'cannot change dimensionality of Colville function!
            Np% = NumProbesPerDimension%*Nd%
            REDIM XiMin(1 TO Nd%), XiMax(1 TO Nd%) : FOR i% = 1 TO Nd% : XiMin(i%) = -10## : XiMax(i%) = 10## : NEXT i%
            REDIM StartingXiMin(1 TO Nd%), StartingXiMax(1 TO Nd%) : FOR i% = 1 TO Nd% : StartingXiMin(i%) = XiMin(i%)
: StartingXiMax(i%) = XiMax(i%) : NEXT i%
            IF PlaceInitialProbes$ = "2D GRID" THEN
                Np% = NumProbesPerDimension%^2 : REDIM R(1 TO Np%, 1 TO Nd%, 0 TO Nt&) 'to create (Np/Nd) x (Np/Nd)
grid
            END IF
        CASE "GRIEWANK"
            Nd%                     = 2
            NumProbesPerDimension% = 4 '14
            Np%                     = NumProbesPerDimension%*Nd%
            Nt&      = 500
            G        = 2##
            Alpha    = 2##
            Beta     = 2##
            DeltaT   = 1##
            Frep     = 0.5##
            PlaceInitialProbes$  = "UNIFORM ON-AXIS"
            InitialAcceleration$ = "ZERO"
            RepositionFactor$    = "VARIABLE"
            CALL
ChangeRunParameters(NumProbesPerDimension%,Np%,Nd%,Nt&,G,Alpha,Beta,DeltaT,Frep,PlaceInitialProbes$,InitialAcceleration
$,RepositionFactor$,FunctionName$)
            Np% = NumProbesPerDimension%*Nd%
            REDIM XiMin(1 TO Nd%), XiMax(1 TO Nd%) : FOR i% = 1 TO Nd% : XiMin(i%) = -600## : XiMax(i%) = 600## : NEXT
i%
            REDIM StartingXiMin(1 TO Nd%), StartingXiMax(1 TO Nd%) : FOR i% = 1 TO Nd% : StartingXiMin(i%) = XiMin(i%)
: StartingXiMax(i%) = XiMax(i%) : NEXT i%
            IF PlaceInitialProbes$ = "2D GRID" THEN
                Np% = NumProbesPerDimension%^2 : REDIM R(1 TO Np%, 1 TO Nd%, 0 TO Nt&) 'to create (Np/Nd) x (Np/Nd)
grid
            END IF
       CASE "HIMMELBLAU"
            Nd%                     = 2
            NumProbesPerDimension% = 4 '14
            Np%                     = NumProbesPerDimension%*Nd%
            Nt&      = 500
            G        = 2##
            Alpha    = 2##
            Beta     = 2##
            DeltaT   = 1##
            Frep     = 0.5##
            PlaceInitialProbes$  = "UNIFORM ON-AXIS"
            InitialAcceleration$ = "ZERO"
            RepositionFactor$    = "VARIABLE"
            CALL
ChangeRunParameters(NumProbesPerDimension%,Np%,Nd%,Nt&,G,Alpha,Beta,DeltaT,Frep,PlaceInitialProbes$,InitialAcceleration
$,RepositionFactor$,FunctionName$)
            Nd% = 2 'cannot change dimensionality of Himmelblau function!
            Np% = NumProbesPerDimension%*Nd%
            REDIM XiMin(1 TO Nd%), XiMax(1 TO Nd%) : FOR i% = 1 TO Nd% : XiMin(i%) = -6## : XiMax(i%) = 6## : NEXT i%
            REDIM StartingXiMin(1 TO Nd%), StartingXiMax(1 TO Nd%) : FOR i% = 1 TO Nd% : StartingXiMin(i%) = XiMin(i%)
: StartingXiMax(i%) = XiMax(i%) : NEXT i%
            IF PlaceInitialProbes$ = "2D GRID" THEN
```

```
                Np% = NumProbesPerDimension%^2 : REDIM R(1 TO Np%, 1 TO Nd%, 0 TO Nt&) 'to create (Np/Nd) x (Np/Nd)
grid
            END IF
        CASE "F1" '(n-D)
            Nd%                     = 30
            NumProbesPerDimension% = 2
            Np%                     = NumProbesPerDimension%*Nd%
            Nt&      = 500
            G        = 2##
            Alpha    = 2##
            Beta     = 2##
            DeltaT   = 1##
            Frep     = 0.5##
            PlaceInitialProbes$  = "UNIFORM ON-AXIS"
            InitialAcceleration$ = "ZERO"
            RepositionFactor$    = "VARIABLE"
            CALL
ChangeRunParameters(NumProbesPerDimension%,Np%,Nd%,Nt&,G,Alpha,Beta,DeltaT,Frep,PlaceInitialProbes$,InitialAcceleration
$,RepositionFactor$,FunctionName$)
            Np% = NumProbesPerDimension%*Nd%
            REDIM XiMin(1 TO Nd%), XiMax(1 TO Nd%) : FOR i% = 1 TO Nd% : XiMin(i%) = -100## : XiMax(i%) = 100## : NEXT
i%
            REDIM StartingXiMin(1 TO Nd%), StartingXiMax(1 TO Nd%) : FOR i% = 1 TO Nd% : StartingXiMin(i%) = XiMin(i%)
: StartingXiMax(i%) = XiMax(i%) : NEXT i%
            IF PlaceInitialProbes$ = "2D GRID" THEN
                Np% = NumProbesPerDimension%^2 : REDIM R(1 TO Np%, 1 TO Nd%, 0 TO Nt&) 'to create (Np/Nd) x (Np/Nd)
grid
            END IF
       CASE "F2" '(n-D)
            Nd%                     = 30
            NumProbesPerDimension% = 2
            Np%                     = NumProbesPerDimension%*Nd%
            Nt&      = 500
            G        = 2##
            Alpha    = 2##
            Beta     = 2##
            DeltaT   = 1##
            Frep     = 0.5##
            PlaceInitialProbes$  = "UNIFORM ON-AXIS"
            InitialAcceleration$ = "ZERO"
            RepositionFactor$    = "VARIABLE"
            CALL
ChangeRunParameters(NumProbesPerDimension%,Np%,Nd%,Nt&,G,Alpha,Beta,DeltaT,Frep,PlaceInitialProbes$,InitialAcceleration
$,RepositionFactor$,FunctionName$)
            Np% = NumProbesPerDimension%*Nd%
            REDIM XiMin(1 TO Nd%), XiMax(1 TO Nd%) : FOR i% = 1 TO Nd% : XiMin(i%) = -10## : XiMax(i%) = 10## : NEXT i%
            REDIM StartingXiMin(1 TO Nd%), StartingXiMax(1 TO Nd%) : FOR i% = 1 TO Nd% : StartingXiMin(i%) = XiMin(i%)
: StartingXiMax(i%) = XiMax(i%) : NEXT i%
            IF PlaceInitialProbes$ = "2D GRID" THEN
                Np% = NumProbesPerDimension%^2 : REDIM R(1 TO Np%, 1 TO Nd%, 0 TO Nt&) 'to create (Np/Nd) x (Np/Nd)
grid
            END IF
      CASE "F3" '(n-D)
            Nd%                     = 30
            NumProbesPerDimension% = 2
            Np%                     = NumProbesPerDimension%*Nd%
            Nt&      = 500
            G        = 2##
            Alpha    = 2##
            Beta     = 2##
            DeltaT   = 1##
            Frep     = 0.5##
            PlaceInitialProbes$  = "UNIFORM ON-AXIS"
            InitialAcceleration$ = "ZERO"
            RepositionFactor$    = "VARIABLE"
            CALL
ChangeRunParameters(NumProbesPerDimension%,Np%,Nd%,Nt&,G,Alpha,Beta,DeltaT,Frep,PlaceInitialProbes$,InitialAcceleration
$,RepositionFactor$,FunctionName$)
            Np% = NumProbesPerDimension%*Nd%
            REDIM XiMin(1 TO Nd%), XiMax(1 TO Nd%) : FOR i% = 1 TO Nd% : XiMin(i%) = -100## : XiMax(i%) = 100## : NEXT
i%
            REDIM StartingXiMin(1 TO Nd%), StartingXiMax(1 TO Nd%) : FOR i% = 1 TO Nd% : StartingXiMin(i%) = XiMin(i%)
: StartingXiMax(i%) = XiMax(i%) : NEXT i%
            IF PlaceInitialProbes$ = "2D GRID" THEN
                Np% = NumProbesPerDimension%^2 : REDIM R(1 TO Np%, 1 TO Nd%, 0 TO Nt&) 'to create (Np/Nd) x (Np/Nd)
grid
            END IF
      CASE "F4" '(n-D)
            Nd%                     = 30
            NumProbesPerDimension% = 2
            Np%                     = NumProbesPerDimension%*Nd%
            Nt&      = 500
            G        = 2##
            Alpha    = 2##
            Beta     = 2##
            DeltaT   = 1##
            Frep     = 0.5##
            PlaceInitialProbes$  = "UNIFORM ON-AXIS"
            InitialAcceleration$ = "ZERO"
            RepositionFactor$    = "VARIABLE"
            CALL
ChangeRunParameters(NumProbesPerDimension%,Np%,Nd%,Nt&,G,Alpha,Beta,DeltaT,Frep,PlaceInitialProbes$,InitialAcceleration
$,RepositionFactor$,FunctionName$)
            Np% = NumProbesPerDimension%*Nd%
```

```
            REDIM XiMin(1 TO Nd%), XiMax(1 TO Nd%) : FOR i% = 1 TO Nd% : XiMin(i%) = -100## : XiMax(i%) = 100## : NEXT
i%
            REDIM StartingXiMin(1 TO Nd%), StartingXiMax(1 TO Nd%) : FOR i% = 1 TO Nd% : StartingXiMin(i%) = XiMin(i%)
: StartingXiMax(i%) = XiMax(i%) : NEXT i%
            IF PlaceInitialProbes$ = "2D GRID" THEN
                Np% = NumProbesPerDimension%^2 : REDIM R(1 TO Np%, 1 TO Nd%, 0 TO Nt&) 'to create (Np/Nd) x (Np/Nd)
grid
            END IF
      CASE "F5" '(n-D)
            Nd%                    = 30
            NumProbesPerDimension% = 2
            Np%                    = NumProbesPerDimension%*Nd%
            Nt&      = 500
            G        = 2##
            Alpha    = 2##
            Beta     = 2##
            DeltaT   = 1##
            Frep     = 0.5##
            PlaceInitialProbes$  = "UNIFORM ON-AXIS"
            InitialAcceleration$ = "ZERO"
            RepositionFactor$    = "VARIABLE"
            CALL
ChangeRunParameters(NumProbesPerDimension%,Np%,Nd%,Nt&,G,Alpha,Beta,DeltaT,Frep,PlaceInitialProbes$,InitialAcceleration
$,RepositionFactor$,FunctionName$)
            Np% = NumProbesPerDimension%*Nd%
            REDIM XiMin(1 TO Nd%), XiMax(1 TO Nd%) : FOR i% = 1 TO Nd% : XiMin(i%) = -30## : XiMax(i%) = 30## : NEXT i%
            REDIM StartingXiMin(1 TO Nd%), StartingXiMax(1 TO Nd%) : FOR i% = 1 TO Nd% : StartingXiMin(i%) = XiMin(i%)
: StartingXiMax(i%) = XiMax(i%) : NEXT i%
            IF PlaceInitialProbes$ = "2D GRID" THEN
                Np% = NumProbesPerDimension%^2 : REDIM R(1 TO Np%, 1 TO Nd%, 0 TO Nt&) 'to create (Np/Nd) x (Np/Nd)
grid
            END IF
       CASE "F6" '(n-D) STEP
            Nd%                    = 30
            NumProbesPerDimension% = 2 '20
            Np%                    = NumProbesPerDimension%*Nd%
            Nt&      = 500
            G        = 2##
            Alpha    = 2##
            Beta     = 2##
            DeltaT   = 1##
            Frep     = 0.5##
            PlaceInitialProbes$  = "UNIFORM ON-AXIS"
            InitialAcceleration$ = "ZERO"
            RepositionFactor$    = "VARIABLE" '"FIXED"
            CALL
ChangeRunParameters(NumProbesPerDimension%,Np%,Nd%,Nt&,G,Alpha,Beta,DeltaT,Frep,PlaceInitialProbes$,InitialAcceleration
$,RepositionFactor$,FunctionName$)
            Np% = NumProbesPerDimension%*Nd%
            REDIM XiMin(1 TO Nd%), XiMax(1 TO Nd%) : FOR i% = 1 TO Nd% : XiMin(i%) = -100## : XiMax(i%) = 100## : NEXT
i%
            REDIM StartingXiMin(1 TO Nd%), StartingXiMax(1 TO Nd%) : FOR i% = 1 TO Nd% : StartingXiMin(i%) = XiMin(i%)
: StartingXiMax(i%) = XiMax(i%) : NEXT i%
            IF PlaceInitialProbes$ = "2D GRID" THEN
                Np% = NumProbesPerDimension%^2 : REDIM R(1 TO Np%, 1 TO Nd%, 0 TO Nt&) 'to create (Np/Nd) x (Np/Nd)
grid
            END IF
       CASE "F7" '(n-D)
            Nd%                    = 30
            NumProbesPerDimension% = 2 '20
            Np%                    = NumProbesPerDimension%*Nd%
            Nt&      = 100          'BECAUSE THIS FUNCTION HAS A RANDOM COMPONENT !!
            G        = 2##
            Alpha    = 2##
            Beta     = 2##
            DeltaT   = 1##
            Frep     = 0.5##
            PlaceInitialProbes$  = "UNIFORM ON-AXIS"
            InitialAcceleration$ = "ZERO"
            RepositionFactor$    = "VARIABLE" '"FIXED"
            CALL
ChangeRunParameters(NumProbesPerDimension%,Np%,Nd%,Nt&,G,Alpha,Beta,DeltaT,Frep,PlaceInitialProbes$,InitialAcceleration
$,RepositionFactor$,FunctionName$)
            Np% = NumProbesPerDimension%*Nd%
            REDIM XiMin(1 TO Nd%), XiMax(1 TO Nd%) : FOR i% = 1 TO Nd% : XiMin(i%) = -1.28## : XiMax(i%) = 1.28## :
NEXT i%
            REDIM StartingXiMin(1 TO Nd%), StartingXiMax(1 TO Nd%) : FOR i% = 1 TO Nd% : StartingXiMin(i%) = XiMin(i%)
: StartingXiMax(i%) = XiMax(i%) : NEXT i%
            IF PlaceInitialProbes$ = "2D GRID" THEN
                Np% = NumProbesPerDimension%^2 : REDIM R(1 TO Np%, 1 TO Nd%, 0 TO Nt&) 'to create (Np/Nd) x (Np/Nd)
grid
            END IF 'F7
       CASE "F8" '(n-D)
            Nd%                    = 30
            NumProbesPerDimension% = 2 '4 '20
            Np%                    = NumProbesPerDimension%*Nd%
            Nt&      = 500
            G        = 2##
            Alpha    = 2##
            Beta     = 2##
            DeltaT   = 1##
            Frep     = 0.5##
            PlaceInitialProbes$  = "UNIFORM ON-AXIS"
            InitialAcceleration$ = "ZERO"
            RepositionFactor$    = "VARIABLE" '"FIXED"
```

```
            CALL
ChangeRunParameters(NumProbesPerDimension%,Np%,Nd%,Nt&,G,Alpha,Beta,DeltaT,Frep,PlaceInitialProbes$,InitialAcceleration
$,RepositionFactor$,FunctionName$)
            Np% = NumProbesPerDimension%*Nd%
            REDIM XiMin(1 TO Nd%), XiMax(1 TO Nd%) : FOR i% = 1 TO Nd% : XiMin(i%) = -500## : XiMax(i%) = 500## : NEXT
i%
            REDIM StartingXiMin(1 TO Nd%), StartingXiMax(1 TO Nd%) : FOR i% = 1 TO Nd% : StartingXiMin(i%) = XiMin(i%)
: StartingXiMax(i%) = XiMax(i%) : NEXT i%
            IF PlaceInitialProbes$ = "2D GRID" THEN
                Np% = NumProbesPerDimension%^2 : REDIM R(1 TO Np%, 1 TO Nd%, 0 TO Nt&) 'to create (Np/Nd) x (Np/Nd)
grid
            END IF 'F8
       CASE "F9" '(n-D)
            Nd%                    = 30
            NumProbesPerDimension% = 2 '4 '20
            Np%                    = NumProbesPerDimension%*Nd%
            Nt&      = 500
            G        = 2##
            Alpha    = 2##
            Beta     = 2##
            DeltaT   = 1##
            Frep     = 0.5##
            PlaceInitialProbes$  = "UNIFORM ON-AXIS"
            InitialAcceleration$ = "ZERO"
            RepositionFactor$    = "VARIABLE" '"FIXED"
            CALL
ChangeRunParameters(NumProbesPerDimension%,Np%,Nd%,Nt&,G,Alpha,Beta,DeltaT,Frep,PlaceInitialProbes$,InitialAcceleration
$,RepositionFactor$,FunctionName$)
            Np% = NumProbesPerDimension%*Nd%
            REDIM XiMin(1 TO Nd%), XiMax(1 TO Nd%) : FOR i% = 1 TO Nd% : XiMin(i%) = -5.12## : XiMax(i%) = 5.12## :
NEXT i%
            REDIM StartingXiMin(1 TO Nd%), StartingXiMax(1 TO Nd%) : FOR i% = 1 TO Nd% : StartingXiMin(i%) = XiMin(i%)
: StartingXiMax(i%) = XiMax(i%) : NEXT i%
            IF PlaceInitialProbes$ = "2D GRID" THEN
                Np% = NumProbesPerDimension%^2 : REDIM R(1 TO Np%, 1 TO Nd%, 0 TO Nt&) 'to create (Np/Nd) x (Np/Nd)
grid
            END IF 'F9
       CASE "F10" '(n-D) Ackley's Function
            Nd%                    = 30
            NumProbesPerDimension% = 2 '4 '20
            Np%                    = NumProbesPerDimension%*Nd%
            Nt&      = 500
            G        = 2##
            Alpha    = 2##
            Beta     = 2##
            DeltaT   = 1##
            Frep     = 0.5##
            PlaceInitialProbes$  = "UNIFORM ON-AXIS"
            InitialAcceleration$ = "ZERO"
            RepositionFactor$    = "VARIABLE" '"FIXED"
            CALL
ChangeRunParameters(NumProbesPerDimension%,Np%,Nd%,Nt&,G,Alpha,Beta,DeltaT,Frep,PlaceInitialProbes$,InitialAcceleration
$,RepositionFactor$,FunctionName$)
            Np% = NumProbesPerDimension%*Nd%
            REDIM XiMin(1 TO Nd%), XiMax(1 TO Nd%) : FOR i% = 1 TO Nd% : XiMin(i%) = -32## : XiMax(i%) = 32## : NEXT i%
            REDIM StartingXiMin(1 TO Nd%), StartingXiMax(1 TO Nd%) : FOR i% = 1 TO Nd% : StartingXiMin(i%) = XiMin(i%)
: StartingXiMax(i%) = XiMax(i%) : NEXT i%
            IF PlaceInitialProbes$ = "2D GRID" THEN
                Np% = NumProbesPerDimension%^2 : REDIM R(1 TO Np%, 1 TO Nd%, 0 TO Nt&) 'to create (Np/Nd) x (Np/Nd)
grid
            END IF 'F10
       CASE "F11" '(n-D)
            Nd%                    = 30
            NumProbesPerDimension% = 2 '4 '20
            Np%                    = NumProbesPerDimension%*Nd%
            Nt&      = 500
            G        = 2##
            Alpha    = 2##
            Beta     = 2##
            DeltaT   = 1##
            Frep     = 0.5##
            PlaceInitialProbes$  = "UNIFORM ON-AXIS"
            InitialAcceleration$ = "ZERO"
            RepositionFactor$    = "VARIABLE" '"FIXED"
            CALL
ChangeRunParameters(NumProbesPerDimension%,Np%,Nd%,Nt&,G,Alpha,Beta,DeltaT,Frep,PlaceInitialProbes$,InitialAcceleration
$,RepositionFactor$,FunctionName$)
            Np% = NumProbesPerDimension%*Nd%
            REDIM XiMin(1 TO Nd%), XiMax(1 TO Nd%) : FOR i% = 1 TO Nd% : XiMin(i%) = -600## : XiMax(i%) = 600## : NEXT
i%
            REDIM StartingXiMin(1 TO Nd%), StartingXiMax(1 TO Nd%) : FOR i% = 1 TO Nd% : StartingXiMin(i%) = XiMin(i%)
: StartingXiMax(i%) = XiMax(i%) : NEXT i%
            IF PlaceInitialProbes$ = "2D GRID" THEN
                Np% = NumProbesPerDimension%^2 : REDIM R(1 TO Np%, 1 TO Nd%, 0 TO Nt&) 'to create (Np/Nd) x (Np/Nd)
grid
            END IF 'F11
       CASE "F12" '(n-D) Penalized #1
            Nd%                    = 30
            NumProbesPerDimension% = 2 '4 '20
            Np%                    = NumProbesPerDimension%*Nd%
            Nt&      = 500
            G        = 2##
            Alpha    = 2##
            Beta     = 2##
            DeltaT   = 1##
```

```
            Frep     = 0.5##
            PlaceInitialProbes$  = "UNIFORM ON-AXIS"
            InitialAcceleration$ = "ZERO"
            RepositionFactor$    = "VARIABLE" '"FIXED"
            CALL
ChangeRunParameters(NumProbesPerDimension%,Np%,Nd%,Nt&,G,Alpha,Beta,DeltaT,Frep,PlaceInitialProbes$,InitialAcceleration
$,RepositionFactor$,FunctionName$)
            Np% = NumProbesPerDimension%*Nd%
            REDIM XiMin(1 TO Nd%), XiMax(1 TO Nd%) : FOR i% = 1 TO Nd% : XiMin(i%) = -50## : XiMax(i%) = 50## : NEXT i%
            REDIM StartingXiMin(1 TO Nd%), StartingXiMax(1 TO Nd%) : FOR i% = 1 TO Nd% : StartingXiMin(i%) = XiMin(i%)
: StartingXiMax(i%) = XiMax(i%) : NEXT i%
            IF PlaceInitialProbes$ = "2D GRID" THEN
                Np% = NumProbesPerDimension%^2 : REDIM R(1 TO Np%, 1 TO Nd%, 0 TO Nt&) 'to create (Np/Nd) x (Np/Nd)
grid
            END IF 'F12
       CASE "F13" '(n-D) Penalized #2
            Nd%                   = 30
            NumProbesPerDimension% = 2 '4 '20
            Np%                   = NumProbesPerDimension%*Nd%
            Nt&      = 500
            G        = 2##
            Alpha    = 2##
            Beta     = 2##
            DeltaT   = 1##
            Frep     = 0.5##
            PlaceInitialProbes$  = "UNIFORM ON-AXIS"
            InitialAcceleration$ = "ZERO"
            RepositionFactor$    = "VARIABLE" '"FIXED"
            CALL
ChangeRunParameters(NumProbesPerDimension%,Np%,Nd%,Nt&,G,Alpha,Beta,DeltaT,Frep,PlaceInitialProbes$,InitialAcceleration
$,RepositionFactor$,FunctionName$)
            Np% = NumProbesPerDimension%*Nd%
            REDIM XiMin(1 TO Nd%), XiMax(1 TO Nd%) : FOR i% = 1 TO Nd% : XiMin(i%) = -50## : XiMax(i%) = 50## : NEXT i%
            REDIM StartingXiMin(1 TO Nd%), StartingXiMax(1 TO Nd%) : FOR i% = 1 TO Nd% : StartingXiMin(i%) = XiMin(i%)
: StartingXiMax(i%) = XiMax(i%) : NEXT i%
            IF PlaceInitialProbes$ = "2D GRID" THEN
                Np% = NumProbesPerDimension%^2 : REDIM R(1 TO Np%, 1 TO Nd%, 0 TO Nt&) 'to create (Np/Nd) x (Np/Nd)
grid
            END IF 'F13
       CASE "F14" '(2-D) Shekel's Foxholes
            Nd%                   = 2
            NumProbesPerDimension% = 4 '20
            Np%                   = NumProbesPerDimension%*Nd%
            Nt&      = 500
            G        = 2##
            Alpha    = 2##
            Beta     = 2##
            DeltaT   = 1##
            Frep     = 0.5##
            PlaceInitialProbes$  = "UNIFORM ON-AXIS"
            InitialAcceleration$ = "ZERO"
            RepositionFactor$    = "VARIABLE" '"FIXED"
            CALL
ChangeRunParameters(NumProbesPerDimension%,Np%,Nd%,Nt&,G,Alpha,Beta,DeltaT,Frep,PlaceInitialProbes$,InitialAcceleration
$,RepositionFactor$,FunctionName$)
            Nd% = 2 'cannot change dimensionality of Shekel's Foxholes function!
            Np% = NumProbesPerDimension%*Nd%
            REDIM XiMin(1 TO Nd%), XiMax(1 TO Nd%) : FOR i% = 1 TO Nd% : XiMin(i%) = -65.536## : XiMax(i%) = 65.536## :
NEXT i%
            REDIM StartingXiMin(1 TO Nd%), StartingXiMax(1 TO Nd%) : FOR i% = 1 TO Nd% : StartingXiMin(i%) = XiMin(i%)
: StartingXiMax(i%) = XiMax(i%) : NEXT i%
            IF PlaceInitialProbes$ = "2D GRID" THEN
                Np% = NumProbesPerDimension%^2 : REDIM R(1 TO Np%, 1 TO Nd%, 0 TO Nt&) 'to create (Np/Nd) x (Np/Nd)
grid
            END IF 'F14
       CASE "F15" '(4-D) Kowalik's Function
            Nd%                   = 4
            NumProbesPerDimension% = 4 '20
            Np%                   = NumProbesPerDimension%*Nd%
            Nt&      = 500
            G        = 2##
            Alpha    = 2##
            Beta     = 2##
            DeltaT   = 1##
            Frep     = 0.5##
            PlaceInitialProbes$  = "UNIFORM ON-AXIS"
            InitialAcceleration$ = "ZERO"
            RepositionFactor$    = "VARIABLE" '"FIXED"
            CALL
ChangeRunParameters(NumProbesPerDimension%,Np%,Nd%,Nt&,G,Alpha,Beta,DeltaT,Frep,PlaceInitialProbes$,InitialAcceleration
$,RepositionFactor$,FunctionName$)
            Nd% = 4 'cannot change dimensionality of Kowalik's Function!
            Np% = NumProbesPerDimension%*Nd%
            REDIM XiMin(1 TO Nd%), XiMax(1 TO Nd%) : FOR i% = 1 TO Nd% : XiMin(i%) = -5## : XiMax(i%) = 5## : NEXT i%
            REDIM StartingXiMin(1 TO Nd%), StartingXiMax(1 TO Nd%) : FOR i% = 1 TO Nd% : StartingXiMin(i%) = XiMin(i%)
: StartingXiMax(i%) = XiMax(i%) : NEXT i%
       CASE "F16" '(2-D) Camel Back
            Nd%                   = 2
            NumProbesPerDimension% = 4 '20
            Np%                   = NumProbesPerDimension%*Nd%
            Nt&      = 500
            G        = 2##
            Alpha    = 2##
            Beta     = 2##
            DeltaT   = 1##
```

```
            Frep     = 0.5##
            PlaceInitialProbes$  = "UNIFORM ON-AXIS"
            InitialAcceleration$ = "ZERO"
            RepositionFactor$    = "VARIABLE" '"FIXED"
            CALL
ChangeRunParameters(NumProbesPerDimension%,Np%,Nd%,Nt&,G,Alpha,Beta,DeltaT,Frep,PlaceInitialProbes$,InitialAcceleration
$,RepositionFactor$,FunctionName$)
            Nd% = 2 'cannot change dimensionality of Camel Back function!
            Np% = NumProbesPerDimension%*Nd%
            REDIM XiMin(1 TO Nd%), XiMax(1 TO Nd%) : FOR i% = 1 TO Nd% : XiMin(i%) = -5## : XiMax(i%) = 5## : NEXT i%
            REDIM StartingXiMin(1 TO Nd%), StartingXiMax(1 TO Nd%) : FOR i% = 1 TO Nd% : StartingXiMin(i%) = XiMin(i%)
: StartingXiMax(i%) = XiMax(i%) : NEXT i%
            IF PlaceInitialProbes$ = "2D GRID" THEN
                Np% = NumProbesPerDimension%^2 : REDIM R(1 TO Np%, 1 TO Nd%, 0 TO Nt&) 'to create (Np/Nd) x (Np/Nd)
grid
            END IF 'F16
       CASE "F17" '(2-D) Branin
            Nd%                   = 2
            NumProbesPerDimension% = 4 '20
            Np%                   = NumProbesPerDimension%*Nd%
            Nt&      = 500
            G        = 2##
            Alpha    = 2##
            Beta     = 2##
            DeltaT   = 1##
            Frep     = 0.5##
            PlaceInitialProbes$  = "UNIFORM ON-AXIS"
            InitialAcceleration$ = "ZERO"
            RepositionFactor$    = "VARIABLE" '"FIXED"
            CALL
ChangeRunParameters(NumProbesPerDimension%,Np%,Nd%,Nt&,G,Alpha,Beta,DeltaT,Frep,PlaceInitialProbes$,InitialAcceleration
$,RepositionFactor$,FunctionName$)
            Nd% = 2 'cannot change dimensionality of Branin function!
            Np% = NumProbesPerDimension%*Nd%
            REDIM XiMin(1 TO Nd%), XiMax(1 TO Nd%) : XiMin(1) = -5## : XiMax(1) = 10## : XiMin(2) = 0## : XiMax(2) =
15##
            REDIM StartingXiMin(1 TO Nd%), StartingXiMax(1 TO Nd%) : FOR i% = 1 TO Nd% : StartingXiMin(i%) = XiMin(i%)
: StartingXiMax(i%) = XiMax(i%) : NEXT i%
            IF PlaceInitialProbes$ = "2D GRID" THEN
                Np% = NumProbesPerDimension%^2 : REDIM R(1 TO Np%, 1 TO Nd%, 0 TO Nt&) 'to create (Np/Nd) x (Np/Nd)
grid
            END IF 'F17
       CASE "F18" '(2-D) Goldstein-Price
            Nd%                   = 2
            NumProbesPerDimension% = 4 '20
            Np%                   = NumProbesPerDimension%*Nd%
            Nt&      = 500
            G        = 2##
            Alpha    = 2##
            Beta     = 2##
            DeltaT   = 1##
            Frep     = 0.5##
            PlaceInitialProbes$  = "UNIFORM ON-AXIS"
            InitialAcceleration$ = "ZERO"
            RepositionFactor$    = "VARIABLE" '"FIXED"
            CALL
ChangeRunParameters(NumProbesPerDimension%,Np%,Nd%,Nt&,G,Alpha,Beta,DeltaT,Frep,PlaceInitialProbes$,InitialAcceleration
$,RepositionFactor$,FunctionName$)
            Nd% = 2 'cannot change dimensionality of Branin function!
            Np% = NumProbesPerDimension%*Nd%
            REDIM XiMin(1 TO Nd%), XiMax(1 TO Nd%) : XiMin(1) = -2## : XiMax(1) = 2## : XiMin(2) = -2## : XiMax(2) =
2##
            REDIM StartingXiMin(1 TO Nd%), StartingXiMax(1 TO Nd%) : FOR i% = 1 TO Nd% : StartingXiMin(i%) = XiMin(i%)
: StartingXiMax(i%) = XiMax(i%) : NEXT i%
            IF PlaceInitialProbes$ = "2D GRID" THEN
                Np% = NumProbesPerDimension%^2 : REDIM R(1 TO Np%, 1 TO Nd%, 0 TO Nt&) 'to create (Np/Nd) x (Np/Nd)
grid
            END IF 'F18
       CASE "F19" '(3-D) Hartman's Family #1
            Nd%                   = 3
            NumProbesPerDimension% = 4 '20
            Np%                   = NumProbesPerDimension%*Nd%
            Nt&      = 500
            G        = 2##
            Alpha    = 2##
            Beta     = 2##
            DeltaT   = 1##
            Frep     = 0.5##
            PlaceInitialProbes$  = "UNIFORM ON-AXIS"
            InitialAcceleration$ = "ZERO"
            RepositionFactor$    = "VARIABLE" '"FIXED"
            CALL
ChangeRunParameters(NumProbesPerDimension%,Np%,Nd%,Nt&,G,Alpha,Beta,DeltaT,Frep,PlaceInitialProbes$,InitialAcceleration
$,RepositionFactor$,FunctionName$)
            Nd% = 3 'cannot change dimensionality of Hartman's Family!
            Np% = NumProbesPerDimension%*Nd%
            REDIM XiMin(1 TO Nd%), XiMax(1 TO Nd%) : FOR i% = 1 TO Nd% : XiMin(i%) = 0## : XiMax(i%) = 1## : NEXT i%
            REDIM StartingXiMin(1 TO Nd%), StartingXiMax(1 TO Nd%) : FOR i% = 1 TO Nd% : StartingXiMin(i%) = XiMin(i%)
: StartingXiMax(i%) = XiMax(i%) : NEXT i%
       CASE "F20" '(6-D) Hartman's Family #2
            Nd%                   = 6
            NumProbesPerDimension% = 4 '20
            Np%                   = NumProbesPerDimension%*Nd%
            Nt&      = 500
            G        = 2##
```

```
            Alpha    = 2##
            Beta     = 2##
            DeltaT   = 1##
            Frep     = 0.5##
            PlaceInitialProbes$  = "UNIFORM ON-AXIS"
            InitialAcceleration$ = "ZERO"
            RepositionFactor$    = "VARIABLE" '"FIXED"
            CALL
ChangeRunParameters(NumProbesPerDimension%,Np%,Nd%,Nt&,G,Alpha,Beta,DeltaT,Frep,PlaceInitialProbes$,InitialAcceleration
$,RepositionFactor$,FunctionName$)
            Nd% = 6 'cannot change dimensionality of Hartman's Family!
            Np% = NumProbesPerDimension%*Nd%
            REDIM XiMin(1 TO Nd%), XiMax(1 TO Nd%) : FOR i% = 1 TO Nd% : XiMin(i%) = 0## : XiMax(i%) = 1## : NEXT i%
            REDIM StartingXiMin(1 TO Nd%), StartingXiMax(1 TO Nd%) : FOR i% = 1 TO Nd% : StartingXiMin(i%) = XiMin(i%)
: StartingXiMax(i%) = XiMax(i%) : NEXT i%
       CASE "F21" '(4-D) Shekel's Family m=5
            Nd%                   = 4
            NumProbesPerDimension% = 4 '20
            Np%                   = NumProbesPerDimension%*Nd%
            Nt&      = 500
            G        = 2##
            Alpha    = 2##
            Beta     = 2##
            DeltaT   = 1##
            Frep     = 0.5##
            PlaceInitialProbes$  = "UNIFORM ON-AXIS"
            InitialAcceleration$ = "ZERO"
            RepositionFactor$    = "VARIABLE" '"FIXED"
            CALL
ChangeRunParameters(NumProbesPerDimension%,Np%,Nd%,Nt&,G,Alpha,Beta,DeltaT,Frep,PlaceInitialProbes$,InitialAcceleration
$,RepositionFactor$,FunctionName$)
            Nd% = 4 'cannot change dimensionality of Shekel's Family!
            Np% = NumProbesPerDimension%*Nd%
            REDIM XiMin(1 TO Nd%), XiMax(1 TO Nd%) : FOR i% = 1 TO Nd% : XiMin(i%) = 0## : XiMax(i%) = 10## : NEXT i%
            REDIM StartingXiMin(1 TO Nd%), StartingXiMax(1 TO Nd%) : FOR i% = 1 TO Nd% : StartingXiMin(i%) = XiMin(i%)
: StartingXiMax(i%) = XiMax(i%) : NEXT i%
       CASE "F22" '(4-D) Shekel's Family m=7
            Nd%                   = 4
            NumProbesPerDimension% = 4 '20
            Np%                   = NumProbesPerDimension%*Nd%
            Nt&      = 500
            G        = 2##
            Alpha    = 2##
            Beta     = 2##
            DeltaT   = 1##
            Frep     = 0.5##
            PlaceInitialProbes$  = "UNIFORM ON-AXIS"
            InitialAcceleration$ = "ZERO"
            RepositionFactor$    = "VARIABLE" '"FIXED"
            CALL
ChangeRunParameters(NumProbesPerDimension%,Np%,Nd%,Nt&,G,Alpha,Beta,DeltaT,Frep,PlaceInitialProbes$,InitialAcceleration
$,RepositionFactor$,FunctionName$)
            Nd% = 4 'cannot change dimensionality of Shekel's Family!
            Np% = NumProbesPerDimension%*Nd%
            REDIM XiMin(1 TO Nd%), XiMax(1 TO Nd%) : FOR i% = 1 TO Nd% : XiMin(i%) = 0## : XiMax(i%) = 10## : NEXT i%
            REDIM StartingXiMin(1 TO Nd%), StartingXiMax(1 TO Nd%) : FOR i% = 1 TO Nd% : StartingXiMin(i%) = XiMin(i%)
: StartingXiMax(i%) = XiMax(i%) : NEXT i%
       CASE "F23" '(4-D) Shekel's Family m=10
            Nd%                   = 4
            NumProbesPerDimension% = 4 '20
            Np%                   = NumProbesPerDimension%*Nd%
            Nt&      = 500
            G        = 2##
            Alpha    = 2##
            Beta     = 2##
            DeltaT   = 1##
            Frep     = 0.5##
            PlaceInitialProbes$  = "UNIFORM ON-AXIS"
            InitialAcceleration$ = "ZERO"
            RepositionFactor$    = "VARIABLE" '"FIXED"
            CALL
ChangeRunParameters(NumProbesPerDimension%,Np%,Nd%,Nt&,G,Alpha,Beta,DeltaT,Frep,PlaceInitialProbes$,InitialAcceleration
$,RepositionFactor$,FunctionName$)
            Nd% = 4 'cannot change dimensionality of Shekel's Family!
            Np% = NumProbesPerDimension%*Nd%
            REDIM XiMin(1 TO Nd%), XiMax(1 TO Nd%) : FOR i% = 1 TO Nd% : XiMin(i%) = 0## : XiMax(i%) = 10## : NEXT i%
            REDIM StartingXiMin(1 TO Nd%), StartingXiMax(1 TO Nd%) : FOR i% = 1 TO Nd% : StartingXiMin(i%) = XiMin(i%)
: StartingXiMax(i%) = XiMax(i%) : NEXT i%
         CASE "PBM_1" '2-D
            Nd%                   = 2
            NumProbesPerDimension% = 2 '4 '20
            Np%                   = NumProbesPerDimension%*Nd%
            Nt&      = 100
            G        = 2##
            Alpha    = 2##
            Beta     = 2##
            DeltaT   = 1##
            Frep     = 0.5##
            PlaceInitialProbes$  = "UNIFORM ON-AXIS"
            InitialAcceleration$ = "ZERO"
            RepositionFactor$    = "VARIABLE" '"FIXED"
            CALL
ChangeRunParameters(NumProbesPerDimension%,Np%,Nd%,Nt&,G,Alpha,Beta,DeltaT,Frep,PlaceInitialProbes$,InitialAcceleration
$,RepositionFactor$,FunctionName$)
            Nd% = 2 'cannot change dimensionality of PBM_1!
```

```
            Np% = NumProbesPerDimension%*Nd%
            REDIM XiMin(1 TO Nd%), XiMax(1 TO Nd%)
            XiMin(1) = 0.5## : XiMax(1) = 3## 'dipole length, L, in Wavelengths
            XiMin(2) = 0##   : XiMax(2) = Pi2 'polar angle, Theta, in Radians
            REDIM StartingXiMin(1 TO Nd%), StartingXiMax(1 TO Nd%) : FOR i% = 1 TO Nd% : StartingXiMin(i%) = XiMin(i%)
: StartingXiMax(i%) = XiMax(i%) : NEXT i%
            NN% = FREEFILE : OPEN "INFILE.DAT" FOR OUTPUT AS #NN% : PRINT #NN%,"PBM1.NEC" : PRINT #NN%,"PBM1.OUT" :
CLOSE #NN% 'NEC Input/Output Files
        CASE "PBM_2" '2-D
            AddNoiseToPBM2$ = "NO" '"YES" '"NO" '"YES"
            Nd%                    = 2
            NumProbesPerDimension% = 4 '20
            Np%                    = NumProbesPerDimension%*Nd%
            Nt&      = 100
            G        = 2##
            Alpha    = 2##
            Beta     = 2##
            DeltaT   = 1##
            Frep     = 0.5##
            PlaceInitialProbes$  = "UNIFORM ON-AXIS"
            InitialAcceleration$ = "ZERO"
            RepositionFactor$    = "VARIABLE" '"FIXED"
            CALL
ChangeRunParameters(NumProbesPerDimension%,Np%,Nd%,Nt&,G,Alpha,Beta,DeltaT,Frep,PlaceInitialProbes$,InitialAcceleration
$,RepositionFactor$,FunctionName$)
            Nd% = 2 'cannot change dimensionality of PBM_2!
            Np% = NumProbesPerDimension%*Nd%
            REDIM XiMin(1 TO Nd%), XiMax(1 TO Nd%)
            XiMin(1) = 5## : XiMax(1) = 15## 'dipole separation, D, in Wavelengths
            XiMin(2) = 0## : XiMax(2) = Pi   'polar angle, Theta, in Radians
            REDIM StartingXiMin(1 TO Nd%), StartingXiMax(1 TO Nd%) : FOR i% = 1 TO Nd% : StartingXiMin(i%) = XiMin(i%)
: StartingXiMax(i%) = XiMax(i%) : NEXT i%
            NN% = FREEFILE : OPEN "INFILE.DAT" FOR OUTPUT AS #NN% : PRINT #NN%,"PBM2.NEC" : PRINT #NN%,"PBM2.OUT" :
CLOSE #NN%
        CASE "PBM_3" '2-D
            Nd%                    = 2
            NumProbesPerDimension% = 4 '20
            Np%                    = NumProbesPerDimension%*Nd%
            Nt&      = 100
            G        = 2##
            Alpha    = 2##
            Beta     = 2##
            DeltaT   = 1##
            Frep     = 0.5##
            PlaceInitialProbes$  = "UNIFORM ON-AXIS"
            InitialAcceleration$ = "ZERO"
            RepositionFactor$    = "VARIABLE" '"FIXED"
            CALL
ChangeRunParameters(NumProbesPerDimension%,Np%,Nd%,Nt&,G,Alpha,Beta,DeltaT,Frep,PlaceInitialProbes$,InitialAcceleration
$,RepositionFactor$,FunctionName$)
            Nd% = 2 'cannot change dimensionality of PBM_3!
            Np% = NumProbesPerDimension%*Nd%
            REDIM XiMin(1 TO Nd%), XiMax(1 TO Nd%)
            XiMin(1) = 0## : XiMax(1) = 4## 'Phase Parameter, Beta (0-4)
            XiMin(2) = 0## : XiMax(2) = Pi  'polar angle, Theta, in Radians
            REDIM StartingXiMin(1 TO Nd%), StartingXiMax(1 TO Nd%) : FOR i% = 1 TO Nd% : StartingXiMin(i%) = XiMin(i%)
: StartingXiMax(i%) = XiMax(i%) : NEXT i%
            NN% = FREEFILE : OPEN "INFILE.DAT" FOR OUTPUT AS #NN% : PRINT #NN%,"PBM3.NEC" : PRINT #NN%,"PBM3.OUT" :
CLOSE #NN%
        CASE "PBM_4" '2-D
            Nd%                    = 2
            NumProbesPerDimension% = 4 '6 '2 '4 '20
            Np%                    = NumProbesPerDimension%*Nd%
            Nt&      = 100
            G        = 2##
            Alpha    = 2##
            Beta     = 2##
            DeltaT   = 1##
            Frep     = 0.5##
            PlaceInitialProbes$  = "UNIFORM ON-AXIS"
            InitialAcceleration$ = "ZERO"
            RepositionFactor$    = "VARIABLE" '"FIXED"
            CALL
ChangeRunParameters(NumProbesPerDimension%,Np%,Nd%,Nt&,G,Alpha,Beta,DeltaT,Frep,PlaceInitialProbes$,InitialAcceleration
$,RepositionFactor$,FunctionName$)
            Nd% = 2 'cannot change dimensionality of PBM_4!
            Np% = NumProbesPerDimension%*Nd%
            REDIM XiMin(1 TO Nd%), XiMax(1 TO Nd%)
            XiMin(1) = 0.5##   : XiMax(1) = 1.5##  'ARM LENGTH (NOT Total Length), wavelengths (0.5-1.5)
            XiMin(2) = Pi/18## : XiMax(2) = Pi/2## 'Inner angle, Alpha, in Radians (Pi/18-Pi/2)
            REDIM StartingXiMin(1 TO Nd%), StartingXiMax(1 TO Nd%) : FOR i% = 1 TO Nd% : StartingXiMin(i%) = XiMin(i%)
: StartingXiMax(i%) = XiMax(i%) : NEXT i%
            NN% = FREEFILE : OPEN "INFILE.DAT" FOR OUTPUT AS #NN% : PRINT #NN%,"PBM4.NEC" : PRINT #NN%,"PBM4.OUT" :
CLOSE #NN%
        CASE "PBM_5"
            NumCollinearElements%  = 6 '30 'EVEN or ODD: 6,7,10,13,16,24 used by PBM
            Nd%                    = NumCollinearElements% - 1
            NumProbesPerDimension% = 4 '20
            Np%                    = NumProbesPerDimension%*Nd%
            Nt&      = 100
            G        = 2##
            Alpha    = 2##
            Beta     = 2##
            DeltaT   = 1##
            Frep     = 0.5##
```

```
            PlaceInitialProbes$  = "UNIFORM ON-AXIS"
            InitialAcceleration$ = "ZERO"
            RepositionFactor$    = "VARIABLE" '"FIXED"
            CALL
ChangeRunParameters(NumProbesPerDimension%,Np%,Nd%,Nt&,G,Alpha,Beta,DeltaT,Frep,PlaceInitialProbes$,InitialAcceleration
$,RepositionFactor$,FunctionName$)
            Nd% = NumCollinearElements% - 1
            Np% = NumProbesPerDimension%*Nd%
            REDIM XiMin(1 TO Nd%), XiMax(1 TO Nd%) : FOR i% = 1 TO Nd% : XiMin(i%) = 0.5## : XiMax(i%) = 1.5## : NEXT
i%
            REDIM StartingXiMin(1 TO Nd%), StartingXiMax(1 TO Nd%) : FOR i% = 1 TO Nd% : StartingXiMin(i%) = XiMin(i%)
: StartingXiMax(i%) = XiMax(i%) : NEXT i%
            NN% = FREEFILE : OPEN "INFILE.DAT" FOR OUTPUT AS #NN% : PRINT #NN%,"PBM5.NEC" : PRINT #NN%,"PBM5.OUT" :
CLOSE #NN%
'   ==========================================================================================
'   NOTE - DON'T FORGET TO ADD NEW TEST FUNCTIONS TO FUNCTION ObjectiveFunction() ABOVE !!
'   ==========================================================================================
    END SELECT
    IF Nd% > 100 THEN Nt& = MIN(Nt&,200) 'to avoid array dimensioning problems
    DiagLength = 0## : FOR i% = 1 TO Nd% : DiagLength = DiagLength + (XiMax(i%)-XiMin(i%))^2 : NEXT i% : DiagLength =
SQR(DiagLength) 'compute length of decision space principal diagonal
END SUB 'GetFunctionRunParameters()
'--------------------------------
FUNCTION ParrottF4(R(),Nd%,p%,j&) 'Parrott F4 (1-D)
'MAXIMUM = 1 AT ~0.0796875... WITH ZERO OFFSET.
'References:
'Beasley, D., D. R. Bull, and R. R. Martin, “A Sequential Niche Technique for Multimodal
'Function Optimization,” Evol. Comp. (MIT Press), vol. 1, no. 2, 1993, pp. 101-125
'(online at http://citeseer.ist.psu.edu/beasley93sequential.html).
'Parrott, D., and X. Li, “Locating and Tracking Multiple Dynamic Optima by a Particle Swarm
'Model Using Speciation,” IEEE Trans. Evol. Computation, vol. 10, no. 4, Aug. 2006, pp. 440-458.
LOCAL Z, x, offset AS EXT
    offset = 0##
    x = R(p%,1,j&)
    Z = EXP(-2##*LOG(2##)*((x-0.08##-offset)/0.854##)^2)*(SIN(5##*Pi*((x-offset)^0.75##-0.05##)))^6 'WARNING! This is a
NATURAL LOG, NOT Log10!!!
    ParrottF4 = Z
END FUNCTION 'ParrottF4()
'-----------------------------
FUNCTION SGO(R(),Nd%,p%,j&) 'SGO Function (2-D)
'MAXIMUM = ~130.8323226... @ ~(-2.8362075...,-2.8362075...) WITH ZERO OFFSET.
'Reference:
'Hsiao, Y., Chuang, C., Jiang, J., and Chien, C., "A Novel Optimization Algorithm: Space
'Gravitational Optimization," Proc. of 2005 IEEE International Conference on Systems, Man,
'and Cybernetics, 3, 2323-2328. (2005)
    LOCAL x1, x2, Z, t1, t2, SGOx1offset, SGOx2offset AS EXT
    SGOx1offset = 0## : SGOx2offset = 0##
'   SGOx1offset = 40## : SGOx2offset = 10##
    x1 = R(p%,1,j&) - SGOx1offset : x2 = R(p%,2,j&) - SGOx2offset
    t1 = x1^4 - 16##*x1^2 + 0.5##*x1 : t2 = x2^4 - 16##*x2^2 + 0.5##*x2
    Z = t1 + t2
    SGO = -Z
END FUNCTION 'SGO()
'------------------
FUNCTION GoldsteinPrice(R(),Nd%,p%,j&) 'Goldstein-Price Function (2-D)
'MAXIMUM = -3 @ (0,-1) WITH ZERO OFFSET.
'Reference:
'Cui, Z., Zeng, J., and Sun, G. (2006) ‘A Fast Particle Swarm Optimization,’ Int’l. J.
'Innovative Computing, Information and Control, vol. 2, no. 6, December, pp. 1365-1380.
    LOCAL Z, x1, x2, offset1, offset2, t1, t2 AS EXT
    offset1 = 0## : offset2 = 0##
'   offset1 = 20## : offset2 = -10##
    x1 = R(p%,1,j&)-offset1 : x2 = R(p%,2,j&)-offset2
    t1 = 1##+(x1+x2+1##)^2*(19##-14##*x1+3##*x1^2-14##*x2+6##*x1*x2+3##*x2^2)
    t2 = 30##+(2##*x1-3##*x2)^2*(18##-32##*x1+12##*x1^2+48##*x2-36##*x1*x2+27##*x2^2)
    Z  = t1*t2
    GoldsteinPrice = -Z
END FUNCTION 'GoldesteinPrice()
'-----------
FUNCTION StepFunction(R(),Nd%,p%,j&) 'Step Function (n-D)
'MAXIMUM VALUE = 0 @ [Offset]^n.
'Reference:
'Yao, X., Liu, Y., and Lin, G., “Evolutionary Programming Made Faster,”
'IEEE Trans. Evolutionary Computation, Vol. 3, No. 2, 82-102, Jul. 1999.
    LOCAL Offset, Z AS EXT
    LOCAL i%
    Z = 0## : Offset = 0## '75.123## '0##
    FOR i% = 1 TO Nd%
        IF Nd% = 2 AND i% = 1 THEN Offset = 75 '75##
        IF Nd% = 2 AND i% = 2 THEN Offset = 35 '30 '35##
        Z = Z + INT((R(p%,i%,j&)-Offset) + 0.5##)^2
    NEXT i%
    StepFunction = -Z
END FUNCTION 'StepFunction()
'-----------
FUNCTION Schwefel226(R(),Nd%,p%,j&) 'Schwefel Problem 2.26 (n-D)
'MAXIMUM = 12,569.5 @ [420.8687]^30 (30-D CASE).
'Reference:
'Yao, X., Liu, Y., and Lin, G., “Evolutionary Programming Made Faster,”
'IEEE Trans. Evolutionary Computation, Vol. 3, No. 2, 82-102, Jul. 1999.
    LOCAL Z, Xi AS EXT
    LOCAL i%
    Z = 0##
    FOR i% = 1 TO Nd%
        Xi = R(p%,i%,j&)
```

```
        Z = Z + Xi*SIN(SQR(ABS(Xi)))
    NEXT i%
    Schwefel226 = Z
END FUNCTION 'SCHWEFEL226()
'-----------
FUNCTION Colville(R(),Nd%,p%,j&) 'Colville Function (4-D)
'MAXIMUM = 0 @ (1,1,1,1) WITH ZERO OFFSET.
'Reference:
'Doo-Hyun, and Se-Young, O., "A New Mutation Rule for Evolutionary Programming Motivated from
'Backpropagation Learning," IEEE Trans. Evolutionary Computation, Vol. 4, No. 2, pp. 188-190,
'July 2000.
    LOCAL Z, x1, x2, x3, x4, offset AS EXT
    offset = 7.123##
    x1 = R(p%,1,j&)-offset : x2 = R(p%,2,j&)-offset : x3 = R(p%,3,j&)-offset : x4 = R(p%,4,j&)-offset
    Z =  100##*(x2-x1^2)^2 + (1##-x1)^2  + _
          90##*(x4-x3^2)^2 + (1##-x3)^2  + _
        10.1##*((x2-1##)^2 + (x4-1##)^2) + _
        19.8##*(x2-1##)*(x4-1##)
    Colville = -Z
END FUNCTION 'Colville()
'-----------
'Max of zero at (0,...,0)
FUNCTION Griewank(R(),Nd%,p%,j&) 'Griewank (n-D)
    LOCAL Offset, Sum, Prod, Z, Xi AS EXT
    LOCAL i%
    Sum = 0## : Prod = 1##
    Offset = 75.123##
    FOR i% = 1 TO Nd%
        Xi = R(p%,i%,j&) - Offset
        Sum = Sum + Xi^2
        Prod = Prod*COS(Xi/SQR(i%))
    NEXT i%
    Z = Sum/4000## - Prod + 1##
    Griewank = -Z
END FUNCTION 'Griewank()
'-----------
FUNCTION Himmelblau(R(),Nd%,p%,j&) 'Himmelblau (2-D)
    LOCAL Z, x1, x2, offset AS EXT
    offset = 0##
    x1 = R(p%,1,j&)-offset : x2 = R(p%,2,j&)-offset
    Z = 200## - (x1^2 + x2 -11##)^2 - (x1+x2^2-7##)^2
    Himmelblau = Z
END FUNCTION 'Himmelblau()
'-----------
FUNCTION F1(R(),Nd%,p%,j&) 'F1 (n-D)
'MAXIMUM = ZERO (n-D CASE).
'Reference:
    LOCAL Z, Xi AS EXT
    LOCAL i%
    Z = 0##
    FOR i% = 1 TO Nd%
        Xi = R(p%,i%,j&)
        Z = Z + Xi^2
    NEXT i%
    F1 = -Z
END FUNCTION 'F1
'-----------
FUNCTION F2(R(),Nd%,p%,j&) 'F2 (n-D)
'MAXIMUM = ZERO (n-D CASE).
'Reference:
    LOCAL Sum, prod, Z, Xi AS EXT
    LOCAL i%
    Z = 0## : Sum = 0## : Prod = 1##
    FOR i% = 1 TO Nd%
        Xi = R(p%,i%,j&)
        Sum  = Sum+ ABS(Xi)
        Prod = Prod*ABS(Xi)
    NEXT i%
    Z = Sum + Prod
    F2 = -Z
END FUNCTION 'F2
'-----------
FUNCTION F3(R(),Nd%,p%,j&) 'F3 (n-D)
'MAXIMUM = ZERO (n-D CASE).
'Reference:
    LOCAL Z, Xk, Sum AS EXT
    LOCAL i%, k%
    Z = 0##
    FOR i% = 1 TO Nd%
        Sum = 0##
        FOR k% = 1 TO i%
            Xk = R(p%,k%,j&)
            Sum = Sum + Xk
        NEXT k%
        Z = Z + Sum^2
    NEXT i%
    F3 = -Z
END FUNCTION 'F3
'-----------
FUNCTION F4(R(),Nd%,p%,j&) 'F4 (n-D)
'MAXIMUM = ZERO (n-D CASE).
'Reference:
    LOCAL Z, Xi, MaxXi AS EXT
    LOCAL i%
    MaxXi = -1E4200
```

```
    FOR i% = 1 TO Nd%
        Xi = R(p%,i%,j&)
        IF ABS(Xi) >= MaxXi THEN MaxXi = ABS(Xi)
    NEXT i%
    F4 = -MaxXi
END FUNCTION 'F4
'-----------
FUNCTION F5(R(),Nd%,p%,j&) 'F5 (n-D)
'MAXIMUM = ZERO (n-D CASE).
'Reference:
    LOCAL Z, Xi, XiPlus1 AS EXT
    LOCAL i%
    Z = 0##
    FOR i% = 1 TO Nd%-1
        Xi      = R(p%,i%,j&)
        XiPlus1 = R(p%,i%+1,j&)
        Z = Z + (100##*(XiPlus1-Xi^2)^2+(Xi-1##))^2
    NEXT i%
    F5 = -Z
END FUNCTION 'F5
'-----------
FUNCTION F6(R(),Nd%,p%,j&) 'F6
'MAXIMUM VALUE = 0 @ [Offset]^n.
'Reference:
'Yao, X., Liu, Y., and Lin, G., "Evolutionary Programming Made Faster,"
'IEEE Trans. Evolutionary Computation, Vol. 3, No. 2, 82-102, Jul. 1999.
    LOCAL Z AS EXT
    LOCAL i%
    Z = 0##
    FOR i% = 1 TO Nd%
        Z = Z + INT(R(p%,i%,j&) + 0.5##)^2
    NEXT i%
    F6 = -Z
END FUNCTION 'F6
'-----------
FUNCTION F7(R(),Nd%,p%,j&) 'F7
'MAXIMUM VALUE = 0 @ [Offset]^n.
'Reference:
'Yao, X., Liu, Y., and Lin, G., "Evolutionary Programming Made Faster,"
'IEEE Trans. Evolutionary Computation, Vol. 3, No. 2, 82-102, Jul. 1999.
    LOCAL Z, Xi AS EXT
    LOCAL i%
    Z = 0##
    FOR i% = 1 TO Nd%
        Xi = R(p%,i%,j&)
        Z = Z + i%*Xi^4
    NEXT i%
    F7 = -Z - RandomNum(0##,1##)
END FUNCTION 'F7
'-----------
FUNCTION F8(R(),Nd%,p%,j&) '(n-D) F8 [Schwefel Problem 2.26]
'MAXIMUM = 12,569.5 @ [420.8687]^30 (30-D CASE).
'Reference:
'Yao, X., Liu, Y., and Lin, G., "Evolutionary Programming Made Faster,"
'IEEE Trans. Evolutionary Computation, Vol. 3, No. 2, 82-102, Jul. 1999.
    LOCAL Z, Xi AS EXT
    LOCAL i%
    Z = 0##
    FOR i% = 1 TO Nd%
        Xi = R(p%,i%,j&)
        Z  = Z - Xi*SIN(SQR(ABS(Xi)))
    NEXT i%
    F8 = -Z
END FUNCTION 'F8
'-----------
FUNCTION F9(R(),Nd%,p%,j&) '(n-D) F9 [Rastrigin]
'MAXIMUM = ZERO (n-D CASE).
'Reference:
'Yao, X., Liu, Y., and Lin, G., "Evolutionary Programming Made Faster,"
'IEEE Trans. Evolutionary Computation, Vol. 3, No. 2, 82-102, Jul. 1999.
    LOCAL Z, Xi AS EXT
    LOCAL i%
    Z = 0##
    FOR i% = 1 TO Nd%
        Xi = R(p%,i%,j&)
        Z  = Z + (Xi^2 - 10##*COS(TwoPi*Xi) + 10##)^2
    NEXT i%
    F9 = -Z
END FUNCTION 'F9
'-----------
FUNCTION F10(R(),Nd%,p%,j&) '(n-D) F10 [Ackley's Function]
'MAXIMUM = ZERO (n-D CASE).
'Reference:
'Yao, X., Liu, Y., and Lin, G., "Evolutionary Programming Made Faster,"
'IEEE Trans. Evolutionary Computation, Vol. 3, No. 2, 82-102, Jul. 1999.
    LOCAL Z, Xi, Sum1, Sum2 AS EXT
    LOCAL i%
    Z = 0## : Sum1 = 0## : Sum2 = 0##
    FOR i% = 1 TO Nd%
        Xi   = R(p%,i%,j&)
        Sum1 = Sum1 + Xi^2
        Sum2 = Sum2 + COS(TwoPi*Xi)
    NEXT i%
    Z = -20##*EXP(-0.2##*SQR(Sum1/Nd%)) - EXP(Sum2/Nd%) + 20## + e
    F10 = -Z
```

```
END FUNCTION 'F10
'-----------
FUNCTION F11(R(),Nd%,p%,j&) '(n-D) F11
'MAXIMUM = ZERO (n-D CASE).
'Reference:
'Yao, X., Liu, Y., and Lin, G., "Evolutionary Programming Made Faster,"
'IEEE Trans. Evolutionary Computation, Vol. 3, No. 2, 82-102, Jul. 1999.
    LOCAL Z, Xi, Sum, Prod AS EXT
    LOCAL i%
    Z = 0## : Sum = 0## : Prod = 1##
    FOR i% = 1 TO Nd%
        Xi   = R(p%,i%,j&)
        Sum  = Sum + (Xi-100##)^2
        Prod = Prod*COS((Xi-100##)/SQR(i%))
    NEXT i%
    Z = Sum/4000## - Prod + 1##
    F11 = -Z
END FUNCTION 'F11
'-----
FUNCTION F12(R(),Nd%,p%,j&) '(n-D) F12, Penalized #1
    LOCAL Offset, Sum1, Sum2, Z, u, Xi, Xn, XiPlus1, Yi, YiPlus1, Yn, X1, Y1, a, k AS EXT
    LOCAL i%, m%
    a = 5## : k = 100## : m% = 4
    X1 = R(p%,1,j&) : Y1 = 1## + (X1+1##)/4##
    Sum1 = 10##*SIN(Pi*Y1)^2 + (Yn-1##)^2
    FOR i% = 1 TO Nd%-1
        Xi = R(p%,i%,j&) : XiPlus1 = R(p%,i%+1,j&) : Xn = R(p%,Nd%,j&)
        Yi = 1## + (Xi+1##)/4## : YiPlus1 = 1## + (XiPlus1+1##)/4## : Yn = 1## + (Xn+1##)/4##
        Sum1 = Sum1 + (Yi-1##)^2*(1##+10##*SIN(Pi*YiPlus1)^2)
    NEXT i%
    Sum1 = Pi*Sum1/Nd%
    Sum2 = 0##
    FOR i% = 1 TO Nd%
        Xi = R(p%,i%,j&)
        u = 0##
        IF Xi >  a THEN u = k*(Xi-a)^m%
        IF Xi < -a THEN u = k*(-Xi-a)^m%
        Sum2 = Sum2 + u
    NEXT i%
    Z = Sum1 + Sum2
    F12 = -Z
END FUNCTION 'F12()
'-----
FUNCTION F13(R(),Nd%,p%,j&) '(n-D) F13, Penalized #2
    LOCAL Offset, Sum1, Sum2, Z, u, Xi, Xn, Xi1, X1, a, k AS EXT
    LOCAL i%, m%
    a = 5## : k = 100## : m% = 4
    X1 = R(p%,1,j&)
    Xn = R(p%,Nd%,j&)
    Sum1 = 0##
    FOR i% = 1 TO Nd%-1
        Xi  = R(p%,i%,j&)
        Xi1 = R(p%,i%+1,j&)
        Sum1 = Sum1 + (Xi-1##)^2*(1##+(SIN(3##*Pi*Xi1))^2)+(Xn-1##)^2*(1##+(SIN(TwoPi*Xn))^2)
    NEXT i%
    Sum2 = 0##
    FOR i% = 1 TO Nd%
        Xi = R(p%,i%,j&)
        u = 0##
        IF Xi >  a THEN u = k*(Xi-a)^m%
        IF Xi < -a THEN u = k*(-Xi-a)^m%
        Sum2 = Sum2 + u
    NEXT i%
    Z = ((SIN(3##*Pi*X1))^2+Sum1)/10## + Sum2
    F13 = -Z
END FUNCTION 'F13()
'-----
SUB FillArrayAij  'needed for function F14, Shekel's Foxholes
    Aij(1,1)=-32##  : Aij(1,2)=-16##  : Aij(1,3)=0##  : Aij(1,4)=16##  : Aij(1,5)=32##
    Aij(1,6)=-32##  : Aij(1,7)=-16##  : Aij(1,8)=0##  : Aij(1,9)=16##  : Aij(1,10)=32##
    Aij(1,11)=-32## : Aij(1,12)=-16## : Aij(1,13)=0## : Aij(1,14)=16## : Aij(1,15)=32##
    Aij(1,16)=-32## : Aij(1,17)=-16## : Aij(1,18)=0## : Aij(1,19)=16## : Aij(1,20)=32##
    Aij(1,21)=-32## : Aij(1,22)=-16## : Aij(1,23)=0## : Aij(1,24)=16## : Aij(1,25)=32##
    Aij(2,1)=-32##  : Aij(2,2)=-32## : Aij(2,3)=-32## : Aij(2,4)=-32## : Aij(2,5)=-32##
    Aij(2,6)=-16##  : Aij(2,7)=-16## : Aij(2,8)=-16## : Aij(2,9)=-16## : Aij(2,10)=-16##
    Aij(2,11)=0##   : Aij(2,12)=0##  : Aij(2,13)=0##  : Aij(2,14)=0##  : Aij(2,15)=0##
    Aij(2,16)=16##  : Aij(2,17)=16## : Aij(2,18)=16## : Aij(2,19)=16## : Aij(2,20)=16##
    Aij(2,21)=32##  : Aij(2,22)=32## : Aij(2,23)=32## : Aij(2,24)=32## : Aij(2,25)=32#
END SUB
'-----
FUNCTION F14(R(),Nd%,p%,j&) 'F14 (2-D) Shekel's Foxholes (INVERTED...)
    LOCAL Sum1, Sum2, Z, Xi AS EXT
    LOCAL i%, jj%
    Sum1 = 0##
    FOR jj% = 1 TO 25
        Sum2 = 0##
        FOR i% = 1 TO 2
            Xi = R(p%,i%,j&)
            Sum2 = Sum2 + (Xi-Aij(i%,jj%))^6
        NEXT i%
        Sum1 = Sum1 + 1##/(jj%+Sum2)
    NEXT j%
    Z = 1##/(0.002##+Sum1)
    F14 = -Z
END FUNCTION 'F14
```

```
'-----------
FUNCTION F16(R(),Nd%,p%,j&) 'F16 (2-D) 6-Hump Camel-Back
    LOCAL x1, x2, Z AS EXT
    x1 = R(p%,1,j&) : x2 = R(p%,2,j&)
    Z = 4##*x1^2 - 2.1##*x1^4 + x1^6/3## + x1*x2 - 4*x2^2 + 4*x2^4
    F16 = -Z
END FUNCTION 'F16
'-----------
FUNCTION F15(R(),Nd%,p%,j&) 'F15 (4-D) Kowalik's Function
'Global maximum = -0.0003075 @ (0.1928,0.1908,0.1231,0.1358)
    LOCAL x1, x2, x3, x4, Num, Denom, Z, Aj(), Bj() AS EXT
    LOCAL jj%
    REDIM Aj(1 TO 11), Bj(1 TO 11)
    Aj(1)  = 0.1957## : Bj(1)  = 1##/0.25##
    Aj(2)  = 0.1947## : Bj(2)  = 1##/0.50##
    Aj(3)  = 0.1735## : Bj(3)  = 1##/1.00##
    Aj(4)  = 0.1600## : Bj(4)  = 1##/2.00##
    Aj(5)  = 0.0844## : Bj(5)  = 1##/4.00##
    Aj(6)  = 0.0627## : Bj(6)  = 1##/6.00##
    Aj(7)  = 0.0456## : Bj(7)  = 1##/8.00##
    Aj(8)  = 0.0342## : Bj(8)  = 1##/10.0##
    Aj(9)  = 0.0323## : Bj(9)  = 1##/12.0##
    Aj(10) = 0.0235## : Bj(10) = 1##/14.0##
    Aj(11) = 0.0246## : Bj(11) = 1##/16.0##
    Z = 0##
    x1 = R(p%,1,j&) : x2 = R(p%,2,j&) : x3 = R(p%,3,j&) : x4 = R(p%,4,j&)
    FOR jj% = 1 TO 11
       Num   = x1*(Bj(jj%)^2+Bj(jj%)*x2)
       Denom = Bj(jj%)^2+Bj(jj%)*x3+x4
       Z = Z + (Aj(jj%)-Num/Denom)^2
    NEXT jj%
    F15 = -Z
END FUNCTION 'F15
'-----------
FUNCTION F17(R(),Nd%,p%,j&) 'F17, (2-D) Branin
'Global maximum = -0.398 @ (-3.142.12.275), (3.142,2.275), (9.425,2.425)
    LOCAL x1, x2, Z AS EXT
    x1 = R(p%,1,j&) : x2 = R(p%,2,j&)
    Z = (x2-5.1##*x1^2/(4##*Pi^2)+5##*x1/Pi-6##)^2 + 10##*(1##-1##/(8##*Pi))*COS(x1) + 10##
    F17 = -Z
END FUNCTION 'F17
'-----------
FUNCTION F18(R(),Nd%,p%,j&) 'Goldstein-Price 2-D Test Function
'Global maximum = -3 @ (0,-1)
    LOCAL Z, x1, x2, t1, t2 AS EXT
    x1 = R(p%,1,j&) : x2 = R(p%,2,j&)
    t1 = 1##+(x1+x2+1##)^2*(19##-14##*x1+3##*x1^2-14##*x2+6##*x1*x2+3##*x2^2)
    t2 = 30##+(2##*x1-3##*x2)^2*(18##-32##*x1+12##*x1^2+48##*x2-36##*x1*x2+27##*x2^2)
    Z  = t1*t2
    F18 = -Z
END FUNCTION 'F18()
'-----------
FUNCTION F19(R(),Nd%,p%,j&) 'F19 (3-D) Hartman's Family #1
'Global maximum = 3.86 @ (0.114,0.556,0.852)
    LOCAL Xi, Z, Sum, Aji(), Cj(), Pji() AS EXT
    LOCAL i%, jj%, m%
    REDIM Aji(1 TO 4, 1 TO 3), Cj(1 TO 4), Pji(1 TO 4, 1 TO 3)
    Aji(1,1) = 3.0## : Aji(1,2) = 10## : Aji(1,3) = 30## : Cj(1) = 1.0##
    Aji(2,1) = 0.1## : Aji(2,2) = 10## : Aji(2,3) = 35## : Cj(2) = 1.2##
    Aji(3,1) = 3.0## : Aji(3,2) = 10## : Aji(3,3) = 30## : Cj(3) = 3.0##
    Aji(4,1) = 0.1## : Aji(4,2) = 10## : Aji(4,3) = 35## : Cj(4) = 3.2##
    Pji(1,1) = 0.36890## : Pji(1,2) = 0.1170## : Pji(1,3) = 0.2673##
    Pji(2,1) = 0.46990## : Pji(2,2) = 0.4387## : Pji(2,3) = 0.7470##
    Pji(3,1) = 0.10910## : Pji(3,2) = 0.8732## : Pji(3,3) = 0.5547##
    Pji(4,1) = 0.03815## : Pji(4,2) = 0.5743## : Pji(4,3) = 0.8828##
    Z = 0##
    FOR jj% = 1 TO 4
        Sum = 0##
        FOR i% = 1 TO 3
            Xi = R(p%,i%,j&)
            Sum = Sum + Aji(jj%,i%)*(Xi-Pji(jj%,i%))^2
        NEXT i%
        Z = Z + Cj(jj%)*EXP(-Sum)
    NEXT jj%
    F19 = Z
END FUNCTION 'F19
'-----------
FUNCTION F20(R(),Nd%,p%,j&) 'F20 (6-D) Hartman's Family #2
'Global maximum = 3.32 @ (0.201,0.150,0.477,0.275,0.311,0.657)
    LOCAL Xi, Z, Sum, Aji(), Cj(), Pji() AS EXT
    LOCAL i%, jj%, m%
    REDIM Aji(1 TO 4, 1 TO 6), Cj(1 TO 4), Pji(1 TO 4, 1 TO 6)
    Aji(1,1) = 10.0## : Aji(1,2) = 3.00## : Aji(1,3) = 17.0## : Cj(1) = 1.0##
    Aji(2,1) = 0.05## : Aji(2,2) = 10.0## : Aji(2,3) = 17.0## : Cj(2) = 1.2##
    Aji(3,1) = 3.00## : Aji(3,2) = 3.50## : Aji(3,3) = 1.70## : Cj(3) = 3.0##
    Aji(4,1) = 17.0## : Aji(4,2) = 8.00## : Aji(4,3) = 0.05## : Cj(4) = 3.2##
    Aji(1,4) = 3.5## : Aji(1,5) = 1.7## : Aji(1,6) =  8##
    Aji(2,4) = 0.1## : Aji(2,5) =   8## : Aji(2,6) = 14##
    Aji(3,4) =  10## : Aji(3,5) =  17## : Aji(3,6) =  8##
    Aji(4,4) =  10## : Aji(4,5) = 0.1## : Aji(4,6) = 14##
    Pji(1,1) = 0.13120## : Pji(1,2) = 0.1696## : Pji(1,3) = 0.5569##
    Pji(2,1) = 0.23290## : Pji(2,2) = 0.4135## : Pji(2,3) = 0.8307##
    Pji(3,1) = 0.23480## : Pji(3,2) = 0.1415## : Pji(3,3) = 0.3522##
    Pji(4,1) = 0.40470## : Pji(4,2) = 0.8828## : Pji(4,3) = 0.8732##
    Pji(1,4) = 0.01240## : Pji(1,5) = 0.8283## : Pji(1,6) = 0.5886##
```

```
    Pji(2,4) = 0.37360## : Pji(2,5) = 0.1004## : Pji(2,6) = 0.9991##
    Pji(3,4) = 0.28830## : Pji(3,5) = 0.3047## : Pji(3,6) = 0.6650##
    Pji(4,4) = 0.57430## : Pji(4,5) = 0.1091## : Pji(4,6) = 0.0381##
    Z = 0##
    FOR jj% = 1 TO 4
        Sum = 0##
        FOR i% = 1 TO 6
            Xi = R(p%,i%,j&)
            Sum = Sum + Aji(jj%,i%)*(Xi-Pji(jj%,i%))^2
        NEXT i%
        Z = Z + Cj(jj%)*EXP(-Sum)
    NEXT jj%
    F20 = Z
END FUNCTION 'F20
'-----------
FUNCTION F21(R(),Nd%,p%,j&) 'F21 (4-D) Shekel's Family m=5
'Global maximum = 10
    LOCAL Xi, Z, Sum, Aji(), Cj() AS EXT
    LOCAL i%, jj%, m%
    m% = 5 : REDIM Aji(1 TO m%, 1 TO 4), Cj(1 TO m%)
    Aji(1,1)  = 4## : Aji(1,2)  =   4## : Aji(1,3)  = 4## : Aji(1,4)  =   4## : Cj(1)  = 0.1##
    Aji(2,1)  = 1## : Aji(2,2)  =   1## : Aji(2,3)  = 1## : Aji(2,4)  =   1## : Cj(2)  = 0.2##
    Aji(3,1)  = 8## : Aji(3,2)  =   8## : Aji(3,3)  = 8## : Aji(3,4)  =   8## : Cj(3)  = 0.2##
    Aji(4,1)  = 6## : Aji(4,2)  =   6## : Aji(4,3)  = 6## : Aji(4,4)  =   6## : Cj(4)  = 0.4##
    Aji(5,1)  = 3## : Aji(5,2)  =   7## : Aji(5,3)  = 3## : Aji(5,4)  =   7## : Cj(5)  = 0.4##
    Z = 0##
    FOR jj% = 1 TO m%  'NOTE:  Index jj% is used to avoid same variable name as j&
        Sum = 0##
        FOR i% = 1 TO 4 'Shekel's family is 4-D only
            Xi = R(p%,i%,j&)
            Sum = Sum + (Xi-Aji(jj%,i%))^2
        NEXT i%
        Z = Z + 1##/(Sum + Cj(jj%))
    NEXT jj%
    F21 = Z
END FUNCTION 'F21
'-----------
FUNCTION F22(R(),Nd%,p%,j&) 'F22 (4-D) Shekel's Family m=7
'Global maximum = 10
    LOCAL Xi, Z, Sum, Aji(), Cj() AS EXT
    LOCAL i%, jj%, m%
    m% = 7 : REDIM Aji(1 TO m%, 1 TO 4), Cj(1 TO m%)
    Aji(1,1)  = 4## : Aji(1,2)  =   4## : Aji(1,3)  = 4## : Aji(1,4)  =   4## : Cj(1)  = 0.1##
    Aji(2,1)  = 1## : Aji(2,2)  =   1## : Aji(2,3)  = 1## : Aji(2,4)  =   1## : Cj(2)  = 0.2##
    Aji(3,1)  = 8## : Aji(3,2)  =   8## : Aji(3,3)  = 8## : Aji(3,4)  =   8## : Cj(3)  = 0.2##
    Aji(4,1)  = 6## : Aji(4,2)  =   6## : Aji(4,3)  = 6## : Aji(4,4)  =   6## : Cj(4)  = 0.4##
    Aji(5,1)  = 3## : Aji(5,2)  =   7## : Aji(5,3)  = 3## : Aji(5,4)  =   7## : Cj(5)  = 0.4##
    Aji(6,1)  = 2## : Aji(6,2)  =   9## : Aji(6,3)  = 2## : Aji(6,4)  =   9## : Cj(6)  = 0.6##
    Aji(7,1)  = 5## : Aji(7,2)  =   5## : Aji(7,3)  = 3## : Aji(7,4)  =   3## : Cj(7)  = 0.3##
    Z = 0##
    FOR jj% = 1 TO m%  'NOTE:  Index jj% is used to avoid same variable name as j&
        Sum = 0##
        FOR i% = 1 TO 4 'Shekel's family is 4-D only
            Xi = R(p%,i%,j&)
            Sum = Sum + (Xi-Aji(jj%,i%))^2
        NEXT i%
        Z = Z + 1##/(Sum + Cj(jj%))
    NEXT jj%
    F22 = Z
END FUNCTION 'F22
'-----------
FUNCTION F23(R(),Nd%,p%,j&) 'F23 (4-D) Shekel's Family m=10
'Global maximum = 10
    LOCAL Xi, Z, Sum, Aji(), Cj() AS EXT
    LOCAL i%, jj%, m%
    m% = 10 : REDIM Aji(1 TO m%, 1 TO 4), Cj(1 TO m%)
    Aji(1,1)  = 4## : Aji(1,2)  =   4## : Aji(1,3)  = 4## : Aji(1,4)  =   4## : Cj(1)  = 0.1##
    Aji(2,1)  = 1## : Aji(2,2)  =   1## : Aji(2,3)  = 1## : Aji(2,4)  =   1## : Cj(2)  = 0.2##
    Aji(3,1)  = 8## : Aji(3,2)  =   8## : Aji(3,3)  = 8## : Aji(3,4)  =   8## : Cj(3)  = 0.2##
    Aji(4,1)  = 6## : Aji(4,2)  =   6## : Aji(4,3)  = 6## : Aji(4,4)  =   6## : Cj(4)  = 0.4##
    Aji(5,1)  = 3## : Aji(5,2)  =   7## : Aji(5,3)  = 3## : Aji(5,4)  =   7## : Cj(5)  = 0.4##
    Aji(6,1)  = 2## : Aji(6,2)  =   9## : Aji(6,3)  = 2## : Aji(6,4)  =   9## : Cj(6)  = 0.6##
    Aji(7,1)  = 5## : Aji(7,2)  =   5## : Aji(7,3)  = 3## : Aji(7,4)  =   3## : Cj(7)  = 0.3##
    Aji(8,1)  = 8## : Aji(8,2)  =   1## : Aji(8,3)  = 8## : Aji(8,4)  =   1## : Cj(8)  = 0.7##
    Aji(9,1)  = 6## : Aji(9,2)  =   2## : Aji(9,3)  = 6## : Aji(9,4)  =   2## : Cj(9)  = 0.5##
    Aji(10,1) = 7## : Aji(10,2) = 3.6## : Aji(10,3) = 7## : Aji(10,4) = 3.6## : Cj(10) = 0.5##
    Z = 0##
    FOR jj% = 1 TO m%  'NOTE:  Index jj% is used to avoid same variable name as j&
        Sum = 0##
        FOR i% = 1 TO 4 'Shekel's family is 4-D only
            Xi = R(p%,i%,j&)
            Sum = Sum + (Xi-Aji(jj%,i%))^2
        NEXT i%
        Z = Z + 1##/(Sum + Cj(jj%))
    NEXT jj%
    F23 = Z
END FUNCTION 'F23
'========================================================= END FUNCTION DEFINITIONS
=============================================================
SUB Plot2DbestProbeTrajectories(NumTrajectories%,M(),R(),XiMin(),XiMax(),Np%,Nd%,LastStep&,FunctionName$)
LOCAL TrajectoryNumber%, ProbeNumber%, StepNumber&, N%, M%, ProcID???
LOCAL MaximumFitness, MinimumFitness AS EXT
LOCAL BestProbeThisStep%()
LOCAL BestFitnessThisStep(), TempFitness() AS EXT
LOCAL Annotation$, xCoord$, yCoord$, GnuPlotEXE$, PlotWithLines$
```

```
    Annotation$    = ""
    PlotWithLines$ = "YES" '"NO"
    NumTrajectories% = MIN(Np%,NumTrajectories%)
    GnuPlotEXE$ = "wgnuplot.exe"
'   ---------------- Get Min/Max Fitnesses -----------------
    MaximumFitness = M(1,0) : MinimumFitness = M(1,0)  'Note:  M(p%,j&)
    FOR StepNumber& = 0 TO LastStep&
        FOR ProbeNumber% = 1 TO Np%
            IF M(ProbeNumber%,StepNumber&) >= MaximumFitness THEN MaximumFitness = M(ProbeNumber%,StepNumber&)
            IF M(ProbeNumber%,StepNumber&) =< MinimumFitness THEN MinimumFitness = M(ProbeNumber%,StepNumber&)
        NEXT ProbeNumber%
    NEXT StepNumber%
'   ------------- Copy Fitness Array M() into TempFitness to Preserve M() ----------------
    REDIM TempFitness(1 TO Np%, 0 TO LastStep&)
    FOR StepNumber& = 0 TO LastStep&
        FOR ProbeNumber% = 1 TO Np%
            TempFitness(ProbeNumber%,StepNumber&) = M(ProbeNumber%,StepNumber&)
        NEXT ProbeNumber%
    NEXT StepNumber%
'   ------------ LOOP ON TRAJECTORIES -----------
    FOR TrajectoryNumber% = 1 TO NumTrajectories%
'       --------------- Get Trajectory Coordinate Data -----------------
        REDIM BestFitnessThisStep(0 TO LastStep&), BestProbeThisStep%(0 TO LastStep&)
        FOR StepNumber& = 0 TO LastStep&
            BestFitnessThisStep(StepNumber&) = TempFitness(1,StepNumber&)
            FOR ProbeNumber% = 1 TO Np%
                IF TempFitness(ProbeNumber%,StepNumber&) >= BestFitnessThisStep(StepNumber&) THEN
                    BestFitnessThisStep(StepNumber&) = TempFitness(ProbeNumber%,StepNumber&)
                    BestProbeThisStep%(StepNumber&)  = ProbeNumber%
                END IF
            NEXT ProbeNumber%
        NEXT StepNumber&
'   ----- Create Plot Data File -----
    N% = FREEFILE
    SELECT CASE TrajectoryNumber%
        CASE 1  : OPEN "t1"  FOR OUTPUT AS #N%
        CASE 2  : OPEN "t2"  FOR OUTPUT AS #N%
        CASE 3  : OPEN "t3"  FOR OUTPUT AS #N%
        CASE 4  : OPEN "t4"  FOR OUTPUT AS #N%
        CASE 5  : OPEN "t5"  FOR OUTPUT AS #N%
        CASE 6  : OPEN "t6"  FOR OUTPUT AS #N%
        CASE 7  : OPEN "t7"  FOR OUTPUT AS #N%
        CASE 8  : OPEN "t8"  FOR OUTPUT AS #N%
        CASE 9  : OPEN "t9"  FOR OUTPUT AS #N%
        CASE 10 : OPEN "t10" FOR OUTPUT AS #N%
    END SELECT
'   ------------ Write Plot File Data ------------
'   M% = freefile : open "BestProbebData" for output as #M% 'debug
'   print #M%, "  Step #  BestProbe#           x1                  x2"
    FOR StepNumber& = 0 TO LastStep&
        PRINT #N%, USING$("######.########
######.########",R(BestProbeThisStep%(StepNumber&),1,StepNumber&),R(BestProbeThisStep%(StepNumber&),2,StepNumber&))
'       PRINT #M%, USING$("#####     #####      ######.########
######.########",StepNumber&,BestProbeThisStep%(StepNumber&),R(BestProbeThisStep%(StepNumber&),1,StepNumber&),R(BestPro
beThisStep%(StepNumber&),2,StepNumber&))
        TempFitness(BestProbeThisStep%(StepNumber&),StepNumber&) = MinimumFitness 'so that same max will not be found
for next trajectory
    NEXT StepNumber%
    CLOSE #N%
'   Close #M%
    NEXT TrajectoryNumber%
'   ------------------------- Plot Trajectories --------------------------
    CALL CreateGNUplotINIfile(0.13##*ScreenWidth&,0.18##*ScreenHeight&,0.7##*ScreenHeight&,0.7##*ScreenHeight&)
    Annotation$ = ""
    N% = FREEFILE
    OPEN "cmd2d.gp" FOR OUTPUT AS #N%
        PRINT #N%, "set xrange ["+REMOVE$(STR$(XiMin(1)),ANY"" )+":"+REMOVE$(STR$(XiMax(1)),ANY" ")+"]"
        PRINT #N%, "set yrange ["+REMOVE$(STR$(XiMin(2)),ANY" ")+":"+REMOVE$(STR$(XiMax(2)),ANY" ")+"]"
        'PRINT #N%, "set label "      + Quote$ + Annotation$ + Quote$ + " at graph " + xCoord$ + "," + yCoord$
        PRINT #N%, "set grid xtics " + "10"
        PRINT #N%, "set grid ytics " + "10"
        PRINT #N%, "set grid mxtics"
        PRINT #N%, "set grid mytics"
        PRINT #N%, "show grid"
        PRINT #N%, "set title "  + Quote$ + "2D "+ FunctionName$+" TRAJECTORIES OF PROBES WITH BEST\nFITNESSES (ORDERED
BY FITNESS)" + "\n" + RunID$ + Quote$
        PRINT #N%, "set xlabel " + Quote$ + "x1\n\n"                                   + Quote$
        PRINT #N%, "set ylabel " + Quote$ + "\nx2"                                     + Quote$
        IF PlotWithLines$ = "YES" THEN
            SELECT CASE NumTrajectories%
                CASE 1  : PRINT #N%, "plot "+Quote$+"t1"+Quote$+" w l lw 3"
                CASE 2  : PRINT #N%, "plot "+Quote$+"t1"+Quote$+" w l lw 3,"+Quote$+"t2"+Quote$+" w l"
                CASE 3  : PRINT #N%, "plot "+Quote$+"t1"+Quote$+" w l lw 3,"+Quote$+"t2"+Quote$+" w
l,"+Quote$+"t3"+Quote$+" w l"
                CASE 4  : PRINT #N%, "plot "+Quote$+"t1"+Quote$+" w l lw 3,"+Quote$+"t2"+Quote$+" w
l,"+Quote$+"t3"+Quote$+" w l,"+Quote$+"t4"+Quote$+" w l"
                CASE 5  : PRINT #N%, "plot "+Quote$+"t1"+Quote$+" w l lw 3,"+Quote$+"t2"+Quote$+" w
l,"+Quote$+"t3"+Quote$+" w l,"+Quote$+"t4"+Quote$+" w l,"+Quote$+"t5"+Quote$+" w l"
                CASE 6  : PRINT #N%, "plot "+Quote$+"t1"+Quote$+" w l lw 3,"+Quote$+"t2"+Quote$+" w
l,"+Quote$+"t3"+Quote$+" w l,"+Quote$+"t4"+Quote$+" w l,"+Quote$+"t5"+Quote$+" w l,"+Quote$+"t6"+Quote$+" w l"
                CASE 7  : PRINT #N%, "plot "+Quote$+"t1"+Quote$+" w l lw 3,"+Quote$+"t2"+Quote$+" w
l,"+Quote$+"t3"+Quote$+" w l,"+Quote$+"t4"+Quote$+" w l,"+Quote$+"t5"+Quote$+" w l,"+Quote$+"t6"+Quote$+" w l,"+_
                                             Quote$+"t7"+Quote$+" w l"
                CASE 8  : PRINT #N%, "plot "+Quote$+"t1"+Quote$+" w l lw 3,"+Quote$+"t2"+Quote$+" w
l,"+Quote$+"t3"+Quote$+" w l,"+Quote$+"t4"+Quote$+" w l,"+Quote$+"t5"+Quote$+" w l,"+Quote$+"t6"+Quote$+" w l,"+_
```

```
                                        Quote$+"t7"+Quote$+" w l,"     +Quote$+"t8"+Quote$+" w l"
                CASE 9  : PRINT #N%, "plot "+Quote$+"t1"+Quote$+" w l lw 3,"+Quote$+"t2"+Quote$+" w
l,"+Quote$+"t3"+Quote$+" w l,"+Quote$+"t4"+Quote$+" w l,"+Quote$+"t5"+Quote$+" w l,"+Quote$+"t6"+Quote$+" w l,"+_
                                        Quote$+"t7"+Quote$+" w l,"     +Quote$+"t8"+Quote$+" w
l,"+Quote$+"t9"+Quote$+" w l"
                CASE 10 : PRINT #N%, "plot "+Quote$+"t1"+Quote$+" w l lw 3,"+Quote$+"t2"+Quote$+" w
l,"+Quote$+"t3"+Quote$+" w l,"+Quote$+"t4"+Quote$+" w l,"+Quote$+"t5"+Quote$+" w l,"+Quote$+"t6"+Quote$+" w l,"+_
                                        Quote$+"t7"+Quote$+" w l,"     +Quote$+"t8"+Quote$+" w
l,"+Quote$+"t9"+Quote$+" w l,"+Quote$+"t10"+Quote$+" w l"
            END SELECT
        ELSE
            SELECT CASE NumTrajectories%
                CASE 1  : PRINT #N%, "plot "+Quote$+"t1"+Quote$+" lw 2"
                CASE 2  : PRINT #N%, "plot "+Quote$+"t1"+Quote$+" lw 2,"+Quote$+"t2"+Quote$
                CASE 3  : PRINT #N%, "plot "+Quote$+"t1"+Quote$+" lw 2,"+Quote$+"t2"+Quote$+" ,"+Quote$+"t3"+Quote$
                CASE 4  : PRINT #N%, "plot "+Quote$+"t1"+Quote$+" lw 2,"+Quote$+"t2"+Quote$+" ,"+Quote$+"t3"+Quote$+"
,"+Quote$+"t4"+Quote$
                CASE 5  : PRINT #N%, "plot "+Quote$+"t1"+Quote$+" lw 2,"+Quote$+"t2"+Quote$+" ,"+Quote$+"t3"+Quote$+"
,"+Quote$+"t4"+Quote$+" ,"+Quote$+"t5"+Quote$
                CASE 6  : PRINT #N%, "plot "+Quote$+"t1"+Quote$+" lw 2,"+Quote$+"t2"+Quote$+" ,"+Quote$+"t3"+Quote$+"
,"+Quote$+"t4"+Quote$+" ,"+Quote$+"t5"+Quote$+" ,"+Quote$+"t6"+Quote$
                CASE 7  : PRINT #N%, "plot "+Quote$+"t1"+Quote$+" lw 2,"+Quote$+"t2"+Quote$+" ,"+Quote$+"t3"+Quote$+"
,"+Quote$+"t4"+Quote$+" ,"+Quote$+"t5"+Quote$+" ,"+Quote$+"t6"+Quote$+" ,"+_
                                        Quote$+"t7"+Quote$
                CASE 8  : PRINT #N%, "plot "+Quote$+"t1"+Quote$+" lw 2,"+Quote$+"t2"+Quote$+" ,"+Quote$+"t3"+Quote$+"
,"+Quote$+"t4"+Quote$+" ,"+Quote$+"t5"+Quote$+" ,"+Quote$+"t6"+Quote$+" ,"+_
                                        Quote$+"t7"+Quote$+" ,"    +Quote$+"t8"+Quote$
                CASE 9  : PRINT #N%, "plot "+Quote$+"t1"+Quote$+" lw 2,"+Quote$+"t2"+Quote$+" ,"+Quote$+"t3"+Quote$+"
,"+Quote$+"t4"+Quote$+" ,"+Quote$+"t5"+Quote$+" ,"+Quote$+"t6"+Quote$+" ,"+_
                                        Quote$+"t7"+Quote$+" ,"    +Quote$+"t8"+Quote$+" ,"+Quote$+"t9"+Quote$
                CASE 10 : PRINT #N%, "plot "+Quote$+"t1"+Quote$+" lw 2,"+Quote$+"t2"+Quote$+" ,"+Quote$+"t3"+Quote$+"
,"+Quote$+"t4"+Quote$+" ,"+Quote$+"t5"+Quote$+" ,"+Quote$+"t6"+Quote$+" ,"+_
                                        Quote$+"t7"+Quote$+" ,"    +Quote$+"t8"+Quote$+" ,"+Quote$+"t9"+Quote$+"
,"+Quote$+"t10"+Quote$
            END SELECT
        END IF
    CLOSE #N%
    ProcID??? = SHELL(GnuPlotEXE$+" cmd2d.gp -") : CALL Delay(0.5##)
END SUB 'Plot2DbestProbeTrajectories()
'----
SUB Plot2DindividualProbeTrajectories(NumTrajectories%,M(),R(),XiMin(),XiMax(),Np%,Nd%,LastStep&,FunctionName$)
LOCAL ProbeNumber%, StepNumber&, N%, ProcID???
LOCAL Annotation$, xCoord$, yCoord$, GnuPlotEXE$, PlotWithLines$
    NumTrajectories% = MIN(Np%,NumTrajectories%)
    Annotation$    = ""
    PlotWithLines$ = "YES" '"NO"
    GnuPlotEXE$ = "wgnuplot.exe"
'   -------------- LOOP ON PROBES ---------------
    FOR ProbeNumber% = 1 TO MIN(NumTrajectories%,Np%)
'   ----- Create Plot Data File -----
    N% = FREEFILE
    SELECT CASE ProbeNumber%
        CASE 1  : OPEN "p1"  FOR OUTPUT AS #N%
        CASE 2  : OPEN "p2"  FOR OUTPUT AS #N%
        CASE 3  : OPEN "p3"  FOR OUTPUT AS #N%
        CASE 4  : OPEN "p4"  FOR OUTPUT AS #N%
        CASE 5  : OPEN "p5"  FOR OUTPUT AS #N%
        CASE 6  : OPEN "p6"  FOR OUTPUT AS #N%
        CASE 7  : OPEN "p7"  FOR OUTPUT AS #N%
        CASE 8  : OPEN "p8"  FOR OUTPUT AS #N%
        CASE 9  : OPEN "p9"  FOR OUTPUT AS #N%
        CASE 10 : OPEN "p10" FOR OUTPUT AS #N%
        CASE 11 : OPEN "p11" FOR OUTPUT AS #N%
        CASE 12 : OPEN "p12" FOR OUTPUT AS #N%
        CASE 13 : OPEN "p13" FOR OUTPUT AS #N%
        CASE 14 : OPEN "p14" FOR OUTPUT AS #N%
        CASE 15 : OPEN "p15" FOR OUTPUT AS #N%
        CASE 16 : OPEN "p16" FOR OUTPUT AS #N%
    END SELECT
'   ------------ Write Plot File Data ------------
    FOR StepNumber& = 0 TO LastStep&
        PRINT #N%, USING$("######.########
######.########",R(ProbeNumber%,1,StepNumber&),R(ProbeNumber%,2,StepNumber&))
    NEXT StepNumber%
    CLOSE #N%
    NEXT ProbeNumber%
'   ------------------------------------------- Plot Trajectories ---------------------------------------------
'usage:  CALL CreateGNUplotINIfile(PlotWindowULC_X%,PlotWindowULC_Y%,PlotWindowWidth%,PlotWindowHeight%)
    CALL CreateGNUplotINIfile(0.17##*ScreenWidth&,0.22##*ScreenHeight&,0.7##*ScreenHeight&,0.7##*ScreenHeight&)
    Annotation$ = ""
    N% = FREEFILE
    OPEN "cmd2d.gp" FOR OUTPUT AS #N%
        PRINT #N%, "set xrange ["+REMOVE$(STR$(XiMin(1)),ANY"" )+":"+REMOVE$(STR$(XiMax(1)),ANY" ")+"]"
        PRINT #N%, "set yrange ["+REMOVE$(STR$(XiMin(2)),ANY" ")+":"+REMOVE$(STR$(XiMax(2)),ANY" ")+"]"
        'PRINT #N%, "set label "      + Quote$ + Annotation$ + Quote$ + " at graph " + xCoord$ + "," + yCoord$
        PRINT #N%, "set grid xtics " + "10"
        PRINT #N%, "set grid ytics " + "10"
        PRINT #N%, "set grid mxtics"
        PRINT #N%, "set grid mytics"
        PRINT #N%, "show grid"
        PRINT #N%, "set title "  + Quote$ + "2D "+ FunctionName$+" INDIVIDUAL PROBE TRAJECTORIES\n(ORDERED BY PROBE #)"
+ "\n" + RunID$ + Quote$
        PRINT #N%, "set xlabel " + Quote$ + "x1\n\n"                                  + Quote$
        PRINT #N%, "set ylabel " + Quote$ + "\nx2"                                    + Quote$
        IF PlotWithLines$ = "YES" THEN
```

```
            SELECT CASE NumTrajectories%
                CASE 1  : PRINT #N%, "plot "+Quote$+"p1"  +Quote$+" w l lw 1"
                CASE 2  : PRINT #N%, "plot "+Quote$+"p1"  +Quote$+" w l lw 1,"+Quote$+"p2"+Quote$+" w l"
                CASE 3  : PRINT #N%, "plot "+Quote$+"p1"  +Quote$+" w l lw 1,"+Quote$+"p2"+Quote$+" w
l,"+Quote$+"p3"+Quote$+" w l"
                CASE 4  : PRINT #N%, "plot "+Quote$+"p1"  +Quote$+" w l lw 1,"+Quote$+"p2"+Quote$+" w
l,"+Quote$+"p3"+Quote$+" w l,"+Quote$+"p4"+Quote$+" w l"
                CASE 5  : PRINT #N%, "plot "+Quote$+"p1"  +Quote$+" w l lw 1,"+Quote$+"p2"+Quote$+" w
l,"+Quote$+"p3"+Quote$+" w l,"+Quote$+"p4"+Quote$+" w l,"+Quote$+"p5"+Quote$+" w l"
                CASE 6  : PRINT #N%, "plot "+Quote$+"p1"  +Quote$+" w l lw 1,"+Quote$+"p2"+Quote$+" w
l,"+Quote$+"p3"+Quote$+" w l,"+Quote$+"p4"+Quote$+" w l,"+Quote$+"p5"+Quote$+" w l,"+Quote$+"p6"+Quote$+" w l"
                CASE 7  : PRINT #N%, "plot "+Quote$+"p1"  +Quote$+" w l lw 1,"+Quote$+"p2"+Quote$+" w
l,"+Quote$+"p3"+Quote$+" w l,"+Quote$+"p4"+Quote$+" w l,"+Quote$+"p5"+Quote$+" w l,"+Quote$+"p6"+Quote$+" w l,"+_
                                            Quote$+"p7"  +Quote$+" w l"
                CASE 8  : PRINT #N%, "plot "+Quote$+"p1"  +Quote$+" w l lw 1,"+Quote$+"p2"+Quote$+" w
l,"+Quote$+"p3"+Quote$+" w l,"+Quote$+"p4"+Quote$+" w l,"+Quote$+"p5"+Quote$+" w l,"+Quote$+"p6"+Quote$+" w l,"+_
                                            Quote$+"p7"  +Quote$+" w l,"     +Quote$+"p8"+Quote$+" w l"
                CASE 9  : PRINT #N%, "plot "+Quote$+"p1"  +Quote$+" w l lw 1,"+Quote$+"p2"+Quote$+" w
l,"+Quote$+"p3"+Quote$+" w l,"+Quote$+"p4"+Quote$+" w l,"+Quote$+"p5"+Quote$+" w l,"+Quote$+"p6"+Quote$+" w l,"+_
                                            Quote$+"p7"  +Quote$+" w l,"     +Quote$+"p8"+Quote$+" w
l,"+Quote$+"p9"+Quote$+" w l"
                CASE 10 : PRINT #N%, "plot "+Quote$+"p1"  +Quote$+" w l lw 1,"+Quote$+"p2" +Quote$+" w l,"+Quote$+"p3"
+Quote$+" w l,"+Quote$+"p4" +Quote$+" w l,"+Quote$+"p5"+Quote$+" w l,"+Quote$+"p6"+Quote$+" w l,"+_
                                            Quote$+"p7"  +Quote$+" w l,"     +Quote$+"p8" +Quote$+" w l,"+Quote$+"p9"
+Quote$+" w l,"+Quote$+"p10"+Quote$+" w l"
                CASE 11 : PRINT #N%, "plot "+Quote$+"p1"  +Quote$+" w l lw 1,"+Quote$+"p2" +Quote$+" w l,"+Quote$+"p3"
+Quote$+" w l,"+Quote$+"p4" +Quote$+" w l,"+Quote$+"p5" +Quote$+" w l,"+Quote$+"p6" +Quote$+" w l,"+_
                                            Quote$+"p7"  +Quote$+" w l,"     +Quote$+"p8" +Quote$+" w l,"+Quote$+"p9"
+Quote$+" w l,"+Quote$+"p10"+Quote$+" w l,"+Quote$+"p11"+Quote$+" w l"
                CASE 12 : PRINT #N%, "plot "+Quote$+"p1"  +Quote$+" w l lw 1,"+Quote$+"p2" +Quote$+" w l,"+Quote$+"p3"
+Quote$+" w l,"+Quote$+"p4" +Quote$+" w l,"+Quote$+"p5" +Quote$+" w l,"+Quote$+"p6" +Quote$+" w l,"+_
                                            Quote$+"p7"  +Quote$+" w l,"     +Quote$+"p8" +Quote$+" w l,"+Quote$+"p9"
+Quote$+" w l,"+Quote$+"p10"+Quote$+" w l,"+Quote$+"p11"+Quote$+" w l,"+Quote$+"p12"+Quote$+" w l"
                CASE 13 : PRINT #N%, "plot "+Quote$+"p1"  +Quote$+" w l lw 1,"+Quote$+"p2" +Quote$+" w l,"+Quote$+"p3"
+Quote$+" w l,"+Quote$+"p4" +Quote$+" w l,"+Quote$+"p5" +Quote$+" w l,"+Quote$+"p6" +Quote$+" w l,"+_
                                            Quote$+"p7"  +Quote$+" w l,"     +Quote$+"p8" +Quote$+" w l,"+Quote$+"p9"
+Quote$+" w l,"+Quote$+"p10"+Quote$+" w l,"+Quote$+"p11"+Quote$+" w l,"+Quote$+"p12"+Quote$+" w l,"+_
                                            Quote$+"p13 "+Quote$+" w l"
                CASE 14 : PRINT #N%, "plot "+Quote$+"p1"  +Quote$+" w l lw 1,"+Quote$+"p2" +Quote$+" w l,"+Quote$+"p3"
+Quote$+" w l,"+Quote$+"p4" +Quote$+" w l,"+Quote$+"p5" +Quote$+" w l,"+Quote$+"p6" +Quote$+" w l,"+_
                                            Quote$+"p7"  +Quote$+" w l,"     +Quote$+"p8" +Quote$+" w l,"+Quote$+"p9"
+Quote$+" w l,"+Quote$+"p10"+Quote$+" w l,"+Quote$+"p11"+Quote$+" w l,"+Quote$+"p12"+Quote$+" w l,"+_
                                            Quote$+"p13" +Quote$+" w l,"     +Quote$+"p14"+Quote$+" w l"
                CASE 15 : PRINT #N%, "plot "+Quote$+"p1"  +Quote$+" w l lw 1,"+Quote$+"p2" +Quote$+" w l,"+Quote$+"p3"
+Quote$+" w l,"+Quote$+"p4" +Quote$+" w l,"+Quote$+"p5" +Quote$+" w l,"+Quote$+"p6" +Quote$+" w l,"+_
                                            Quote$+"p7"  +Quote$+" w l,"     +Quote$+"p8" +Quote$+" w l,"+Quote$+"p9"
+Quote$+" w l,"+Quote$+"p10"+Quote$+" w l,"+Quote$+"p11"+Quote$+" w l,"+Quote$+"p12"+Quote$+" w l,"+_
                                            Quote$+"p13" +Quote$+" w l,"     +Quote$+"p14"+Quote$+" w
l,"+Quote$+"p15"+Quote$+" w l"
                CASE 16 : PRINT #N%, "plot "+Quote$+"p1"  +Quote$+" w l lw 1,"+Quote$+"p2" +Quote$+" w l,"+Quote$+"p3"
+Quote$+" w l,"+Quote$+"p4" +Quote$+" w l,"+Quote$+"p5" +Quote$+" w l,"+Quote$+"p6" +Quote$+" w l,"+_
                                            Quote$+"p7"  +Quote$+" w l,"     +Quote$+"p8" +Quote$+" w l,"+Quote$+"p9"
+Quote$+" w l,"+Quote$+"p10"+Quote$+" w l,"+Quote$+"p11"+Quote$+" w l,"+Quote$+"p12"+Quote$+" w l,"+_
                                            Quote$+"p13" +Quote$+" w l,"     +Quote$+"p14"+Quote$+" w
l,"+Quote$+"p15"+Quote$+" w l,"+Quote$+"p16"+Quote$+" w l"
            END SELECT
        ELSE
            SELECT CASE NumTrajectories%
                CASE 1  : PRINT #N%, "plot "+Quote$+"p1"+Quote$+" lw 1"
                CASE 2  : PRINT #N%, "plot "+Quote$+"p1"+Quote$+" lw 1,"+Quote$+"p2"+Quote$
                CASE 3  : PRINT #N%, "plot "+Quote$+"p1"+Quote$+" lw 1,"+Quote$+"p2"+Quote$+" ,"+Quote$+"p3"+Quote$
                CASE 4  : PRINT #N%, "plot "+Quote$+"p1"+Quote$+" lw 1,"+Quote$+"p2"+Quote$+" ,"+Quote$+"p3"+Quote$+"
,"+Quote$+"p4"+Quote$
                CASE 5  : PRINT #N%, "plot "+Quote$+"p1"+Quote$+" lw 1,"+Quote$+"p2"+Quote$+" ,"+Quote$+"p3"+Quote$+"
,"+Quote$+"p4"+Quote$+" ,"+Quote$+"p5"+Quote$
                CASE 6  : PRINT #N%, "plot "+Quote$+"p1"+Quote$+" lw 1,"+Quote$+"p2"+Quote$+" ,"+Quote$+"p3"+Quote$+"
,"+Quote$+"p4"+Quote$+" ,"+Quote$+"p5"+Quote$+" ,"+Quote$+"p6"+Quote$
                CASE 7  : PRINT #N%, "plot "+Quote$+"p1"+Quote$+" lw 1,"+Quote$+"p2"+Quote$+" ,"+Quote$+"p3"+Quote$+"
,"+Quote$+"p4"+Quote$+" ,"+Quote$+"p5"+Quote$+" ,"+Quote$+"p6"+Quote$+" ,"+_
                                            Quote$+"p7"+Quote$
                CASE 8  : PRINT #N%, "plot "+Quote$+"p1"+Quote$+" lw 1,"+Quote$+"p2"+Quote$+" ,"+Quote$+"p3"+Quote$+"
,"+Quote$+"p4"+Quote$+" ,"+Quote$+"p5"+Quote$+" ,"+Quote$+"p6"+Quote$+" ,"+_
                                            Quote$+"p7"+Quote$+" ,"    +Quote$+"p8"+Quote$
                CASE 9  : PRINT #N%, "plot "+Quote$+"p1"+Quote$+" lw 1,"+Quote$+"p2"+Quote$+" ,"+Quote$+"p3"+Quote$+"
,"+Quote$+"p4"+Quote$+" ,"+Quote$+"p5"+Quote$+" ,"+Quote$+"p6"+Quote$+" ,"+_
                                            Quote$+"p7"+Quote$+" ,"    +Quote$+"p8"+Quote$+" ,"+Quote$+"p9"+Quote$
                CASE 10 : PRINT #N%, "plot "+Quote$+"p1"+Quote$+" lw 1," +Quote$+"p2" +Quote$+" ,"+Quote$+"p3"+Quote$+"
,"+Quote$+"p4"  +Quote$+" ,"+Quote$+"p5"+Quote$+" ,"+Quote$+"p6"+Quote$+" ,"+_
                                            Quote$+"p7"+Quote$+" ,"     +Quote$+"p8" +Quote$+" ,"+Quote$+"p9"+Quote$+"
,"+Quote$+"p10" +Quote$
                CASE 11 : PRINT #N%, "plot "+Quote$+"p1"+Quote$+" lw 1," +Quote$+"p2" +Quote$+" ,"+Quote$+"p3"+Quote$+"
,"+Quote$+"p4"  +Quote$+" ,"+Quote$+"p5" +Quote$+" ,"+Quote$+"p6"+Quote$+" ,"+_
                                            Quote$+"p7"+Quote$+" ,"     +Quote$+"p8" +Quote$+" ,"+Quote$+"p9"+Quote$+"
,"+Quote$+"p10" +Quote$+" ,"+Quote$+"p11"+Quote$
                CASE 12 : PRINT #N%, "plot "+Quote$+"p1"+Quote$+" lw 1," +Quote$+"p2" +Quote$+" ,"+Quote$+"p3"+Quote$+"
,"+Quote$+"p4"  +Quote$+" ,"+Quote$+"p5" +Quote$+" ,"+Quote$+"p6" +Quote$+" ,"+_
                                            Quote$+"p7"+Quote$+" ,"     +Quote$+"p8" +Quote$+" ,"+Quote$+"p9"+Quote$+"
,"+Quote$+"p10" +Quote$+" ,"+Quote$+"p11"+Quote$+" ,"+Quote$+"p12"+Quote$
                CASE 13 : PRINT #N%, "plot "+Quote$+"p1" +Quote$+" lw 1,"+Quote$+"p2" +Quote$+" ,"+Quote$+"p3"+Quote$+"
,"+Quote$+"p4"  +Quote$+" ,"+Quote$+"p5" +Quote$+" ," +Quote$+"p6" +Quote$+" ,"+_
                                            Quote$+"p7" +Quote$+" ,"    +Quote$+"p8" +Quote$+" ,"+Quote$+"p9"+Quote$+"
,"+Quote$+"p10" +Quote$+" ,"+Quote$+"p11"+Quote$+" ," +Quote$+"p12"+Quote$+" ,"+_
                                            Quote$+"p13"+Quote$
                CASE 14 : PRINT #N%, "plot "+Quote$+"p1" +Quote$+" lw 1,"+Quote$+"p2" +Quote$+" ,"+Quote$+"p3"+Quote$+"
,"+Quote$+"p4"  +Quote$+" ,"+Quote$+"p5" +Quote$+" ," +Quote$+"p6" +Quote$+" ,"+_
```

```
                                             Quote$+"p7" +Quote$+" ,"    +Quote$+"p8" +Quote$+" ,"+Quote$+"p9"+Quote$+"
,"+Quote$+"p10" +Quote$+" ,"+Quote$+"p11"+Quote$+" ," +Quote$+"p12"+Quote$+" ,"+_
                                             Quote$+"p13"+Quote$+" ,"    +Quote$+"p14"+Quote$
                CASE 15 : PRINT #N%, "plot "+Quote$+"p1" +Quote$+" lw 1,"+Quote$+"p2" +Quote$+" ,"+Quote$+"p3"
+Quote$+" ,"+Quote$+"p4" +Quote$+" ,"+Quote$+"p5" +Quote$+" ," +Quote$+"p6" +Quote$+" ,"+_
                                             Quote$+"p7" +Quote$+" ,"    +Quote$+"p8" +Quote$+" ,"+Quote$+"p9"
+Quote$+" ,"+Quote$+"p10"+Quote$+" ,"+Quote$+"p11"+Quote$+" ," +Quote$+"p12"+Quote$+" ,"+_
                                             Quote$+"p13"+Quote$+" ,"    +Quote$+"p14"+Quote$+" ,"+Quote$+"p15"+Quote$
                CASE 16 : PRINT #N%, "plot "+Quote$+"p1" +Quote$+" lw 1,"+Quote$+"p2" +Quote$+" ,"+Quote$+"p3"
+Quote$+" ,"+Quote$+"p4" +Quote$+" ,"+Quote$+"p5" +Quote$+" ," +Quote$+"p6" +Quote$+" ,"+_
                                             Quote$+"p7" +Quote$+" ,"    +Quote$+"p8" +Quote$+" ,"+Quote$+"p9"
+Quote$+" ,"+Quote$+"p10"+Quote$+" ,"+Quote$+"p11"+Quote$+" ," +Quote$+"p12"+Quote$+" ,"+_
                                             Quote$+"p13"+Quote$+" ,"    +Quote$+"p14"+Quote$+"
,"+Quote$+"p15"+Quote$+" ,"+Quote$+"p16"+Quote$
            END SELECT
        END IF
    CLOSE #N%
    ProcID??? = SHELL(GnuPlotEXE$+" cmd2d.gp -") : CALL Delay(0.5##)
END SUB 'Plot2DindividualProbeTrajectories()
'----
SUB Plot3DbestProbeTrajectories(NumTrajectories%,M(),R(),XiMin(),XiMax(),Np%,Nd%,LastStep&,FunctionName$) 'XYZZY
LOCAL TrajectoryNumber%, ProbeNumber%, StepNumber&, N%, M%, ProcID???
LOCAL MaximumFitness, MinimumFitness AS EXT
LOCAL BestProbeThisStep%()
LOCAL BestFitnessThisStep(), TempFitness() AS EXT
LOCAL Annotation$, xCoord$, yCoord$, zCoord$, GnuPlotEXE$, PlotWithLines$
    Annotation$      = ""
    PlotWithLines$   = "NO" '"YES" '"NO"
    NumTrajectories% = MIN(Np%,NumTrajectories%)
    GnuPlotEXE$ = "wgnuplot.exe"
'   ---------------- Get Min/Max Fitnesses -----------------
    MaximumFitness = M(1,0) : MinimumFitness = M(1,0)  'Note:  M(p%,j&)
    FOR StepNumber& = 0 TO LastStep&
        FOR ProbeNumber% = 1 TO Np%
            IF M(ProbeNumber%,StepNumber&) >= MaximumFitness THEN MaximumFitness = M(ProbeNumber%,StepNumber&)
            IF M(ProbeNumber%,StepNumber&) =< MinimumFitness THEN MinimumFitness = M(ProbeNumber%,StepNumber&)
        NEXT ProbeNumber%
    NEXT StepNumber%
'   ------------- Copy Fitness Array M() into TempFitness to Preserve M() ----------------
    REDIM TempFitness(1 TO Np%, 0 TO LastStep&)
    FOR StepNumber& = 0 TO LastStep&
        FOR ProbeNumber% = 1 TO Np%
            TempFitness(ProbeNumber%,StepNumber&) = M(ProbeNumber%,StepNumber&)
        NEXT ProbeNumber%
    NEXT StepNumber%
'   ------------ LOOP ON TRAJECTORIES -----------
    FOR TrajectoryNumber% = 1 TO NumTrajectories%
'       --------------- Get Trajectory Coordinate Data -----------------
        REDIM BestFitnessThisStep(0 TO LastStep&), BestProbeThisStep%(0 TO LastStep&)
        FOR StepNumber& = 0 TO LastStep&
            BestFitnessThisStep(StepNumber&) = TempFitness(1,StepNumber&)
            FOR ProbeNumber% = 1 TO Np%
                IF TempFitness(ProbeNumber%,StepNumber&) >= BestFitnessThisStep(StepNumber&) THEN
                    BestFitnessThisStep(StepNumber&) = TempFitness(ProbeNumber%,StepNumber&)
                    BestProbeThisStep%(StepNumber&)  = ProbeNumber%
                END IF
            NEXT ProbeNumber%
        NEXT StepNumber&
'   ----- Create Plot Data File -----
    N% = FREEFILE
    SELECT CASE TrajectoryNumber%
        CASE 1  : OPEN "t1"  FOR OUTPUT AS #N%
        CASE 2  : OPEN "t2"  FOR OUTPUT AS #N%
        CASE 3  : OPEN "t3"  FOR OUTPUT AS #N%
        CASE 4  : OPEN "t4"  FOR OUTPUT AS #N%
        CASE 5  : OPEN "t5"  FOR OUTPUT AS #N%
        CASE 6  : OPEN "t6"  FOR OUTPUT AS #N%
        CASE 7  : OPEN "t7"  FOR OUTPUT AS #N%
        CASE 8  : OPEN "t8"  FOR OUTPUT AS #N%
        CASE 9  : OPEN "t9"  FOR OUTPUT AS #N%
        CASE 10 : OPEN "t10" FOR OUTPUT AS #N%
    END SELECT
'   ------------ Write Plot File Data ------------
    FOR StepNumber& = 0 TO LastStep&
        PRINT #N%, USING$("######.######## ######.########
######.########",R(BestProbeThisStep%(StepNumber&),1,StepNumber&),R(BestProbeThisStep%(StepNumber&),2,StepNumber&),R(Be
stProbeThisStep%(StepNumber&),3,StepNumber&))+CHR$(13)
        TempFitness(BestProbeThisStep%(StepNumber&),StepNumber&) = MinimumFitness 'so that same max will not be found
for next trajectory
    NEXT StepNumber%
    CLOSE #N%
    NEXT TrajectoryNumber%
'   ------------------------- Plot Trajectories --------------------------
   'CALL CreateGNUplotINIfile(0.1##*ScreenWidth&,0.25##*ScreenHeight&,0.6##*ScreenHeight&,0.6##*ScreenHeight&)
    Annotation$ = ""
    N% = FREEFILE
    OPEN "cmd3d.gp" FOR OUTPUT AS #N%
    PRINT #N%, "set pm3d"
    PRINT #N%, "show pm3d"
    PRINT #N%, "set hidden3d"
    PRINT #N%, "set view 45, 45, 1, 1"
    PRINT #N%, "unset colorbox"
    PRINT #N%, "set xrange [" + REMOVE$(STR$(XiMin(1)),ANY"" ) + ":" + REMOVE$(STR$(XiMax(1)),ANY"" ) + "]"
    PRINT #N%, "set yrange [" + REMOVE$(STR$(XiMin(2)),ANY"" ) + ":" + REMOVE$(STR$(XiMax(2)),ANY"" ) + "]"
    PRINT #N%, "set zrange [" + REMOVE$(STR$(XiMin(3)),ANY"" ) + ":" + REMOVE$(STR$(XiMax(3)),ANY"" ) + "]"
```

```
'   PRINT #N%, "set label "   + Quote$  + Annotation$ + Quote$+" at graph "+xCoord$+","+yCoord$+","+zCoord$
'   PRINT #N%, "show label"
    PRINT #N%, "set grid xtics ytics ztics"
    PRINT #N%, "show grid"
    PRINT #N%, "set title "  + Quote$ + "3D " + FunctionName$ + " PROBE TRAJECTORIES" + "\n" + RunID$ + Quote$
    PRINT #N%, "set xlabel " + Quote$ + "x1"                                          + Quote$
    PRINT #N%, "set ylabel " + Quote$ + "x2"                                          + Quote$
    PRINT #N%, "set zlabel " + Quote$ + "x3"                                          + Quote$
    IF PlotWithLines$ = "YES" THEN
        SELECT CASE NumTrajectories%
            CASE 1  : PRINT #N%, "splot "+Quote$+"t1"+Quote$+" w l lw 3"
            CASE 2  : PRINT #N%, "splot "+Quote$+"t1"+Quote$+" w l lw 3,"+Quote$+"t2"+Quote$+" w l"
            CASE 3  : PRINT #N%, "splot "+Quote$+"t1"+Quote$+" w l lw 3,"+Quote$+"t2"+Quote$+" w
l,"+Quote$+"t3"+Quote$+" w l"
            CASE 4  : PRINT #N%, "splot "+Quote$+"t1"+Quote$+" w l lw 3,"+Quote$+"t2"+Quote$+" w
l,"+Quote$+"t3"+Quote$+" w l,"+Quote$+"t4"+Quote$+" w l"
            CASE 5  : PRINT #N%, "splot "+Quote$+"t1"+Quote$+" w l lw 3,"+Quote$+"t2"+Quote$+" w
l,"+Quote$+"t3"+Quote$+" w l,"+Quote$+"t4"+Quote$+" w l,"+Quote$+"t5"+Quote$+" w l"
            CASE 6  : PRINT #N%, "splot "+Quote$+"t1"+Quote$+" w l lw 3,"+Quote$+"t2"+Quote$+" w
l,"+Quote$+"t3"+Quote$+" w l,"+Quote$+"t4"+Quote$+" w l,"+Quote$+"t5"+Quote$+" w l,"+Quote$+"t6"+Quote$+" w l"
            CASE 7  : PRINT #N%, "splot "+Quote$+"t1"+Quote$+" w l lw 3,"+Quote$+"t2"+Quote$+" w
l,"+Quote$+"t3"+Quote$+" w l,"+Quote$+"t4"+Quote$+" w l,"+Quote$+"t5"+Quote$+" w l,"+Quote$+"t6"+Quote$+" w l,"+_
                                          Quote$+"t7"+Quote$+" w l"
            CASE 8  : PRINT #N%, "splot "+Quote$+"t1"+Quote$+" w l lw 3,"+Quote$+"t2"+Quote$+" w
l,"+Quote$+"t3"+Quote$+" w l,"+Quote$+"t4"+Quote$+" w l,"+Quote$+"t5"+Quote$+" w l,"+Quote$+"t6"+Quote$+" w l,"+_
                                          Quote$+"t7"+Quote$+" w l,"     +Quote$+"t8"+Quote$+" w l"
            CASE 9  : PRINT #N%, "splot "+Quote$+"t1"+Quote$+" w l lw 3,"+Quote$+"t2"+Quote$+" w
l,"+Quote$+"t3"+Quote$+" w l,"+Quote$+"t4"+Quote$+" w l,"+Quote$+"t5"+Quote$+" w l,"+Quote$+"t6"+Quote$+" w l,"+_
                                          Quote$+"t7"+Quote$+" w l,"     +Quote$+"t8"+Quote$+" w
l,"+Quote$+"t9"+Quote$+" w l"
            CASE 10 : PRINT #N%, "splot "+Quote$+"t1"+Quote$+" w l lw 3,"+Quote$+"t2"+Quote$+" w
l,"+Quote$+"t3"+Quote$+" w l,"+Quote$+"t4"+Quote$+" w l,"+Quote$+"t5"+Quote$+" w l,"+Quote$+"t6"+Quote$+" w l,"+_
                                          Quote$+"t7"+Quote$+" w l,"     +Quote$+"t8"+Quote$+" w
l,"+Quote$+"t9"+Quote$+" w l,"+Quote$+"t10"+Quote$+" w l"
        END SELECT
    ELSE
        SELECT CASE NumTrajectories%
            CASE 1  : PRINT #N%, "splot "+Quote$+"t1"+Quote$+" lw 2"
            CASE 2  : PRINT #N%, "splot "+Quote$+"t1"+Quote$+" lw 2,"+Quote$+"t2"+Quote$
            CASE 3  : PRINT #N%, "splot "+Quote$+"t1"+Quote$+" lw 2,"+Quote$+"t2"+Quote$+" ,"+Quote$+"t3"+Quote$
            CASE 4  : PRINT #N%, "splot "+Quote$+"t1"+Quote$+" lw 2,"+Quote$+"t2"+Quote$+" ,"+Quote$+"t3"+Quote$+"
,"+Quote$+"t4"+Quote$
            CASE 5  : PRINT #N%, "splot "+Quote$+"t1"+Quote$+" lw 2,"+Quote$+"t2"+Quote$+" ,"+Quote$+"t3"+Quote$+"
,"+Quote$+"t4"+Quote$+" ,"+Quote$+"t5"+Quote$
            CASE 6  : PRINT #N%, "splot "+Quote$+"t1"+Quote$+" lw 2,"+Quote$+"t2"+Quote$+" ,"+Quote$+"t3"+Quote$+"
,"+Quote$+"t4"+Quote$+" ,"+Quote$+"t5"+Quote$+" ,"+Quote$+"t6"+Quote$
            CASE 7  : PRINT #N%, "splot "+Quote$+"t1"+Quote$+" lw 2,"+Quote$+"t2"+Quote$+" ,"+Quote$+"t3"+Quote$+"
,"+Quote$+"t4"+Quote$+" ,"+Quote$+"t5"+Quote$+" ,"+Quote$+"t6"+Quote$+" ,"+_
                                          Quote$+"t7"+Quote$
            CASE 8  : PRINT #N%, "splot "+Quote$+"t1"+Quote$+" lw 2,"+Quote$+"t2"+Quote$+" ,"+Quote$+"t3"+Quote$+"
,"+Quote$+"t4"+Quote$+" ,"+Quote$+"t5"+Quote$+" ,"+Quote$+"t6"+Quote$+" ,"+_
                                          Quote$+"t7"+Quote$+" ,"    +Quote$+"t8"+Quote$
            CASE 9  : PRINT #N%, "splot "+Quote$+"t1"+Quote$+" lw 2,"+Quote$+"t2"+Quote$+" ,"+Quote$+"t3"+Quote$+"
,"+Quote$+"t4"+Quote$+" ,"+Quote$+"t5"+Quote$+" ,"+Quote$+"t6"+Quote$+" ,"+_
                                          Quote$+"t7"+Quote$+" ,"    +Quote$+"t8"+Quote$+" ,"+Quote$+"t9"+Quote$
            CASE 10 : PRINT #N%, "splot "+Quote$+"t1"+Quote$+" lw 2,"+Quote$+"t2"+Quote$+" ,"+Quote$+"t3"+Quote$+"
,"+Quote$+"t4"+Quote$+" ,"+Quote$+"t5"+Quote$+" ,"+Quote$+"t6"+Quote$+" ,"+_
                                          Quote$+"t7"+Quote$+" ,"    +Quote$+"t8"+Quote$+" ,"+Quote$+"t9"+Quote$+"
,"+Quote$+"t10"+Quote$
        END SELECT
    END IF
    CLOSE #N%
    ProcID??? = SHELL(GnuPlotEXE$+" cmd3d.gp -") : CALL Delay(0.5##)
END SUB 'Plot3DbestProbeTrajectories()
'----
FUNCTION HasFITNESSsaturated$(Nsteps&,j&,Np%,Nd%,M(),R(),DiagLength)
LOCAL A$
LOCAL k&, p%
LOCAL BestFitness, SumOfBestFitnesses, BestFitnessStepJ, FitnessSatTOL AS EXT
    A$ = "NO"
    FitnessSatTOL = 0.000001## 'tolerance for FITNESS saturation
    IF j& < Nsteps& + 10 THEN GOTO ExitHasFITNESSsaturated 'execute at least 10 steps after averaging interval before
performing this check
    SumOfBestFitnesses = 0##
    FOR k& = j&-Nsteps&+1 TO j&
        BestFitness = M(k&,1)
        FOR p% = 1 TO Np%
            IF M(p%,k&) >= BestFitness THEN BestFitness = M(p%,k&)
        NEXT p%
        IF k& = j& THEN BestFitnessStepJ = BestFitness
        SumOfBestFitnesses = SumOfBestFitnesses + BestFitness
    NEXT k&
    IF ABS(SumOfBestFitnesses/Nsteps&-BestFitnessStepJ) =< FitnessSatTOL THEN A$ = "YES" 'saturation if (avg value -
last value) are within TOL
ExitHasFITNESSsaturated:
    HasFITNESSsaturated$ = A$
END FUNCTION 'HasFITNESSsaturated$()
'-----------
FUNCTION HasDAVGsaturated$(Nsteps&,j&,Np%,Nd%,M(),R(),DiagLength)
LOCAL A$
LOCAL k&
LOCAL SumOfDavg, DavgStepJ AS EXT
LOCAL DavgSatTOL AS EXT
    A$ = "NO"
    DavgSatTOL = 0.0005## 'tolerance for DAVG saturation
```

```
    IF j& < Nsteps& + 10 THEN GOTO ExitHasDAVGsaturated 'execute at least 10 steps after averaging interval before
performing this check
    DavgStepJ = DavgThisStep(j&,Np%,Nd%,M(),R(),DiagLength)
    SumOfDavg = 0##
    FOR k& = j&-Nsteps&+1 TO j& 'check this step and previous (Nsteps&-1) steps
        SumOfDavg = SumOfDavg + DavgThisStep(k&,Np%,Nd%,M(),R(),DiagLength)
    NEXT k&
    IF ABS(SumOfDavg/Nsteps&-DavgStepJ) =< DavgSatTOL THEN A$ = "YES" 'saturation if (avg value - last value) are
within TOL
ExitHasDAVGsaturated:
    HasDAVGsaturated$ = A$
END FUNCTION 'HasDAVGsaturated$()
'-----------
FUNCTION OscillationInDavg$(j&,Np%,Nd%,M(),R(),DiagLength)
LOCAL A$
LOCAL k&, NumSlopeChanges%
    A$ = "NO"
    NumSlopeChanges% = 0
    IF j& < 15 THEN GOTO ExitDavgOscillation 'wait at least 15 steps
    FOR k& = j&-10 TO j&-1 'check previous ten steps
        IF (DavgThisStep(k&,Np%,Nd%,M(),R(),DiagLength)-DavgThisStep(k&-1,Np%,Nd%,M(),R(),DiagLength))* _
           (DavgThisStep(k&+1,Np%,Nd%,M(),R(),DiagLength)-DavgThisStep(k&,Np%,Nd%,M(),R(),DiagLength)) < 0## THEN INCR
NumSlopeChanges%
    NEXT j&
    IF NumSlopeChanges% >= 3 THEN A$ = "YES"
ExitDavgOscillation:
    OscillationInDavg$ = A$
END FUNCTION 'OscillationInDavg()
'------
FUNCTION DavgThisStep(j&,Np%,Nd%,M(),R(),DiagLength)
LOCAL BestFitness, TotalDistanceAllProbes, SumSQ AS EXT
LOCAL p%, k&, N%, i%, BestProbeNumber%, BestTimeStep&
'   ----------- Best Probe #, etc. -----------
    FOR k& = 0 TO j&
        BestFitness = M(1,k&)
        FOR p% = 1 TO Np%
            IF M(p%,k&) >= BestFitness THEN
                BestFitness = M(p%,k&) : BestProbeNumber% = p% : BestTimeStep& = k&
            END IF
        NEXT p% 'probe #
    NEXT k& 'time step
'   --------- Average Distance to Best Probe -----------
    TotalDistanceAllProbes = 0##
    FOR p% = 1 TO Np%
        SumSQ = 0##
        FOR i% = 1 TO Nd%
            SumSQ = SumSQ + (R(BestProbeNumber%,i%,BestTimeStep&)-R(p%,i%,j&))^2 'do not exclude
p%=BestProbeNumber%(j&) from sum because it adds zero
         NEXT i%
        TotalDistanceAllProbes = TotalDistanceAllProbes + SQR(SumSQ)
    NEXT p%
    DavgThisStep = TotalDistanceAllProbes/(DiagLength*(Np%-1)) 'but exclude best prove from average
END FUNCTION 'DavgThisStep()
'-----------
SUB
PlotBestFitnessEvolution(Nd%,Np%,LastStep&,G,DeltaT,Alpha,Beta,Frep,M(),PlaceInitialProbes$,InitialAcceleration$,Reposi
tionFactor$,FunctionName$,Gamma)
LOCAL BestFitness(), GlobalBestFitness AS EXT
LOCAL PlotAnnotation$, PlotTitle$
LOCAL p%, j&, N%
    REDIM BestFitness(0 TO LastStep&)
    CALL
GetPlotAnnotation(PlotAnnotation$,Nd%,Np%,LastStep&,G,DeltaT,Alpha,Beta,Frep,M(),PlaceInitialProbes$,InitialAcceleratio
n$,RepositionFactor$,FunctionName$,Gamma)
    GlobalBestFitness = M(1,0)
    FOR j& = 0 TO LastStep&
        BestFitness(j&) = M(1,j&)
        FOR p% = 1 TO Np%
            IF M(p%,j&) >= BestFitness(j&)   THEN BestFitness(j&)   = M(p%,j&)
            IF M(p%,j&) >= GlobalBestFitness THEN GlobalBestFitness = M(p%,j&)
        NEXT p% 'probe #
    NEXT j& 'time step
    N% = FREEFILE
    OPEN "Fitness" FOR OUTPUT AS #N%
        FOR j& = 0 TO LastStep&
            PRINT #N%, USING$("###### #######.#######",j&,BestFitness(j&))
        NEXT j&
    CLOSE #N%
    PlotAnnotation$ = PlotAnnotation$ + "Best Fitness = " + REMOVE$(STR$(ROUND(GlobalBestFitness,8)),ANY" ")
    PlotTitle$ = "Best Fitness vs Time Step\n" + "[" + REMOVE$(STR$(Np%),ANY" ") + " probes,
"+REMOVE$(STR$(LastStep&),ANY" ")+" time steps]"
    CALL CreateGNUplotINIfile(0.1##*ScreenWidth&,0.1##*ScreenHeight&,0.6##*ScreenWidth&,0.6##*ScreenHeight&)
    CALL TwoDplot("Fitness","Best Fitness","0.7","0.7","Time Step\n\n.",".\n\nBest Fitness(X)", _
                "","","","","","","","","wgnuplot.exe"," with lines linewidth 2",PlotAnnotation$)
END SUB 'PlotBestFitnessEvolution()
'------
SUB
PlotAverageDistance(Nd%,Np%,LastStep&,G,DeltaT,Alpha,Beta,Frep,M(),PlaceInitialProbes$,InitialAcceleration$,RepositionF
actor$,FunctionName$,R(),DiagLength,Gamma)
LOCAL Davg(), BestFitness(), TotalDistanceAllProbes, SumSQ AS EXT
LOCAL PlotAnnotation$, PlotTitle$
LOCAL p%, j&, N%, i%, BestProbeNumber%(), BestTimeStep&()
    REDIM Davg(0 TO LastStep&), BestFitness(0 TO LastStep&), BestProbeNumber%(0 TO LastStep&), BestTimeStep&(0 TO
LastStep&)
```

```
    CALL
GetPlotAnnotation(PlotAnnotation$,Nd%,Np%,LastStep&,G,DeltaT,Alpha,Beta,Frep,M(),PlaceInitialProbes$,InitialAcceleratio
n$,RepositionFactor$,FunctionName$,Gamma)
'   ----------- Best Probe #, etc. -----------
    FOR j& = 0 TO LastStep&
        BestFitness(j&) = M(1,j&)
        FOR p% = 1 TO Np%
            IF M(p%,j&) >= BestFitness(j&) THEN
                BestFitness(j&) = M(p%,j&) : BestProbeNumber%(j&) = p% : BestTimeStep&(j&) = j& 'only probe number is
used at this time, but other data are computed for possible future use.
            END IF
        NEXT p% 'probe #
    NEXT j& 'time step
    N% = FREEFILE
'   --------- Average Distance to Best Probe -----------
    FOR j& = 0 TO LastStep&
        TotalDistanceAllProbes = 0##
        FOR p% = 1 TO Np%
            SumSQ = 0##
            FOR i% = 1 TO Nd%
                SumSQ = SumSQ + (R(BestProbeNumber%(j&),i%,j&)-R(p%,i%,j&))^2 'do not exclude p%=BestProbeNumber%(j&)
from sum because it adds zero
            NEXT i%
            TotalDistanceAllProbes = TotalDistanceAllProbes + SQR(SumSQ)
        NEXT p%
        Davg(j&) = TotalDistanceAllProbes/(DiagLength*(Np%-1)) 'but exclude best prove from average
    NEXT j&
'   ------------ Create Plot Data File -----------
    OPEN "Davg" FOR OUTPUT AS #N%
        FOR j& = 0 TO LastStep&
            PRINT #N%, USING$("###### #######.#######",j&,Davg(j&))
        NEXT j&
    CLOSE #N%
    PlotTitle$ = "Average Distance of " + REMOVE$(STR$(Np%-1),ANY" ") + " Probes to Best Probe\nNormalized to Size of
Decision Space\n" + _
                 "[" + REMOVE$(STR$(Np%),ANY" ") + " probes, " + REMOVE$(STR$(LastStep&),ANY" ") + " time steps]"
    CALL CreateGNUplotINIfile(0.2##*ScreenWidth&,0.2##*ScreenHeight&,0.6##*ScreenWidth&,0.6##*ScreenHeight&)
    CALL TwoDplot("Davg",PlotTitle$,"0.7","0.9","Time Step\n\n.",".\n\n<D>/Ldiag", _
                  "","","","","","","","","wgnuplot.exe"," with lines linewidth 2",PlotAnnotation$)
END SUB 'PlotAverageDistance()
'------
SUB
GetPlotAnnotation(PlotAnnotation$,Nd%,Np%,LastStep&,G,DeltaT,Alpha,Beta,Frep,M(),PlaceInitialProbes$,InitialAcceleratio
n$,RepositionFactor$,FunctionName$,Gamma)
LOCAL A$
    A$ = "" : IF PlaceInitialProbes$ = "UNIFORM ON-AXIS" AND Nd% > 1 THEN A$ = " ("+REMOVE$(STR$(Np%/Nd%),ANY" ") +
"/axis)"
    PlotAnnotation$ = RunID$ + "\n" + _
                      FunctionName$ + " Function" + " ("+ FormatInteger$(Nd%) + "-D) \n"    +_
                      FormatInteger$(Np%) + " probes"       + A$ + "\n" +_
                      "G = " + FormatFP$(G,2)               + "\n" +_
                      "Alpha = "     + FormatFP$(Alpha,1)   + "\n" +_
                      "Beta = "      + FormatFP$(Beta,1)    + "\n" +_
                      "DelT = "      + FormatFP$(DeltaT,1)  + "\n" +_
                      "Gamma = "     + FormatFP$(Gamma,3)   + "\n" +_
                      "Init Probes " + PlaceInitialProbes$  + "\n" +_
                      "Init Accel "  + InitialAcceleration$ + "\n" +_
                      "Frep = "      + FormatFP$(Frep,3)    + " (" + RepositionFactor$ + ")\n"
'    SELECT CASE RepositionFactor$
 '       CASE "FIXED" : PlotAnnotation$ = PlotAnnotation$ + "Frep = " + FormatFP$(Frep,3) + " fixed" + "\n"
  '  END SELECT
END SUB
'------
SUB
PlotBestProbeVsTimeStep(Nd%,Np%,LastStep&,G,DeltaT,Alpha,Beta,Frep,M(),PlaceInitialProbes$,InitialAcceleration$,Reposit
ionFactor$,FunctionName$,Gamma)
LOCAL BestFitness AS EXT
LOCAL PlotAnnotation$, PlotTitle$
LOCAL p%, j&, N%, BestProbeNumber%()
    REDIM BestProbeNumber%(0 TO LastStep&)
    CALL
GetPlotAnnotation(PlotAnnotation$,Nd%,Np%,LastStep&,G,DeltaT,Alpha,Beta,Frep,M(),PlaceInitialProbes$,InitialAcceleratio
n$,RepositionFactor$,FunctionName$,Gamma)
    FOR j& = 0 TO LastStep&
        Bestfitness = M(1,j&)
        FOR p% = 1 TO Np%
            IF M(p%,j&) >= BestFitness THEN
                BestFitness = M(p%,j&) : BestProbeNumber%(j&) = p%
            END IF
        NEXT p% 'probe #
    NEXT j& 'time step
    N% = FREEFILE
    OPEN "Best Probe" FOR OUTPUT AS #N%
        FOR j& = 0 TO LastStep&
            PRINT #N%, USING$("###### #####",j&,BestProbeNumber%(j&))
        NEXT j&
    CLOSE #N%
    PlotTitle$ = "Best Probe Number vs Time Step\n" + "[" +REMOVE$(STR$(Np%),ANY" ") + " probes, " +
REMOVE$(STR$(LastStep&),ANY" ") + " time steps]"
    CALL CreateGNUplotINIfile(0.15#*ScreenWidth&,0.15##*ScreenHeight&,0.6##*ScreenWidth&,0.6##*ScreenHeight&)
'USAGE: CALL
TwoDplot(PlotFileName$,PlotTitle$,xCoord$,yCoord$,XaxisLabel$,YaxisLabel$,LogXaxis$,LogYaxis$,xMin$,xMax$,yMin$,yMax$,x
Tics$,yTics$,GnuPlotEXE$,LineType$,Annotation$)
```

```
    CALL TwoDplot("Best Probe",PlotTitle$,"0.7","0.7","Time Step\n\n.",".\n\nBest Probe
#","","","","","0",NoSpaces$(Np%+1,0),"","","wgnuplot.exe"," pt 8 ps .5 lw 1",PlotAnnotation$) 'pt, pointtype; ps,
pointsize; lw, linewidth
END SUB 'PlotBestProbeVsTimeStep()
'------
SUB DisplayBestFitness(Np%,Nd%,LastStep&,M(),R(),BestFitnessProbeNumber%,BestFitnessTimeStep&,FunctionName$)
LOCAL A$, B$, p%, i%, j&
LOCAL BestFitness AS EXT
    B$ = "" : IF Nd% > 1 THEN B$ = "s"
    BestFitness = M(1,0)
    FOR j& = 0 TO LastStep&
        FOR p% = 1 TO Np%
            IF M(p%,j&) >= BestFitness THEN
                BestFitness = M(p%,j&) : BestFitnessProbeNumber% = p% : BestFitnessTimeStep& = j&
            END IF
        NEXT p%
    NEXT j&
    A$ = FunctionName$ + CHR$(13) +_
         "Best Fitness = " + REMOVE$(STR$(ROUND(BestFitness,8)),ANY" ") + " returned by" + CHR$(13)        +_
         "Probe # "        + REMOVE$(STR$(BestFitnessProbeNumber%),ANY" ") +_
         " at Time Step "  + REMOVE$(STR$(BestFitnessTimeStep&),ANY" ") + CHR$(13) + CHR$(13) + "P" +
REMOVE$(STR$(BestFitnessProbeNumber%),ANY" ") + " coordinate" + B$ + ":" + CHR$(13)
    FOR i% = 1 TO Nd% : A$ = A$ + STR$(i%)+"
"+REMOVE$(STR$(ROUND(R(BestFitnessProbeNumber%,i%,BestFitnessTimeStep&),8)),ANY" ")+CHR$(13) : NEXT i%
    MSGBOX(A$)
END SUB 'DisplayBestFitness()
'------
FUNCTION FormatInteger$(M%) : FormatInteger$ = REMOVE$(STR$(M%),ANY" ") : END FUNCTION
'------
FUNCTION FormatFP$(X,Ndigits%)
LOCAL A$
    IF X = 0## THEN
        A$ = "0." : GOTO ExitFormatFP
    END IF
    A$ = REMOVE$(STR$(ROUND(ABS(X),Ndigits%)),ANY" ")
    IF ABS(X) < 1## THEN
        IF X > 0## THEN
            A$ = "0" + A$
        ELSE
            A$ = "-0" + A$
        END IF
    ELSE
        IF X < 0## THEN A$ = "-" + A$
    END IF
ExitFormatFP:
    FormatFP$ = A$
END FUNCTION
'----------
SUB InitialProbeDistribution(Np%,Nd%,Nt&,XiMin(),XiMax(),R(),PlaceInitialProbes$,Gamma)
LOCAL DeltaXi, DelX1, DelX2, Di AS EXT
LOCAL NumProbesPerAxis%, p%, i%, k%, NumX1points%, NumX2points%, x1pointNum%, x2pointNum%, A$
    SELECT CASE PlaceInitialProbes$
        CASE "UNIFORM ON-AXIS"
            IF Nd% > 1 THEN
                NumProbesPerAxis% = Np%\Nd% 'even #
            ELSE
                NumProbesPerAxis% = Np%
            END IF
            FOR i% = 1 TO Nd%
                FOR p% = 1 TO Np%
                    R(p%,i%,0) = XiMin(i%) + Gamma*(XiMax(i%)-XiMin(i%))
                NEXT Np%
            NEXT i%
            FOR i% = 1 TO Nd% 'place probes axis-by-axis (i% is axis [dimension] number)
                DeltaXi = (XiMax(i%)-XiMin(i%))/(NumProbesPerAxis%-1)
                FOR k% = 1 TO NumProbesPerAxis%
                    p% = k% + NumProbesPerAxis%*(i%-1) 'probe #
                    R(p%,i%,0) = XiMin(i%) + (k%-1)*DeltaXi
                NEXT k%
                'DeltaXi = (XiMax(i%)-XiMin(i%))/(NumProbesPerAxis%+1)
                'FOR k% = 1 TO NumProbesPerAxis%
                '    p% = k% + NumProbesPerAxis%*(i%-1) 'probe #
                '    R(p%,i%,0) = XiMin(i%) + k%*DeltaXi
                'NEXT k%
            NEXT i%
        CASE "UNIFORM ON-DIAGONAL"
            FOR p% = 1 TO Np%
                FOR i% = 1 TO Nd%
                    DeltaXi = (XiMax(i%)-XiMin(i%))/(Np%-1)
                    R(p%,i%,0) = XiMin(i%) + (p%-1)*DeltaXi
                NEXT i%
            NEXT p%
        CASE "2D GRID"
            NumProbesPerAxis% = SQR(Np%) : NumX1points% = NumProbesPerAxis% : NumX2points% = NumX1points% 'broken down
for possible future use
            DelX1 = (XiMax(1)-XiMin(1))/(NumX1points%-1)
            DelX2 = (XiMax(2)-XiMin(2))/(NumX2points%-1)
            FOR x1pointNum% = 1 TO NumX1points%
                FOR x2pointNum% = 1 TO NumX2points%
                    p% = NumX1points%*(x1pointNum%-1)+x2pointNum% 'probe #
                    R(p%,1,0) = XiMin(1) + DelX1*(x1pointNum%-1) 'x1 coord
                    R(p%,2,0) = XiMin(2) + DelX2*(x2pointNum%-1) 'x2 coord
                NEXT x2pointNum%
            NEXT x1pointNum%
        CASE "RANDOM"
```

```
            FOR p% = 1 TO Np%
                FOR i% = 1 TO Nd%
                    R(p%,i%,0) = XiMin(i%) + RandomNum(0##,1##)*(XiMax(i%)-XiMin(i%))
                NEXT i%
            NEXT p%
    END SELECT
END SUB 'InitialProbeDistribution()
'------
SUB
InitialProbeAccelerations(Np%,Nd%,A(),InitialAcceleration$,MaxInitialRandomAcceleration,MaxInitialFixedAcceleration)
LOCAL p%, i%
LOCAL A$
LOCAL FixedInitialAcceleration AS EXT
    SELECT CASE InitialAcceleration$
        CASE "ZERO"
            FOR p% = 1 TO Np%
                FOR i% = 1 TO Nd%
                    A(p%,i%,0) = 0##
                NEXT i% 'coordinate #
            NEXT p% 'probe #
        CASE "FIXED"
            A$ = INPUTBOX$("Fixed Initial Probe"+CHR$(13)+"Acceleration? [0.001-
"+REMOVE$(STR$(MaxInitialFixedAcceleration),ANY" ")+"]","PROBES' INITIAL ACCELERATION","0.5")
            FixedInitialAcceleration = VAL(A$)
            IF FixedInitialAcceleration < 0.001## OR FixedInitialAcceleration > MaxInitialFixedAcceleration THEN
FixedInitialAcceleration = 0.5##
            FOR p% = 1 TO Np%
                FOR i% = 1 TO Nd%
                    A(p%,i%,0) = FixedInitialAcceleration
                NEXT i% 'coordinate
            NEXT p% 'probe #
        CASE "RANDOM"
            FOR p% = 1 TO Np%
                FOR i% = 1 TO Nd%
                    A(p%,i%,0) = RandomNum(0##,1##)*MaxInitialRandomAcceleration
                NEXT i% 'coordinate
            NEXT p% 'probe #
    END SELECT
END SUB 'InitialProbeAccelerations()
'------
SUB RetrieveErrantProbes(Np%,Nd%,j&,XiMin(),XiMax(),R(),M(),RepositionFactor$,Frep) 'note: M(), RepositionFcator$
passed but not used in thsi version
LOCAL p%, i%
    FOR p% = 1 TO Np%
        FOR i% = 1 TO Nd%
            IF R(p%,i%,j&) < XiMin(i%) THEN R(p%,i%,j&) = XiMin(i%) + Frep*(R(p%,i%,j&-1)-XiMin(i%))
            IF R(p%,i%,j&) > XiMax(i%) THEN R(p%,i%,j&) = XiMax(i%) - Frep*(XiMax(i%)-R(p%,i%,j&-1))
        NEXT i%
    NEXT p%
END SUB 'RetrieveErrantProbes()
'------
SUB
ChangeRunParameters(NumProbesPerDimension%,Np%,Nd%,Nt&,G,Alpha,Beta,DeltaT,Frep,PlaceInitialProbes$,InitialAcceleration
$,RepositionFactor$,FunctionName$)
LOCAL A$, DefaultValue$
    A$ = INPUTBOX$("# dimensions?","Change # Dimensions ("+FunctionName$+")",NoSpaces$(Nd%+0,0)) : Nd%    = VAL(A$) :
IF Nd% < 1 OR Nd% > 500 THEN Nd% = 2
    A$ = INPUTBOX$("# probes/dimension?","Change # Probes/Axis
("+FunctionName$+")",NoSpaces$(NumProbesPerDimension%+0,0)) : NumProbesPerDimension% = VAL(A$)
    IF NumProbesPerDimension% < 2 OR NumProbesPerDimension% > 500 THEN NumProbesPerDimension% = 10 'restrict values to
reasonable (?) ranges
    IF Nd% > 1 THEN NumProbesPerDimension% = 2*((NumProbesPerDimension%+1)\2) 'require an even # probes on each axis to
avoid overlapping at origin (in symmetrical spaces at least...)
    IF Nd% = 1 THEN NumProbesPerDimension% = MAX(NumProbesPerDimension%,3)    'at least 3 probes on x-axis for 1-D
functions
    Np% = NumProbesPerDimension%*Nd%
    A$ = INPUTBOX$("# time steps?","Change # Steps ("+FunctionName$+")",NoSpaces$(Nt&+0,0)) : Nt&    = VAL(A$) : IF Nt&
< 3                      THEN Nt& = 50
    A$ = INPUTBOX$("Grav Const G?","Change G ("+FunctionName$+")",NoSpaces$(G,2))           : G      = VAL(A$) : IF G <
-100##    OR G > 100##    THEN G   = 2##
    A$ = INPUTBOX$("Alpha?","Change Alpha ("+FunctionName$+")",NoSpaces$(Alpha,2))          : Alpha  = VAL(A$) : IF
Alpha < -50## OR Alpha > 50## THEN Alpha = 2##
    A$ = INPUTBOX$("Beta?","Change Beta ("+FunctionName$+")",NoSpaces$(Beta,2))             : Beta   = VAL(A$) : IF
Beta  < -50## OR Beta  > 50## THEN Beta  = 2##'
    A$ = INPUTBOX$("Delta T","Change Delta-T ("+FunctionName$+")",NoSpaces$(DeltaT,2))      : DeltaT = VAL(A$) : IF
DeltaT =< 0##             THEN DeltaT = 1##
    A$ = INPUTBOX$("Frep [0-1]?","Change Frep ("+FunctionName$+")",NoSpaces$(Frep,3))       : Frep   = VAL(A$) : IF
Frep < 0##    OR Frep > 1##   THEN Frep = 0.5##
'   ------------ Initial Probe Distribution ------------
    SELECT CASE PlaceInitialProbes$
        CASE "UNIFORM ON-AXIS"     : DefaultValue$ = "1"
        CASE "UNIFORM ON-DIAGONAL" : DefaultValue$ = "2"
        CASE "2D GRID"             : DefaultValue$ = "3"
        CASE "RANDOM"              : DefaultValue$ = "4"
    END SELECT
    A$ = INPUTBOX$("Initial Probes?"+CHR$(13)+"1 - UNIFORM ON-AXIS"+CHR$(13)+"2 - UNIFORM ON-DIAGONAL"+CHR$(13)+"3 - 2D
GRID"+CHR$(13)+"4 - RANDOM","Initial Probe Distribution ("+FunctionName$+")",DefaultValue$)
    IF VAL(A$) < 1 OR VAL(A$) > 4 THEN A$ = "1"
    SELECT CASE VAL(A$)
        CASE 1 : PlaceInitialProbes$ = "UNIFORM ON-AXIS"
        CASE 2 : PlaceInitialProbes$ = "UNIFORM ON-DIAGONAL"
        CASE 3 : PlaceInitialProbes$ = "2D GRID"
        CASE 4 : PlaceInitialProbes$ = "RANDOM"
    END SELECT
```

```
    IF Nd% = 1  AND PlaceInitialProbes$ = "UNIFORM ON-DIAGONAL" THEN PlaceInitialProbes$ = "UNIFORM ON-AXIS" 'cannot do
diagonal in 1-D space
    IF Nd% <> 2 AND PlaceInitialProbes$ = "2D GRID" THEN PlaceInitialProbes$ = "UNIFORM ON-AXIS" '2D grid is available
only in 2 dimensions!
'   ------------- Initial Acceleration -----------------
    SELECT CASE InitialAcceleration$
        CASE "ZERO"   : DefaultValue$ = "1"
        CASE "FIXED"  : DefaultValue$ = "2"
        CASE "RANDOM" : DefaultValue$ = "3"
    END SELECT
    A$ = INPUTBOX$("Initial Acceleration?"+CHR$(13)+"1 - ZERO"+CHR$(13)+"2 - FIXED"+CHR$(13)+"3 - RANDOM","Initial
Acceleration ("+FunctionName$+")",DefaultValue$)
   IF VAL(A$) < 1 OR VAL(A$) > 3 THEN A$ = "1"
    SELECT CASE VAL(A$)
        CASE 1 : InitialAcceleration$ = "ZERO"
        CASE 2 : InitialAcceleration$ = "FIXED"
        CASE 3 : InitialAcceleration$ = "RANDOM"
    END SELECT
'   ----------- Reposition Factor ---------------
    SELECT CASE RepositionFactor$
        CASE "FIXED"    : DefaultValue$ = "1"
        CASE "VARIABLE" : DefaultValue$ = "2"
        CASE "RANDOM"   : DefaultValue$ = "3"
    END SELECT
    A$ = INPUTBOX$("Reposition Factor?"+CHR$(13)+"1 - FIXED"+CHR$(13)+"2 - VARIABLE"+CHR$(13)+"3 - RANDOM","Retrieve
Probes ("+FunctionName$+")",DefaultValue$)
    IF VAL(A$) < 1 OR VAL(A$) > 3 THEN A$ = "1"
    SELECT CASE VAL(A$)
        CASE 1 : RepositionFactor$ = "FIXED"
        CASE 2 : RepositionFactor$ = "VARIABLE"
        CASE 3 : RepositionFactor$ = "RANDOM"
    END SELECT
END SUB 'ChangeRunParameters()
'------
FUNCTION NoSpaces$(X,NumDigits%) :  NoSpaces$ = REMOVE$(STR$(X,NumDigits%),ANY" ") : END FUNCTION
'-----------
FUNCTION TerminateNowForSaturation$(j&,Nd%,Np%,Nt&,G,DeltaT,Alpha,Beta,R(),A(),M())
LOCAL A$, i&, p%, NumStepsForAveraging&
LOCAL BestFitness, AvgFitness, FitnessTOL AS EXT 'terminate if avg fitness does not change over NumStepsForAveraging&
time steps
    FitnessTOL = 0.00001## : NumStepsForAveraging& = 10
    A$ = "NO"
    IF j& >= NumStepsForAveraging+10 THEN 'wait until step 10 to start checking for fitness saturation
        AvgFitness = 0##
        FOR i& = j&-NumStepsForAveraging&+1 TO j& 'avg fitness over current step & previous NumStepsForAveraging&-1
steps
            BestFitness = M(1,i&)
            FOR p% = 1 TO Np%
                IF M(p%,i&) >= BestFitness THEN BestFitness = M(p%,i&)
            NEXT p%
            AvgFitness = AvgFitness + BestFitness
        NEXT i&
        AvgFitness = AvgFitness/NumStepsForAveraging&
        IF ABS(AvgFitness-BestFitness) < FitnessTOL THEN A$ = "YES" 'compare avg fitness to best fitness at this step
    END IF
    TerminateNowForSaturation$ = A$
END FUNCTION 'TerminateNowForSaturation$()
'-----------
FUNCTION MagVector(V(),N%) 'returns magnitude of Nx1 column vector V
LOCAL SumSq AS EXT
LOCAL i%
    SumSQ = 0## : FOR i% = 1 TO N% : SumSQ = SumSQ + V(i%)^2 : NEXT i% : MagVector = SQR(SumSQ)
END FUNCTION 'MagVector()
'---
FUNCTION UnitStep(X)
LOCAL Z AS EXT
    IF X < 0## THEN
        Z = 0##
    ELSE
        Z = 1##
    END IF
    UnitStep = Z
END FUNCTION 'UnitStep()
'---
SUB Plot1Dfunction(FunctionName$,XiMin(),XiMax(),R()) 'plots 1D function on-screen
LOCAL NumPoints%, i%, N%
LOCAL DeltaX, X AS EXT
    NumPoints% = 32001
    DeltaX = (XiMax(1)-XiMin(1))/(NumPoints%-1)
    N% = FREEFILE
    SELECT CASE FunctionName$
        CASE "ParrottF4" 'PARROTT F4 FUNCTION
            OPEN "ParrottF4" FOR OUTPUT AS #N%
                FOR i% = 1 TO NumPoints%
                    R(1,1,0) = XiMin(1) + (i%-1)*DeltaX
                    PRINT #N%, USING$("#.######  #.######",R(1,1,0),ParrottF4(R(),1,1,0))
                NEXT i%
            CLOSE #N%
            CALL CreateGNUplotINIfile(0.2##*ScreenWidth&,0.2##*ScreenHeight&,0.6##*ScreenWidth&,0.6##*ScreenHeight&)
            CALL TwoDplot("ParrottF4","Parrott F4 Function","0.7","0.7","X\n\n.",".\n\nParrott
F4(X)","","","0","1","0","1","","","wgnuplot.exe"," with lines linewidth 2","")
    END SELECT
END SUB
'------
SUB CLEANUP 'probe coordinate plot files
```

```
    IF DIR$("P1")  <> "" THEN KILL "P1"
    IF DIR$("P2")  <> "" THEN KILL "P2"
    IF DIR$("P3")  <> "" THEN KILL "P3"
    IF DIR$("P4")  <> "" THEN KILL "P4"
    IF DIR$("P5")  <> "" THEN KILL "P5"
    IF DIR$("P6")  <> "" THEN KILL "P6"
    IF DIR$("P7")  <> "" THEN KILL "P7"
    IF DIR$("P8")  <> "" THEN KILL "P8"
    IF DIR$("P9")  <> "" THEN KILL "P9"
    IF DIR$("P10") <> "" THEN KILL "P10"
    IF DIR$("P11") <> "" THEN KILL "P11"
    IF DIR$("P12") <> "" THEN KILL "P12"
    IF DIR$("P13") <> "" THEN KILL "P13"
    IF DIR$("P14") <> "" THEN KILL "P14"
    IF DIR$("P15") <> "" THEN KILL "P15"
END SUB
'------
SUB Plot2Dfunction(FunctionName$,XiMin(),XiMax(),R())
LOCAL A$
LOCAL NumPoints%, i%, k%, N%
LOCAL DelX1, DelX2, Z AS EXT
    SELECT CASE FunctionName$
        CASE "PBM_1","PBM_2","PBM_3","PBM_4","PBM_5" : NumPoints% = 25
        CASE ELSE : NumPoints% = 100
    END SELECT
    N% = FREEFILE : OPEN "TwoDplot.DAT" FOR OUTPUT AS #N%
    DelX1 = (XiMax(1)-XiMin(1))/(NumPoints%-1) : DelX2 = (XiMax(2)-XiMin(2))/(NumPoints%-1)
    FOR i% = 1 TO NumPoints%
        R(1,1,0) = XiMin(1) + (i%-1)*DelX1 'x1 value
        FOR k% = 1 TO NumPoints%
            R(1,2,0) = XiMin(2) + (k%-1)*DelX2 'x2 value
            Z = ObjectiveFunction(R(),2,1,0,FunctionName$)
            PRINT #N%, USING$("######.###### ######.###### #######.######^^^^",R(1,1,0),R(1,2,0),Z)
        NEXT k%
        PRINT #N%, ""
    NEXT i%
    CLOSE #N%
    CALL CreateGNUplotINIfile(0.1##*ScreenWidth&,0.1##*ScreenHeight&,0.6##*ScreenWidth&,0.6##*ScreenHeight&)
    A$ = "" : IF INSTR(FunctionName$,"PBM_") > 0 THEN A$ = "Coarse "
    CALL ThreeDplot2("TwoDplot.DAT",A$+"Plot of "+FunctionName$+" Function","","0.6","0.6","1.2", _
                     "x1","x2","z=F(x1,x2)","","","wgnuplot.exe","","","","","")
END SUB
'------
    SUB
TwoDplot3curves(NumCurves%,PlotFileName1$,PlotFileName2$,PlotFileName3$,PlotTitle$,Annotation$,xCoord$,yCoord$,XaxisLab
el$,YaxisLabel$, _
                        LogXaxis$,LogYaxis$,xMin$,xMax$,yMin$,yMax$,xTics$,yTics$,GnuPlotEXE$)
        LOCAL N%
        LOCAL LineSize$
        LineSize$ = "2"
        N% = FREEFILE
        OPEN "cmd2d.gp" FOR OUTPUT AS #N%
            IF LogXaxis$ = "YES" AND LogYaxis$ = "NO"  THEN PRINT #N%, "set logscale x"
            IF LogXaxis$ = "NO"  AND LogYaxis$ = "YES" THEN PRINT #N%, "set logscale y"
            IF LogXaxis$ = "YES" AND LogYaxis$ = "YES" THEN PRINT #N%, "set logscale xy"
            IF xMin$ <> "" AND xMax$ <> "" THEN  PRINT #N%, "set xrange ["+xMin$+":"+xMax$+"]"
            IF yMin$ <> "" AND yMax$ <> "" THEN  PRINT #N%, "set yrange ["+yMin$+":"+yMax$+"]"
            PRINT #N%, "set label "+Quote$+AnnoTation$+Quote$+" at graph "+xCoord$+","+yCoord$
            PRINT #N%, "set grid xtics"
            PRINT #N%, "set grid ytics"
            PRINT #N%, "set xtics "+xTics$
            PRINT #N%, "set ytics "+yTics$
            PRINT #N%, "set grid mxtics"
            PRINT #N%, "set grid mytics"
            PRINT #N%, "set title " +Quote$+PlotTitle$+Quote$
            PRINT #N%, "set xlabel "+Quote$+XaxisLabel$+Quote$
            PRINT #N%, "set ylabel "+Quote$+YaxisLabel$+Quote$
            SELECT CASE NumCurves%
            CASE 1
            PRINT #N%, "plot " + Quote$ + PlotFileName1$ + Quote$ + " with lines linewidth " + LineSize$
            CASE 2
            PRINT #N%, "plot " + Quote$ + PlotFileName1$ + Quote$ + " with lines linewidth " + LineSize$+", " + _
                                 Quote$ + PlotFileName2$ + Quote$ + " with lines linewidth " + LineSize$
            CASE 3
            PRINT #N%, "plot " + Quote$ + PlotFileName1$ + Quote$ + " with lines linewidth " + LineSize$+", " + _
                                 Quote$ + PlotFileName2$ + Quote$ + " with lines linewidth " + LineSize$+", " + _
                                 Quote$ + PlotFileName3$ + Quote$ + " with lines linewidth " + LineSize$
            END SELECT
        CLOSE #N%
        SHELL(GnuPlotEXE$+" cmd2d.gp -")
        CALL Delay(0.3)
    END SUB 'TwoDplot3Curves()
'---
FUNCTION Fibonacci&&(N%) 'RETURNS Nth FIBONACCI NUMBER
LOCAL i%, Fn&&, Fn1&&, Fn2&&
LOCAL A$
    IF N% > 91 OR N% < 0 THEN
        MSGBOX("ERROR!  Fibonacci argument"+STR$(N%)+" > 91.  Out of range or < 0...") : EXIT FUNCTION
    END IF
    SELECT CASE N%
        CASE 0: Fn&& = 1
        CASE ELSE
            Fn&& = 0 : Fn2&& = 1 : i% = 0
            FOR i% = 1 TO N%
                Fn&& = Fn1&& + Fn2&&
```

```
                Fn1&& = Fn2&&
                Fn2&& = Fn&&
            NEXT i% 'LOOP
    END SELECT
    Fibonacci&& = Fn&&
END FUNCTION 'Fibonacci&&()
'-----------
FUNCTION RandomNum(a,b) 'Returns random number X, a=< X < b.
    RandomNum = a + (b-a)*RND
END FUNCTION 'RandomNum()
'-----------
FUNCTION GaussianDeviate(Mu,Sigma) 'returns NORMAL (Gaussian) random deviate with mean Mu and standard deviation Sigma
(variance = Sigma^2)
'Refs: (1) Press, W.H., Flannery, B.P., Teukolsky, S.A., and Vetterling, W.T., "Numerical Recipes: The Art of
Scientific Computing,"
'          §7.2, Cambridge University Press, Cambridge, UK, 1986.
'      (2) Shinzato, T., "Box Muller Method," 2007, http://www.sp.dis.titech.ac.jp/~shinzato/boxmuller.pdf
LOCAL s, t, Z AS EXT
    s = RND : t = RND
    Z = Mu + Sigma*SQR(-2##*LOG(s))*COS(TwoPi*t)
    GaussianDeviate = Z
END FUNCTION 'GaussianDeviate()
'-----------
    SUB ContourPlot(PlotFileName$,PlotTitle$,Annotation$,xCoord$,yCoord$,zCoord$, _
                XaxisLabel$,YaxisLabel$,ZaxisLabel$,zMin$,zMax$,GnuPlotEXE$,A$)
        LOCAL N%
        N% = FREEFILE
        OPEN "cmd3d.gp" FOR OUTPUT AS #N%
            PRINT #N%, "show surface"
            PRINT #N%, "set hidden3d"
            IF zMin$ <> "" AND zMax$ <> "" THEN  PRINT #N%, "set zrange ["+zMin$+":"+zMax$+"]"
            PRINT #N%, "set label "+Quote$+AnnoTation$+Quote$+" at graph "+xCoord$+","+yCoord$+","+zCoord$
            PRINT #N%, "show label"
            PRINT #N%, "set grid xtics ytics ztics"
            PRINT #N%, "show grid"
            PRINT #N%, "set title "+Quote$+PlotTitle$+Quote$
            PRINT #N%, "set xlabel "+Quote$+XaxisLabel$+Quote$
            PRINT #N%, "set ylabel "+Quote$+YaxisLabel$+Quote$
            PRINT #N%, "set zlabel "+Quote$+ZaxisLabel$+Quote$
            PRINT #N%, "splot "+Quote$+PlotFileName$+Quote$+A$  '" notitle with linespoints" 'A$'" notitle with lines"
        CLOSE #N%
        SHELL(GnuPlotEXE$+" cmd3d.gp -")
    END SUB 'ContourPlot()
'---
    SUB ThreeDplot(PlotFileName$,PlotTitle$,Annotation$,xCoord$,yCoord$,zCoord$, _
               XaxisLabel$,YaxisLabel$,ZaxisLabel$,zMin$,zMax$,GnuPlotEXE$,A$)
        LOCAL N%, ProcessID???
        N% = FREEFILE
        OPEN "cmd3d.gp" FOR OUTPUT AS #N%
            PRINT #N%, "set pm3d"
            PRINT #N%, "show pm3d"
            IF zMin$ <> "" AND zMax$ <> "" THEN  PRINT #N%, "set zrange ["+zMin$+":"+zMax$+"]"
            PRINT #N%, "set label "+Quote$+AnnoTation$+Quote$+" at graph "+xCoord$+","+yCoord$+","+zCoord$
            PRINT #N%, "show label"
            PRINT #N%, "set grid xtics ytics ztics"
            PRINT #N%, "show grid"
            PRINT #N%, "set title "+Quote$+PlotTitle$+Quote$
            PRINT #N%, "set xlabel "+Quote$+XaxisLabel$+Quote$
            PRINT #N%, "set ylabel "+Quote$+YaxisLabel$+Quote$
            PRINT #N%, "set zlabel "+Quote$+ZaxisLabel$+Quote$
            PRINT #N%, "splot "+Quote$+PlotFileName$+Quote$+A$+" notitle"' with lines"
        CLOSE #N%
        SHELL(GnuPlotEXE$+" cmd3d.gp -") : CALL Delay(0.5##)
    END SUB 'ThreeDplot()
'---
    SUB ThreeDplot2(PlotFileName$,PlotTitle$,Annotation$,xCoord$,yCoord$,zCoord$, _
                XaxisLabel$,YaxisLabel$,ZaxisLabel$,zMin$,zMax$,GnuPlotEXE$,A$,xStart$,xStop$,yStart$,yStop$)
        LOCAL N%
        N% = FREEFILE
        OPEN "cmd3d.gp" FOR OUTPUT AS #N%
            PRINT #N%, "set pm3d"
            PRINT #N%, "show pm3d"
            PRINT #N%, "set hidden3d"
            PRINT #N%, "set view 45, 45, 1, 1"
            IF zMin$ <> "" AND zMax$ <> "" THEN  PRINT #N%, "set zrange ["+zMin$+":"+zMax$+"]"
            PRINT #N%, "set xrange [" + xStart$ + ":" + xStop$ + "]"
            PRINT #N%, "set yrange [" + yStart$ + ":" + yStop$ + "]"
            PRINT #N%, "set label "   + Quote$  + AnnoTation$ + Quote$+" at graph "+xCoord$+","+yCoord$+","+zCoord$
            PRINT #N%, "show label"
            PRINT #N%, "set grid xtics ytics ztics"
            PRINT #N%, "show grid"
            PRINT #N%, "set title "  + Quote$+PlotTitle$    + Quote$
            PRINT #N%, "set xlabel " + Quote$+XaxisLabel$   + Quote$
            PRINT #N%, "set ylabel " + Quote$+YaxisLabel$   + Quote$
            PRINT #N%, "set zlabel " + Quote$+ZaxisLabel$   + Quote$
            PRINT #N%, "splot "      + Quote$+PlotFileName$ + Quote$ + A$ + " notitle with lines"
        CLOSE #N%
        SHELL(GnuPlotEXE$+" cmd3d.gp -")
    END SUB 'ThreeDplot2()
'---
     SUB TwoDplot2Curves(PlotFileName1$,PlotFileName2$,PlotTitle$,Annotation$,xCoord$,yCoord$,XaxisLabel$,YaxisLabel$,
_
                       LogXaxis$,LogYaxis$,xMin$,xMax$,yMin$,yMax$,xTics$,yTics$,GnuPlotEXE$,LineSize)
        LOCAL N%, ProcessID???
        N% = FREEFILE
```

```
        OPEN "cmd2d.gp" FOR OUTPUT AS #N%
            'print #N%, "set output "+Quote$+"test.plt"+Quote$ 'tried this 3/11/06, didn't work...
            IF LogXaxis$ = "YES" AND LogYaxis$ = "NO"  THEN PRINT #N%, "set logscale x"
            IF LogXaxis$ = "NO"  AND LogYaxis$ = "YES" THEN PRINT #N%, "set logscale y"
            IF LogXaxis$ = "YES" AND LogYaxis$ = "YES" THEN PRINT #N%, "set logscale xy"
            IF xMin$ <> "" AND xMax$ <> "" THEN  PRINT #N%, "set xrange ["+xMin$+":"+xMax$+"]"
            IF yMin$ <> "" AND yMax$ <> "" THEN  PRINT #N%, "set yrange ["+yMin$+":"+yMax$+"]"
            PRINT #N%, "set label "+Quote$+AnnoTation$+Quote$+" at graph "+xCoord$+","+yCoord$
            PRINT #N%, "set grid xtics"
            PRINT #N%, "set grid ytics"
            PRINT #N%, "set xtics "+xTics$
            PRINT #N%, "set ytics "+yTics$
            PRINT #N%, "set grid mxtics"
            PRINT #N%, "set grid mytics"
            PRINT #N%, "set title "+Quote$+PlotTitle$+Quote$
            PRINT #N%, "set xlabel "+Quote$+XaxisLabel$+Quote$
            PRINT #N%, "set ylabel "+Quote$+YaxisLabel$+Quote$
            PRINT #N%, "plot "+Quote$+PlotFileName1$+Quote$+" with lines linewidth "+REMOVE$(STR$(LineSize),ANY"
")+","+ _
                           Quote$+PlotFileName2$+Quote$+" with points pointsize 0.05"'+REMOVE$(STR$(LineSize),ANY"
")
        CLOSE #N%
        ProcessID??? = SHELL(GnuPlotEXE$+" cmd2d.gp -") : CALL Delay(0.5##)
    END SUB 'TwoDplot2Curves()
'---
    SUB Probe2Dplots(ProbePlotsFileList$,PlotTitle$,Annotation$,xCoord$,yCoord$,XaxisLabel$,YaxisLabel$, _
                     LogXaxis$,LogYaxis$,xMin$,xMax$,yMin$,yMax$,xTics$,yTics$,GnuPlotEXE$)
        LOCAL N%, ProcessID???
        N% = FREEFILE
        OPEN "cmd2d.gp" FOR OUTPUT AS #N%
            IF LogXaxis$ = "YES" AND LogYaxis$ = "NO"  THEN PRINT #N%, "set logscale x"
            IF LogXaxis$ = "NO"  AND LogYaxis$ = "YES" THEN PRINT #N%, "set logscale y"
            IF LogXaxis$ = "YES" AND LogYaxis$ = "YES" THEN PRINT #N%, "set logscale xy"
            IF xMin$ <> "" AND xMax$ <> "" THEN  PRINT #N%, "set xrange ["+xMin$+":"+xMax$+"]"
            IF yMin$ <> "" AND yMax$ <> "" THEN  PRINT #N%, "set yrange ["+yMin$+":"+yMax$+"]"
            PRINT #N%, "set label "+Quote$+AnnoTation$+Quote$+" at graph "+xCoord$+","+yCoord$
            PRINT #N%, "set grid xtics"
            PRINT #N%, "set grid ytics"
            PRINT #N%, "set xtics "+xTics$
            PRINT #N%, "set ytics "+yTics$
            PRINT #N%, "set grid mxtics"
            PRINT #N%, "set grid mytics"
            PRINT #N%, "set title "+Quote$+PlotTitle$+Quote$
            PRINT #N%, "set xlabel "+Quote$+XaxisLabel$+Quote$
            PRINT #N%, "set ylabel "+Quote$+YaxisLabel$+Quote$
            PRINT #N%, ProbePlotsFileList$
        CLOSE #N%
        ProcessID??? = SHELL(GnuPlotEXE$+" cmd2d.gp -") : CALL Delay(0.5##)
    END SUB 'Probe2Dplots()
'---
SUB
Show2Dprobes(R(),Np%,Nt&,j&,XiMin(),XiMax(),Frep,BestFitness,BestProbeNumber%,BestTimeStep&,FunctionName$,RepositionFac
tor$,Gamma)
    LOCAL N%, p%
    LOCAL A$, PlotFileName$, PlotTitle$, Symbols$
    LOCAL xMin$, xMax$, yMin$, yMax$
    LOCAL s1, s2, s3, s4 AS EXT
    PlotFileName$ = "Probes("+REMOVE$(STR$(j&),ANY" ")+")"
    IF j& > 0 THEN 'PLOT PROBES AT THIS TIME STEP
        PlotTitle$ = "\nLOCATIONS OF "+REMOVE$(STR$(Np%),ANY" ") + " PROBES AT TIME STEP" + STR$(j&) + " / " +
REMOVE$(STR$(Nt&),ANY" ") + "\n" + _
                     "Fitness = "+REMOVE$(STR$(ROUND(BestFitness,3)),ANY" ") + ", Probe #" +
REMOVE$(STR$(BestProbeNumber%),ANY" ") + " at Step #" + REMOVE$(STR$(BestTimeStep&),ANY" ") + _
                     "  [Frep = "+REMOVE$(STR$(Frep,4),ANY" ") + " " + RepositionFactor$ + "]\n"
    ELSE 'PLOT INITIAL PROBE DISTRIBUTION
        PlotTitle$ = "\nLOCATIONS OF "+REMOVE$(STR$(Np%),ANY" ") + " INITIAL PROBES FOR " + FunctionName$ + "
FUNCTION\n[gamma = "+STR$(ROUND(Gamma,3))+"]\n"
    END IF
    N% = FREEFILE : OPEN PlotFileName$ FOR OUTPUT AS #N%
        FOR p% = 1 TO Np% : PRINT #N%, USING$("######.#####    ######.#####",R(p%,1,j&),R(p%,2,j&)) : NEXT p%
    CLOSE #N%
    s1 = 1.1## : s2 = 1.1## : s3 = 1.1## : s4 = 1.1## 'expand plots axes by 10%
    IF XiMin(1) > 0## THEN s1 = 0.9##
    IF XiMax(1) < 0## THEN s2 = 0.9##
    IF XiMin(2) > 0## THEN s3 = 0.9##
    IF XiMax(2) < 0## THEN s4 = 0.9##
    xMin$ = REMOVE$(STR$(s1*XiMin(1),2),ANY" ")
    xMax$ = REMOVE$(STR$(s2*XiMax(1),2),ANY" ")
    yMin$ = REMOVE$(STR$(s3*XiMin(2),2),ANY" ")
    yMax$ = REMOVE$(STR$(s4*XiMax(2),2),ANY" ")
    CALL
TwoDplot(PlotFileName$,PlotTitle$,"0.6","0.7","x1\n\n","\nx2","NO","NO",xMin$,xMax$,yMin$,yMax$,"5","5","wgnuplot.exe",
" pointsize 1 linewidth 2","")
    KILL PlotFileName$ 'erase plot data file after probes have been displayed
END SUB 'ShowProbes()
'----
    SUB TwoDplot(PlotFileName$,PlotTitle$,xCoord$,yCoord$,XaxisLabel$,YaxisLabel$, _
                 LogXaxis$,LogYaxis$,xMin$,xMax$,yMin$,yMax$,xTics$,yTics$,GnuPlotEXE$,LineType$,Annotation$)
        LOCAL N%, ProcessID???
        N% = FREEFILE
        OPEN "cmd2d.gp" FOR OUTPUT AS #N%
            IF LogXaxis$ = "YES" AND LogYaxis$ = "NO"  THEN PRINT #N%, "set logscale x"
            IF LogXaxis$ = "NO"  AND LogYaxis$ = "YES" THEN PRINT #N%, "set logscale y"
            IF LogXaxis$ = "YES" AND LogYaxis$ = "YES" THEN PRINT #N%, "set logscale xy"
            IF xMin$ <> "" AND xMax$ <> "" THEN  PRINT #N%, "set xrange ["+xMin$+":"+xMax$+"]"
```

```
            IF yMin$ <> "" AND yMax$ <> "" THEN  PRINT #N%, "set yrange ["+yMin$+":"+yMax$+"]"
            PRINT #N%, "set label "      + Quote$ + Annotation$ + Quote$ + " at graph " + xCoord$ + "," + yCoord$
            PRINT #N%, "set grid xtics " + XTics$
            PRINT #N%, "set grid ytics " + yTics$
            PRINT #N%, "set grid mxtics"
            PRINT #N%, "set grid mytics"
            PRINT #N%, "show grid"
            PRINT #N%, "set title "  + Quote$+PlotTitle$+Quote$
            PRINT #N%, "set xlabel " + Quote$+XaxisLabel$+Quote$
            PRINT #N%, "set ylabel " + Quote$+YaxisLabel$+Quote$
            PRINT #N%, "plot "+Quote$+PlotFileName$+Quote$+" notitle"+LineType$
        CLOSE #N%
        ProcessID??? = SHELL(GnuPlotEXE$+" cmd2d.gp -") : CALL Delay(0.5##)
    END SUB 'TwoDplot()
'-----
    SUB CreateGNUplotINIfile(PlotWindowULC_X%,PlotWindowULC_Y%,PlotWindowWidth%,PlotWindowHeight%)
    LOCAL N%, WinPath$, A$, B$, WindowsDirectory$
    WinPath$ = UCASE$(ENVIRON$("Path"))'DIR$("C:\WINDOWS",23)
    DO
        B$ = A$
        A$ = EXTRACT$(WinPath$,";")
        WinPath$ = REMOVE$(WinPath$,A$+";")
        IF RIGHT$(A$,7) = "WINDOWS" OR A$ = B$ THEN EXIT LOOP
        IF RIGHT$(A$,5) = "WINNT"   OR A$ = B$ THEN EXIT LOOP
    LOOP
    WindowsDirectory$ = A$
    N% = FREEFILE
'   ----------- WGNUPLOT.INPUT FILE -----------
    OPEN WindowsDirectory$+"\wgnuplot.ini" FOR OUTPUT AS #N%
        PRINT #N%,"[WGNUPLOT]"
        PRINT #N%,"TextOrigin=0 0"
        PRINT #N%,"TextSize=640 150"
        PRINT #N%,"TextFont=Terminal,9"
        PRINT #N%,"GraphOrigin="+REMOVE$(STR$(PlotWindowULC_X%),ANY" ")+" "+REMOVE$(STR$(PlotWindowULC_Y%),ANY" ")
        PRINT #N%,"GraphSize="  +REMOVE$(STR$(PlotWindowWidth%),ANY" ")+" "+REMOVE$(STR$(PlotWindowHeight%),ANY" ")
        PRINT #N%,"GraphFont=Arial,10"
        PRINT #N%,"GraphColor=1"
        PRINT #N%,"GraphToTop=1"
        PRINT #N%,"GraphBackground=255 255 255"
        PRINT #N%,"Border=0 0 0 0 0"
        PRINT #N%,"Axis=192 192 192 2 2"
        PRINT #N%,"Line1=0 0 255 0 0"
        PRINT #N%,"Line2=0 255 0 0 1"
        PRINT #N%,"Line3=255 0 0 0 2"
        PRINT #N%,"Line4=255 0 255 0 3"
        PRINT #N%,"Line5=0 0 128 0 4"
    CLOSE #N%
    END SUB 'CreateGNUplotINIfile()
'------
    SUB Delay(NumSecs)
        LOCAL StartTime, StopTime AS EXT
        StartTime = TIMER
        DO UNTIL (StopTime-StartTime) >= NumSecs
            StopTime = TIMER
        LOOP
    END SUB 'Delay()
'-----
SUB MathematicalConstants
    EulerConst    = 0.577215664901532860606512##
    Pi            = 3.141592653589793238462643##
    Pi2           = Pi/2##
    Pi4           = Pi/4##
    TwoPi         = 2##*Pi
    FourPi        = 4##*Pi
    e             = 2.718281828459045235360287##
    Root2         = 1.4142135623730950488##
END SUB
'-----
SUB AlphabetAndDigits
    Alphabet$   = "ABCDEFGHIJKLMNOPQRSTUVWXYZabcdefghijklmnopqrstuvwxyz"
    Digits$     = "0123456789"
    RunID$      = DATE$ + ", " + TIME$
END SUB
'------
SUB SpecialSymbols
    Quote$             = CHR$(34) 'Quotation mark "
    SpecialCharacters$ = "'(),#:;/_"
END SUB
'-----
SUB EMconstants
    Mu0  = 4E-7##*Pi     'hy/meter
    Eps0 = 8.854##*1E-12 'fd/meter
    c    = 2.998E8##     'velocity of light, 1##/SQR(Mu0*Eps0) 'meters/sec
    eta0 = SQR(Mu0/Eps0) 'impedance of free space, ohms
END SUB
'------
SUB ConversionFactors
    Rad2Deg       = 180##/Pi
    Deg2Rad       = 1##/Rad2Deg
    Feet2Meters   = 0.3048##
    Meters2Feet   = 1##/Feet2Meters
    Inches2Meters = 0.0254##
    Meters2Inches = 1##/Inches2Meters
    Miles2Meters  = 1609.344##
    Meters2Miles  = 1##/Miles2Meters
```

```
    NautMi2Meters = 1852##
    Meters2NautMi = 1##/NautMi2Meters
END SUB
'------
SUB ShowConstants 'puts up msgbox showing all constants
LOCAL A$
A$ = _
"Mathematical Constants:"+CHR$(13)+_
"Euler const="+STR$(EulerConst)+CHR$(13)+_
"Pi="+STR$(Pi)+CHR$(13)+_
"Pi/2="+STR$(Pi2)+CHR$(13)+_
"Pi/4="+STR$(Pi4)+CHR$(13)+_
"2Pi="+STR$(TwoPi)+CHR$(13)+_
"4Pi="+STR$(FourPi)+CHR$(13)+_
"e="+STR$(e)+CHR$(13)+_
"Sqr2="+STR$(Root2)+CHR$(13)+CHR$(13)+_
"Alphabet, Digits & Special Characters:"+CHR$(13)+_
"Alphabet="+Alphabet$+CHR$(13)+_
"Digits="+Digits$+CHR$(13)+_
"quote="+Quote$+CHR$(13)+_
"Spec chars="+SpecialCharacters$+CHR$(13)+CHR$(13)+_
"E&M Constants:"+CHR$(13)+_
"Mu0="+STR$(Mu0)+CHR$(13)+_
"Eps0="+STR$(Eps0)+CHR$(13)+_
"c="+STR$(c)+CHR$(13)+_
"Eta0="+STR$(eta0)+CHR$(13)+CHR$(13)+_
"Conversion Factors:"+CHR$(13)+_
"Rad2Deg="+STR$(Rad2Deg)+CHR$(13)+_
"Deg2Rad="+STR$(Deg2Rad)+CHR$(13)+_
"Ft2meters="+STR$(Feet2Meters)+CHR$(13)+_
"Meters2Ft="+STR$(Meters2Feet)+CHR$(13)+_
"Inches2Meters="+STR$(Inches2Meters)+CHR$(13)+_
"Meters2Inches="+STR$(Meters2Inches)+CHR$(13)+_
"Miles2Meters="+STR$(Miles2Meters)+CHR$(13)+_
"Meters2Miles="+STR$(Meters2Miles)+CHR$(13)+_
"NautMi2Meters="+STR$(NautMi2Meters)+CHR$(13)+_
"Meters2NautMi="+STR$(Meters2NautMi)+CHR$(13)+CHR$(13)
MSGBOX(A$)
END SUB
'------
SUB DisplayRmatrix(Np%,Nd%,Nt&,R())
LOCAL p%, i%, j&, A$
    A$ = "Position Vector Matrix R()"+CHR$(13)
    FOR p% = 1 TO Np%
        FOR i% = 1 TO Nd%
            FOR j& = 0 TO Nt&
                A$ = A$ + "R("+STR$(p%)+", "+STR$(i%)+", "+STR$(j&)+ ") ="+STR$(R(p%,i%,j&)) + CHR$(13)
            NEXT j&
        NEXT i%
    NEXT p%
    MSGBOX(A$)
END SUB
'------
SUB DisplayRmatrixThisTimeStep(Np%,Nd%,j&,R())
LOCAL p%, i%, A$
    A$ = "Position Vector Matrix R() at step "+STR$(j&)+":"+CHR$(13)
    FOR p% = 1 TO Np%
        FOR i% = 1 TO Nd%
            A$ = A$ + "R("+STR$(p%)+", "+STR$(i%)+", "+STR$(j&)+ ") ="+STR$(R(p%,i%,j&)) + CHR$(13)
        NEXT i%
    NEXT p%
    MSGBOX(A$)
END SUB
'------
SUB DisplayAmatrix(Np%,Nd%,Nt&,A())
LOCAL p%, i%, j&, A$
    A$ = "Acceleration Vector Matrix A()"+CHR$(13)
    FOR p% = 1 TO Np%
        FOR i% = 1 TO Nd%
            FOR j& = 0 TO Nt&
                A$ = A$ + "A("+STR$(p%)+", "+STR$(i%)+", "+STR$(j&)+ ") ="+STR$(A(p%,i%,j&)) + CHR$(13)
            NEXT j&
        NEXT i%
    NEXT p%
    MSGBOX(A$)
END SUB
'------
SUB DisplayAmatrixThisTimeStep(Np%,Nd%,j&,A())
LOCAL p%, i%, A$
    A$ = "Acceleration matrix A() at step "+STR$(j&)+":"+CHR$(13)
    FOR p% = 1 TO Np%
        FOR i% = 1 TO Nd%
            A$ = A$ + "A("+STR$(p%)+", "+STR$(i%)+", "+STR$(j&)+ ") ="+STR$(A(p%,i%,j&)) + CHR$(13)
        NEXT i%
    NEXT p%
    MSGBOX(A$)
END SUB
'------
SUB DisplayMmatrix(Np%,Nt&,M())
LOCAL p%, j&, A$
    A$ = "Fitness Matrix M()"+CHR$(13)
    FOR p% = 1 TO Np%
       FOR j& = 0 TO Nt&
          A$ = A$ + "M("+STR$(p%)+", "+STR$(j&)+ ") ="+STR$(M(p%,j&)) + CHR$(13)
       NEXT j&
```

```
    NEXT p%
    MSGBOX(A$)
END SUB
'------
SUB DisplayMmatrixThisTimeStep(Np%,j&,M())
LOCAL p%, A$
    A$ = "Fitness matrix M() at step "+STR$(j&)+":"+CHR$(13)
    FOR p% = 1 TO Np%
          A$ = A$ + "M("+STR$(p%)+", "+STR$(j&)+ ") ="+STR$(M(p%,j&)) + CHR$(13)
    NEXT p%
    MSGBOX(A$)
END SUB
'------
SUB DisplayXiMinMax(Nd%,XiMin(),XiMax())
LOCAL i%, A$
    A$ = ""
    FOR i% = 1 TO Nd%
        A$ = A$ + "XiMin("+STR$(i%)+" ) = "+STR$(XiMin(i%))+"   XiMax("+STR$(i%)+" ) = "+STR$(XiMax(i%)) + CHR$(13)
    NEXT i%
    MSGBOX(A$)
END SUB
'------
SUB
DisplayRunParameters2(FunctionName$,Nd%,Np%,Nt&,G,DeltaT,Alpha,Beta,Frep,PlaceInitialProbes$,InitialAcceleration$,Repos
itionFactor$)
LOCAL A$
    A$ = "Function = "+ FunctionName$+CHR$(13)+_
         "Nd = "+STR$(Nd%)+CHR$(13)+_
         "Np = "+STR$(Np%)+CHR$(13)+_
         "Nt = "+STR$(Nt&)+CHR$(13)+_
         "G  = "+STR$(G)+CHR$(13)+_
         "DeltaT = "+STR$(DeltaT)+CHR$(13)+_
         "Alpha = "+STR$(Alpha)+CHR$(13)+_
         "Beta  = "+STR$(Beta)+CHR$(13)+_
         "Frep  = "+STR$(Frep)+CHR$(13)+_
         "Init Probes: "+PlaceInitialProbes$+CHR$(13)+_
         "Init Accel:  "+InitialAcceleration$+CHR$(13)+_
         "Retrive Method: "+RepositionFactor$+CHR$(13)
    MSGBOX(A$)
END SUB
'------
SUB
Tabulate1DprobeCoordinates(Max1DprobesPlotted%,Nd%,Np%,LastStep&,G,DeltaT,Alpha,Beta,Frep,R(),M(),PlaceInitialProbes$,I
nitialAcceleration$,RepositionFactor$,FunctionName$,Gamma)
LOCAL N%, ProbeNum%, FileHeader$, A$, B$, C$, D$, E$, F$, H$, StepNum&, FieldNumber% 'kludgy, yes, but it accomplishes
its purpose...
        CALL
GetPlotAnnotation(FileHeader$,Nd%,Np%,LastStep&,G,DeltaT,Alpha,Beta,Frep,M(),PlaceInitialProbes$,InitialAcceleration$,R
epositionFactor$,FunctionName$,Gamma)
        REPLACE "\n" WITH ", " IN FileHeader$
        FileHeader$ = LEFT$(FileHeader$,LEN(FileHeader$)-2)
        FileHeader$ = "PROBE COORDINATES" + CHR$(13) +_
                      "-----------------" + CHR$(13) + FileHeader$
        N% = FREEFILE : OPEN "ProbeCoordinates.DAT" FOR OUTPUT AS #N%
            A$ = "   Step #     " : B$ = "   ------     " : C$ = ""
            FOR ProbeNum% = 1 TO Np% 'create out data file header
                SELECT CASE ProbeNum%
                    CASE   1 TO   9 : E$ = ""   : F$ = "            " : H$ = "            "
                    CASE  10 TO  99 : E$ = "-"  : F$ = "           "  : H$ = "           "
                    CASE 100 TO 999 : E$ = "--" : F$ = "          "   : H$ = "          "
                END SELECT
                A$ = A$ + "P" + NoSpaces$(ProbeNum%+0,0) + F$ 'note: adding zero to ProbeNum% necessary to convert to
floating point...
                B$ = B$ + E$ + "--" + H$
                C$ = C$ + "######.###    "
'                C$ = C$ + "##.#######"
            NEXT ProbeNum%
            PRINT #N%, FileHeader$ + CHR$(13) : PRINT #N%, A$ : PRINT #N%, B$
            FOR StepNum& = 0 TO LastStep&
                D$ = USING$("######   ",StepNum&)
                FOR ProbeNum% = 1 TO Np% : D$ = D$ + USING$(C$,R(ProbeNum%,1,StepNum&)) : NEXT ProbeNum%
                PRINT #N%, D$
            NEXT StepNum&
        CLOSE #N%
END SUB 'Tabulate1DprobeCoordinates()
'------
SUB
Plot1DprobePositions(Max1DprobesPlotted%,Nd%,Np%,LastStep&,G,DeltaT,Alpha,Beta,Frep,R(),M(),PlaceInitialProbes$,Initial
Acceleration$,RepositionFactor$,FunctionName$,Gamma)
   'plots on-screen 1D function probe positions vs time step if Np =< 10
LOCAL ProcessID???, N%, n1%, n2%, n3%, n4%, n5%, n6%, n7%, n8%, n9%, n10%, n11%, n12%, n13%, n14%, n15%, ProbeNum%,
StepNum&, A$
LOCAL PlotAnnotation$
    IF Np% > Max1DprobesPlotted% THEN EXIT SUB
    CALL CLEANUP 'delete old "Px" plot files, if any
    ProbeNum% = 0
    DO 'create output data files, probe-by-probe
        INCR ProbeNum% : n1%  = FREEFILE : OPEN "P"+REMOVE$(STR$(ProbeNum%),ANY" ") FOR OUTPUT AS #n1%  : IF ProbeNum%
= Np% THEN EXIT LOOP
        INCR ProbeNum% : n2%  = FREEFILE : OPEN "P"+REMOVE$(STR$(ProbeNum%),ANY" ") FOR OUTPUT AS #n2%  : IF ProbeNum%
= Np% THEN EXIT LOOP
        INCR ProbeNum% : n3%  = FREEFILE : OPEN "P"+REMOVE$(STR$(ProbeNum%),ANY" ") FOR OUTPUT AS #n3%  : IF ProbeNum%
= Np% THEN EXIT LOOP
        INCR ProbeNum% : n4%  = FREEFILE : OPEN "P"+REMOVE$(STR$(ProbeNum%),ANY" ") FOR OUTPUT AS #n4%  : IF ProbeNum%
= Np% THEN EXIT LOOP
```

```
        INCR ProbeNum% : n5%  = FREEFILE : OPEN "P"+REMOVE$(STR$(ProbeNum%),ANY" ") FOR OUTPUT AS #n5%  : IF ProbeNum%
= Np% THEN EXIT LOOP
        INCR ProbeNum% : n6%  = FREEFILE : OPEN "P"+REMOVE$(STR$(ProbeNum%),ANY" ") FOR OUTPUT AS #n6%  : IF ProbeNum%
= Np% THEN EXIT LOOP
        INCR ProbeNum% : n7%  = FREEFILE : OPEN "P"+REMOVE$(STR$(ProbeNum%),ANY" ") FOR OUTPUT AS #n7%  : IF ProbeNum%
= Np% THEN EXIT LOOP
        INCR ProbeNum% : n8%  = FREEFILE : OPEN "P"+REMOVE$(STR$(ProbeNum%),ANY" ") FOR OUTPUT AS #n8%  : IF ProbeNum%
= Np% THEN EXIT LOOP
        INCR ProbeNum% : n9%  = FREEFILE : OPEN "P"+REMOVE$(STR$(ProbeNum%),ANY" ") FOR OUTPUT AS #n9%  : IF ProbeNum%
= Np% THEN EXIT LOOP
        INCR ProbeNum% : n10% = FREEFILE : OPEN "P"+REMOVE$(STR$(ProbeNum%),ANY" ") FOR OUTPUT AS #n10% : IF ProbeNum%
= Np% THEN EXIT LOOP
        INCR ProbeNum% : n11% = FREEFILE : OPEN "P"+REMOVE$(STR$(ProbeNum%),ANY" ") FOR OUTPUT AS #n11% : IF ProbeNum%
= Np% THEN EXIT LOOP
        INCR ProbeNum% : n12% = FREEFILE : OPEN "P"+REMOVE$(STR$(ProbeNum%),ANY" ") FOR OUTPUT AS #n12% : IF ProbeNum%
= Np% THEN EXIT LOOP
        INCR ProbeNum% : n13% = FREEFILE : OPEN "P"+REMOVE$(STR$(ProbeNum%),ANY" ") FOR OUTPUT AS #n13% : IF ProbeNum%
= Np% THEN EXIT LOOP
        INCR ProbeNum% : n14% = FREEFILE : OPEN "P"+REMOVE$(STR$(ProbeNum%),ANY" ") FOR OUTPUT AS #n14  : IF ProbeNum%
= Np% THEN EXIT LOOP
        INCR ProbeNum% : n15% = FREEFILE : OPEN "P"+REMOVE$(STR$(ProbeNum%),ANY" ") FOR OUTPUT AS #n15% : IF ProbeNum%
= Np% THEN EXIT LOOP
    LOOP
    ProbeNum% = 0
    DO 'output probe positions as a function of time step
        INCR ProbeNum% : FOR StepNum& = 0 TO LastStep& : PRINT #n1%,  USING$("######
######.########",StepNum&,R(ProbeNum%,1,StepNum&)) : NEXT StepNum& : IF ProbeNum% = Np% THEN EXIT LOOP
        INCR ProbeNum% : FOR StepNum& = 0 TO LastStep& : PRINT #n2%,  USING$("######
######.########",StepNum&,R(ProbeNum%,1,StepNum&)) : NEXT StepNum& : IF ProbeNum% = Np% THEN EXIT LOOP
        INCR ProbeNum% : FOR StepNum& = 0 TO LastStep& : PRINT #n3%,  USING$("######
######.########",StepNum&,R(ProbeNum%,1,StepNum&)) : NEXT StepNum& : IF ProbeNum% = Np% THEN EXIT LOOP
        INCR ProbeNum% : FOR StepNum& = 0 TO LastStep& : PRINT #n4%,  USING$("######
######.########",StepNum&,R(ProbeNum%,1,StepNum&)) : NEXT StepNum& : IF ProbeNum% = Np% THEN EXIT LOOP
        INCR ProbeNum% : FOR StepNum& = 0 TO LastStep& : PRINT #n5%,  USING$("######
######.########",StepNum&,R(ProbeNum%,1,StepNum&)) : NEXT StepNum& : IF ProbeNum% = Np% THEN EXIT LOOP
        INCR ProbeNum% : FOR StepNum& = 0 TO LastStep& : PRINT #n6%,  USING$("######
######.########",StepNum&,R(ProbeNum%,1,StepNum&)) : NEXT StepNum& : IF ProbeNum% = Np% THEN EXIT LOOP
        INCR ProbeNum% : FOR StepNum& = 0 TO LastStep& : PRINT #n7%,  USING$("######
######.########",StepNum&,R(ProbeNum%,1,StepNum&)) : NEXT StepNum& : IF ProbeNum% = Np% THEN EXIT LOOP
        INCR ProbeNum% : FOR StepNum& = 0 TO LastStep& : PRINT #n8%,  USING$("######
######.########",StepNum&,R(ProbeNum%,1,StepNum&)) : NEXT StepNum& : IF ProbeNum% = Np% THEN EXIT LOOP
        INCR ProbeNum% : FOR StepNum& = 0 TO LastStep& : PRINT #n9%,  USING$("######
######.########",StepNum&,R(ProbeNum%,1,StepNum&)) : NEXT StepNum& : IF ProbeNum% = Np% THEN EXIT LOOP
        INCR ProbeNum% : FOR StepNum& = 0 TO LastStep& : PRINT #n10%, USING$("######
######.########",StepNum&,R(ProbeNum%,1,StepNum&)) : NEXT StepNum& : IF ProbeNum% = Np% THEN EXIT LOOP
        INCR ProbeNum% : FOR StepNum& = 0 TO LastStep& : PRINT #n11,  USING$("######
######.########",StepNum&,R(ProbeNum%,1,StepNum&)) : NEXT StepNum& : IF ProbeNum% = Np% THEN EXIT LOOP
        INCR ProbeNum% : FOR StepNum& = 0 TO LastStep& : PRINT #n12%, USING$("######
######.########",StepNum&,R(ProbeNum%,1,StepNum&)) : NEXT StepNum& : IF ProbeNum% = Np% THEN EXIT LOOP
        INCR ProbeNum% : FOR StepNum& = 0 TO LastStep& : PRINT #n13%, USING$("######
######.########",StepNum&,R(ProbeNum%,1,StepNum&)) : NEXT StepNum& : IF ProbeNum% = Np% THEN EXIT LOOP
        INCR ProbeNum% : FOR StepNum& = 0 TO LastStep& : PRINT #n14%, USING$("######
######.########",StepNum&,R(ProbeNum%,1,StepNum&)) : NEXT StepNum& : IF ProbeNum% = Np% THEN EXIT LOOP
        INCR ProbeNum% : FOR StepNum& = 0 TO LastStep& : PRINT #n15%, USING$("######
######.########",StepNum&,R(ProbeNum%,1,StepNum&)) : NEXT StepNum& : IF ProbeNum% = Np% THEN EXIT LOOP
    LOOP
    ProbeNum% = 0
    DO 'close output data files
        INCR ProbeNum% : CLOSE #n1%  : IF ProbeNum% = Np% THEN EXIT LOOP
        INCR ProbeNum% : CLOSE #n2%  : IF ProbeNum% = Np% THEN EXIT LOOP
        INCR ProbeNum% : CLOSE #n3%  : IF ProbeNum% = Np% THEN EXIT LOOP
        INCR ProbeNum% : CLOSE #n4%  : IF ProbeNum% = Np% THEN EXIT LOOP
        INCR ProbeNum% : CLOSE #n5%  : IF ProbeNum% = Np% THEN EXIT LOOP
        INCR ProbeNum% : CLOSE #n6%  : IF ProbeNum% = Np% THEN EXIT LOOP
        INCR ProbeNum% : CLOSE #n7%  : IF ProbeNum% = Np% THEN EXIT LOOP
        INCR ProbeNum% : CLOSE #n8%  : IF ProbeNum% = Np% THEN EXIT LOOP
        INCR ProbeNum% : CLOSE #n9%  : IF ProbeNum% = Np% THEN EXIT LOOP
        INCR ProbeNum% : CLOSE #n10% : IF ProbeNum% = Np% THEN EXIT LOOP
        INCR ProbeNum% : CLOSE #n11% : IF ProbeNum% = Np% THEN EXIT LOOP
        INCR ProbeNum% : CLOSE #n12% : IF ProbeNum% = Np% THEN EXIT LOOP
        INCR ProbeNum% : CLOSE #n13% : IF ProbeNum% = Np% THEN EXIT LOOP
        INCR ProbeNum% : CLOSE #n14% : IF ProbeNum% = Np% THEN EXIT LOOP
        INCR ProbeNum% : CLOSE #n15% : IF ProbeNum% = Np% THEN EXIT LOOP
    LOOP
    ProbeNum% = 0 : A$ = ""
    DO 'create file string for plot command file
        INCR ProbeNum% : A$ = A$ + Quote$ + "P"+REMOVE$(STR$(ProbeNum%),ANY" ") + Quote$ + " w l lw 2, " : IF ProbeNum%
= Np% THEN EXIT LOOP
        INCR ProbeNum% : A$ = A$ + Quote$ + "P"+REMOVE$(STR$(ProbeNum%),ANY" ") + Quote$ + " w l lw 2, " : IF ProbeNum%
= Np% THEN EXIT LOOP
        INCR ProbeNum% : A$ = A$ + Quote$ + "P"+REMOVE$(STR$(ProbeNum%),ANY" ") + Quote$ + " w l lw 2, " : IF ProbeNum%
= Np% THEN EXIT LOOP
        INCR ProbeNum% : A$ = A$ + Quote$ + "P"+REMOVE$(STR$(ProbeNum%),ANY" ") + Quote$ + " w l lw 2, " : IF ProbeNum%
= Np% THEN EXIT LOOP
        INCR ProbeNum% : A$ = A$ + Quote$ + "P"+REMOVE$(STR$(ProbeNum%),ANY" ") + Quote$ + " w l lw 2, " : IF ProbeNum%
= Np% THEN EXIT LOOP
        INCR ProbeNum% : A$ = A$ + Quote$ + "P"+REMOVE$(STR$(ProbeNum%),ANY" ") + Quote$ + " w l lw 2, " : IF ProbeNum%
= Np% THEN EXIT LOOP
        INCR ProbeNum% : A$ = A$ + Quote$ + "P"+REMOVE$(STR$(ProbeNum%),ANY" ") + Quote$ + " w l lw 2, " : IF ProbeNum%
= Np% THEN EXIT LOOP
        INCR ProbeNum% : A$ = A$ + Quote$ + "P"+REMOVE$(STR$(ProbeNum%),ANY" ") + Quote$ + " w l lw 2, " : IF ProbeNum%
= Np% THEN EXIT LOOP
        INCR ProbeNum% : A$ = A$ + Quote$ + "P"+REMOVE$(STR$(ProbeNum%),ANY" ") + Quote$ + " w l lw 2, " : IF ProbeNum%
= Np% THEN EXIT LOOP
```

```
        INCR ProbeNum% : A$ = A$ + Quote$ + "P"+REMOVE$(STR$(ProbeNum%),ANY" ") + Quote$ + " w l lw 2, " : IF ProbeNum%
= Np% THEN EXIT LOOP
        INCR ProbeNum% : A$ = A$ + Quote$ + "P"+REMOVE$(STR$(ProbeNum%),ANY" ") + Quote$ + " w l lw 2, " : IF ProbeNum%
= Np% THEN EXIT LOOP
        INCR ProbeNum% : A$ = A$ + Quote$ + "P"+REMOVE$(STR$(ProbeNum%),ANY" ") + Quote$ + " w l lw 2, " : IF ProbeNum%
= Np% THEN EXIT LOOP
        INCR ProbeNum% : A$ = A$ + Quote$ + "P"+REMOVE$(STR$(ProbeNum%),ANY" ") + Quote$ + " w l lw 2, " : IF ProbeNum%
= Np% THEN EXIT LOOP
        INCR ProbeNum% : A$ = A$ + Quote$ + "P"+REMOVE$(STR$(ProbeNum%),ANY" ") + Quote$ + " w l lw 2, " : IF ProbeNum%
= Np% THEN EXIT LOOP
        INCR ProbeNum% : A$ = A$ + Quote$ + "P"+REMOVE$(STR$(ProbeNum%),ANY" ") + Quote$ + " w l lw 2, " : IF ProbeNum%
= Np% THEN EXIT LOOP
    LOOP
    A$ = LEFT$(A$,LEN(A$)-2)
    CALL
GetPlotAnnotation(PlotAnnotation$,Nd%,Np%,LastStep&,G,DeltaT,Alpha,Beta,Frep,M(),PlaceInitialProbes$,InitialAcceleratio
n$,RepositionFactor$,FunctionName$,Gamma)
    N% = FREEFILE
    OPEN "cmd2d.gp" FOR OUTPUT AS #N%
        PRINT #N%, "set label "      + Quote$ + PlotAnnotation$ + Quote$ + " at graph 0.5,0.95"
        PRINT #N%, "set grid xtics"
        PRINT #N%, "set grid ytics"
        PRINT #N%, "set title "  + Quote$ + "Evolution of "    + FunctionName$ + " Probe Positions"+ "\n" + RunID$ +
Quote$
        PRINT #N%, "set xlabel " + Quote$ + "Time Step"        + Quote$
        PRINT #N%, "set ylabel " + Quote$ + "Probe Coordinate" + Quote$
        PRINT #N%, "plot "       + A$
    CLOSE #N%
    CALL CreateGNUplotINIfile(0.2##*ScreenWidth&,0.2##*ScreenHeight&,0.6##*ScreenWidth&,0.6##*ScreenHeight&) 'USAGE:
CALL CreateGNUplotINIfile(PlotWindowULC_X%,PlotWindowULC_Y%,PlotWindowWidth%,PlotWindowHeight%)
    ProcessID??? = SHELL("wgnuplot.exe"+" cmd2d.gp -") : CALL Delay(5##) 'before SUB Cleanup is called
END SUB
'------
SUB
DisplayRunParameters(FunctionName$,Nd%,Np%,Nt&,G,DeltaT,Alpha,Beta,Frep,R(),A(),M(),PlaceInitialProbes$,InitialAccelera
tion$,RepositionFactor$,RunCFO$,ShrinkDS$,CheckForEarlyTermination$)
LOCAL A$, B$, YN&
    ShrinkDS$ = "NO"                  : YN& = MSGBOX("Adaptively Shrink DS?",%MB_YESNO,"ADAPTIVE DS?")              : IF
YN& = %IDYES THEN ShrinkDS$ = "YES"
    CheckForEarlyTermination$ = "NO" : YN& = MSGBOX("Check for Early Termination?",%MB_YESNO,"EARLY TERMINATION?") : IF
YN& = %IDYES THEN CheckForEarlyTermination$ = "YES"
    B$ = "" : IF PlaceInitialProbes$ = "UNIFORM ON-AXIS" AND Nd% > 1 THEN B$ = "  ["+REMOVE$(STR$(Np%/Nd%),ANY" ") +
"/axis]"
    RunCFO$ = "NO"
    A$ = "RUN CFO WITH THE" + CHR$(13) +_
         "FOLLOWING PARAMETERS?"                           + CHR$(13) + CHR$(13) +_
         "Function "        + FunctionName$                + " (" + REMOVE$(STR$(Nd%),ANY" ") + "-D)" + CHR$(13) +
CHR$(13) +_
         "# probes = "      + REMOVE$(STR$(Np%),ANY" ")      + B$ + CHR$(13) + _
         "# time steps = "  + REMOVE$(STR$(Nt&),ANY" ")      + CHR$(13) + _
         "Grav Const G = "  + REMOVE$(STR$(G,2),ANY" ")      + CHR$(13) + _
         "Delta-T = "       + REMOVE$(STR$(DeltaT,3),ANY" ") + CHR$(13) + _
         "Exp Alpha = "     + REMOVE$(STR$(Alpha,3),ANY" ")  + CHR$(13) + _
         "Exp Beta = "      + REMOVE$(STR$(Beta,3),ANY" ")   + CHR$(13) + _
         "Frep = "          + REMOVE$(STR$(Frep,4),ANY" ")   + " ("+RepositionFactor$ + ")" + CHR$(13) + _
         "Initial Probes: " + PlaceInitialProbes$            + CHR$(13) + _
         "Initial Accel: "  + InitialAcceleration$           + CHR$(13) + CHR$(13)
'   lResult& = MSGBOX(txt$ [, [style&], title$])
    YN& = MSGBOX(A$,%MB_YESNO,"CONFIRM RUN")
    IF YN& = %IDYES THEN RunCFO$ = "YES"
END SUB
'------
SUB StatusWindow(FunctionName$,StatusWindowHandle???)
    GRAPHIC WINDOW "Run Progress,
"+FunctionName$,0.08##*ScreenWidth&,0.08##*ScreenHeight&,0.25##*ScreenWidth&,0.1##*ScreenHeight& TO
StatusWindowHandle???
    GRAPHIC ATTACH StatusWindowHandle???,0,REDRAW
    GRAPHIC FONT "Lucida Console",8,0 '"Courier New",8,0 'Fixed width fonts
    GRAPHIC SET PIXEL (35,15) : GRAPHIC PRINT "  Initializing...      " : GRAPHIC REDRAW
END SUB
'------
SUB GetTestFunctionNumber(FunctionName$)
  LOCAL hDlg AS DWORD
  LOCAL N%, M%
  LOCAL FrameWidth&, FrameHeight&, BoxWidth&, BoxHeight&
  BoxWidth& = 276 : BoxHeight& = 300 : FrameWidth& = 80 : FrameHeight& = BoxHeight&-5
  DIALOG NEW 0, "CENTRAL FORCE OPTIMIZATION TEST FUNCTIONS",,, BoxWidth&, BoxHeight&, %WS_CAPTION OR %WS_SYSMENU, 0 TO
hDlg
'-------------------------------------------------------------------
  CONTROL ADD FRAME,  hDlg, %IDC_FRAME1,  "Test Functions",      5,  2, FrameWidth&, FrameHeight&
  CONTROL ADD FRAME,  hDlg, %IDC_FRAME2,  "GSO Test Functions",  95, 2, FrameWidth&, 255
  CONTROL ADD OPTION, hDlg, %IDC_Function_Number1,  "Parrott F4",10,  14, 60, 10, %WS_GROUP OR %WS_TABSTOP
  CONTROL ADD OPTION, hDlg, %IDC_Function_Number2,  "SGO"      , 10,  24, 60, 10
  CONTROL ADD OPTION, hDlg, %IDC_Function_Number3,  "Goldstein-Price", 10,  34, 60, 10
  CONTROL ADD OPTION, hDlg, %IDC_Function_Number4,  "Step"     , 10,  44, 60, 10
  CONTROL ADD OPTION, hDlg, %IDC_Function_Number5,  "Schwefel 2.26", 10,  54, 60, 10
  CONTROL ADD OPTION, hDlg, %IDC_Function_Number6,  "Colville", 10,  64, 60, 10
  CONTROL ADD OPTION, hDlg, %IDC_Function_Number7,  "Griewank", 10,  74, 60, 10
  CONTROL ADD OPTION, hDlg, %IDC_Function_Number31, "PBM #1",    10,  84, 60, 10
  CONTROL ADD OPTION, hDlg, %IDC_Function_Number32, "PBM #2",    10,  94, 60, 10
  CONTROL ADD OPTION, hDlg, %IDC_Function_Number33, "PBM #3",    10, 104, 60, 10
  CONTROL ADD OPTION, hDlg, %IDC_Function_Number34, "PBM #4",    10, 114, 60, 10
  CONTROL ADD OPTION, hDlg, %IDC_Function_Number35, "PBM #5",    10, 124, 60, 10
  CONTROL ADD OPTION, hDlg, %IDC_Function_Number36, "Himmelblau",10, 134, 60, 10
  CONTROL ADD OPTION, hDlg, %IDC_Function_Number37, "Reserved",  10, 144, 60, 10
```

```
  CONTROL ADD OPTION, hDlg, %IDC_Function_Number38, "Reserved",  10, 154, 60, 10
  CONTROL ADD OPTION, hDlg, %IDC_Function_Number39, "Reserved",  10, 164, 60, 10
  CONTROL ADD OPTION, hDlg, %IDC_Function_Number40, "Reserved",  10, 174, 60, 10
  CONTROL ADD OPTION, hDlg, %IDC_Function_Number41, "Reserved",  10, 184, 60, 10
  CONTROL ADD OPTION, hDlg, %IDC_Function_Number42, "Reserved",  10, 194, 60, 10
  CONTROL ADD OPTION, hDlg, %IDC_Function_Number43, "Reserved",  10, 204, 60, 10
  CONTROL ADD OPTION, hDlg, %IDC_Function_Number44, "Reserved",  10, 214, 60, 10
  CONTROL ADD OPTION, hDlg, %IDC_Function_Number45, "Reserved",  10, 224, 60, 10
  CONTROL ADD OPTION, hDlg, %IDC_Function_Number46, "Reserved",  10, 234, 60, 10
  CONTROL ADD OPTION, hDlg, %IDC_Function_Number47, "Reserved",  10, 244, 60, 10
  CONTROL ADD OPTION, hDlg, %IDC_Function_Number48, "Reserved",  10, 254, 60, 10
  CONTROL ADD OPTION, hDlg, %IDC_Function_Number49, "Reserved",  10, 264, 60, 10
  CONTROL ADD OPTION, hDlg, %IDC_Function_Number50, "Reserved",  10, 274, 60, 10
' --------------------- Test Functions from GSO Paper ---------------------
  CONTROL ADD OPTION, hDlg, %IDC_Function_Number8,  "f1" , 120,  14, 40, 10
  CONTROL ADD OPTION, hDlg, %IDC_Function_Number9,  "f2" , 120,  24, 40, 10
  CONTROL ADD OPTION, hDlg, %IDC_Function_Number10, "f3" , 120,  34, 40, 10
  CONTROL ADD OPTION, hDlg, %IDC_Function_Number11, "f4" , 120,  44, 40, 10
  CONTROL ADD OPTION, hDlg, %IDC_Function_Number12, "f5" , 120,  54, 40, 10
  CONTROL ADD OPTION, hDlg, %IDC_Function_Number13, "f6" , 120,  64, 40, 10
  CONTROL ADD OPTION, hDlg, %IDC_Function_Number14, "f7" , 120,  74, 40, 10
  CONTROL ADD OPTION, hDlg, %IDC_Function_Number15, "f8" , 120,  84, 40, 10
  CONTROL ADD OPTION, hDlg, %IDC_Function_Number16, "f9" , 120,  94, 40, 10
  CONTROL ADD OPTION, hDlg, %IDC_Function_Number17, "f10", 120, 104, 40, 10
  CONTROL ADD OPTION, hDlg, %IDC_Function_Number18, "f11", 120, 114, 40, 10
  CONTROL ADD OPTION, hDlg, %IDC_Function_Number19, "f12", 120, 124, 40, 10
  CONTROL ADD OPTION, hDlg, %IDC_Function_Number20, "f13", 120, 134, 40, 10
  CONTROL ADD OPTION, hDlg, %IDC_Function_Number21, "f14", 120, 144, 40, 10
  CONTROL ADD OPTION, hDlg, %IDC_Function_Number22, "f15", 120, 154, 40, 10
  CONTROL ADD OPTION, hDlg, %IDC_Function_Number23, "f16", 120, 164, 40, 10
  CONTROL ADD OPTION, hDlg, %IDC_Function_Number24, "f17", 120, 174, 40, 10
  CONTROL ADD OPTION, hDlg, %IDC_Function_Number25, "f18", 120, 184, 40, 10
  CONTROL ADD OPTION, hDlg, %IDC_Function_Number26, "f19", 120, 194, 40, 10
  CONTROL ADD OPTION, hDlg, %IDC_Function_Number27, "f20", 120, 204, 40, 10
  CONTROL ADD OPTION, hDlg, %IDC_Function_Number28, "f21", 120, 214, 40, 10
  CONTROL ADD OPTION, hDlg, %IDC_Function_Number29, "f22", 120, 224, 40, 10
  CONTROL ADD OPTION, hDlg, %IDC_Function_Number30, "f23", 120, 234, 40, 10
  CONTROL SET OPTION  hDlg, %IDC_Function_Number1, %IDC_Function_Number1, %IDC_Function_Number3 'default to Parrott F4
'-----------------------------------------------------------------------
  CONTROL ADD BUTTON, hDlg, %IDOK, "&OK", 200, 0.45##*BoxHeight&, 50, 14
'-----------------------------------------------------------------------
  DIALOG SHOW MODAL hDlg CALL DlgProc
  CALL Delay(0.5##)
  IF FunctionNumber% < 1 OR FunctionNumber% > 36 THEN
    FunctionNumber% = 1 : MSGBOX("Error in function number...")
  END IF
' MSGBOX("Test Function is #"+STR$(FunctionNumber%))
    SELECT CASE FunctionNumber%
        CASE 1 : FunctionName$ = "ParrottF4"
        CASE 2 : FunctionName$ = "SGO"
        CASE 3 : FunctionName$ = "GP"
        CASE 4 : FunctionName$ = "STEP"
        CASE 5 : FunctionName$ = "SCHWEFEL_226"
        CASE 6 : FunctionName$ = "COLVILLE"
        CASE 7 : FunctionName$ = "GRIEWANK"
        CASE 8 : FunctionName$ = "F1"
        CASE 9 : FunctionName$ = "F2"
        CASE 10: FunctionName$ = "F3"
        CASE 11: FunctionName$ = "F4"
        CASE 12: FunctionName$ = "F5"
        CASE 13: FunctionName$ = "F6"
        CASE 14: FunctionName$ = "F7"
        CASE 15: FunctionName$ = "F8"
        CASE 16: FunctionName$ = "F9"
        CASE 17: FunctionName$ = "F10"
        CASE 18: FunctionName$ = "F11"
        CASE 19: FunctionName$ = "F12"
        CASE 20: FunctionName$ = "F13"
        CASE 21: FunctionName$ = "F14"
        CASE 22: FunctionName$ = "F15"
        CASE 23: FunctionName$ = "F16"
        CASE 24: FunctionName$ = "F17"
        CASE 25: FunctionName$ = "F18"
        CASE 26: FunctionName$ = "F19"
        CASE 27: FunctionName$ = "F20"
        CASE 28: FunctionName$ = "F21"
        CASE 29: FunctionName$ = "F22"
        CASE 30: FunctionName$ = "F23"
        CASE 31: FunctionName$ = "PBM_1"
        CASE 32: FunctionName$ = "PBM_2"
        CASE 33: FunctionName$ = "PBM_3"
        CASE 34: FunctionName$ = "PBM_4"
        CASE 35: FunctionName$ = "PBM_5"
        CASE 36: FunctionName$ = "HIMMELBLAU"
    END SELECT
END SUB
'-----------
CALLBACK FUNCTION DlgProc() AS LONG
  '--------------------------------------------------------------------
  ' Callback procedure for the main dialog
  '------------------------------------------------------------------
  LOCAL c, lRes AS LONG, sText AS STRING
  SELECT CASE AS LONG CBMSG
  CASE %WM_INITDIALOG' %WM_INITDIALOG is sent right before the dialog is shown.
  CASE %WM_COMMAND              ' <- a control is calling
```

```
     SELECT CASE AS LONG CBCTL  ' <- look at control's id
     CASE %IDOK                 ' <- OK button or Enter key was pressed
         IF CBCTLMSG = %BN_CLICKED THEN
             '---------------------------------------
             ' Loop through the Function_Number controls
             ' to see which one is selected
             '---------------------------------------
             FOR c = %IDC_Function_Number1 TO %IDC_Function_Number50
                 CONTROL GET CHECK CBHNDL, c TO lRes
                 IF lRes THEN EXIT FOR
             NEXT 'c holds the id for selected test function.
             FunctionNumber% = c-120
'DEBUG       sText = FORMAT$(c - %IDC_Function_Number1 + 1) + $CRLF
'            MSGBOX sText, %MB_TASKMODAL, "Selected Function"
            DIALOG END CBHNDL
         END IF
     END SELECT
  END SELECT
END FUNCTION
'---------------------------- PBM ANTENNA BENCHMARK FUNCTIONS ---------------------------
'Reference for benchmarks PBM_1 through PBM_5:
'Pantoja, M F., Bretones, A. R., Martin, R. G., "Benchmark Antenna Problems for Evolutionary
'Optimization Algorithms," IEEE Trans. Antennas & Propagation, vol. 55, no. 4, April 2007,
'pp. 1111-1121
FUNCTION PBM_1(R(),Nd%,p%,j&) 'PBM Benchmark #1: Max D for Variable-Length CF Dipole
    LOCAL Z, LengthWaves, ThetaRadians AS EXT
    LOCAL N%, Nsegs%, FeedSegNum%
    LOCAL NumSegs$, FeedSeg$, HalfLength$, Radius$, ThetaDeg$, Lyne$, GainDB$
    LengthWaves  = R(p%,1,j&)
    ThetaRadians = R(p%,2,j&)
    ThetaDeg$ = REMOVE$(STR$(ROUND(ThetaRadians*Rad2Deg,2)),ANY" ")
    IF TALLY(ThetaDeg$,".") = 0 THEN ThetaDeg$ = ThetaDeg$+"."
    Nsegs% = 2*(INT(100*LengthWaves)\2)+1 '100 segs per wavelength, must be an odd #, VOLTAGE SOURCE
    FeedSegNum% = Nsegs%\2 + 1 'center segment number, VOLTAGE SOURCE
    'Nsegs% = 2*(INT(100*LengthWaves)\2) '100 segs per wavelength, must be an even # for BICONE SOURCE
    'FeedSegNum% = Nsegs%\2 'center segment number for BICONE SOURCE
    NumSegs$    = REMOVE$(STR$(Nsegs%),ANY" ")
    FeedSeg$    = REMOVE$(STR$(FeedSegNum%),ANY" ")
    HalfLength$ = REMOVE$(STR$(ROUND(LengthWaves/2##,6)),ANY" ")
    IF TALLY(HalfLength$,".") = 0 THEN HalfLength$ = HalfLength$+"."
    Radius$     = "0.00001" 'REMOVE$(STR$(ROUND(LengthWaves/1000##,6)),ANY" ")
    N% = FREEFILE
    OPEN "PBM1.NEC" FOR OUTPUT AS #N%
        PRINT #N%,"CM File: PBM1.NEC"
        PRINT #N%,"CM Run ID "+DATE$+" "+TIME$
        PRINT #N%,"CM Nd="+STR$(Nd%)+", p="STR$(p%)+", j="+STR$(j&)
        PRINT #N%,"CM R(p,1,j)="+STR$(R(p%,1,j&))+", R(p,2,j)="+STR$(R(p%,2,j&))
        PRINT #N%,"CE"
        PRINT #N%,"GW 1,"+NumSegs$+",0.,0.,-"+HalfLength$+",0.,0.,"+HalfLength$+","+Radius$
        PRINT #N%,"GE"
        'PRINT #N%,"EX 5,1,"+FeedSeg$+",0,1.,0." 'BICONE SOURCE
        PRINT #N%,"EX 0,1,"+FeedSeg$+",0,1.,0." 'VOLTAGE SOURCE
        PRINT #N%,"FR 0,1,0,0,299.79564,0."
        PRINT #N%,"RP 0,1,1,1001,"+ThetaDeg$+",0.,0.,0.,1000." 'gain at 1000 wavelengths range
        PRINT #N%,"XQ"
        PRINT #N%,"EN"
    CLOSE #N%
'      - - ANGLES - -          - POWER GAINS -        - - - POLARIZATION - - -    - - - E(THETA) - - -    - - - E(PHI)
- - -
'  THETA     PHI        VERT.   HOR.    TOTAL      AXIAL     TILT   SENSE     MAGNITUDE    PHASE      MAGNITUDE
PHASE
' DEGREES  DEGREES       DB      DB      DB        RATIO     DEG.              VOLTS/M    DEGREES      VOLTS/M
DEGREES
'   90.00     0.00       3.91 -999.99    3.91    0.00000     0.00  LINEAR    1.29504E-04     5.37    0.00000E+00    -
5.24
'123456789x123456789x123456789x123456789x123456789x123456789x123456789x123456789x123456789x123456789x123456789x12345678
9x
'        10        20        30        40        50        60        70        80        90       100       110
120
    SHELL "n41_2k1.exe",0
    N% = FREEFILE
    OPEN "PBM1.OUT" FOR INPUT AS #N%
        WHILE NOT EOF(N%)
            LINE INPUT #N%, Lyne$
            IF INSTR(Lyne$,"DEGREES  DEGREES") > 0 THEN EXIT LOOP
        WEND 'position at next data line
        LINE INPUT #N%, Lyne$
    CLOSE #N%
    GainDB$ = REMOVE$(MID$(Lyne$,37,8),ANY" ")
    PBM_1 = 10^(VAL(GainDB$)/10##) 'Directivity
END FUNCTION 'PBM_1()
'----
FUNCTION PBM_2(R(),Nd%,p%,j&) 'PBM Benchmark #2: Max D for Variable-Separation Array of CF Dipoles
    LOCAL Z, DipoleSeparationWaves, ThetaRadians AS EXT
    LOCAL N%, i%
    LOCAL NumSegs$, FeedSeg$, Radius$, ThetaDeg$, Lyne$, GainDB$, Xcoord$, WireNum$
    DipoleSeparationWaves = R(p%,1,j&)
    ThetaRadians          = R(p%,2,j&)
    ThetaDeg$ = REMOVE$(STR$(ROUND(ThetaRadians*Rad2Deg,2)),ANY" ")
    IF TALLY(ThetaDeg$,".") = 0 THEN ThetaDeg$ = ThetaDeg$+"."
    NumSegs$ = "49"
    FeedSeg$ = "25"
    Radius$  = "0.00001"
    N% = FREEFILE
    OPEN "PBM2.NEC" FOR OUTPUT AS #N%
```

```
        PRINT #N%,"CM File: PBM2.NEC"
        PRINT #N%,"CM Run ID "+DATE$+" "+TIME$
        PRINT #N%,"CM Nd="+STR$(Nd%)+", p="STR$(p%)+", j="+STR$(j&)
        PRINT #N%,"CM R(p,1,j)="+STR$(R(p%,1,j&))+", R(p,2,j)="+STR$(R(p%,2,j&))
        PRINT #N%,"CE"
        FOR i% = -9 TO 9 STEP 2
            WireNum$ = REMOVE$(STR$((i%+11)\2),ANY" ")
            Xcoord$  = REMOVE$(STR$(i%*DipoleSeparationWaves/2##),ANY" ")
            PRINT #N%,"GW "+WireNum$+","+NumSegs$+","+Xcoord$+",0.,-0.25,"+Xcoord$+",0.,0.25,"+Radius$
        NEXT i%
        PRINT #N%,"GE"
        FOR i% = 1 TO 10
            PRINT #N%,"EX 0,"+REMOVE$(STR$(i%),ANY" ")+","+FeedSeg$+",0,1.,0." 'VOLTAGE SOURCE
        NEXT i%
        PRINT #N%,"FR 0,1,0,0,299.79564,0."
        PRINT #N%,"RP 0,1,1,1001,"+ThetaDeg$+",90.,0.,0.,1000." 'gain at 1000 wavelengths range
        PRINT #N%,"XQ"
        PRINT #N%,"EN"
    CLOSE #N%
'      - - ANGLES - -          - POWER GAINS -        - - - POLARIZATION - - -    - - - E(THETA) - - -    - - - E(PHI)
- - -
'  THETA     PHI        VERT.   HOR.    TOTAL      AXIAL     TILT   SENSE     MAGNITUDE    PHASE      MAGNITUDE
PHASE
' DEGREES  DEGREES       DB      DB      DB        RATIO     DEG.              VOLTS/M    DEGREES      VOLTS/M
DEGREES
'   90.00     0.00       3.91 -999.99    3.91    0.00000     0.00  LINEAR    1.29504E-04     5.37    0.00000E+00    -
5.24
'123456789x123456789x123456789x123456789x123456789x123456789x123456789x123456789x123456789x123456789x123456789x12345678
9x
'        10        20        30        40        50        60        70        80        90        100       110
120
    SHELL "n41_2k1.exe",0
    N% = FREEFILE
    OPEN "PBM2.OUT" FOR INPUT AS #N%
        WHILE NOT EOF(N%)
            LINE INPUT #N%, Lyne$
            IF INSTR(Lyne$,"DEGREES  DEGREES") > 0 THEN EXIT LOOP
        WEND 'position at next data line
        LINE INPUT #N%, Lyne$
    CLOSE #N%
    GainDB$ = REMOVE$(MID$(Lyne$,37,8),ANY" ")
    IF AddNoiseToPBM2$ = "YES" THEN
        Z = 10^(VAL(GainDB$)/10##) + GaussianDeviate(0##,0.4472##) 'Directivity with Gaussian noise (zero mean, 0.2
variance)
    ELSE
        Z = 10^(VAL(GainDB$)/10##) 'Directivity without noise
    END IF
    PBM_2 = Z
END FUNCTION 'PBM_2()
'----
FUNCTION PBM_3(R(),Nd%,p%,j&) 'PBM Benchmark #3: Max D for Circular Dipole Array
    LOCAL Beta, ThetaRadians, Alpha, ReV, ImV AS EXT
    LOCAL N%, i%
    LOCAL NumSegs$, FeedSeg$, Radius$, ThetaDeg$, Lyne$, GainDB$, Xcoord$, Ycoord$, WireNum$, ReEX$, ImEX$
    Beta         = R(p%,1,j&)
    ThetaRadians = R(p%,2,j&)
    ThetaDeg$ = REMOVE$(STR$(ROUND(ThetaRadians*Rad2Deg,2)),ANY" ")
    IF TALLY(ThetaDeg$,".") = 0 THEN ThetaDeg$ = ThetaDeg$+"."
    NumSegs$ = "49"
    FeedSeg$ = "25"
    Radius$  = "0.00001"
    N% = FREEFILE
    OPEN "PBM3.NEC" FOR OUTPUT AS #N%
        PRINT #N%,"CM File: PBM3.NEC"
        PRINT #N%,"CM Run ID "+DATE$+" "+TIME$
        PRINT #N%,"CM Nd="+STR$(Nd%)+", p="STR$(p%)+", j="+STR$(j&)
        PRINT #N%,"CM R(p,1,j)="+STR$(R(p%,1,j&))+", R(p,2,j)="+STR$(R(p%,2,j&))
        PRINT #N%,"CE"
        FOR i% = 1 TO 8
            WireNum$ = REMOVE$(STR$(i%),ANY" ")
            SELECT CASE i%
                CASE 1 : Xcoord$ = "1"        : Ycoord$ = "0"
                CASE 2 : Xcoord$ = "0.70711"  : Ycoord$ = "0.70711"
                CASE 3 : Xcoord$ = "0"        : Ycoord$ = "1"
                CASE 4 : Xcoord$ = "-0.70711" : Ycoord$ = "0.70711"
                CASE 5 : Xcoord$ = "-1"       : Ycoord$ = "0"
                CASE 6 : Xcoord$ = "-0.70711" : Ycoord$ = "-0.70711"
                CASE 7 : Xcoord$ = "0"        : Ycoord$ = "-1"
                CASE 8 : Xcoord$ = "0.70711"  : Ycoord$ = "-0.70711"
            END SELECT
            PRINT #N%,"GW "+WireNum$+","+NumSegs$+","+Xcoord$+","+Ycoord$+",-0.25,"+Xcoord$+","Ycoord$+",0.25,"+Radius$
        NEXT i%
        PRINT #N%,"GE"
        FOR i% = 1 TO 8
            Alpha = -COS(TwoPi*Beta*(i%-1))
            ReV   = COS(Alpha)
            ImV   = SIN(Alpha)
            ReEX$ = REMOVE$(STR$(ROUND(ReV,6)),ANY" ")
            ImEX$ = REMOVE$(STR$(ROUND(ImV,6)),ANY" ")
            IF TALLY(ReEX$,".") = 0 THEN ReEX$ = ReEX$+"."
            IF TALLY(ImEX$,".") = 0 THEN ImEX$ = ImEX$+"."
            PRINT #N%,"EX 0,"+REMOVE$(STR$(i%),ANY" ")+","+FeedSeg$+",0,"+ReEX$+","+ImEX$ 'VOLTAGE SOURCE
        NEXT i%
        PRINT #N%,"FR 0,1,0,0,299.79564,0."
        PRINT #N%,"RP 0,1,1,1001,"+ThetaDeg$+",0.,0.,0.,1000." 'gain at 1000 wavelengths range
```

```
        PRINT #N%,"XQ"
        PRINT #N%,"EN"
    CLOSE #N%
'      - - ANGLES - -          - POWER GAINS -        - - - POLARIZATION - - -    - - - E(THETA) - - -    - - - E(PHI)
- - -
'  THETA     PHI        VERT.   HOR.    TOTAL      AXIAL     TILT   SENSE     MAGNITUDE    PHASE      MAGNITUDE
PHASE
' DEGREES  DEGREES       DB      DB      DB        RATIO     DEG.              VOLTS/M    DEGREES      VOLTS/M
DEGREES
'   90.00     0.00       3.91 -999.99    3.91    0.00000     0.00  LINEAR    1.29504E-04     5.37    0.00000E+00    -
5.24
'123456789x123456789x123456789x123456789x123456789x123456789x123456789x123456789x123456789x123456789x123456789x12345678
9x
'        10        20        30        40        50        60        70        80        90        100       110
120
    SHELL "n41_2k1.exe",0
    N% = FREEFILE
    OPEN "PBM3.OUT" FOR INPUT AS #N%
        WHILE NOT EOF(N%)
            LINE INPUT #N%, Lyne$
            IF INSTR(Lyne$,"DEGREES  DEGREES") > 0 THEN EXIT LOOP
        WEND 'position at next data line
        LINE INPUT #N%, Lyne$
    CLOSE #N%
    GainDB$ = REMOVE$(MID$(Lyne$,37,8),ANY" ")
    PBM_3 = 10^(VAL(GainDB$)/10##) 'Directivity
END FUNCTION 'PBM_3()
'----
FUNCTION PBM_4(R(),Nd%,p%,j&) 'PBM Benchmark #4: Max D for Vee Dipole
    LOCAL TotalLengthWaves, AlphaRadians, ArmLength, Xlength, Zlength, Lfeed AS EXT
    LOCAL N%, i%, Nsegs%, FeedZcoord$
    LOCAL NumSegs$, Lyne$, GainDB$, Xcoord$, Zcoord$
    TotalLengthWaves = 2##*R(p%,1,j&)
    AlphaRadians     = R(p%,2,j&)
    Lfeed            = 0.01##
    FeedZcoord$      = REMOVE$(STR$(Lfeed),ANY" ")
    ArmLength = (TotalLengthWaves-2##*Lfeed)/2##
    Xlength  = ROUND(ArmLength*COS(AlphaRadians),6)
    Xcoord$  = REMOVE$(STR$(Xlength),ANY" ") : IF TALLY(Xcoord$,".") = 0 THEN Xcoord$ = Xcoord$+"."
    Zlength  = ROUND(ArmLength*SIN(AlphaRadians),6)
    Zcoord$  = REMOVE$(STR$(Zlength+Lfeed),ANY" ") : IF TALLY(Zcoord$,".") = 0 THEN Zcoord$ = Zcoord$+"."
    Nsegs%   = 2*(INT(TotalLengthWaves*100)\2) 'even number, total # segs
    NumSegs$ = REMOVE$(STR$(Nsegs%\2),ANY" ") '# segs per arm
    N% = FREEFILE
    OPEN "PBM4.NEC" FOR OUTPUT AS #N%
        PRINT #N%,"CM File: PBM4.NEC"
        PRINT #N%,"CM Run ID "+DATE$+" "+TIME$
        PRINT #N%,"CM Nd="+STR$(Nd%)+", p="STR$(p%)+", j="+STR$(j&)
        PRINT #N%,"CM R(p,1,j)="+STR$(R(p%,1,j&))+", R(p,2,j)="+STR$(R(p%,2,j&))
        PRINT #N%,"CE"
        PRINT #N%,"GW 1,5,0.,0.,-"+FeedZcoord$+",0.,0.,"+FeedZcoord$+",0.00001" 'feed wire, 1 segment, 0.01 wvln
        PRINT #N%,"GW 2,"+NumSegs$+",0.,0.,"+FeedZcoord$+","+Xcoord$+",0.,"+Zcoord$+",0.00001" 'upper arm
        PRINT #N%,"GW 3,"+NumSegs$+",0.,0.,-"+FeedZcoord$+","+Xcoord$+",0.,-"+Zcoord$+",0.00001" 'lower arm
        PRINT #N%,"GE"
        PRINT #N%,"EX 0,1,3,0,1.,0." 'VOLTAGE SOURCE
        PRINT #N%,"FR 0,1,0,0,299.79564,0."
        PRINT #N%,"RP 0,1,1,1001,90.,0.,0.,0.,1000." 'ENDFIRE gain at 1000 wavelengths range
        PRINT #N%,"XQ"
        PRINT #N%,"EN"
    CLOSE #N%
'      - - ANGLES - -          - POWER GAINS -        - - - POLARIZATION - - -    - - - E(THETA) - - -    - - - E(PHI)
- - -
'  THETA     PHI        VERT.   HOR.    TOTAL      AXIAL     TILT   SENSE     MAGNITUDE    PHASE      MAGNITUDE
PHASE
' DEGREES  DEGREES       DB      DB      DB        RATIO     DEG.              VOLTS/M    DEGREES      VOLTS/M
DEGREES
'   90.00     0.00       3.91 -999.99    3.91    0.00000     0.00  LINEAR    1.29504E-04     5.37    0.00000E+00    -
5.24
'123456789x123456789x123456789x123456789x123456789x123456789x123456789x123456789x123456789x123456789x123456789x12345678
9x
'        10        20        30        40        50        60        70        80        90        100       110
120
    SHELL "n41_2k1.exe",0
    N% = FREEFILE
    OPEN "PBM4.OUT" FOR INPUT AS #N%
        WHILE NOT EOF(N%)
            LINE INPUT #N%, Lyne$
            IF INSTR(Lyne$,"DEGREES  DEGREES") > 0 THEN EXIT LOOP
        WEND 'position at next data line
        LINE INPUT #N%, Lyne$
    CLOSE #N%
    GainDB$ = REMOVE$(MID$(Lyne$,37,8),ANY" ")
    PBM_4 = 10^(VAL(GainDB$)/10##) 'Directivity
END FUNCTION 'PBM_4()
'----
FUNCTION PBM_5(R(),Nd%,p%,j&) 'PBM Benchmark #5: N-element collinear array (Nd=N-1)
    LOCAL TotalLengthWaves, Di(), Ystart, Y1, Y2, SumDi AS EXT
    LOCAL N%, i%, q%
    LOCAL Lyne$, GainDB$
    REDIM Di(1 TO Nd%)
    FOR i% = 1 TO Nd%
        Di(i%) = R(p%,i%,j&) 'dipole separation, wavelengths
        'MSGBOX("R="+STR$(R(p%,i%,j&))+"  p="+STR$(p%)+"   i="+STR$(i%)+"    j="+STR$(j&)+"    Nd="+STR$(Nd%))
    NEXT i%
    TotalLengthWaves = 0##
```

```
    FOR i% = 1 TO Nd%
        TotalLengthwaves = TotalLengthWaves + Di(i%)
    NEXT i%
    TotalLengthWaves = TotalLengthWaves + 0.5## 'add half-wavelength of 1 meter at 299.8 MHz
    Ystart = -TotalLengthWaves/2##
    N% = FREEFILE
    OPEN "PBM5.NEC" FOR OUTPUT AS #N%
        PRINT #N%,"CM File: PBM5.NEC"
        PRINT #N%,"CM Run ID "+DATE$+" "+TIME$
        PRINT #N%,"CM Nd="+STR$(Nd%)+", p="STR$(p%)+", j="+STR$(j&)
        PRINT #N%,"CM R(p,1,j)="+STR$(R(p%,1,j&))+", R(p,2,j)="+STR$(R(p%,2,j&))
        PRINT #N%,"CE"
        FOR i% = 1 TO Nd%+1
            SumDi = 0##
            FOR q% = 1 TO i%-1
                SumDi = SumDi + Di(q%)
            NEXT q%
            Y1 = ROUND(Ystart + SumDi,6)
            Y2 = ROUND(Y1+0.5##,6) 'add one-half wavelength for other end of dipole
            PRINT #N%,"GW "+REMOVE$(STR$(i%),ANY" ")+",49,0.,"+REMOVE$(STR$(Y1),ANY" ")+",0.,0.,"+REMOVE$(STR$(Y2),ANY"
")+",0.,0.00001"
        NEXT i%
        PRINT #N%,"GE"
        FOR i% = 1 TO Nd%+1
            PRINT #N%,"EX 0,"+REMOVE$(STR$(i%),ANY" ")+",25,0,1.,0." 'VOLTAGE SOURCES
        NEXT i%
        PRINT #N%,"FR 0,1,0,0,299.79564,0."
        PRINT #N%,"RP 0,1,1,1001,90.,0.,0.,0.,1000." 'gain at 1000 wavelengths range
        PRINT #N%,"XQ"
        PRINT #N%,"EN"
    CLOSE #N%
'      - - ANGLES - -          - POWER GAINS -        - - - POLARIZATION - - -    - - - E(THETA) - - -    - - - E(PHI)
- - -
'  THETA     PHI        VERT.   HOR.    TOTAL      AXIAL     TILT   SENSE     MAGNITUDE    PHASE      MAGNITUDE
PHASE
' DEGREES  DEGREES       DB      DB      DB        RATIO     DEG.             VOLTS/M    DEGREES      VOLTS/M
DEGREES
'   90.00     0.00       3.91 -999.99    3.91    0.00000     0.00  LINEAR    1.29504E-04     5.37    0.00000E+00    -
5.24
'123456789x123456789x123456789x123456789x123456789x123456789x123456789x123456789x123456789x123456789x123456789x12345678
9x
'        10        20        30        40        50        60        70        80        90       100       110
120
    SHELL "n41_2k1.exe",0
    N% = FREEFILE
    OPEN "PBM5.OUT" FOR INPUT AS #N%
        WHILE NOT EOF(N%)
            LINE INPUT #N%, Lyne$
            IF INSTR(Lyne$,"DEGREES  DEGREES") > 0 THEN EXIT LOOP
        WEND 'position at next data line
        LINE INPUT #N%, Lyne$
    CLOSE #N%
    GainDB$ = REMOVE$(MID$(Lyne$,37,8),ANY" ")
    PBM_5 = 10^(VAL(GainDB$)/10##) 'Directivity
END FUNCTION 'PBM_5()
'*********************************  END PROGRAM 'CFO_11-26-09.BAS'  *********************************
```